\documentclass[
	fontsize=12pt, 
	twoside=true, 
	numbers=noenddot, 
]{kaobook}

\usepackage[latin,greek,english]{babel} 
\usepackage[english=british]{csquotes} 
\usepackage[T1]{fontenc}

\usepackage[style=alphabetic,backref=true,maxbibnames=99]{biblatex}
\usepackage{blindtext}
\usepackage{glossaries}
\usepackage{xcolor}
\usepackage{amsmath}
\usepackage{algpseudocode}
\usepackage{epigraph}
\usepackage{textgreek}
\usepackage{diagbox}
\usepackage[normalem]{ulem}
\hypersetup{
	colorlinks = true,
	citecolor = RedOrange,
	linkcolor = MidnightBlue,
	urlcolor = ForestGreen,
	hyperindex=true,
}
\usepackage{rotating}
\usepackage{makeidx}
\usepackage{etoolbox}
\usepackage{braket}

\definecolor{codegreen}{rgb}{0,0.6,0}
\definecolor{codegray}{rgb}{0.5,0.5,0.5}
\definecolor{codepurple}{rgb}{0.58,0,0.82}
\definecolor{backcolour}{rgb}{0.95,0.95,0.92}

\lstdefinestyle{mystyle}{
    backgroundcolor=\color{backcolour},   
    commentstyle=\color{codegreen},
    keywordstyle=\color{magenta},
    numberstyle=\tiny\color{codegray},
    stringstyle=\color{codepurple},
    basicstyle=\ttfamily\footnotesize,
    breakatwhitespace=false,         
    breaklines=true,                 
    captionpos=b,                    
    keepspaces=true,                 
    numbers=left,                    
    numbersep=5pt,                  
    showspaces=false,                
    showstringspaces=false,
    showtabs=false,                  
    tabsize=2
}

\newcommand\mydots{\hbox to 1em{.\,.\,.}}

\newcommand{\dt}{\ensuremath{d_t}\xspace}
\newcommand{\dr}{\ensuremath{d_r}\xspace}
\newcommand{\vth}{\ensuremath{v_{\mathrm{th}}}\xspace}
\newcommand{\oth}{\ensuremath{\omega_{\mathrm{th}}}\xspace}
\newcommand{\Kn}{\ensuremath{\mathrm{Kn}}\xspace}
\newcommand{\rr}{\mathbf{r}}
\newcommand{\cc}{\mathbf{c}}
\newcommand{\ww}{\mathbf{w}}
\newcommand{\vvel}{\mathbf{v}}
\newcommand{\HCS}{{\mathrm{H}}}
\newcommand{\Ttr}{T_t}
\newcommand{\Trot}{T_r}
\newcommand{\een}{\alpha}
\newcommand{\eet}{\beta}
\newcommand{\en}{\overline{\alpha}}
\newcommand{\et}{\overline{\beta}}
\newcommand{\eq}{\mathrm{eq}}
\newcommand{\kb}{k_{\mathrm{B}}}
\newcommand{\vab}{v_{12}}
\newcommand{\vvab}{\vvel_{12}}
\newcommand{\ggab}{\mathbf{g}_{12}}
\newcommand{\Qqab}{\mathbf{Q}_{12}}
\newcommand{\oo}{\boldsymbol{\omega}}
\newcommand{\ssab}{\widehat{\boldsymbol{\sigma}}}
\newcommand{\vom}{\boldsymbol{\Gamma}}
\newcommand{\cw}{\widetilde{\boldsymbol{\Gamma}}}
\newcommand{\dif}{\mathrm{d}}
\newcommand{\pprime}{\prime\prime}
\newcommand{\llangle}{\langle\!\langle}
\newcommand{\rrangle}{\rangle\!\rangle}
\newcommand{\Twn}{T^{\mathrm{wn}}}
\newcommand{\nuwn}{\nu^{\mathrm{wn}}}
\newcommand{\KLD}{\mathcal{D}_{\mathrm{KL}}}

\newcommand{\ancho}{0.14}
\usepackage{lscape}
\usepackage{rotating}

\newcommand\gobbleone[1]{}

\usepackage{tabularx}%
\usepackage[intoc]{nomencl}
\usepackage{styles/kaobiblio}
\usepackage{styles/mdftheorems}
\DeclareMathAlphabet{\mathcal}{OMS}{zplm}{m}{n}
\SetMathAlphabet{\mathcal}{bold}{OMS}{zplm}{b}{n}
\DeclareMathAlphabet{\mathbfcal}{OMS}{zplm}{b}{n}
\graphicspath{{examples/documentation/images/}{images/}} 

\makeglossaries 
\newacronym{VDF}{VDF}{Velocity distribution function}
\newacronym{HCS}{HCS}{Homogeneous Cooling State}
\newacronym{HVT}{HVT}{High-velocity tail}
\newacronym{HS}{HS}{Hard sphere}
\newacronym{HD}{HD}{Hard disk}
\newacronym{BBGKY}{BBGKY}{Bogoliubov--Born--Green--Kirkwood--Yvon}
\newacronym{EHS}{EHS}{Elastic hard sphere}
\newacronym{IHS}{IHS}{Inelastic hard sphere}
\newacronym{IRHS}{IRHS}{Inelastic and rough hard sphere}
\newacronym{BE}{BE}{Boltzmann equation}
\newacronym{CE}{CE}{Chapman--Enskog}
\newacronym{SA}{SA}{Sonine approximation}
\newacronym{ST}{ST}{Splitting thermostat}
\newacronym{MA}{MA}{Maxwellian approximation}
\newacronym{ME}{ME}{Mpemba effect}
\newacronym{TME}{TME}{Thermal Mpemba effect}
\newacronym{EME}{EME}{Entropic Mpemba effect}
\newacronym{OME}{OME}{Overshoot Mpemba effect}
\newacronym{IME}{IME}{Inverse Mpemba effect}
\newacronym{DME}{DME}{Direct Mpemba effect}
\newacronym{KE}{KE}{Kovacs effect}
\newacronym{MD}{MD}{Molecular dynamics}
\newacronym{DSMC}{DSMC}{Direct simulation Monte Carlo}
\newacronym{EDMD}{EDMD}{Event-driven molecular dynamics}
\newacronym{TDMD}{TDMD}{Time-driven molecular dynamics}
\newacronym{KLD}{KLD}{Kullback--Leibler divergence}
\newacronym{LE}{LE}{Langevin equation}
\newacronym{BSA}{BSA}{Basic Sonine approximation}
\newacronym{ESA}{ESA}{Extended Sonine approximation}
\newacronym{FSA}{FSA}{First Sonine approximation}
\newacronym{LBSA}{LBSA}{Linearized basic Sonine approximation}
\newacronym{AGF}{AGF}{Approximate Green function}
\newacronym{SDE}{SDE}{Stochastic differential equation}
\newacronym{NSF}{NSF}{Navier--Stokes--Fourier}
\newacronym{FDT}{FDT}{Fluctuation-dissipation theorem}
\newacronym{BFPE}{BFPE}{Boltzmann--Fokker--Planck equation}
\newacronym{FPE}{FPE}{Fokker--Planck equation}
\newacronym{NNL}{NNL}{Nearest neighbor list}
\newacronym{HUR}{HUR}{Highly unstable region}
\newacronym{QR}{QR}{Quasi-Rayleigh}
\newglossaryentry{maths}
{
    name=mathematics,
    description={Mathematics is what mathematicians do}
}

\makeatletter
\renewcommand{\part}{%
  \if@openright
    \cleardoublepage
  \else
    \clearpage
  \fi%
  \thispagestyle{empty}%
  \if@twocolumn
    \onecolumn
    \@tempswatrue
  \else
    \@tempswafalse
  \fi%
  \null\vfil%
  \secdef\@part\@spart%
}
\makeatother

\begin{document}

\title{}
\author{}
\date{}
\makeatletter
\thispagestyle{empty}
\begin{center}
{\includegraphics[width=0.9\textwidth]{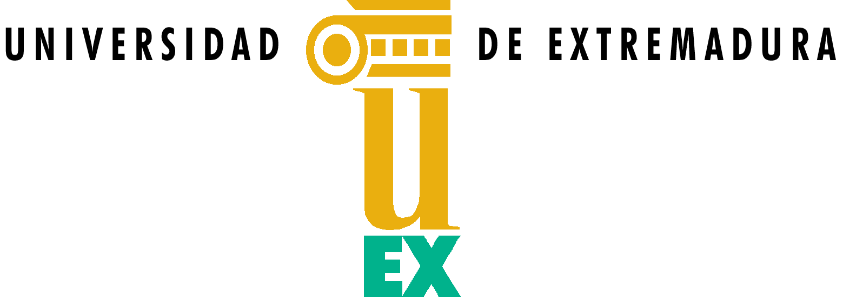}\par}
\vspace{2cm}
{\scshape\Huge\textbf{Influence of roughness on the dynamical properties of granular gases} \par}
\vspace{1cm}
{\itshape\large A thesis submitted in fulfilment of the requirements for the degree of Doctor of Philosophy \par}
\vspace{3.5cm}
{\Large Author: \par}
{\Large Alberto Meg\'ias Fern\'andez \par}
\vspace{0.5cm}
{\Large Supervisor: \par}
{\Large Prof. Dr. Andr\'es Santos\par}
\end{center}

%


\frontmatter 




\makeatletter
\thispagestyle{empty}
\uppertitleback{\@titlehead} 

\lowertitleback{
	
	\medskip
	
	\textbf{Creative Commons License}\\
	\ccbyncsa\ This license allows reusers to distribute, remix, adapt, and build upon the material in any medium or format for noncommercial purposes only, and only so long as attribution is given to the creator. If you remix, adapt, or build upon the material, you must license the modified material under identical terms. 
	
	To view a copy of the CC BY-NC-SA code, visit: \\\url{https://creativecommons.org/licenses/by-nc-sa/4.0/}
	
%
%
%
%
	
}
\makeatother

\thispagestyle{empty}
\dedication{
\flushright {\Large
	\emph{A mis padres y a mi hermano}}\\
\flushright {\Large
	\emph{A Blanca}}	
}

\makeatother

	

\maketitle
\thispagestyle{empty}
\vspace*{5cm}
{\flushleft {
	\emph{There are very, very simple problems that are very hard to understand}\\
	Giorgio Parisi, \href{https://www.nobelprize.org/prizes/physics/2021/parisi/interview/}{Nobel Prize in Physics 2021}}}

\newpage
\blankpage
\newpage	
\phantomsection
\addcontentsline{toc}{chapter}{Resumen}
 \thispagestyle{plain}
\begin{center}
\vspace{10pt}
\textcolor{marineblue}{\rule{\textwidth}{1mm}}

\vspace{20pt}
    {\Huge{\textcolor{marineblue}{\textbf{Resumen}}}}
\vspace{10pt}

\textcolor{marineblue}{\rule{\textwidth}{1mm}}
\vspace{10pt}
\end{center}

El estudio físico-estadístico de la materia granular es esencial para entender desde un punto de vista fundamental los fenómenos que presentan estos sistemas clásicos de muchas partículas. En condiciones de flujo rápido, los materiales granulares exhiben un comportamiento que recuerda al de un gas y que puede estudiarse desde el punto de vista de la teoría cinética de gases. A diferencia de un gas molecular, un gas granular disipa energía debido a las colisiones entre sus partículas. Por ello, mientras que el modelo más sencillo de un gas molecular consiste en una colección de esferas duras elásticas, para un gas granular, éstas se consideran inelásticas y su extensión más realista consiste en considerar también su rugosidad superficial. Además, un gas granular puede encontrarse tanto en libre evolución, como forzado o sumergido en un fluido intersticial. En este trabajo, se ha tratado de describir la dinámica de gases granulares bajo distintas ligaduras con el objetivo de extraer propiedades específicas del caso rugoso. Para ello, se ha utilizado la ecuación de Boltzmann, añadiendo, en el caso forzado y de suspensiones, términos de tipo Fokker--Planck. Los principales objetivos de la tesis han sido describir teóricamente los estados homogéneos de los gases considerados, el estudio de efectos de memoria en ellos, y la caracterización de las propiedades de transporte e inestabilidades en la descripción hidrodinámica de un gas granular de partículas rugosas. Todo este análisis ha sido complementado con simulaciones por ordenador bajo algoritmos de tipo Monte Carlo y de dinámica molecular.

\newpage
\blankpage
\newpage
\phantomsection
\addcontentsline{toc}{chapter}{Abstract}
 \thispagestyle{plain}
\begin{center}
\vspace{10pt}
\textcolor{marineblue}{\rule{\textwidth}{1mm}}

\vspace{20pt}
    {\Huge{\textcolor{marineblue}{\textbf{Abstract}}}}
\vspace{10pt}

\textcolor{marineblue}{\rule{\textwidth}{1mm}}
\vspace{10pt}
\end{center}

The statistical-physical study of granular matter is essential to understand, from a fundamental point of view, the many different phenomena emerging in these classical many-body systems. Under rapid-flow conditions, granular materials exhibit a gas-like behavior, which can be described from the kinetic theory of gases. However, unlike molecular gases, a granular gas dissipates energy upon particle collisions, thus being completely out of equilibrium. Then, whereas the simplest model for a molecular gas consists in a collection of elastic hard spheres, a granular gas is simply described by inelastic hard spheres, which is usually improved by the consideration of surface roughness in the particles. Moreover, a granular gas can be found in free evolution, externally driven, or even immersed in an interstitial fluid. In this work, the description of the dynamics of granular gases has been carried out with the aim of identifying the specific properties emerging from the model by considering the surface roughness of granular particles. For that end, we have used the Boltzmann equation, adding in the driven and suspension cases a Fokker--Planck-like term. The main objectives of this thesis have been the theoretical description of the homogeneous states of the considered models, as well as the study of memory effects emerging from out-of-equilibrium states, and, finally, the transport properties and the appearance of instabilities in the hydrodynamic description of a granular gas of inelastic and rough particles. All these kinetic-theory analyses have been supplemented with computer simulations from Monte Carlo and molecular dynamics algorithms.

\newpage
\blankpage
\newpage


\addchap[Agradecimientos]{Agradecimientos}

Como ya se explicará, uno de los resultados de esta tesis es que muchos ``granitos'' pueden llevar a resultados muy satisfactorios. La propia consecución de este trabajo es una analogía de ello. Y es que, su realización ha sido posible gracias a la aportación de muchos de estos ``granitos'' por parte de muchas personas, tanto en el ámbito científico como en el personal.

Como es evidente, en unos simples y cortos párrafos es imposible expresar todo, ni siquiera mencionar a todas esas personas que me han aportado algo durante esta etapa. Sin embargo, me gustaría destacar ciertas personas y momentos que han sido cruciales durante el desarrollo de mi tesis doctoral. Aunque también a usted, lector, debo agradecerle que se haya parado a leer esta memoria y espero que, de alguna forma, aunque no sea en su contenido, algo le resulte interesante o le sea de utilidad, incluso si ha llegado aquí por mera casualidad.

Primero de todo, y como no podría ser de otra forma, quería agradecer todo su tiempo, su trabajo y su ayuda a mi director de tesis, Andrés Santos. Esta historia comenzó con unas clases de Física Cuántica I en inglés, en las que por primera vez disfruté de su labor docente. Ello me llevó a querer (y conseguir) realizar la ``Beca de Colaboración'' y el TFG con él, que supondría alcanzar mi primer sueño, el de ser graduado en Física. Esa primera época llena de apoyo y consejos conllevó al comienzo y desarrollo de una etapa dura, pero muy bonita: la realización de esta tesis doctoral. Muchas gracias, Andrés, por guiarme, comprenderme, apoyarme y sobre todo, confiar en mí en todo momento, en los mejores y en los de más agobio. Gracias, sobre todo, por querer siempre lo mejor para mí y darme la oportunidad de buscarlo. Llevaré toda mi vida conmigo tus consejos y los conocimientos (en el sentido más amplio de la palabra) que me has transmitido durante estos años.

Quería agradecer también a Antonio Prados su ayuda, su tiempo y su trabajo, del que he podido aprender mucho y conseguir diferentes perspectivas durante el desarrollo de la investigación. 

A pesar de que la tesis se ha desarrollado en la Universidad de Extremadura, he podido de disfrutar de estancias de investigación en el extranjero. Primero, me gustaría agradecer a Nikolai V. Brilliantov su gran acogida durante mi estancia en ``Skolkovo Institute of Science and Technology'' en Moscú, Rusia. Gracias a Nikolai pude asistir a un curso durante mi primer año de doctorado que ha sido fundamental durante mi desarrollo como investigador y, además, me permitió comenzar mi primera experiencia de una colaboración científica en un centro de alto prestigio. Por otro lado, me gustaría agradecer también su gran hospitalidad a Hans J. Herrmann durante mi estancia en el ``Laboratoire de Physique et Mécanique des Milieux Hétérogènes'' (PMMH) de la ESPCI en París, Francia. Hans, no solo me acogió con los brazos abiertos, sino que me integró estupendamente desde el primer día en el laboratorio. Le estaré siempre agradecido por lo todo anterior y lo aprendido con él sobre el mundo granular durante esos meses. Finalmente, no podría olvidarme de mis compañeros Ernesto, Fernando, Javier y María de Skoltech y Benjamín, Chloé, Francisco, Juan, Lars, Manon y Samantha de PMMH, entre otros muchos, que me han hecho disfrutar mucho más de estas estancias, me ayudaron y me acogieron con mucho cariño. Así como de Jasmin, Khalil, Mohammed, Matrona, Soheil, Tarik y Tim que me hicieron disfrutar aún más de la escuela MOLSIM-2020 en Ámsterdam, Países Bajos.

Tampoco podría olvidarme de todos los miembros del grupo SPhinX de la Universidad de Extremadura, con los que he trabajado estos cuatro años de tesis doctoral y que muchos de ellos han participado en mi formación y motivación previa como estudiante de Física. Gracias al resto de los investigadores ``seniors'' Enrique Abad, Antonio Astillero, Santos Bravo, Vicente Garzó, Antonio Gordillo, Juan José Meléndez, Álvaro Rodríguez, Juan Jesús Ruiz y Francisco Vega que me han dado consejo y ayuda siempre, y que han favorecido un ambiente de trabajo muy agradable y cercano. Quería mencionar especialmente a Santos Bravo y Juan Jesús Ruiz por permitirme compartir con ellos labores docentes, así como sus enseñanzas y consejos en este aspecto, que me han hecho crecer mucho más como profesional y sobre todo como persona. También quería destacar a Enrique Abad y Vicente Garzó por sus conversaciones futboleras, pero, sobre todo, por sus labores en la coordinación del grupo, mirando siempre por los más jóvenes y por ayudarnos a alcanzar todos nuestros objetivos. Me siento también orgulloso de haber conocido a mis compañeros de viajes y penurias, con muchas risas y buenos momentos en esos maravillosos ``Coffee Breaks'': Ana, Andrés, Beatriz, Costantino, Felipe, Javier, Jesús M., Juan, Juanfran, Lucía, Miguel, Miguel Ángel, Moisés y Rubén. También, debo destacar y valorar la labor de Francisco Naranjo y Nuria María García del Moral como gestores del grupo durante este período. Más aún, quería agradecer lo enriquecedor que ha sido contar con las visitas de Mariano López de Haro y S\l{}awomir Pieprzyk, entre otros.

Además, quería expresar mi profundo agradecimiento a los revisores externos de esta tesis doctoral, Meheboob Alam, Stefan Luding y Satoshi Takada, por su disposición a realizar esta labor, así como por leerla de forma cuidada y, también, por las muy interesantes aportaciones que han contribuido a la mejora de esta memoria.

Mi pasión por el mundo investigador es fruto de muchas personas que me han acompañado en este camino. Primero, gracias a mis profesoras de las asignaturas de Química, Física y Matemáticas: Mari Ángeles, Lola y Fátima. Sin vuestra labor, yo no estaría aquí. Por supuesto, gracias a los amigos que he hecho durante mis estudios de grado y máster, cuya amistad espero preservar toda la vida: Álvaro, Blanca, Celia, Cristina, David, Germán, Jesús T., Toribio, Virginia y Álex. Además, quería mencionar al personal del Fis \& Kids. En especial a Marisa Cancillo por su trabajo, pero aun más, por su cariño al frente de este proyecto tan bonito y darme la oportunidad de haber sido espectador y partícipe de él. Así como a los distintos componentes ``seniors'' del mismo: Javier Acero, María Cruz Gallego, Agustín García, Antonio Serrano y José Vaquero, entre muchos otros. También a todos mis compañeros del Fis \& Kids (muchos), investigadores y alumnos, que me habéis acompañado en esta bonita labor de la difusión científica. Una labor que ha sido complementada estos dos últimos años gracias a Guillermo Mena y Laura Morala, y el resto del equipo de la Real Sociedad Extremeña de Amigos del País de Badajoz, por las charlas del ``Café Con'', y a mi amigo, anteriormente mencionado, Álex (García-Quismondo), por pensar siempre en mí para todo esto.

También quería destacar y agradecer a la Asociación de Doctorandos de la Universidad de Extremadura (ADUEx) su labor al frente de la búsqueda del bienestar de los estudiantes de doctorado y ayudar a la creación de vínculos entre nosotros. Me siento afortunado de haber formado parte de la Junta Directiva como secretario y haber compartido trabajo con magníficas personas como Juan, Luz, Marta, Manu, Paula y Sara, así como a las antiguas y futuras Juntas. Quiero añadir también, en este agradecimiento, a la asociación InvestigaEx por su lucha constante por la mejora y dignificación de la labor investigadora en Extremadura, sobre todo, para el personal más precario.

En el ámbito más personal yo no habría conseguido nada sin el apoyo constante de mi familia. En especial, mis padres, Marisa y Francisco. No solo me dieron la vida, también me educaron, cuidaron, aconsejaron, apoyaron en todo momento y me permitieron buscar mis objetivos cuales fueran, siempre desde un amor incondicional, que es mutuo, y del que les estaré eternamente agradecido. Estas líneas son muy pocas con respecto a todo lo que os quiero y os debo. También, a mi hermano mayor, Francisco, por ser mi compañero de vida y quererme, apoyarme y defenderme siempre, te quiero. Gracias también a mi sobrino, Francisco Manuel, por llenar de luz, cariño y alegría nuestras vidas. Así como a las peques, María y Julia, que nos aportan continuamente felicidad. Desearía tener unas palabras de agradecimiento para mis tíos, Dioni, Félix y Paco y para mi prima (y hermana) María por vuestro cariño y apoyo constante, así como el del resto de nuestra gran familia. También, gracias a todos mis amigos, del colegio y fuera de él, por vuestro cariño y apoyo incansable. 

Por supuesto, no me olvido de mis abuelos, Francisco y Eugenia, que no han podido disfrutar de la consecución de este trabajo, pero sé, que allá dónde estéis con Adela y Blas, me cuidáis y me mandáis toda la fuerza que necesito día a día. Estoy seguro de que los cuatro estáis muy felices por mí.

Y cómo no hablar de Blanca, con la que he compartido estos años de mi vida y espero que sean muchos más. Gracias por ser mi apoyo incondicional, mi sostén emocional y mi felicidad diaria. Gracias por ser mi teorema. No podría haber hecho nada sin ti y sin tu cariño constante. Te quiero. También, gracias a Nati y Miguel por su apoyo y ayuda durante todo este tiempo.

Finalmente, agradezco la financiación por parte del Ministerio de Ciencia, Innovación y Universidades por la ayuda FPU18-3503 que ha hecho posible la realización de esta tesis doctoral, y por la ayuda EST22-00145 que me ha permitido realizar la segunda estancia de investigación en el ``Laboratoire de Physique et Mécanique des Milieux Hétérogènes'' (PMMH) de la ESPCI en París, Francia.

\newpage




\setglossarystyle{list}
\printglossary[title=Acronyms, toctitle=Acronyms] 

\addchap[Symbols]{Symbols}

\vspace*{4cm}
\emph{The next list describes several symbols that will be later used within the body of the document (not necessary in the publications, whose notation is independent) and the first page of their occurrence. The independence of the publications produces that some symbols could be duplicated. Then, if it is needed, all the possible meanings will be explained below. Notice that, if the symbol is firstly introduced in a publication, the referenced page would correspond to the first page of the section where the publication is inserted.}

\vspace*{0.6cm}
\begin{description}[leftmargin=!,labelwidth=1.5cm]
\item[$a_k$] $k^{\mathrm{th}}$ cumulant of a velocity distribution function (p.~\pageref{eq:sonine_expansion_c}).
\item[$a_{kA}$] $k^{\mathrm{th}}$ cumulant of a velocity distribution function of a sample A initially further from equilibrium than another sample B (p.~\pageref{sym:a2A_a2B}).
\item[$a_{kB}$] $k^{\mathrm{th}}$ cumulant of a velocity distribution function of a sample B initially closer to equilibrium than another sample A (p.~\pageref{sym:a2A_a2B}).
\item[$a_{k}^0$] $\equiv a_k(t_0)$, initial value of $a_k$ (p.~\pageref{sym:a2A_a2B}).
\item[$a_{kA}^0$] $\equiv a_k(t_0)$, initial value of $a_k$ of a sample A initially closer to the steady-state temperature than another sample B (p.~\pageref{sym:a2A_a2B}).
\item[$a_{kB}^0$] $\equiv a_k(t_0)$, initial value of $a_k$ of a sample B initially closer to the steady-state temperature than another sample A (p.~\pageref{sym:a2A_a2B}).
\item[$a_{pq}^{(r)}$] Cumulant of order $(p,q,r)$ (p.~\pageref{eq:Sonine_expansion_IRHS_a}).
\item[$a_{pq}$] $=a_{pq}^{(0)}$, cumulant of order $(p,q,0)$ (p.~\pageref{sym:apq}).
\item[$a_k^{\HCS}$] Value of the $k^{\mathrm{th}}$ cumulant at the homogeneous cooling state (p.~\pageref{sym:akHCS}).
\item[$a_k^{\mathrm{st}}$] Steady-state value of the $k^{\mathrm{th}}$ cumulant (p.~\pageref{sym:a2st}).
\item[$\mathbf{a}$] Arbitrary vectorial function (p.~\pageref{eq:general_form_LE}).
\item[$\mathcal{A}$] Arbitrary normalization constant of a velocity distribution function (p.~\pageref{eq:HVT_IHS_KTGG}).
\item[$\boldsymbol{\mathcal{A}}$] Vectorial function of the velocities coupled with the temperature gradient in the first-order distribution $f^{(1)}$ (p.~\pageref{sym:ABCE_quant}).
\item[$\mathsf{b}$] Arbitrary tensorial function (p.~\pageref{eq:general_form_LE}).
\item[$B(\cdot,\cdot)$] Differential cross-section (p.~\pageref{sym:Bgpprime}).
\item[$\boldsymbol{\mathcal{B}}$]  Vectorial function of the velocities coupled with the density gradient in the first-order distribution $f^{(1)}$ (p.~\pageref{sym:ABCE_quant}).
\item[$\mathfrak{B}_{ij,\ssab}$] Postcollisional operator for a collision between a pair $ij$ of particles with unit vector $\ssab$ (p.~\pageref{sym:B_op}).
\item[$\mathfrak{B}^{\mathrm{put}}_{ij,\ssab}$] Postcollisional operator for a collision between a pair $ij$ of particles with unit vector $\ssab$ using putative velocities (p.~\pageref{eq:rv_final_Lan_EDMD}).
\item[$\mathfrak{B}^{\mathrm{fic}}_{ij,\ssab}$] Postcollisional operator for a collision between a pair $ij$ of particles with unit vector $\ssab$ using fictive velocities (p.~\pageref{eq:rv_final_Lan_EDMD}).
\item[$c$] Modulus of the reduced translational velocity of a particle (p.~\pageref{eq:vb_cb}).
\item[$c_b$] Modulus of the reduced translational velocity of a thermal-bath particle (p.~\pageref{eq:vb_cb}).
\item[$\cc$] $=\vvel/\vth$, reduced translational velocity of a particle (p.~\pageref{eq:vb_cb}).
\item[$\cc_{ij}$] $=\cc_i-\cc_j$, relative reduced translational velocity between particles $i$ and $j$ (p.~\pageref{eq:I_coll_def}).
\item[$\cc_b$] $=\vvel/\vth$, reduced translational velocity of a thermal-bath particle (p.~\pageref{eq:vb_cb}).
\item[$\mathcal{C}_{ij}$] Tensorial function of the velocities coupled with $\partial_j u_i$ in the first-order velocity distribution function $f^{(1)}$ (p.~\pageref{sym:ABCE_quant}).
\item[$\dt$] Translational degrees of freedom (p.~\pageref{sym:dt}).
\item[$\dr$] Rotational degrees of freedom (p.~\pageref{sym:dr}).
\item[$\dif \mathbf{S}$] Cross-section of a differential collisional cylinder (p.~\pageref{fig:collision_cylinder}).
\item[$D$] Diffusion coefficient (p.~\pageref{sym:diff_coeff}).
\item[$\mathbf{D}^{(1)}$] Drift vector (p.~\pageref{eq:app_FPE}).
\item[$\mathsf{D}^{(2)}$] Diffusion matrix (p.~\pageref{eq:app_FPE}).
\item[$\mathcal{D}_t$] $=\partial_t+\mathbf{u}\cdot\nabla$, material time derivative (p.~\pageref{eq:mat_time_der}).
\item[$\mathcal{D}_t^{(k)}$] $k^{\mathrm{th}}$-order term of $\mathcal{D}_t$ in a Chapman--Enskog expansion (p.~\pageref{eq:ders_expansion}).
\item[$\KLD$]  Kullback--Leibler divergence (p.~\pageref{eq:KLD_def}).
\item[$\mathcal{D}_A$] Nonequilibrium entropy of a sample A initially further from the equilibrium (steady) state as compared with another sample B (p.~\pageref{sec:Art1}).
\item[$\mathcal{D}_B$] Nonequilibrium entropy of a sample B initially closer to the equilibrium (steady) state as compared with another sample A (p.~\pageref{sec:Art1}).
\item[$\mathcal{D}^0$] Initial value of the nonequilibrium entropy of a system (p.~\pageref{sec:Art1}).
\item[$\mathcal{D}^0_A$] Initial value of the nonequilibrium entropy of a sample A initially further from the equilibrium (steady) state as compared with another sample B (p.~\pageref{sec:Art1}).
\item[$\mathcal{D}^0_B$] Initial value of the nonequilibrium entropy of a sample B initially closer to the equilibrium (steady) state as compared with another sample A (p.~\pageref{sec:Art1}).
\item[$\mathcal{D}^{\mathrm{kin}}$] Kinetic counterpart of the nonequilibrium entropy (p.~\pageref{sym:kin_LE}).
\item[$\mathcal{D}^{\mathrm{LE}}$] Local-equilibrium counterpart of the nonequilibrium entropy (p.~\pageref{sym:kin_LE}).
\item[$\mathcal{D}^{\mathrm{LE}}_{A}$] $\mathcal{D}^{\mathrm{LE}}$ curve of a sample A initially further from the equilibrium (steady) state than another sample B (p.~\pageref{sym:KLD_LE_A_B}).
\item[$\mathcal{D}^{\mathrm{LE}}_{B}$] $\mathcal{D}^{\mathrm{LE}}$ curve of a sample B initially closer to the equilibrium (steady) state than another sample B (p.~\pageref{sym:KLD_LE_A_B}).
\item[$\mathfrak{D}$] Thermal distance (p.~\pageref{tab:phenom_ME}).
\item[$\mathfrak{D}_{A}$] Thermal distance of a sample A initially further from the steady-state temperature as compared with another sample B (p.~\pageref{sym:DfrakA_B}).
\item[$\mathfrak{D}_B$] Thermal distance of a sample B initially closer to the steady-state temperature as compared with another sample A (p.~\pageref{sym:DfrakA_B}).
\item[$E$] Kinetic energy of a particle (p.~\pageref{eq:cons_E_IHS}).
\item[$E^t$] Translational kinetic energy of a particle (p.~\pageref{eq:Energy_t_r}).
\item[$E^r$] Rotational kinetic energy of a particle (p.~\pageref{eq:Energy_t_r}).
\item[$E_{12}$] Kinetic energy of particles 1 and 2 (p.~\pageref{eq:cons_E_IHS}).
\item[$E^{\prime}_{12}$] Postcollisional kinetic energy of particles 1 and 2 (p.~\pageref{eq:cons_E_IHS}).
\item[$\mathcal{E}$] Scalar function of the velocities coupled with the divergence
of the flow velocity field in the first-order distribution $f^{(1)}$ (p.~\pageref{sym:ABCE_quant}).
\item[$f$] One-body velocity distribution function (p.~\pageref{sym:f}).
\item[$f_b$] One-body velocity distribution function for thermal-bath particles (p.~\pageref{eq:equilibrium_VDF_fb}).
\item[$f_{\HCS}$] One-body velocity distribution function at the homogeneous cooling state (p.~\pageref{eq:scaling_HCS}).
\item[$f_{\mathrm{M}}$] Maxwellian velocity distribution function (p.~\pageref{eq:MDF_EHS}).
\item[$f_{\mathrm{ref}}$] Reference velocity distribution function (p.~\pageref{sym:fref}).
\item[$f^{\pprime}_i$] $\equiv f(\vom^{\pprime}_i)$ with $i=1,2$, one-body velocity distribution function of the precollisional velocity variables (p.~\pageref{eq:BCO}).
\item[$f^{\mathrm{LE}}$] Local-equilibrium velocity distribution function (p.~\pageref{fig:EME_MSP22}).
\item[$f^{(k)}$] $k^{\mathrm{th}}$-order of $f$ in a Chapman--Enskog (or Hilbert) expansion (p.~\pageref{eq:Hilbert_expansion}).
\item[$\mathbf{F}^{\mathrm{ext}}$] External force applied to a system (p.~\pageref{sym:rr_ast}).
\item[$\mathbf{F}_{\mathrm{coll}}$] Effective force due to collisions (p.~\pageref{eq:LE_NLD_KTGG_def}).
\item[$\mathbf{F}_{\mathrm{drag}}$] Drag force (p.~\pageref{sym:Fdrag}).
\item[$F_{X}$] Rate of change of a macrostate variable of the system $X$ (p.~\pageref{eq:memory_descr}).
\item[$F_{Y}$] Rate of change of an inner variable of the system $Y$ (p.~\pageref{eq:memory_descr}).
\item[$\mathfrak{F}$] Function that returns the numerical solution of a Langevin-like system of equations (p.~\pageref{eq:vels_pos_FrakF}).
\item[$g$] Modulus of $\mathbf{g}$ (p.~\pageref{sym:Bgpprime}).
\item[$g_{12}$] Modulus of $\mathbf{g}_{12}$ (p.~\pageref{sym:Bgpprime}).
\item[$g_{12}^{\pprime}$] Modulus of the precollisional version of $\mathbf{g}_{12}$ (p.~\pageref{sym:Bgpprime}).
\item[$\mathbf{g}$] $=\vvel_b-\vvel$, relative velocity between thermal-bath and Brownian particles (p.~\pageref{sym:Qbar}).
\item[$\mathbf{g}^{\prime}$] Postcollisional version of $\mathbf{g}$ (p.~\pageref{sym:Bgpprime}).
\item[$\ggab$] Relative velocity of the points at contact in a binary collision between particles 1 and 2 (p.~\pageref{eq:CTR}).
\item[$\ggab^{\pprime}$] Precollisional version of $\ggab$ (p.~\pageref{eq:CTR}).
\item[$\mathbf{G}$] $=\{\mathbf{F}^{\mathrm{ext}}/m,\boldsymbol{\tau}^{\mathrm{ext}}/I\}$, generalized force per unit mass (p.~\pageref{eq:jac_BE}).
\item[$h$] Hydrodynamic length scale (p.~\pageref{sym:h}).
\item[$H$] Boltzmann's $H$-functional (p.~\pageref{sym:ShannonH}).
\item[$H_{\mathrm{eq}}$] Value of $H$ at equilibrium (p.~\pageref{eq:S_with_H_functional}).
\item[$H^*$] Boltzmann's $H$-functional for the reduced velocity distribution function (p.~\pageref{sym:Hast}).
\item[$I$] Moment of inertia of a particle (p.~\pageref{sym:oo_ast}).
\item[$I{[ \cdot ]}$] Reduced collisional operator (p.~\pageref{eq:red_BE_IHS_free}).
\item[$\mathcal{I}$] Collisional integral (p.~\pageref{eq:coll_integral_I_cal}).
\item[$\mathcal{I}_0$] $\equiv 2\pi^{\frac{\dt-1}{2}}/(\dt+1)$, constant introduced in the computation of the collisional integrals (p.~\pageref{tab:EPR_2body}).
\item[$J$] Boltzmann collisional operator (p.~\pageref{eq:general_BE}).
\item[$J_G$] Gain term of the Boltzmann collisional operator (p.~\pageref{sym:JG}).
\item[$J_L$] Loss term of the Boltzmann collisional operator (p.~\pageref{sym:JL}).
\item[$J^{(k)}$] $k^{\mathrm{th}}$-order term of $J$ in a Chapman--Enskog (or Hilbert) expansion (p.~\pageref{eq:Hilbert_expansion}).
\item[$\mathcal{J}$] Collisional integral (p.~\pageref{eq:balance}).
\item[$\mathfrak{J}$] Jacobian of the transformation $(\vom_1^{\prime},\vom_2^{\prime})\to (\vom_1,\vom_2)$ or, equivalently, $(\vom_1,\vom_2)\to (\vom_1^{\pprime},\vom_2^{\pprime})$ (p.~\pageref{eq:JG_dif_vom1_J}).
\item[$\ell$] Mean free path (p.~\pageref{eq:mean_free_path}).
\item[$\kb$] Boltzmann constant (p.~\pageref{sym:kb}).
\item[$k_{\parallel}$] Critical wave number associated with the longitudinal modes (p.~\pageref{sym:ins_modes}).
\item[$k_{\perp}$] Critical wave number associated with the transverse modes (p.~\pageref{sym:ins_modes}).
\item[$L$] Characteristic length of a system (p.~\pageref{sym:L}).
\item[$L_c$] Critical length for instabilities (p.~\pageref{sym:Lc}).
\item[$L^{\mathrm{cell}}$] Characteristic length of a square cell (p.~\pageref{alg:CA2D}).
\item[$L^{\mathrm{cell}}_{x,y}$] Length of the $x$ or $y$ component of a rectangular cell (p.~\pageref{alg:CA2D}).
\item[$L_k^{(n)}$] Associated Laguerre polynomial (p.~\pageref{eq:Sk_sonine_def}).
\item[$K_d$] Constant appearing in the definition of the collisional frequency in $d$ dimensions (p.~\pageref{eq:BFPE_epsilon_ST}).
\item[$\mathrm{Kn}$] Knudsen number (p.~\pageref{sym:Kn}).
\item[$n$] Number density (p.~\pageref{sym:n}).
\item[$n_b$] Number density of thermal-bath particles (p.~\pageref{sym:nb}).
\item[$N$] Number of particles in a system (p.~\pageref{sym:N}).
\item[$N_{x,y}^{\mathrm{cell}}$] Number of cells in a row ($x$) or a column ($y$) of the simulation box (p.~\pageref{alg:CA2D}).
\item[$\mathcal{N}_k$] Normalization constant of $S_k$ (p.~\pageref{eq:Sk_sonine_def}).
\item[$\mathcal{N}_{pq}^{(r)}$] Normalization constant of $\Psi_{pq}^{(r)}$ (p.~\pageref{eq:Sonine_expansion_IRHS_a}).
\item[$m$] Mass of a particle (p.~\pageref{sym:rr_ast}).
\item[$m_b$] Mass of a thermal-bath particle (p.~\pageref{sym:mb}).
\item[$P_k$] Legendre polynomial of order $k$ (p.~\pageref{eq:Sonine_expansion_IRHS_a}).
\item[$\mathsf{P}$, $P_{ij}$] Pressure tensor (p.~\pageref{sym:Ptensor}).
\item[$\mathbf{q}$] (Total) heat flux (p.~\pageref{sym:qflux}).
\item[$\mathbf{q}_t$] Translational counterpart of the heat flux (p.~\pageref{sec:Art7}).
\item[$\mathbf{q}_r$] Rotational counterpart of the heat flux (p.~\pageref{sec:Art7}).
\item[$\overline{Q}$] Total cross section (p.~\pageref{sym:Qbar}).
\item[$\mathbf{Q}_{12}$] Impulse exerted by particle 1 on particle 2 in a binary collision (p.~\pageref{sym:Q12}).
\item[$\mathbf{Q}_{12}^{-}$] Impulse acting on the precollisional form of a binary collision (p.~\pageref{eq:coll_rules_IRHS_rest}).
\item[$\mathbf{Q}^{\parallel}_{12}$] Component of $\mathbf{Q}_{12}$ parallel to the unit vector $\ssab$ (p.~\pageref{eq:Qab_par_perp}).
\item[$\mathbf{Q}^{\perp}_{12}$] Component of $\mathbf{Q}_{12}$ perpendicular to the unit vector $\ssab$ (p.~\pageref{eq:Qab_par_perp}).
\item[$\rr$] Position of a particle (p.~\pageref{sym:rij}).
\item[$\rr^*$] Position of a particle after a small time interval of width $\delta t$ (p.~\pageref{sym:rr_ast}).
\item[$\rr^{\mathrm{fic}}$] Fictive positions in an event-driven molecular dynamics algorithm with Langevin-like dynamics (p.~\pageref{sym:rput}).
\item[$\rr^{\mathrm{put}}$] Putative positions in an event-driven molecular dynamics algorithm with Langevin-like dynamics (p.~\pageref{sym:rput}).
\item[$r_{ij}$] $=|\rr_i-\rr_j|$, distance between a pair $ij$ of particles (p.~\pageref{sym:rij}).
\item[$s$] Dimensionless time scale that represents the (local-equilibrium) accumulated average number of collisions per particle up to a certain time (p.~\pageref{eq:s_variable}).
\item[$s_0$] Accumulated average number of collisions per particle at $t_0$ (p.~\pageref{eq:Haffs_law_coll}).
\item[$s_w$] Waiting number of collisions per particle in the velocity-inversion computational experiment (p.~\pageref{sym:sw}).
\item[$S$] (Nonequilibrium) entropy (p.~\pageref{sym:SQ}).
\item[$S_{\mathrm{eq}}$] Equilibrium value of the entropy (p.~\pageref{eq:S_with_H_functional}).
\item[$S_k$] $k^{\mathrm{th}}$-order Sonine polynomial (p.~\pageref{eq:Sk_sonine_def}).
\item[$\mathbf{S}_{12}$] $=\sigma(\oo_1+\oo_2)/2$, mean angular velocity vector of particles 1 and 2 in units of their diameter, $\sigma$ (p.~\pageref{eq:ggab}).
\item[$\mathrm{Sr}$] Strouhal number (p.~\pageref{sym:Sr}).
\item[$t$] Time variable (p.~\pageref{sym:t}).
\item[$t_0$] Reference initial time value (p.~\pageref{eq:Haffs_law}).
\item[$t_{\mathrm{coll}}$] Time when a collision event occurs (p.~\pageref{sym:tcoll}).
\item[$t_\theta$] Thermal crossing time (p.~\pageref{sym:t_theta}).
\item[$t_{\mathcal{D}}$] Entropic crossing time (p.~\pageref{sym:t_KLD}).
\item[$t_{\mathcal{D}^{\mathrm{LE}}}$] Thermal-distance crossing time (p.~\pageref{sym:t_KLD_LE}).
\item[$t_{O}$] Overshoot crossing time (p.~\pageref{sym:t_O}).
\item[$t_w$] Waiting time (p.~\pageref{fig:Kovacs_protocol}).
\item[$t^*$] $=t+\delta t$, time translated by a small time interval $\delta t$ (from p.~\pageref{sym:t_ast}).\\
$\equiv t\nu_b$ or $\equiv t\nuwn$, reduced time variable for a system under the action of a thermal bath (introduced in p.~\pageref{sec:Art1}).
\item[$t^*_K$] $\equiv \nu_b t_K$, Kovacs waiting time (p.~\pageref{sym:t_K_ast}).
\item[$T$] Temperature of a system (p.~\pageref{eq:molecular_temp}).
\item[$T_A$] Temperature of a sample A initially further from the equilibrium or steady-state temperature as compared with another sample B (p.~\pageref{sym:TA_TB}).
\item[$T_b$] Temperature associated with a thermal bath (p.~\pageref{sym:Tb}).
\item[$T_B$] Temperature of a sample B initially closer to the equilibrium or steady-state temperature as compared with another sample A (p.~\pageref{sym:TA_TB}).
\item[$T_{\HCS}$] Temperature of a system at the homogeneous cooling state (p.~\pageref{sym:zetaHCS}).
\item[$T_{\mathrm{ref}}$] Reference bath-temperature (p.~\pageref{fig:Kovacs_protocol}).
\item[$\Ttr$] Translational temperature (p.~\pageref{eq:tr_rot_temp}).
\item[$\Trot$] Rotational temperature (p.~\pageref{eq:tr_rot_temp}).
\item[$\Twn$] Temperature associated with a thermostat (p.~\pageref{eq:def_Twn_ST}).
\item[$T^{\mathrm{st}}$] Steady-state value of the temperature (p.~\pageref{sym:Tst}).
\item[$T^{0}$] $\equiv T(t_0)$, initial temperature (p.~\pageref{sym:TA_TB}).
\item[$T^{0}_A$ ($T_{0A}$)] Initial temperature of a sample A initially further from the equilibrium or steady-state temperature as compared with another sample B (p.~\pageref{sym:TA_TB}).
\item[$T^{0}_B$ ($T_{0B}$)] Initial temperature of a sample B initially closer to the equilibrium or steady-state temperature as compared with another sample A (p.~\pageref{sym:TA_TB}).
\item[$\mathbf{u}$] Flow field of the system (p.~\pageref{eq:u_def}).
\item[$U_{ij}$] Potential energy of a pair $ij$ of particles (p.~\pageref{eq:hard_potential}).
\item[$v$] Velocity modulus of a paticle (p.~\pageref{eq:vb_cb}).
\item[$v_{x,y,z}$] $x$, $y$ or $z$ velocity component of $\vvel$ (p.~\pageref{alg:EDMD_BC_x}).
\item[$v_b$] Velocity modulus of a thermal-bath particle (p.~\pageref{eq:vb_cb}).
\item[$\vth$] Thermal velocity (p.~\pageref{eq:red_velocity}).
\item[$v_{\mathrm{th},b}$] Thermal velocity of thermal-bath particles (p.~\pageref{eq:red_velocity}).
\item[$\vvel$] Translational velocity of a particle (p.~\pageref{sym:vvel}).
\item[$\vvel^{\prime}$] Postcollisional translational velocity of a particle (p.~\pageref{eq:dir_rules_elastic}).
\item[$\vvel^{\pprime}$] Precollisional translational velocity of a particle (p.~\pageref{eq:JG_dif_vom1}).
\item[$\vvel^*$] Translational velocity of a particle after a small time interval $\delta t$ (p.~\pageref{sym:vvel_ast}).
\item[$\vvel_b$] Translational velocity of a thermal-bath particle (p.~\pageref{sym:vb}).
\item[$\vvel_{ij}$] $=\vvel_i-\vvel_j$, relative translational velocity of a pair $ij$ of particles (p.~\pageref{sym:vvel_12}).
\item[$\vvel_{ij}^{\prime}$] Postcollisional relative translational velocity of a pair $ij$ of particles (p.~\pageref{eq:dir_rules_elastic}).
\item[$\vvel^{\prime\prime}_{ij}$] Precollisional relative translational velocity of a pair $ij$ of particles (p.~\pageref{eq:JG_dif_vom1}).
\item[$\vvel^{\mathrm{fic}}$] Fictive velocities in an event-driven molecular dynamics algorithm with Langevin-like dynamics (p.~\pageref{sym:rput}).
\item[$\vvel^{\mathrm{put}}$] Putative velocities in an event-driven molecular dynamics algorithm with Langevin-like dynamics (p.~\pageref{sym:rput}).
\item[$V$] Modulus of the peculiar velocity (p.~\pageref{eq:T_CE_IRHS}).
\item[$V_{\mathrm{eq}}$] Volume at equilibrium (p.~\pageref{fig:Kovacs_protocol}).
\item[$\mathbf{V}$] $=\vvel-\mathbf{u}$, peculiar velocity (p.~\pageref{eq:T_CE_IRHS}).
\item[$\ww$] $=\oo/\oth$, reduced angular velocity of a particle (p.~\pageref{eq:f_resc_IRHS}).
\item[$\overline{W}$] Number of microstates compatible with a given macrostate of the system (p.~\pageref{sym:overlineW}).
\item[$\mathbf{W}_{12}$] $\equiv(\ww_1+\ww_2)/2$, mean reduced angular velocity of particles 1 and 2 (p.~\pageref{sym:W12}).
\item[$\mathbf{W}_t$] Wiener process (p.~\pageref{eq:general_form_LE}).
\item[$\boldsymbol{\mathcal{W}},\bar{\boldsymbol{\mathcal{W}}}$] Random Gaussian variables emerging from the solution of Langevin-like equations (p.~\pageref{eq:W_rand_def_Lang_LE}).
\item[$X$] Arbitrary macrostate quantity in the context of nonequilibrium processes (p.~\pageref{fig:Kovacs_protocol}).
\item[$\widetilde{X}$] $\equiv I\bar{\Omega}^2/\dr\Trot$, nondimensional rotational quantity (p.~\pageref{sym:X_rot}).
\item[$\widetilde{X}^{\HCS}$] Value of $I\bar{\Omega}^2/\dr\Trot$ at the homogeneous cooling state (p.~\pageref{sym:XHCS}).
\item[$\mathsf{X}$] Noise-intensity diagonal matrix (p.~\pageref{eq:ST_LE_first}).
\item[$Y$] Inner arbitrary dynamic quantity of the system in the context of nonequilibrium processes (p.~\pageref{eq:memory_descr}).
\item[$\boldsymbol{\mathcal{Y}},\bar{\boldsymbol{\mathcal{Y}}}$] Random Gaussian variables of unit variance and zero mean appearing in the definition of the variables $\boldsymbol{\mathcal{W}},\bar{\boldsymbol{\mathcal{W}}}$ (p.~\pageref{eq:W_rand_def_Lang_LE}).
\item[$\een$] Coefficient of normal restitution (p.~\pageref{sym:een}).
\item[$\en$] Reduced coefficient of normal restitution (p.~\pageref{eq:red_coef_rest}).
\item[$\eet$] Coefficient of tangential restitution (p.~\pageref{sym:een}).
\item[$\et$] Reduced coefficient of tangential restitution (p.~\pageref{eq:red_coef_rest}).
\item[$\gamma$] Control parameter of the nonlinearity in a drag coefficient (p.~\pageref{eq:xi_QR_gamma}).
\item[$\gamma_c$] Exponent of the high-velocity tail of $\phi^{\HCS}_{\cc}$ (p.~\pageref{eq:HVT_IHS_KTGG}).\\
           Critical value of the nonlinear parameter $\gamma$ (p.~\pageref{sym:gammac_IHS}).
\item[$\gamma_c^{\mathrm{IHS}}$] Value of the exponent $\gamma_c$ for the inelastic hard-sphere model (p.~\pageref{eq:HVT_IHS_KTGG}).
\item[$\gamma_c^{\mathrm{IRHS}}$] Value of the exponent $\gamma_c$ for the inelastic and rough hard-sphere model (p.~\pageref{sym:gammacIRHS}).
\item[$\gamma_c^{\mathrm{wn}}$] Value of the exponent $\gamma_c$ for the externally driven case (p.~\pageref{sym:phiwn}).
\item[$\gamma_{cw}$] Exponent of the high-velocity tail of the $\phi^{\HCS}_{cw}$ (p.~\pageref{sym:gammacw}).
\item[$\gamma_w$] Exponent of the high-velocity tail of $\phi^{\HCS}_{\cc}$ (p.~\pageref{sym:gammaw}).
\item[$\bar{\gamma}$] Set of noise parameters (p.~\pageref{sym:bargamma}).
\item[$\Gamma(\cdot)$] Gamma function (p.~\pageref{eq:nuwn}).
\item[$\vom$] Generalized velocity vector of a particle (p.~\pageref{sym:vom}).
\item[$\vom^{\prime}$] Postcollisional generalized velocity vector of a particle (p.~\pageref{sym:vom_prime}).
\item[$\vom^{\pprime}$] Precollisional generalized velocity vector of a particle (p.~\pageref{sym:vom_pprime}).
\item[$\vom^{\mathrm{put}}$] Putative generalized velocity vector of a particle (p.~\pageref{alg:AGF_collision}).
\item[$\vom^{\mathrm{fic}}$] Fictive generalized velocity vector of a particle (p.~\pageref{alg:AGF_collision}).
\item[$\cw$] Generalized reduced-velocity vector of a particle (p.~\pageref{eq:f_resc_IRHS}).
\item[$\delta(\cdot)$] Dirac delta (p.~\pageref{eq:white_noise_corr}).
\item[$\delta_{ij}$] Kronecker delta (p.~\pageref{eq:white_noise_corr}).
\item[$\delta t$] Small perturbative time interval (p.~\pageref{sym:delta_t}).
\item[$\delta Q$] Heat transferred in a reversible process (p.~\pageref{sym:SQ}).
\item[$\Delta t$] Time step (or width of a time interval) (p.~\pageref{sym:timestep}).
\item[$\Delta t_{\mathrm{coll}}$] $\equiv t_{\mathrm{coll}}-t$, width of a time interval between a time of a collision event and current time (p.~\pageref{eq:rv_final_Lan_EDMD}).
\item[$\varepsilon$] $=\frac{\dr}{\dt}\frac{I}{m}\frac{\chi_r^2}{\chi^2}$, rotational-to-total noise intensity ratio (p.~\pageref{eq:total_and_varepsilon_ST}).
\item[$\varepsilon_{\mathrm{cr}}$] Critical value of $\varepsilon$ that controls the emergence of overshoot (p.~\pageref{sym:epsi_cr}).
\item[$\varepsilon_{\mathrm{ref}}$] Reference value of $\varepsilon$ in a Mpemba-like effect protocol (p.~\pageref{sym:epsi_ref}).
\item[$\epsilon$] Bookkeeping parameter (p.~\pageref{sym:epsilon_book}).
\item[$\zeta$] Cooling rate (p.~\pageref{eq:diff_ev_T}).
\item[$\zeta^*$] Reduced cooling rate (p.~\pageref{eq:Haffs_law_coll}).
\item[$\zeta_{\HCS}$] Cooling rate at the homogeneous cooling state (p.~\pageref{sym:zetaHCS}).
\item[$\zeta_{\HCS}^*$] Reduced cooling rate at the homogeneous cooling state (p.~\pageref{eq:Haffs_law_coll}).
\item[$\zeta_{\HCS}^{*\mathrm{MA}}$] Reduced cooling rate at the homogeneous cooling state in the Maxwellian approximation (p.~\pageref{eq:zeta_HCS_red_MA}).
\item[$\zeta_{\bar{\Omega}}$] Production rate associated with the mean angular velocity evolution (p.~\pageref{eq:ev_eq_Omega_zeta}).
\item[$\zeta^{\HCS}_{\bar{\Omega}}$] Production rate associated with the mean angular velocity evolution (p.~\pageref{sym:omega_zeta_HCS}).
\item[$\zeta^*_{\bar{\Omega}}$] $\equiv \zeta_{\bar{\Omega}}/\nu$, reduced production rate associated with the mean angular velocity evolution (p.~\pageref{sym:zetaOmega_ast}).
\item[$\eta$] Shear viscosity (p.~\pageref{sym:viscosities}).
\item[$\eta_b$] Bulk viscosity (p.~\pageref{sym:viscosities}).
\item[$\boldsymbol{\eta}$] Random Gaussian-distributed force (p.~\pageref{eq:white_noise}).
\item[$\bar{\boldsymbol{\eta}}$] Random Gaussian-distributed force of unit variance (p.~\pageref{eq:white_noise_eta_bar}).
\item[$\bar{\boldsymbol{\eta}}_d$] Random Gaussian-distributed force of unit variance of $d$ dimensions (p.~\pageref{eq:ST_LE_first}).
\item[$\theta$] $=\Trot/\Ttr$, rotational-to-translational temperature ratio (p.~\pageref{eq:theta_equip_def}).
\item[$\theta^{\mathrm{st}}$] Steady-state value of $\theta$ (p.~\pageref{sym:thetast}).
\item[$\theta^{\HCS}$] Value of $\theta$ at the homogeneous cooling state (p.~\pageref{sym:thetaHCS}).
\item[$\theta^{\HCS}_{\mathrm{M}}$] $\theta^{\HCS}$ for the Maxwellian approximation (p.~\pageref{eq:expr_theta_M}).
\item[$\theta^0$] Initial condition for $\theta$ (p.~\pageref{sym:theta0}).
\item[$\theta^0_A$ ($\theta_{0A}$)] Initial condition for $\theta$ of a sample A initially further from the steady-state temperature as compared with another sample B (p.~\pageref{sym:theta0A_B}).
\item[$\theta^0_B$ ($\theta_{0B}$)] Initial condition for $\theta$ of a sample B initially closer to the steady-state temperature as compared with another sample A (p.~\pageref{sym:theta0A_B}).
\item[$\vartheta$] Angle formed by $\vvel_{12}$ and $\dif \mathbf{S}$ (p.~\pageref{fig:collision_cylinder}).\\
 		Impact angle in a binary collision (from p.~\pageref{fig:COL_EHS}).
\item[$\vartheta^{\pprime}$] Precollisional impact angle in a binary collision (p.~\pageref{fig:COL_EHS}).
\item[$\overline{\vartheta}$] Angle between $\mathbf{g}^{\prime}$ and $\mathbf{g}$ (p.~\pageref{sym:Qbar}).
\item[$\Theta$] Heaviside step-function (p.~\pageref{eq:JL_dif_vom1}).
\item[$\kappa$] $=4I/m\sigma^2$, reduced moment of inertia (p.~\pageref{eq:red_coef_rest}).
\item[$\lambda$] Thermal conductivity (p.~\pageref{sym:lambda}).
\item[$\mu$] Dufour-like transport coefficient (p.~\pageref{sym:Dufour}).
\item[$\mu_k$] Collisional moment of order $k$ (p.~\pageref{eq:red_BE_IHS_free}).
\item[$\mu_k^{\HCS}$] Collisional moment of order $k$ at the homogeneous cooling state (p.~\pageref{eq:eq_steady_HCS_IHS}).
\item[$\mu_k^{(p)}$] Coefficient of the $k^{\mathrm{th}}$ collisional moment accompanying $a_p$ from a linearization in cumulants of the Sonine approximation (p.~\pageref{sec:Art4}).
\item[$\mu_{pq}$] $\equiv \mu_{pq}^{(0)}$, collisional moment of order $(p,q,0)$ (p.~\pageref{sec:Art5}).
\item[$\mu_{pq}^{\HCS}$] $\equiv \mu_{pq}^{(0)\HCS}$, collisional moment of order $(p,q,0)$ at the homogeneous cooling state (p.~\pageref{sec:Art5}).
\item[$\mu_{pq}^{(r)}$] Collisional moment of order $(p,q,r)$ (p.~\pageref{eq:red_BE_IRHS_free}).
\item[$\mu_{pq}^{(r)\HCS}$] Collisional moment of order $(p,q,r)$ at the homogeneous cooling state (p.~\pageref{eq:red_BE_IRHS_free}).
\item[$\nu$] Collision frequency (p.~\pageref{eq:s_variable}).
\item[$\nu_b$] Collision frequency of a system at $T_b$ (p.~\pageref{sym:nub}).
\item[$\nuwn$] Collision frequency associated with the thermostat energy injection (p.~\pageref{eq:BFPE_epsilon_ST}).
\item[$\xi$] Velocity-divergence transport coefficient (p.~\pageref{sec:Art7}).
\item[$\xi(v)$] Nonlinear drag coefficient (p.~\pageref{eq:LE_NLD}).
\item[$\xi_0$] Constant drag coefficient (p.~\pageref{eq:white_noise}).
\item[$\xi^*_0$] $\equiv \xi_0/\nu_b$ reduced constant drag coefficient (p.~\pageref{eq:T_Langevin_ev}).
\item[$\xi_{\mathrm{eff}}$] Effective drag coefficient in a system with Langevin-like dynamics under nonlinear drag (p.~\pageref{sec:Art1}).
\item[$\xi_{\mathrm{QR}}$] Drag coefficient in the quasi-Rayleigh regime (p.~\pageref{eq:mom_drag}).
\item[$\xi_t$] Translational energy production rate (p.~\pageref{eq:P_Rates_t}).
\item[$\xi_r$] Rotational energy production rate (p.~\pageref{eq:P_Rates_t}).
\item[$\xi^{\HCS}_{t,r}$] Energy production rates at the homogeneous cooling state (p.~\pageref{sym:xi_prod_HCS}).
\item[$\xi^{*}_{t,r}$] $\equiv \xi_{t,r}/\nu$, reduced energy production rates (p.~\pageref{sym:zetaOmega_ast}).
\item[$\xi^{*\HCS}_{t,r}$] Value of $\xi^{*}_{t,r}$ at the homogeneous cooling state (p.~\pageref{sym:xitrHCS}).
\item[$\varpi_{ij}$] Estimator of the collision rate (p.~\pageref{sym:varpi_ij}).
\item[$\varpi_{\max}$] Upper bound estimate of the collision rate (p.~\pageref{sym:varpi_ij}).
\item[$\rho$] Mass density (p.~\pageref{sym:rho_den}).
\item[$\sigma$] Diameter of a particle (p.~\pageref{sym:sigma}).
\item[$\sigma_b$] Diameter of a thermal-bath particle (p.~\pageref{sym:sb}).
\item[$\sigma_{ij}$] $=(\sigma_i+\sigma_j)/2$, mean diameter of a pair $ij$ of particles (p.~\pageref{sym:sigma}).
\item[$\ssab$ ($\ssab_{ij}$)] $=(\rr_i-\rr_j)/r_{ij}$, unit intercenter vector of a pair $ij$ of particles. (p.~\pageref{sym:ssab})
\item[$\ssab_{ij}^{\mathrm{fic}}$] Fictive unit intercenter vector in an event-driven molecular dynamics algorithm with Langevin-like dynamics (p.~\pageref{eq:rv_final_Lan_EDMD}).
\item[$\tau$] Mean free time (p.~\pageref{sym:tau0}).
\item[$\tau_0$] Typical time scale of a problem (p.~\pageref{sym:tau0}).
\item[$\tau_b$] Mean free time for a dilute gas at $T_b$ (p.~\pageref{sym:taub}).
\item[$\tau_c$] Collision duration (p.~\pageref{sym:tauc}).
\item[$\tau_h$] Hydrodynamic time scale (p.~\pageref{sym:tauh}).
\item[$\tau_t$] $=\Ttr/T$, translational temperature ratio (p.~\pageref{eq:f_resc_IRHS}).
\item[$\tau_t^{\HCS}$] $\tau_t$ evaluated at the homogeneous cooling state (p.~\pageref{eq:f_resc_IRHS}).
\item[$\tau_r$] $=\Trot/T$, rotational temperature ratio (p.~\pageref{eq:f_resc_IRHS}).
\item[$\tau_r^{\HCS}$] $\tau_r$ evaluated at the homogeneous cooling state (p.~\pageref{eq:f_resc_IRHS}).
\item[$\boldsymbol{\tau}^{\mathrm{ext}}$] External torque applied to a system (p.~\pageref{sym:oo_ast}).
\item[$\phi$] Reduced one-body velocity distribution function (p.~\pageref{sym:phi}).
\item[$\phi_i^{\pprime}$] $\equiv\phi(\cw_i^{\pprime})$, precollisional version of the reduced one-body velocity distribution function (p.~\pageref{eq:I_coll_def}).
\item[$\phi_{\HCS}$] Reduced one-body velocity distribution function at the homogeneous cooling state (p.~\pageref{eq:scaling_HCS}).
\item[$\phi_{\mathrm{M}}$] Maxwellian reduced velocity distribution function (p.~\pageref{eq:phiM_def}).
\item[$\phi^{\mathrm{wn}}$] Reduced one-body steady-state velocity distribution function of an externally driven granular gas (p.~\pageref{sym:phiwn}).
\item[$\phi_{\cc}$] Translational marginal velocity distribution function (p.~\pageref{eq:MA_VDF_IRHS}).
\item[$\phi_{\ww}$] Rotational marginal velocity distribution function (p.~\pageref{eq:MA_VDF_IRHS}).
\item[$\phi_{cw}$] Marginal velocity distribution function for the variable $c^2w^2$ (p.~\pageref{sym:phi_marg_HCS}).
\item[$\phi^{\HCS}_{\cc}$] $\phi_{\cc}$ at the homogeneous cooling state (p.~\pageref{sym:phi_marg_HCS}).
\item[$\phi^{\HCS}_{\ww}$] $\phi_{\ww}$ at the homogeneous cooling state (p.~\pageref{sym:phi_marg_HCS}).
\item[$\phi^{\HCS}_{cw}$] $\phi_{cw}$ at the homogeneous cooling state (p.~\pageref{sym:phi_marg_HCS}).
\item[$\Phi$] Slope for the temperature evolution in an homogeneous granular gas of inelastic and rough hard disks under the action of the splitting thermostat (p.~\pageref{sym:Phi_slope}).
\item[$\Phi_{0,A}$] Initial value of the slope $\Phi$ of a sample A initially further from the steady-state temperature as compared with another sample B (p.~\pageref{sec:Art6}).
\item[$\Phi_{0,B}$] Initial value of the slope $\Phi$ of a sample B initially closer to the steady-state temperature as compared with another sample A (p.~\pageref{sec:Art6}).
\item[$\chi^2$] (Total) noise intensity (p.~\pageref{eq:white_noise_corr}).
\item[$\chi_t^2$] Noise intensity of the translational degrees of freedom (p.~\pageref{eq:ext_stoc_force}).
\item[$\chi_r^2$] Noise intensity of the rotational degrees of freedom (p.~\pageref{eq:ext_stoc_force}).
\item[$\psi$] Generic velocity-related quantity (p.~\pageref{eq:cool_rate_def}).
\item[$\Psi_{pq}^{(r)}$] Polynomial of order $(p,q,r)$ forming an orthogonal basis (p.~\pageref{eq:Sonine_expansion_IRHS_a}).
\item[$\oth$] Thermal angular velocity (p.~\pageref{eq:f_resc_IRHS}).
\item[$\oo$] Angular velocity of a particle (p.~\pageref{sym:oo}).
\item[$\oo^{\prime}$] Postcollisional angular velocity of a particle (p.~\pageref{eq:coll_rules_IRHS}).
\item[$\oo^*$] Angular velocity of a particle after a small time interval $\delta t$ (p.~\pageref{sym:oo_ast}).
\item[$\Omega_d$] Solid angle in $d$ dimensions (p.~\pageref{eq:def_Twn_ST}).
\item[$\boldsymbol{\Omega}$] Solid angle vector (p.~\pageref{fig:collision_cylinder}).
\item[$\bar{\boldsymbol{\Omega}}$] $\equiv\langle \oo\rangle$, mean angular velocity of a system (p.~\pageref{sym:avOmega}).
\item[$\bar{\boldsymbol{\Omega}}_{\HCS}$] Value of $\bar{\boldsymbol{\Omega}}$ at the homogeneous cooling state (p.~\pageref{sym:avOmega}).
\item[$\langle\cdot\rangle_{\HCS}$] Average at the homogeneous cooling state (p.~\pageref{eq:hierarchy_moms_IHS}).
\item[$\langle\cdot\rangle_{\mathrm{M}}$] Average under the Maxwellian approximation (p.~\pageref{eq:comp_moments_ck_Max}).
\item[$\mathbb{1}_d$] Unit matrix in $d$ dimenstions (p.~\pageref{eq:white_noise_corr}).
\item[$\lfloor \cdot \rfloor$] Largest integer smaller than a given real number (p.~\pageref{sym:floor_op}).
\item[$\int_{+}\dif\ssab$] $\equiv \int\dif\ssab\,\Theta(\vvel_{12}\cdot\ssab)$, integration in the semiplane $\vvel_{12}\cdot\ssab\geq 0$ (p.~\pageref{eq:BCO}).
\end{description}
\newpage

\cleardoublepage
\phantomsection
\addcontentsline{toc}{chapter}{\listfigurename}
\begingroup 

\setlength{\textheight}{23cm} 

\etocstandarddisplaystyle 
\etocstandardlines 

\listoffigures 


\phantomsection
\addcontentsline{toc}{chapter}{\listtablename}
\listoftables 

\endgroup
\tableofcontents


\setchapterstyle{kao}{} 
\pagelayout{wide}

\checkoddpage
\ifoddpage{
	\afterpage{\blankpage}
}\fi

\checkoddpage
\ifoddpage{
	\afterpage{\blankpage}
}\fi

\oldmainmatter 

\setlength{\overflowingheadlen}{\linewidth}
\addtolength{\overflowingheadlen}{\marginparsep}
\addtolength{\overflowingheadlen}{\marginparwidth}

\pagestyle{scrheadings} 

\addpart{Introduction}

\thispagestyle{empty}
\begin{center}
\vspace*{0.5cm}
{\large{\textbf{The Granular Gas Poem}}}\\
\vspace*{0.5cm}
\textit{Invisible and swirling,}\\
\textit{Particles collide and mix,}\\
\textit{A dance of frenzied motion,}\\
\textit{In a world of tiny tricks.}

\textit{Granular gases they are called,}\\
\textit{A realm beyond our sight,}\\
\textit{But their impact is felt,}\\
\textit{As they move with all their might.}

\textit{From powders to sand dunes,}\\
\textit{These granules have a way,}\\
\textit{Of behaving like a fluid,}\\
\textit{But they never stay at bay.}

\textit{Their energy is chaotic,}\\
\textit{Like a wild and whirling storm,}\\
\textit{And yet they find a way,}\\
\textit{To maintain their curious form.}

\textit{They flow and bounce and scatter,}\\
\textit{In ways we can't explain,}\\
\textit{A never-ending story,}\\
\textit{Of movement that sustains.}

\textit{So when you next hold sand,}\\
\textit{Or pour sugar in a jar,}\\
\textit{Remember the granular gases,}\\
\textit{That swirl both near and far.}

\textit{For in these tiny particles,}\\
\textit{There's a world of mystery,}\\
\textit{A hidden universe,}\\
\textit{That we can only see through history.}

{\small{\href{https://chat.openai.com/}{ChatGPT - OpenAI}}}
\end{center}

\setcounter{chapter}{0}
\chapter{General introduction}
\labch{general_introduction}
{\flushleft{
\emph{For dust you are, and to dust you shall return}\\
\emph{Genesis 3:19}
}}

{\flushleft{\emph{We’re All Made of Dust—
But It’s Star Dust!}\\
William E. Barton, \emph{The Evening News (Michigan, 1921)}
}}

{\flushleft{
\emph{The cosmos is within us.\\
We are made of star-stuff.\\
We are a way for the universe to know itself.}\\
    Carl Sagan, \emph{Cosmos: A Personal Voyage (1980)}
}}

{\flushleft{
	\emph{We are stardust brought to life, then empowered by the universe to figure itself out – and we have only just begun.}\\
 Neil deGrasse Tyson, \emph{Astrophysics for People in a Hurry (2017)}}}
\vspace{0.5cm}

Science and Christianity agree, \emph{we're all made of dust}. This quote must be remembered and highlighted. It is definitively rare that (old) religious writings and (contemporary) science could agree in something so important as the origin of human beings, of course, both under their own words, meanings, and purposes.

Far from the Christian sense, in the early 20$^{\mathrm{th}}$ Century, astronomers and astrophysicists began to understand the problem of star and planet formation and so the beginning of living beings' birthplace. Whereas this popular phrase existed before Carl Sagan, it was him (and more recently, Neil deGrasse Tyson, picking up Sagan's baton), in 1980, in a magnificent effort to communicate the mysteries of the Universe in the TV series \emph{Cosmos: A Personal Voyage}, that made this sentence common in literature, cinema, and daily language. Everyone, captured by Sagan's science communication, understood that we were---and still we are---stardust trying to understand the origin of stars. However, are we trying to comprehend dust itself? The answer to the latter question is ``yes,'' but certainly the study of dust, sand, powders, in general, ``granular materials'' is something \emph{modern} in science, and particularly, in physics. 

The 20$^{\mathrm{th}}$ Century was the ``worst'' epoch for Newton's description of Nature. All the classical theories were questioned, and modern concepts and new discoveries arose. In the first decades, most of attention by physicists was paid to quantum, special and general relativity theories, statistical mechanics (actually present since late 1800s), and solid state physical description. However, something apparently simpler, like granular matter, did not become a real hot topic until 1980s, but its importance and presence in physical, chemical, mathematical, and engineering research increased ever since, becoming the study of granular materials not only an essential research branch nowadays, but one of the starting paradigms in the world of complex systems.

But, how do we define granular materials? A widely-spread definition is that granular materials correspond to large collections of discrete and macroscopic particles. Those particles are the \emph{grains} of such conglomerations. The sizes of such grains vary in a huge range of length scales, since orders of magnitude of $1\,\mu\mathrm{m}$ (ultrafine grains), up to hundreds---or even thousands---of kilometers (like asteroids), with very different morphologies~\cite{AFP13}. The latter implies that the variety of systems that can be considered a granular material is enormous. In addition, they are involved constantly in a huge diversity of natural processes like geological phenomena, such as avalanches, erosion, dune formation, earthquakes, or landslides, among others~\cite{R90,S91,H95,H06,H07,LOMKR08,BMB21,ZH21}. But the action of granular materials is not restricted to Earth, but, also, granular matter participates in astrophysical processes in the form of dust, asteroids, or protoplanetary disks~\cite{SSHT17,SMKSKJDTW21}, as some examples. Moreover, research in granular materials about the latter systems has reported quite good predictions, like the size distribution of asteroids in Saturn's rings~\cite{BKBSHSS15}.

Granular materials are also very present in industrial engineering. In fact, it is estimated that, in the last decades, around 10\% of the human energy consumption is due to processes linked to granular matter~\cite{D12}, implying a huge impact in the economy and, more importantly, in global warming. Moreover, granular matter is one of the most used type of material in industry, and it is being situated by some authors to be ranked as the second one, only overcome by water~\cite{B06c,D12}. It can be highlighted its implication in industries like pharmaceutics, agriculture, construction materials, mining, or fossil-fuel treatments, as some examples~\cite{D12,OFDHMH14,S15b,HM16,CWGJZ18,WGBHSZ20}. Therefore, the importance of the understanding about granular processes could solve several natural and industrial issues.

To address the comprehension of granular matter phenomena, one must look at the interactions between grains, which are macroscopically determined by dissipative collisions. In fact, it is important to note that ordinary \index{Temperature}temperature does not play any role in their description, since its associated thermal energy scale of the grains is negligible compared to their kinetic energy~\cite{JNB96a,BP04,G19}. The loss of kinetic energy in collisions can be characterized by several causes like friction, internal vibration, or plastic deformation~\cite{BP04}. The phenomena underlying granular matter ensembles are plenty and, together with the dissipative character of particles, unveil certain complexity, such that a fully understanding of this topic is a challenge. The dissipative mechanisms and the mechanical properties of the grains, like size, shapes, or masses, as well as possible external interactions, determine important and different effects, consequences of their collective motion, like the Brazil nut effect~\cite{W76,H95,HQL01,G08}, mixing and demixing~\cite{M02}, or aggregation and fragmentation processes~\cite{BKBSHSS15,BFP18}, among others. In addition, the interest on granular materials does not only reside in their vast applications in different natural and industrial processes, but, also, despite its supposedly simplicity, granular systems have intriguing behaviors differing from those of more usual forms of matter. In fact, whereas some granular flows could recall the classical three states of matter, solid, liquid, and gas~\cite{vW08}, granular matter can be consider itself as a different state of matter exhibiting different phase behaviors in between solids and fluids~\cite{EP97}. 

The macroscopic feature of the grains invites to solve simply their single Newton's equations of motion to obtain the real dynamics of the system. However, the number of grains in the composition of a granular material is usually very large. Then, solving this system of coupled second order differential equations is almost impossible from an analytical point of view, and time- and energy-consuming in computational terms. The quite large number of constituents invites to base the theoretical approach essentially on statistical physics, fluid, and continuum mechanics~\cite{dG99,BP04,G19}, and, of course, simulation methods are the best complement for that~\cite{H99,PS05}. The density and characteristics of a granular flow determine which approach is more appropriate, as well as the expected mechanisms and effects that could be observed \emph{a priori} in the granular system. Roughly speaking, we could classify the granular dynamics into \emph{quasi-static} regimes or \emph{rapid} flows~\cite{G19}. In the former case, systems are usually dense, and solid- and liquid-like behaviors can be observed, with clogging and jamming~\cite{AD06,BCNO16,BC19}, crystal-type transitions~\cite{KCD22}, or frustration~\cite{LRLHR19} being some examples of underlying phenomena. On the other hand, the rapid flow scheme corresponds to high-speed flows. These fluidized granular systems implies that the grain motions are independent of even its nearest neighbors. Thus, collisions between granular particles are nearly instantaneous---if interactions are noncohesive---as compared with the mean free time, recalling the molecular gas case~\cite{C90,G03,G19}. Therefore, these systems are called \emph{granular gases}. This consideration is analogous to the kinetic-theory picture of molecular gases, and, in fact, the mean-square value of their random velocities is commonly known as the \emph{granular temperature}~\cite{OUO80}. These fluidized states can be achieved by strong forcing like shearing, shaking, vibration, or other methods. The way of generating the rapid flow is important to address a proper description of the flow. Mainly, granular gases can be classified into \emph{dry} granular gases of \emph{gas-solid} flows. In the former description, the interaction of the granular gas with the surrounding medium is negligible or absent, and they can be classified into \emph{driven} (externally forced) granular gases, in which there is an external energy injection playing an important role in the dynamics, and \emph{freely evolving} systems, in which there is no external energy sources, and the gas is continuously dissipating energy. Furthermore, when the granular gas is interacting with an interstitial fluid, it is said that we are in conditions of a gas-solid flow or a \emph{suspension}. From a kinetic-theory description, there are similarities between driven granular gases and gas-solid flows, where a Fokker--Planck term into the kinetic equation appears, coming from a stochastic energy injection in the former case, or a Langevin-like interaction in the latter one. Here, these two types are going to be considered as \emph{thermostatted} systems.

Throughout this thesis, we have focused on rapid granular flows in free evolution and thermostatted states. In order to complete the kinetic-theory description of the system, we must infer a model for the dissipative granular interactions. First, for the sake of simplicity, a common description of granular particles, employed in most theoretical models and computer simulations, as well as in many experiments, consists in considering them as hard spheres (\acrshort{HS}) in three dimensions or hard disks (\acrshort{HD}) in two-dimensional setups. These hard-core repulsive interactions are, by definition, noncohesive, and the instantaneous approach introduced above is fulfilled. In general, the nature of dissipation is quite complex and usually very dependent on the material of the grain. However, the easiest paradigm, from which one could extract properly the dynamical and statistical features of these rapid flows, consists in considering the hard particles to be inelastic, this feature being accounted for by a coefficient of normal restitution\index{Coefficient of normal restitution}, \phantomsection\label{sym:een}$\een$, considered to be constant~\cite{C90,G03,G19}. Moreover, in this thesis, based on what experimental observations claim~\cite{YSS20}, we have improved this simple model by considering the effect of surface roughness in particle-particle collisions. Here, it is accounted for by a coefficient of tangential restitution\index{Coefficient of tangential restitution}, \phantomsection\label{sym:eet}$\eet$, which in the simplest case is considered to be constant as well~\cite{BP04,G19}. 

In addition, one expects, as occurs with molecular gases, that rapid granular flows may admit a \index{Hydrodynamic!description}hydrodynamic description. Then, Navier--Stokes--Fourier (\acrshort{NSF}) equations could be extended to the granular case. For a usual fluid, \phantomsection\label{sym:dt}$\dt+2$ conservation or \emph{balance} equations form this description, with \dt being the dimension of the vector space for velocities; one of them corresponds to mass conservation, $\dt$ equations for linear momentum conservation, and another one for energy \index{Temperature!granular}(temperature) conservation. On the other hand, the dissipation linked to the granular interactions suggests the addition of a sink term in the energy balance equation. Moreover, to complete this description, usually phenomenological expressions for the constitutive equations are proposed, establishing a form of the fluxes in terms of gradients. For an ordinary fluid, the stress tensor and the heat flux are the fluxes appearing in the \acrshort{NSF} scheme, providing, in their phenomenological forms, three main transport coefficients\index{Transport coefficient}: the shear\index{Transport coefficient!shear viscosity} and bulk viscosities\index{Transport coefficient!bulk viscosity} from the former, and the thermal conductivity\index{Transport coefficient!thermal conductivity} from the latter. However, in the granular gaseous case, two important modifications are observed. First, dissipation implies that a density gradient contributes to the heat flux, which comes together with a proper transport (Dufour-like) coefficient\index{Transport coefficient!Dufour-like coefficient}. Moreover, the sink term in the \index{Temperature!granular}temperature equation can be affected by transport, complicating in fact the description. Furthermore, a particular feature of granular gases, when one studies its hydrodynamics, is the appearance of \index{Instability}instabilities by means of \index{Instability!clustering}clustering and \index{Instability!vortices}vortices, which exhibit a time-evolving growth~\cite{LH99,PS05} in freely cooling granular gases. Then, in order to infer the proper description arising from the peculiarities of granular interactions, a many-body approach of the problem is needed. Boltzmann or Enskog kinetic theories are applied to granular gaseous flows in order to address those issues~\cite{G19}. 

We have tried to describe the latter features with special attention to the effect of the surface roughness of the granular particles on the dynamics. However, in order to answer some questions that will be raised throughout the memoir, we have also studied some properties and effects in molecular gases or inelastic but smooth granular gases. In fact, the nonequilibrium characteristic of the granular gases, in free evolution, externally driven or as a granular suspension with a background fluid, is the perfect scenario to observe memory effects. This approach is good to be understood in simpler models to get a full perspective of the possible phenomenology that one could expect. And not only these effects, but the proper freely evolving granular gas description is not even trivial in the simplest collisional model, where traditional results of Boltzmann equation for molecular gases are completely lost. Then, before entering into the details of the implementation of the surface roughness in the granular interactions, we try to get a more general scope of those topics. This has been crucial in the structure of this thesis which is detailed in the following section.

\section{Structure of the thesis}

 The theoretical framework of this thesis is then based on kinetic-theory tools. Moreover, we exclusively work on the dilute regime and, then, the proper kinetic equation for this picture is the so-called Boltzmann equation (\acrshort{BE}). In addition, as it will be detailed, we study a certain model of molecular gases, whose knowledge is useful to the description of a certain type of gas-solid flow. We also extract information for certain memory effects arising in granular gases. In fact, we consider three types of particle interaction in our kinetic-theory-based description: elastic, inelastic and smooth, and inelastic and rough hard particles. Whereas the first one is applied to the mentioned molecular system, the other two models are used for freely evolving and thermostatted granular gaseous models.
 
The thesis is the result of a compendium of eight scientific publications, which are part of a coherent research line. They are, in order of appearance in the thesis, the following:
\begin{itemize}
\item \cite{MSP22}~\textbf{Article 1}. A. Meg\'ias, A. Santos and A. Prados. `Thermal versus entropic Mpemba effect in molecular gases with nonlinear drag'. \emph{Phys. Rev. E} 105 (2022), p.~054140. DOI:~\href{https://doi.org/10.1103/PhysRevE.105.054140}{10.1103/PhysRevE.105.054140}

\item \cite{MS20}~\textbf{Article 2}. A. Meg\'ias and A. Santos. `Kullback--Leibler divergence of a freely cooling granular gas'. \emph{Entropy} 22 (2020), p.~1308. DOI:~\href{https://doi.org/10.3390/e22111308}{10.3390/e22111308}

\item \cite{MS21}~\textbf{Article 3}. A. Meg\'ias and A. Santos. `Relative entropy of freely cooling granular gases. A molecular dynamics study'. \emph{EPJ Web Conf.} 249 (2021), p.~04006. DOI:~\href{https://doi.org/10.1051/epjconf/202124904006 }{{10.1051/epjconf/202124904006}}

\item \cite{MS22c}~\textbf{Article 4}. A. Meg\'ias and A. Santos. `Kinetic Theory and Memory Effects of Homogeneous Inelastic Granular Gases under Nonlinear Drag'. \emph{Entropy} 24 (2022), p.~1436. DOI:~\href{https://doi.org/10.3390/e24101436}{10.3390/e24101436}

\item \cite{MS23}~\textbf{Article 5}. A. Meg\'ias and A. Santos. `Translational and rotational non-Gaussianities in homogeneous freely evolving granular gases'. \emph{Phys. Rev. E} (2023). \href{https://journals.aps.org/pre/accepted/f1073R5eP7e14b25f62a3df806d9f883fdc37b54c}{Accepted paper}.

\item \cite{MS22b}~\textbf{Article 6}. A. Megías and A. Santos. `Mpemba-like effect protocol for granular gases of inelastic and rough hard disks'. \emph{Front. Phys.} 10 (2022), p.~971671. DOI:~\href{https://doi.org/10.3389/fphy.2022.971671}{{10.3389/fphy.2022.971671}}

\item \cite{MS21a}~\textbf{Article 7}. A. Meg\'ias and A. Santos. `Hydrodynamics of granular gases of inelastic and rough hard disks or spheres. I. Transport coefficients'. \emph{Phys. Rev. E} \textbf{104} (2021), p.~034901. DOI:~\href{https://doi.org/10.1103/PhysRevE.104.034901}{10.1103/PhysRevE.104.034901}

\item \cite{MS21b}~\textbf{Article 8}. A. Meg\'ias and A. Santos. `Hydrodynamics of granular gases of inelastic and rough hard disks or spheres. II. Stability analysis'. \emph{Phys. Rev. E} \textbf{104} (2021), p.~034902. DOI:~\href{https://doi.org/10.1103/PhysRevE.104.034902}{10.1103/PhysRevE.104.034902}
\end{itemize}

This thesis is organized in six parts containing different chapters, which are detailed as follows. 

First, there is an \textbf{Introduction} part, which contains this and other three chapters with the aim of presenting the basic and state-of-the-art knowledge about the investigated topics. After this general introduction, we introduce the necessary features of kinetic theory to understand and put in context all the articles forming part of the thesis in \refch{KTGG}. After that, the main concepts and definitions, related to some memory effects studied in this thesis, are described in \refch{ME_KTGG}, particularly, the Mpemba effect (\acrshort{ME}) and the Kovacs effect (\acrshort{KE}) are conceptually explained. Moreover, in \refch{SM}, we detail the simulation methods implemented to confirm or discard the theoretical developments of this thesis. 

Then, three intermediate parts are classified in order of difficulty of the collisional model used in the kinetic-theory description. The second part of the thesis is dedicated to \textbf{Molecular gases}. In this part, we exclusively study a single model of molecular fluid of identical particles, which is developed in \refch{molecular_gases} (\textbf{Article 1}, reproduced in \refsec{Art1}). Here, we describe a molecular gas as a collection of elastic hard \dt-spheres in contact with an interstitial fluid. This background fluid is assumed to be in equilibrium at temperature\index{Temperature!of the thermal bath}\phantomsection\label{sym:Tb}$T_b$ and it is considered to act as a thermal bath on the molecular gas. Therefore, the interactions between the gas molecules and the bath particles are modeled by a coarse-grained Langevin-like description. That is, apart from the collisional-driven evolution of the system, the thermal bath acts on the gas via two forces, a drag force accounting for a viscous action on the molecules and a stochastic force, which models the energy injection due to the thermal action of the bath into the gas particles. In fact, it is imposed that the drag coefficient and the noise intensity of the stochastic force are related via a fluctuation-dissipation relation. Moreover, as a more realistic model than the original Langevin's one~\cite{L08,LG97}, we set the drag coefficient to be nonlinear, i.e., it is treated as a function of the gas molecule velocity and not as a constant. The system is studied in terms of a Boltzmann--Fokker--Planck equation (\acrshort{BFPE}), in which the collisional contribution comes from the Boltzmann collisional operator, and the Fokker--Planck term describes the action of the bath. This system is known to exhibit the \acrshort{ME}~\cite{SP20,PSP21}, which is a memory effect that is also present in granular fluids~\cite{LVPS17}. The aim of this work is to study the thermal and entropic descriptions of the memory effect in the particular case of the molecular gas under nonlinear drag. We identify the similarities and differences between those two pictures and elaborate a full description of the phenomenology arising. The theoretical predictions are compared with simulation results to check our approximate theoretical schemes and arguments.

The third part of this thesis, \textbf{Granular gases of inelastic and smooth hard particles}, consists in the study of certain aspects related to the simplest collisional model of a granular gas. That is, we model the granular particles by inelastic and smooth hard \dt-spheres. This inelasticity is described by a coefficient of normal restitution\index{Coefficient of normal restitution}, $\een$, such that for $0\leq \een < 1$ particles are inelastic, the molecular gas case being recovered when $\een=1$. This part is divided into two chapters. First, in \refch{KLD_IHS}, we focus on the extension to the granular gaseous realm of a very important feature of Boltzmann kinetic theory of molecular gases, the $H$-theorem. In this chapter, \textbf{Article 2} and \textbf{Article 3} are introduced in \refsec{Art2} and \refsec{Art3}, respectively. In them, we detail the problems that the original $H$-theorem exhibits in the inelastic version of the \acrshort{BE} for a monodisperse granular gas in free evolution and provide theoretical arguments in favor of a conjecture about a functional for describing the nonequilibrium granular entropy for this system. Moreover, in \textbf{Article 2} and \textbf{Article 3}, we show computer simulation results that reinforce our theoretical schemes. In order to address this problem, we analyze first the homogeneous states of this model of granular gas that, under the \acrshort{BE} framework, reaches a state in which the one-particle velocity distribution function \index{VDF|see{Velocity distribution function}}\index{Velocity distribution function!one-particle}(\acrshort{VDF}) of the system adopts a scaling form. In this state, the time evolution of the system is completely captured by the \index{Temperature!granular}temperature continuous cooling. Such a state is called \emph{homogeneous cooling state} (\acrshort{HCS}), and it has been widely studied in the literature~\cite{G19}. We base our theoretical description on the evolution of the system toward the \acrshort{HCS} under certain approximations, which are supported by simulation results. Then, we study how the conjecture proposed in Refs.~\cite{BPV13,GMMMRT15} solves the encountered issues in the $H$-theorem. This proposal consists in considering the \emph{relative entropy}, or \emph{Kullback--Leibler divergence} (\acrshort{KLD})~\cite{KL51} of the time-dependent \index{Velocity distribution function!time-dependent}\acrshort{VDF} with respect to the \index{Velocity distribution function!of the HCS}\acrshort{HCS} \acrshort{VDF}, to be the nonequilibrium entropy of the system (up to a minus sign). Under some approximations, we elaborate a toy model for such functional, which turns out to behave as the expected nonequilibrium entropy. Computer simulations support these results and discard other possible candidates. Finally, in \textbf{Article 3} (\refsec{Art3}), and for the sake of completeness, we carry out a velocity-inversion computational experiment, based on the original work of Orban and Bellemans~\cite{OB67,OB69}, in which we show the role of inelasticity when the system is forced to turn back to its initial state.

After this first study about a granular gas of inelastic and smooth granular particles, we consider in \refch{IHS_ND} (\textbf{Article 4}, reproduced in \refsec{Art4}) the granular extension to the system studied in \textbf{Article 1} (\refsec{Art1}). Here, a monodisperse granular gas of inelastic hard \dt-spheres is surrounded by a background fluid which acts on the granular particles as a thermal bath at \index{Temperature!of the thermal bath}temperature $T_b$ and via a drag force, with nonlinear drag coefficient, and a stochastic force, both related by a fluctuation-dissipation relation. The goal of this study is to analyze the effect of the nonlinearities of the drag coefficient into the transient and steady homogeneous states of this gas-solid flow. In order to address this, we introduce similar approaches already elaborated in \textbf{Article 1} (\refsec{Art1}) and \textbf{Article 2} (\refsec{Art2}). This analysis allows us to study the emergence of some memory effects in this system, which are known to appear in the molecular case. Particularly, we predict the emergence of the \acrshort{ME} and the \acrshort{KE}. All the theoretical expressions, related to the dynamical description of the system and the appearance of the memory effects, are tested by computer simulation outcomes.

Then, after the study of the molecular and inelastic and smooth case, we investigate the influence of the addition of surface roughness on the collision mechanism of a granular gas in a fourth part of this thesis, called \textbf{Granular gases of inelastic and rough particles}. This part splits into three chapters, two of them dedicated to homogeneous states and a last third one focused on the study of the transport properties of a granular gaseous system with inelastic and rough hard particles in free evolution. In \refch{IRHS_HS} (\textbf{Article 5}, reproduced in \refsec{Art5}), we detail an analysis about the homogeneous states of a freely evolving granular gas of inelastic and rough \acrshort{HD} and \acrshort{HS} from a unified framework in terms of the rotational and translational degrees of freedom, $\dt$ and \phantomsection\label{sym:dr}$\dr$, respectively, and already introduced in Refs.~\cite{MS19,MS19b}. Whereas transient states are also introduced, we have focused on the description of the \acrshort{HCS} features of this system, with the aim of characterizing the non-Gaussian properties of the \acrshort{VDF} at this particular state. We characterize these non-Gaussianities in terms of the first order nontrivial cumulants from a well-known Sonine expansion, extending previous results of the \acrshort{HS} case~\cite{VSK14}, and, also, in terms of the high-velocity tails (\acrshort{HVT}) of some relevant marginal \index{Velocity distribution function!marginal}\acrshort{VDF} of the system, this being a novel contribution for the rough particle case. All these findings are supported by simulation results too. 

Nonetheless, the study of homogeneous states in this model of granular collisions is not only restricted to freely evolving granular particles, but also to another dry and driven setup, in which we assume an injection of energy into inelastic and rough \acrshort{HD} via their translational and rotational degrees of freedom. A description of this system is introduced in \textbf{Article 6} (\refsec{Art6}). Contrarily to the freely evolving case, the system reaches a steady state, which is studied together with transient states, describing the evolution of the system toward the stationary. For that, and expecting that the non-Gaussianities for this type of system are less important than in free evolution, we assume that the \acrshort{VDF} stays close to the Maxwellian at every stage of system evolution and in the steady state. This allows us to solve explicitly the transient and the steady homogeneous states of the system, validating the approximation with computer simulations. Furthermore, in Ref.~\cite{TLLVPS19}, it was found that the \acrshort{ME} appears from a single mechanism of injection restricted to the translational degrees of freedom. Then, it is expected that the \acrshort{ME} is present in these thermostatted states too. Indeed, we apply the theoretical approximate description to satisfactorily infer a theoretical protocol for the preparation of the initial states of the samples involved in a Mpemba-like experiment, in agreement with computer simulation outcomes.

Finally, the last works forming part of this thesis are \textbf{Article 7} and \textbf{Article 8}, appearing in \refch{IRHS_IS}, and reproduced in \refsec{Art7} and in \refsec{Art8}, respectively. In them, we study the transport properties of a \index{Hydrodynamic!description}hydrodynamic description of a monodisperse and dilute granular gas of inelastic and rough \acrshort{HD} or \acrshort{HS} in free evolution, and using the degrees-of-freedom framework. The main goals of these works are the computation of the \acrshort{NSF} \index{Transport coefficient}transport coefficients, in \textbf{Article 7} (\refsec{Art7}), treating the \acrshort{HCS} as the base state in the application of the perturbative \index{Chapman--Enskog method}Chapman--Enskog (\acrshort{CE}) method, and, in \textbf{Article 8}, the determination of the stability of the \acrshort{HCS} via linear inhomogeneous perturbations of the \acrshort{HCS} as an application of the results of \textbf{Article 7}. For the former problem, we use the simplest approximation to develop the corresponding analysis for the computation of such coefficients. In fact, we recover the results of Ref.~\cite{KSG14} for \acrshort{HS} ($\dt=\dr=3$) and obtain the form of the \index{Transport coefficient}transport coefficients for certain limiting cases of interest. All the results derived in \textbf{Article 7} are purely theoretical. Afterwards, the knowledge of those outcomes are applied to the mentioned linear stability analysis of the \acrshort{HCS}, validating previous results for the \acrshort{HS} reported in Ref.~\cite{GSK18}. As an important result, we observe a region of the parameter space where the \index{Hydrodynamic!description}hydrodynamic description of the system seems to break down. However, a discussion from computer simulation results clarifies the situation and shows that a region of permanent unstable systems does not strictly exist. 

To clarify the structure of the main part of the memoir, the different models addressed in this thesis and their corresponding chapters and publications are summarized in \reftab{sum_thesis}.

\begin{table}[h!]
\centering
\caption{Summary of the models addressed in this thesis and their corresponding occurrences.}
\labtab{sum_thesis}
\begin{tabularx}{\textwidth}{c|c|c|c}
\hline\hline
	\diagbox[width=0.33\textwidth]{\multicolumn{1}{c}{\begin{tabular}[c]{@{}c@{}}Free-streaming\\ model\end{tabular}}}{\multicolumn{1}{c}{\begin{tabular}[c]{@{}c@{}}Collisional\\ model\end{tabular}}} & \multicolumn{1}{c|}{\begin{tabular}[c]{@{}c@{}}Molecular\\ gas\end{tabular}} & \multicolumn{1}{c|}{\begin{tabular}[c]{@{}c@{}}Granular gas of\\ inelastic particles\end{tabular}} & \multicolumn{1}{c}{\begin{tabular}[c]{@{}c@{}}Granular gas of\\ inelastic and rough\\ particles\end{tabular}}\\
	\hline
	Free evolution & & \multicolumn{1}{c|}{\begin{tabular}[c]{@{}c@{}}\refch{KLD_IHS}\\ (\textbf{Articles 2 \& 3})\end{tabular}} & \multicolumn{1}{c}{\begin{tabular}[c]{@{}c@{}}\refch{IRHS_HS} \& \refch{IRHS_IS}\\ (\textbf{Articles 5, 7 \& 8})\end{tabular}}\\\hline
	\multicolumn{1}{c|}{\begin{tabular}[c]{@{}c@{}}Translational and\\ rotational driving\end{tabular}}& & & \multicolumn{1}{c}{\begin{tabular}[c]{@{}c@{}}\refch{IRHS_ME}\\ (\textbf{Article 6})\end{tabular}}\\\hline
	\multicolumn{1}{c|}{\begin{tabular}[c]{@{}c@{}}Suspension model\\ with nonlinear drag\end{tabular}} & \multicolumn{1}{c|}{\begin{tabular}[c]{@{}c@{}}\refch{molecular_gases}\\ (\textbf{Article 1})\end{tabular}} & \multicolumn{1}{c|}{\begin{tabular}[c]{@{}c@{}}\refch{IHS_ND}\\ (\textbf{Article 4})\end{tabular}} & \\ \hline\hline
\end{tabularx}
\end{table}

Once all the publications forming part of this compendium are described and explicitly presented, we discuss, in a fifth part, made of \refch{R_D} and \refch{C_O}, the main \textbf{Results and conclusions} of the thesis. 

As a colophon, a final part is dedicated to some \textbf{Appendices}, in which some computations and additional information are detailed.


\chapter{Kinetic-theoretical description of molecular and granular gaseous flows}
\labch{KTGG}
\vspace*{-2cm}
\begin{minipage}[t]{.6\textwidth}
\vspace{0pt}
\begin{center}
\includegraphics[width=0.8\linewidth]{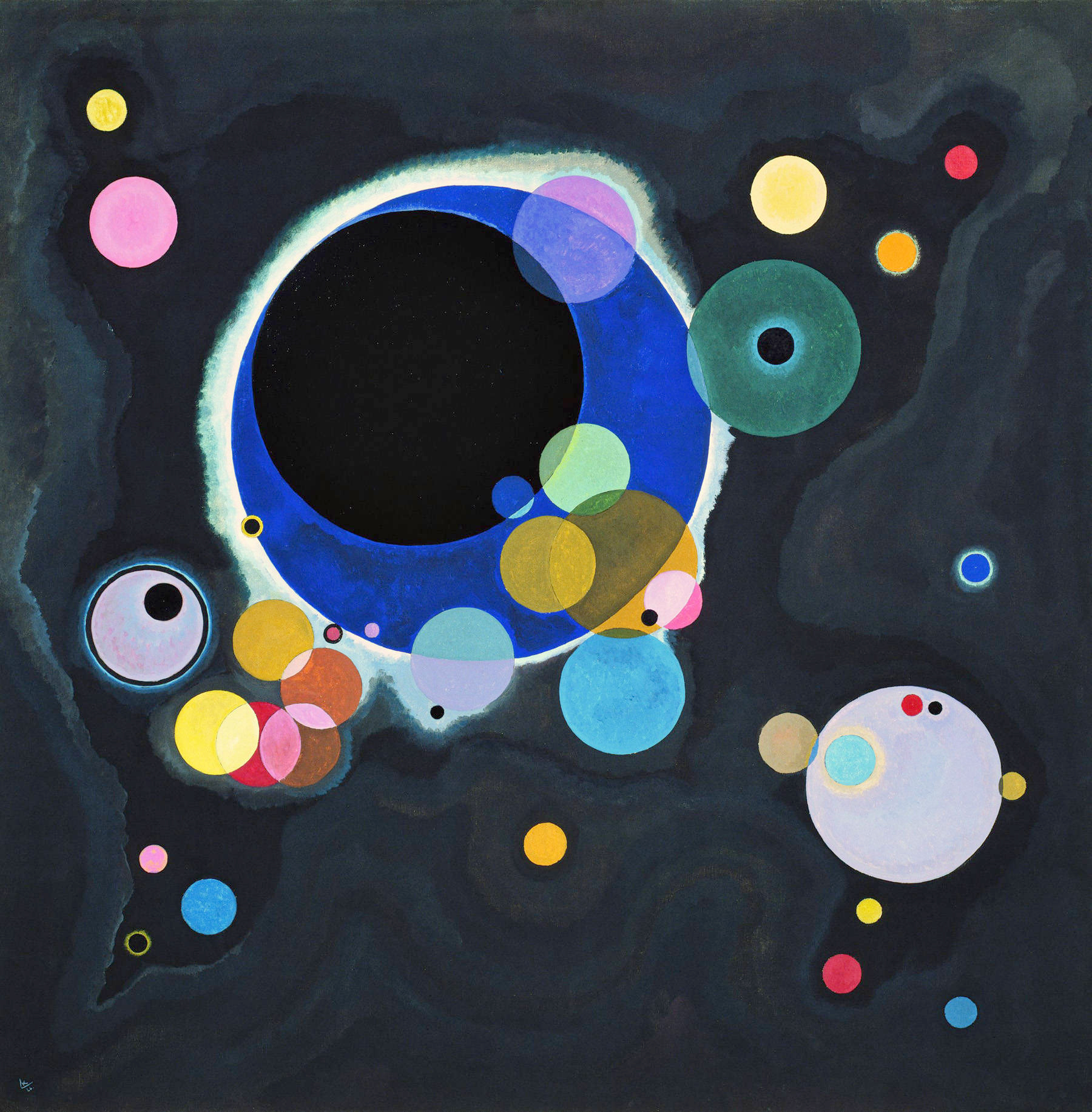}
\\
{\tiny{\copyright2018 Artists Rights Society (ARS), New York/ADAGP, Paris}}\\
{Vasily Kandinsky, \emph{Einige Kreise} (1926). \href{https://www.guggenheim.org/artwork/1992}{Guggenheim Museum, New York}.}
\end{center}
\end{minipage}

\vspace*{0.5cm}

\section{Introduction}\labsec{intro_KTGG}

The kinetic theory of gases became an essential theory in the understanding of nonequilibrium properties of fluids in the late 19$^{\mathrm{th}}$ Century, reaching its most powerful results by James C.~Maxwell and Ludwig Boltzmann~\cite{B86,B86b}. The theory was supported by experimental results and afterwards by the development of computer simulation techniques. Whereas it was introduced prior to statistical mechanics, it is nowadays considered to be a branch of it. The idea behind kinetic theory is to fully describe the state of a large physical system---originally a gas---made of \phantomsection\label{sym:N}$N$ constituents by a time-dependent probability density, where the positions and velocities of the gas particles (or generalized coordinates and their conjugate momenta) are assumed to be random variables of this probability density, which follows integro-differential \emph{kinetic} equations for its time-evolution.

In 1872, Boltzmann derived its original kinetic equation~\cite{F73,C98}, being in more modern terms the very first equation of the well-known Bogoliubov--Born--Green--Kirkwood--Yvon (\acrshort{BBGKY}) hierarchy~\cite{Y35,B46,K46,BG46,K47} after an adequate closure. The closure behind the \acrshort{BE} consists in approximating the two-particle \index{Velocity distribution function!two-particle}\acrshort{VDF} in the collisional operator by the product of two one-particle distribution functions. This is based on the hypothesis introduced by J.~C.~Maxwell~\cite{M67} known as \emph{molecular chaos} or, in the original German word, \emph{Stosszahlansatz} (introduced by Paul Ehrenfest \cite{EE02}), that is, the assumption of statistical uncorrelation of the velocities of the colliding particles. Whereas this might seem quite strong an assumption, it has been widely accepted and justified in the context of dilute systems.

The concepts of kinetic theory have been applied to granular gaseous flows since the late 20$^{\mathrm{th}}$ Century. Here, the thermodynamic \index{Temperature}temperature does not play any significant role in the granular dynamics since it is remarkably smaller than the average kinetic energy in usual experiments at room \index{Temperature}temperature~\cite{BP04,G19}. Moreover, experiments have observed evidence of molecular chaos in granular low-density systems~\cite{BO07}, so that the kinetic-theory approach seems to be applicable. 

The aim of this chapter is to introduce, for the sake of self-consistency, the main aspects and results related to kinetic theory of molecular and granular gases that have been used throughout this thesis.

\section{Derivation of the Boltzmann equation for hard spherical particles}\labsec{der_BE_KTGG}

Let us consider an arbitrary dilute monodisperse gas made of a very large number $N$ of identical hard hyperspheres in \dt translational dimensions, such that the dynamics of each individual particle is determined by its general velocities \phantomsection\label{sym:vom}$\vom$. In the absence of rotations, \phantomsection\label{sym:vvel}$\vom=\vvel$ is identical to the translational velocity, while, if rotations are known to be present, one has $\vom=\{\vvel,\oo\}$, \phantomsection\label{sym:oo}$\oo$ being the angular velocity. For the moment, we are not going to specify whether it is a molecular or a granular gas. Each specific collisional model will give us information about how particles modify their velocities in a collision, as will be specified in next \refsec{collisional_models_KTGG}.

The aim of this section is to derive the general form of the \acrshort{BE} for hard interactions between the particles of the described gas. The derivation will be based on heuristic arguments rather than as the first equation of the \acrshort{BBGKY} hierarchy derived from the Liouville equation.

The hard-sphere interaction is defined by the following discontinuous \emph{potential},
\begin{align}\labeq{hard_potential}
    U_{ij}(r_{ij})=\left\{\begin{array}{cc}
        \infty &  \text{if }r_{ij}<\sigma_{ij}\\
        0 & \text{if }r_{ij}\geq\sigma_{ij}
    \end{array} \right.,
\end{align}
with \phantomsection\label{sym:sigma}$\sigma_{ij}=(\sigma_i+\sigma_j)/2$ the distance of the centers of the spheres at contact, $\sigma_i$ being the diameter of the $i^{\mathrm{th}}$ particle and \phantomsection\label{sym:rij}$r_{ij}=|\rr_i-\rr_j|$ the distance between their centers. This represents the interaction between impenetrable spheres. All of this can be generalized straightforwardly to hyperspheres with an arbitrary number $\dt$ of translational dimensions. From now on, as we are working with identical particles, it will be considered that $\sigma_i\equiv \sigma$, $\forall i\in\{1,\dots,N\}$.

Let us consider that all the relevant information about the system is contained in its one-body probability distribution function, \phantomsection\label{sym:f}$f(\rr,\vom;t)$, such that $f(\rr,\vom;t)\dif\rr\dif\vom$ is the expected number of particles at time \phantomsection\label{sym:t}$t$ residing in the differential volume element of the phase space $\dif\rr\dif\vom$, that is, the average number of particles localized in the volume element $\dif\rr$ centered at position $\rr$, and moving with velocities valued between $\vom$ and $\vom+\dif\vom$. The knowledge of this \acrshort{VDF} can be employed to obtain macroscopic properties of the gas. Therefore, it is essential to infer its time evolution. The assumption of diluteness allows us to decouple the evolution of the VDF into two terms: a free streaming evolution, where only external forces might apply, and, on the other hand, an interaction-driven evolution, which in the case of \acrshort{HS} is equivalent to a collisional-driven change. Hence,
\begin{equation}\label{eq:dfdt_str_coll}
    \frac{\partial f}{\partial t} = \left( \frac{\partial f}{\partial t}\right)_{\mathrm{str}} + \left( \frac{\partial f}{\partial t}\right)_{\mathrm{coll}},
\end{equation}
where the first term on the right-hand side corresponds to the free-streaming evolution and the second one to the collisional rate of change. Let us infer these two different contributions.

First of all, to derive the free-streaming term, a collisionless gas will be assumed, namely the Knudsen gas~\cite{C98}. Thus, in a differential time interval \phantomsection\label{sym:delta_t}$\delta t$, a particle located at $\rr$ with velocities $\vom=\{\vvel,\oo\}$ will move to \phantomsection\label{sym:rr_ast}$\rr^*=\rr+\vvel\delta t$. Moreover, if external forces and torques are present, then \phantomsection\label{sym:vvel_ast}$\vvel^*=\vvel+m^{-1}\mathbf{F}^{\mathrm{ext}}\delta t$ and \phantomsection\label{sym:oo_ast}$\oo^* = \oo +I^{-1}\boldsymbol{\tau}^{\mathrm{ext}}\delta t$ for spinning particles.  Here, $m$ and $I$ are the mass and moment of inertia of a particle, respectively. The superscript $*$ denotes quantities evaluated at \phantomsection\label{sym:t_ast}$t^*=t+\delta t$. Hence,
\begin{equation}\label{eq:ConsProb_BE}
    f(\rr,\vom;t)\dif\rr\dif\vom = f(\rr^*,\vom^*;t^*)\dif\rr^*\dif\vom^*,
\end{equation}
implying that the Jacobian associated with the transformation $(\rr,\vom)\rightarrow (\rr^*,\vom^*)$ is given by,
\begin{equation}\labeq{jac_BE}
    \dif\rr^*\dif\vom^* = \left[1+\frac{\partial}{\partial \vom}\cdot \mathbf{G}\delta t +\mathcal{O}(\delta t^2) \right]\dif\rr\dif\vom,
\end{equation}
where $\mathbf{G}=\{\mathbf{F}^{\mathrm{ext}}/m,\boldsymbol{\tau}^{\mathrm{ext}}/I \}$ are the \emph{generalized forces per unit mass}. Moreover, the Taylor expansion of $f(\rr^*,\vom^*;t^*)$ in powers of $\delta t$ reads, up to first order,
\begin{equation}\label{eq:TE_BE}
    f(\rr^*,\vom^*;t^*)=f(\rr,\vom;t)+\left[\frac{\partial f(\rr,\vom;t)}{\partial t}\right]_{\mathrm{str}}\delta t+\vvel\cdot\nabla f(\rr,\vom;t)\delta t+ \mathbf{G}\cdot \frac{\partial f(\rr,\vom;t)}{\partial \vom}\delta t+\mathcal{O}(\delta t^2),
\end{equation}
with $\nabla \equiv \partial_\rr$. Using \refeqs{ConsProb_BE}--\eqref{eq:TE_BE}, we finally get
\begin{equation}
    \left(\frac{\partial f}{\partial t}\right)_{\mathrm{str}} = -\vvel\cdot\nabla f - \frac{\partial}{\partial\vom}\cdot(f\mathbf{G}).
\end{equation}
In \refsec{therm_models_KTGG}, we will study different types of $\mathbf{G}$.

In order to complete the time evolution of the one-particle \index{Velocity distribution function!one-particle}VDF we need to specify its rate of change due to collisions exclusively. Formally, the change due to the interactions would depend on the two-body \index{Velocity distribution function!two-particle}VDF. However, from the assumption of diluteness, as introduced in \refsec{intro_KTGG}, the average distance between particles, \phantomsection\label{sym:n}$n^{-1/\dt}$, is much smaller than the interaction range, with $n$ being the particle number density. In our case, due to the hard-core interaction, the average distance is much smaller than the diameter of the (hyper)spheres, $\sigma$, that is $n\sigma^{\dt}\ll 1$. Therefore, particles move most of the time due to the free streaming and eventually binary collision occur (collisions involving more than two particles are very improbable and so they are neglected). The latter statement implies that the \emph{mean free path} is much larger than $\sigma$, whereas the former establishes that the \emph{mean free time} is much higher than the duration of a collision.

From these considerations, we are going to derive the collisional-driven change of $f$. That is, we want to consider the change due to collisions of the number of particles $f(\rr,\vom_1;t)\dif\rr\dif\vom_1$ in the differential volume $\dif\rr\dif\boldsymbol{\Gamma}$ of the phase space around the point $(\rr,\vom_1)$ during a small time interval $\delta t$. To that end, we split the collisional mechanism into two processes. In the first one, there are particles in our initial phase space volume that collide during the time interval $\delta t$, and, then, they get postcollisional velocities \phantomsection\label{sym:vom_prime}$\vom_1^\prime$, such that do not contribute to the number of particles $f(\rr,\vom_1;t+\delta t)\dif\rr\dif\vom_1$. This process defines a loss term, \phantomsection\label{sym:JL}$J_L$, in the time evolution of $f$. Second, we must consider those particles with initial precollisional velocities \phantomsection\label{sym:vom_pprime}$\vom_1^{\prime\prime}$ involved in a collision such that they are counted in $f(\rr,\vom_1;t+\delta t)\dif\rr\dif\vom_1$. This latter process defines a gain term, \phantomsection\label{sym:JG}$J_G$, in this collisional-driven change of $f$. That is,
\begin{equation}
    \left(\frac{\partial f}{\partial t}\right)_{\mathrm{coll}}=J_G-J_L.
\end{equation}

\begin{figure}[t]
    \centering
    \includegraphics[width=0.8\textwidth]{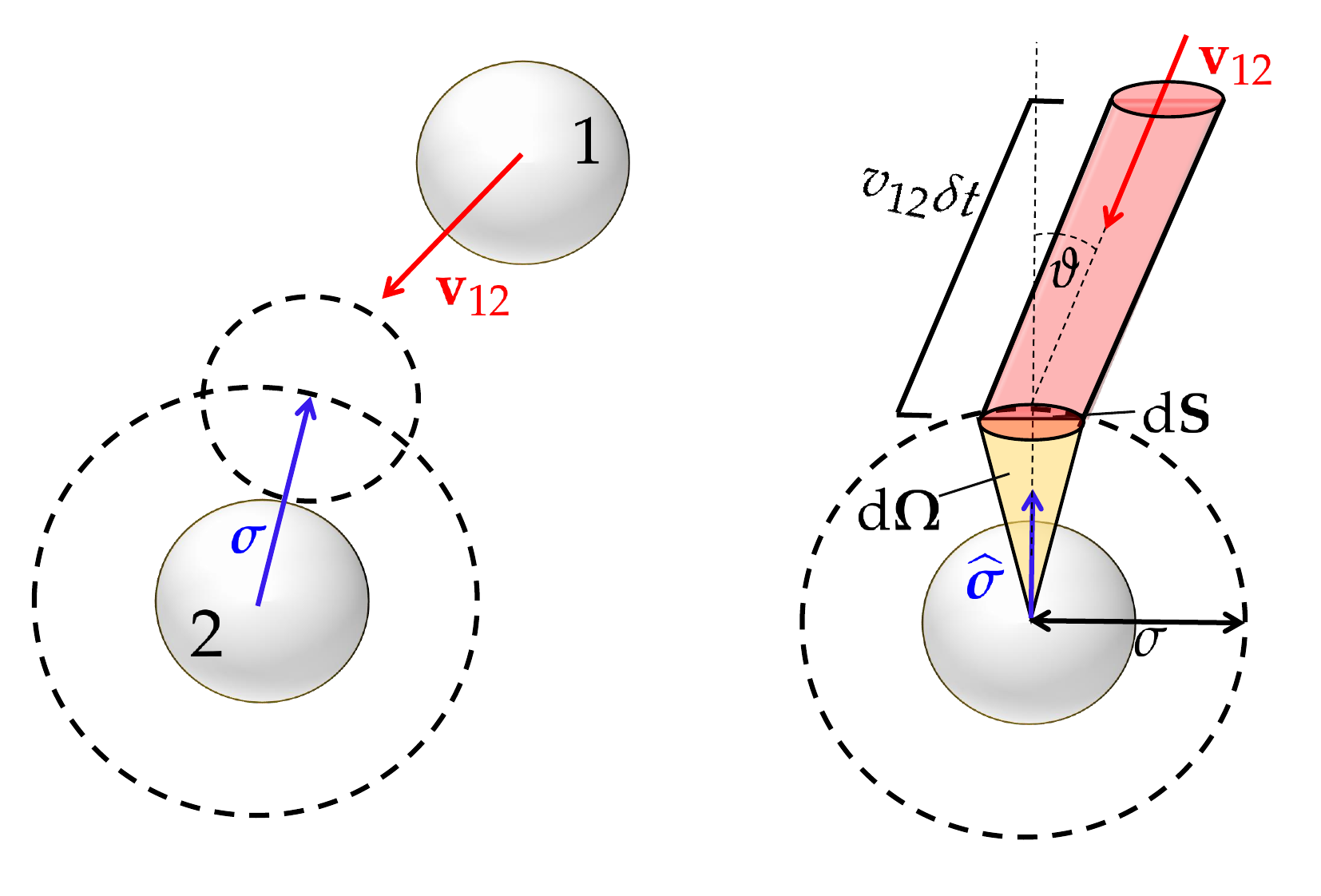}
    \caption{On the left, we show a sketch of a binary collision between a projectile particle (1) and a target particle (2). The big dashed circle represents the distance at $\sigma$ from the center of particle  2 where a collision takes place if the center of another sphere is located there. The smaller dashed circle represents the location of particle 1 at the time of collision. On the right part of the figure, there is a scheme of the collision cylinder represented by a reddish cylinder, where $\vartheta$ is the angle formed by the vector $\vvel_{12}$ and $\dif \mathbf{S}$, which is the differential area at contact shared by the dashed sphere of radius $\sigma$ centered at particle 2 and the collision cylinder. In addition, the differential solid angle $\dif\boldsymbol{\Omega}$ is depicted by a yellowish cone.}
    \labfig{collision_cylinder}
\end{figure}

We will distinguish between target particles moving with translational velocities $\vvel_2$, and incident particles moving with $\vvel_1$. Working in the frame of reference of particle 2, incident particles move with translational velocities \phantomsection\label{sym:vvel_12}$\vvel_{12}=\vvel_1-\vvel_2$. Therefore, for each target there exists an individual collision cylinder (see \reffig{collision_cylinder}), which accounts for all possible collisions that may occur within the infinitesimal solid angle $\dif\boldsymbol{\Omega}$ in the direction of the unit vector \phantomsection\label{sym:ssab}$\ssab=(\rr_2-\rr_1)/| \rr_2-\rr_1|$. This cylinder has a cross-section $\dif\mathbf{S}= \sigma^{\dt-1}\dif\boldsymbol{\Omega}$ and length $\vab \delta t$. Therefore, the number of particles scattered per unit time due to collisions of type $(\vom_1,\vom_2)$ into the element of solid angle $\dif\ssab$ around the unit intercenter vector $\ssab$ is~\cite{GS03,G19}
\begin{equation}
    \sigma^{\dt-1}|\vvel_{12}\cdot\ssab|f(\vom_1)f(\vom_2)\dif\ssab\dif\vom_1\dif\vom_2.
\end{equation}
From \reffig{collision_cylinder}, one can deduce that only those velocities with $\vvel_{12}\cdot\ssab\geq 0$ lead to a true impact. Then, the rate of loss per unit time of particles with velocities between $\vom_1$ and $\vom_1+\dif\vom_1$ is
\begin{equation}\labeq{JL_dif_vom1}
    J_L\dif\vom_1 = \sigma^{\dt-1}\int\dif\vom_2\int\dif\ssab\, \Theta(\vvel_{12}\cdot\ssab)(\vvel_{12}\cdot\ssab)f(\vom_1)f(\vom_2)\dif\vom_1,
\end{equation}
$\Theta$ being the Heaviside step-function. Equivalently, for the rate of gain we obtain that
\begin{equation}\labeq{JG_dif_vom1}
    J_G\dif\vom_1 = \sigma^{\dt-1}\int\dif\vom^{\prime\prime}_2\int\dif\ssab\, \Theta(\vvel^{\prime\prime}_{12}\cdot\ssab)(\vvel^{\prime\prime}_{12}\cdot\ssab)f(\vom^{\prime\prime}_1)f(\vom^{\prime\prime}_2)\dif\vom^{\prime\prime}_1.
\end{equation}

In this thesis, we will only consider cases in which $(\vvel^{\prime\prime}_{12}\cdot\ssab)= -\een^{-1}(\vvel_{12}\cdot\ssab)$, with $0\leq \een\leq 1$ a constant coefficient of normal restitution\index{Coefficient of normal restitution}, as it will be specified in \refsec{collisional_models_KTGG}. Hence, conveniently changing $\ssab\to-\ssab$, it is obtained that
\begin{equation}\labeq{JG_dif_vom1_J}
    J_G\dif\vom_1 = \sigma^{\dt-1}\int\dif\vom_2\int\dif\ssab\, \Theta(\vvel_{12}\cdot\ssab)(\vvel_{12}\cdot\ssab)(\een \mathfrak{J})^{-1}f(\vom^{\prime\prime}_1)f(\vom^{\prime\prime}_2)\dif\vom_1,
\end{equation}
$\mathfrak{J}$ being the Jacobian of the transformation $(\vom_1^{\prime},\vom_2^{\prime})\to(\vom_1,\vom_2)$ or, equivalently, $(\vom_1,\vom_2)\to(\vom_1^{\prime\prime},\vom_2^{\prime\prime})$, that is
\begin{equation}\labeq{jac_def}
    \mathfrak{J} = \left| \frac{\partial(\vom_1^{\prime},\vom_2^{\prime})}{\partial(\vom_1,\vom_2)}\right| = \left| \frac{\partial(\vom_1,\vom_2)}{\partial(\vom_1^{\prime\prime},\vom_2^{\prime\prime})}\right|.
\end{equation}
It can be explicitly computed from the chosen collisional model. Thus,
\begin{align}\labeq{BCO}
    \left(\frac{\partial f}{\partial t}\right)_{\mathrm{coll}}\equiv J[\vom_1|f,f] =& \sigma^{\dt-1}\int\dif\vom_2\int_{+}\dif\ssab\, 
    (\vvel_{12}\cdot\ssab)\left[(\een \mathfrak{J})^{-1}f^{\pprime}_1 f^{\pprime}_2-f_1 f_2\right],
\end{align}
where the notation $\int_{+}\dif\ssab\equiv \int\dif\ssab\,\Theta(\vvel_{12}\cdot\ssab)$ is adopted, $J$ is the integro-differential operator acting on $f$ (usually known as \index{Boltzmann collisional operator}Boltzmann collisional operator), and we defined $f_i\equiv f(\vom_i)$ and $f^{\pprime}_i\equiv f(\vom^{\pprime}_i)$ with $i=1,2$. Therefore, the \acrshort{BE} for hard $\dt$-hyperspheres reads
\begin{equation}\labeq{general_BE}
    \frac{\partial f}{\partial t}+\vvel\cdot\nabla f+\frac{\partial}{\partial \vom}\cdot(f\mathbf{G}) = J[\vom|f,f].
\end{equation}
In next sections, the \acrshort{BE} for different collisional models and external forces will be explicitly written.

\section{Collisional models}\labsec{collisional_models_KTGG}

Throughout this thesis, different collisional models for, in general, hard \dt-spheres are considered, \dt being the number of translational degrees of freedom. This will be crucial in the final results of the model. The derivation of the different properties of a specific model has been done under the assumptions of the \acrshort{BE}. In fact, from the molecular chaos ansatz, only binary collisions are considered and their binary collisional rules are essential, not only to specify the form of the \index{Boltzmann collisional operator}Boltzmann collisional operator in Eq.~\eqref{eq:BCO}, but also to derive properly the theoretical predictions.

Let us imagine two identical hard \dt-spheres colliding. Then, given two precollisional velocities, the binary direct collisional rules provide the expressions for the postcollisional velocities with respect to the precollisional ones and, the other way around, for the inverse collisional rules. Mathematically, let $\vom_1$ and $\vom_2$ be the precollisional velocities of particles 1 and 2, respectively. Therefore, the direct collisional rules establish the form of the postcollisional velocities \phantomsection\label{sym:B_op}$\vom_i^\prime=\mathfrak{B}_{12,\ssab}\vom_i$, with $i\in\{1,2\}$, where the operator $\mathfrak{B}_{12,\ssab}$ gives the postcollisional version of a certain velocity-related quantity. Moreover, if we now express $\vom_1$ and $\vom_2$ as the postcollisional velocities of particles 1 and 2, respectively, the binary inverse collisional rules are expressions of the form $\vom_i^{\prime\prime}=\mathfrak{B}_{12,\ssab}^{-1}\vom_i$, with double primed quantities denoting their precollisional values.

Three collisional models will be introduced attending to the systems considered. First, elastic collisions for hard \dt-spheres to describe molecular interaction (see \refsubsec{EHS_KTGG}). Next, granular interactions will be represented by means of two models: first, by introducing inelastic collisions in \refsubsec{IHS_KTGG} and, then, adding surface roughness to the inelastic collisions in \refsubsec{IRHS_KTGG}.

\subsection{Molecular gas of elastic hard spheres}\labsubsec{EHS_KTGG}

In this model, it is assumed that a molecular gas is composed by hard \dt-spheres interacting with elastic collisions. In what follows, this representation will be named as elastic hard-sphere (\acrshort{EHS}) collisional model. Whereas the tangential component of the relative velocity is preserved in a binary collision, the normal components in a collision between particle 1 and 2 accomplish the following reflection rules for the direct collisions,
\begin{equation}\labeq{dir_rules_elastic}
    \vvel_{1,2}^\prime = \mathfrak{B}_{12,\ssab}\vvel_{1,2}\equiv \vvel_{1,2}\mp (\vvel_{12}\cdot\ssab)\ssab,
\end{equation}
where $(\vvel_{12}^\prime\cdot\ssab)=-(\vvel_{12}\cdot\ssab)$ and, then, $(\vvel_{12}^{\prime\prime}\cdot\ssab)=-(\vvel_{12}\cdot\ssab)$. Hence, the inverse or restituting collisional rules become
\begin{equation}\labeq{inv_rules_elastic}
    \vvel_{1,2}^{\prime\prime} = \mathfrak{B}^{-1}_{12,\ssab}\vvel_{1,2}\equiv \vvel_{1,2}\mp (\vvel_{12}\cdot\ssab)\ssab.
\end{equation}
These rules satisfy conservation of linear momentum, i.e.,
\begin{align}\labeq{cons_lin_mom}
   m\vvel_1^{\prime\prime}+m\vvel_2^{\prime\prime} = m\vvel_1+m\vvel_2 =  m\vvel_1^\prime+m\vvel_2^\prime,
\end{align}
and conservation of energy, that is,
\begin{align}
   \frac{1}{2}m{v_1^{\prime\prime}}^2+\frac{1}{2}m{v_2^{\prime\prime}}^2 = \frac{1}{2}m{v_1}^2+\frac{1}{2}m{v_2}^2 =  \frac{1}{2}m{v_1^\prime}^2+\frac{1}{2}m{v_2^\prime}^2.
\end{align}

\begin{figure}[t]
    \centering
    \includegraphics[width=0.49\textwidth]{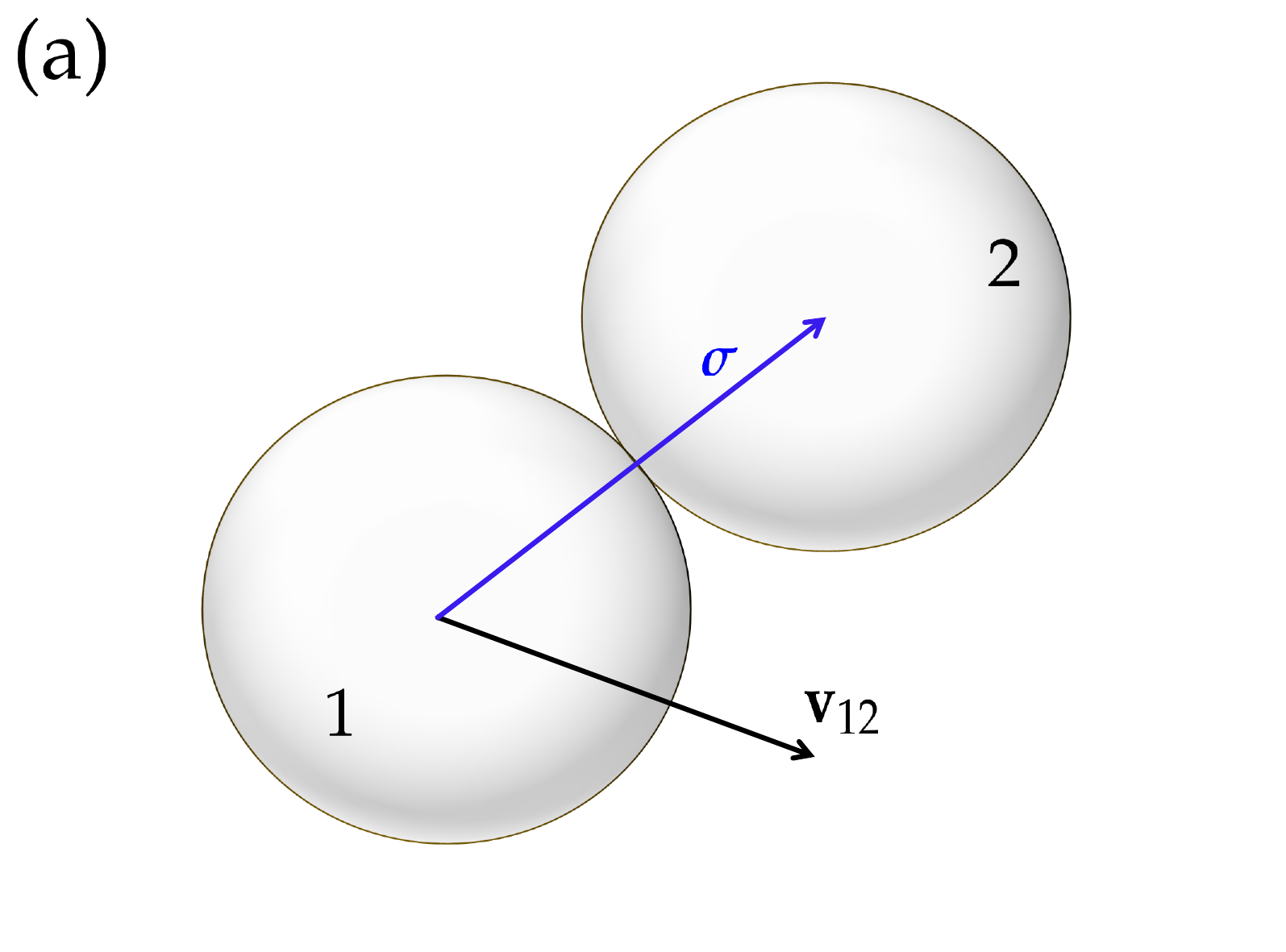}
    \includegraphics[width=0.49\textwidth]{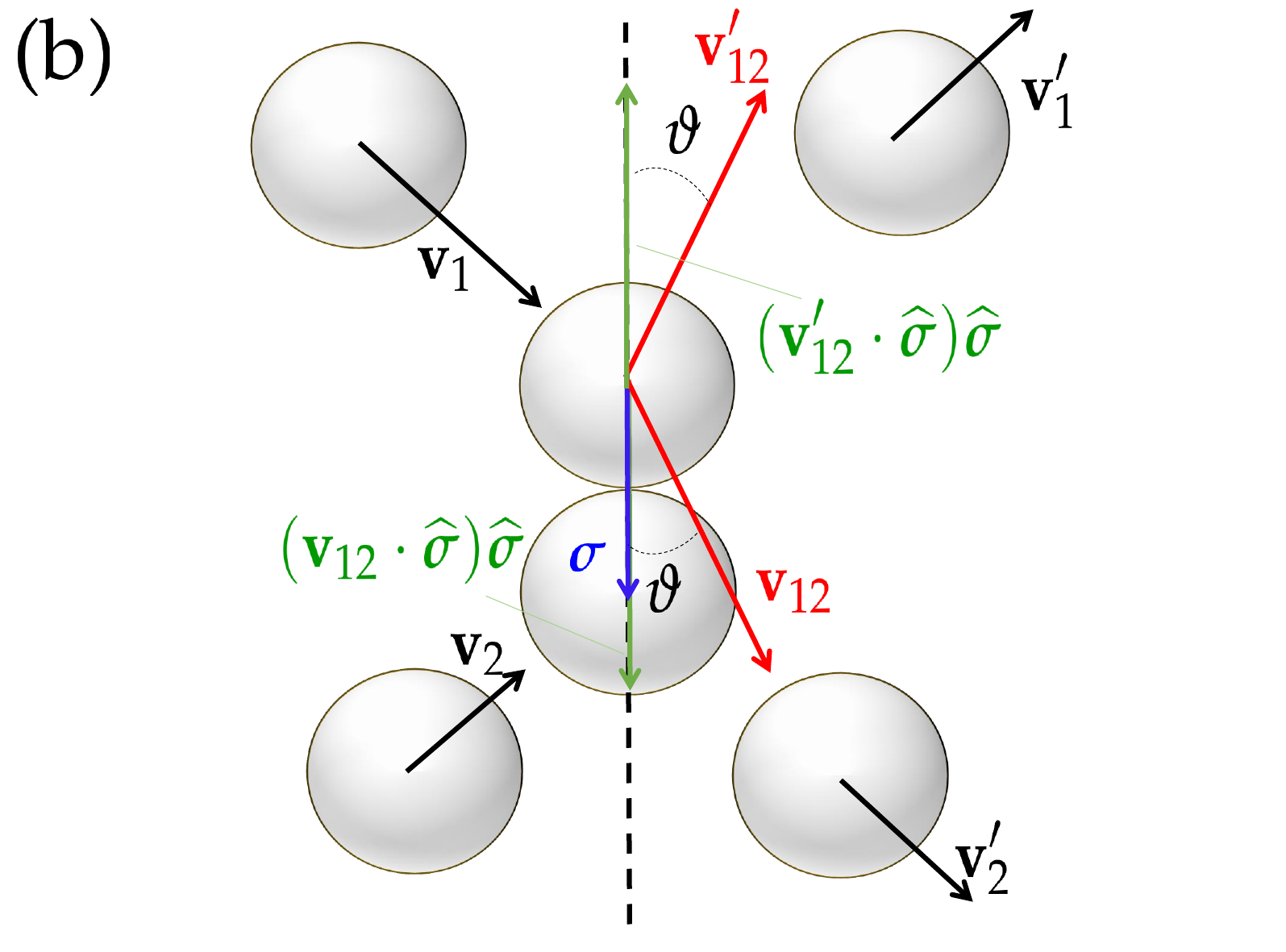}
    \caption{Sketch of (a) a collision between two hard spheres and (b) the collisional process in the \acrshort{EHS} model with the notation of the direct collisional rules for the spatial and velocity quantities. Here, $\vartheta=\arccos(|\vvel_{12}\cdot\ssab|/v_{12})$ is the impact angle, which, due to reflective elastic collisions, coincides with its postcollisional version, $\vartheta^{\prime}=\arccos(|\vvel^{\prime}_{12}\cdot\ssab|/v^{\prime}_{12})$. }
    \labfig{COL_EHS}
\end{figure}

A sketch of the collisional scheme can be observed in \reffig{COL_EHS}, where the reflective behavior in the collision is put into manifest. 

From the collisional rules in Eq.~\eqref{eq:dir_rules_elastic}, one obtains that the Jacobian of the transformation $(\vvel_1^{\prime},\vvel_2^{\prime})\to (\vvel_1,\vvel_2)$ defined in \refeq{jac_def} is $\mathfrak{J}=1$. Thus, the \index{Boltzmann collisional operator!for EHS}Boltzmann collisional operator introduced in \refeq{BCO} reads for the \acrshort{EHS} model,
\begin{align}\labeq{coll_operator_EHS}
    J[\vvel_1|f,f] =& \sigma^{\dt-1}\int\dif\vvel_2\int_{+}\dif\ssab\, 
    (\vvel_{12}\cdot\ssab)\left(f_1^{\pprime}f_2^{\pprime}-f_1f_2\right),
\end{align}
where the condition for elastic collisions, $(\vvel_{12}\cdot\ssab)=-(\vvel^{\pprime}_{12}\cdot\ssab)$, has been used.

\subsection{Inelastic hard spheres}\labsubsec{IHS_KTGG}

The inelastic hard sphere (\acrshort{IHS}) model is the simplest approximation of a granular interaction of impenetrable grains. It is assumed that the granular particles collide inelastically yielding a loss of energy. The reflective collision characteristic of the molecular case is broken down by a decrement of the normal component of the relative velocity parameterized by the coefficient of normal restitution\index{Coefficient of normal restitution} $\een$,
\begin{equation}\labeq{CNR}
    (\vvel_{12}^{\prime}\cdot\ssab)=-\een(\vvel_{12}\cdot\ssab),
\end{equation}
which is assumed to be constant in this model, and throughout this thesis. There are other models that slightly modify this by the introduction of a dependence of the coefficient of restitution on the impact velocity, such as the viscoelastic model~\cite{BSHP96,BP04,SP08}. However, despite the fact that the coefficient of restitution must actually be a function of the impact velocity, its form might not be universal, and the viscoelastic approach is quite specific. The constant coefficient of restitution approximation allows one to extract the properties of granular gases in a simple way. Moreover, recent experiments~\cite{YSS20} conclude that, in fact, this is a better approach than the viscoelastic model. The coefficient of normal restitution is defined such that $0\leq \een\leq 1$, where $\een=1$ applies for an elastic collision and $\een=0$ describes a completely inelastic collision. As in the molecular case, the tangential component of the relative velocity is preserved in the \acrshort{IHS} model. Thus, within this collisional model, the direct collisional rules are
\begin{equation}\labeq{dir_rules_inelastic}
    \vvel_{1,2}^\prime = \mathfrak{B}_{12,\ssab}\vvel_{1,2}\equiv \vvel_{1,2}\mp \frac{1+\een}{2}(\vvel_{12}\cdot\ssab)\ssab.
\end{equation}
On the other hand, the inverse collisional rules are
\begin{equation}\labeq{inv_rules_inelastic}
    \vvel_{1,2}^{\prime\prime} = \mathfrak{B}^{-1}_{12,\ssab}\vvel_{1,2}\equiv \vvel_{1,2}\mp \frac{1+\een^{-1}}{2}(\vvel_{12}\cdot\ssab)\ssab.
\end{equation}
A sketch of a binary collision for the \acrshort{IHS} model is represented in \reffig{COL_IHS}, in which the lack of the reflective collision behavior is depicted. 
\begin{figure}[t]
    \centering
    \includegraphics[width=0.49\textwidth]{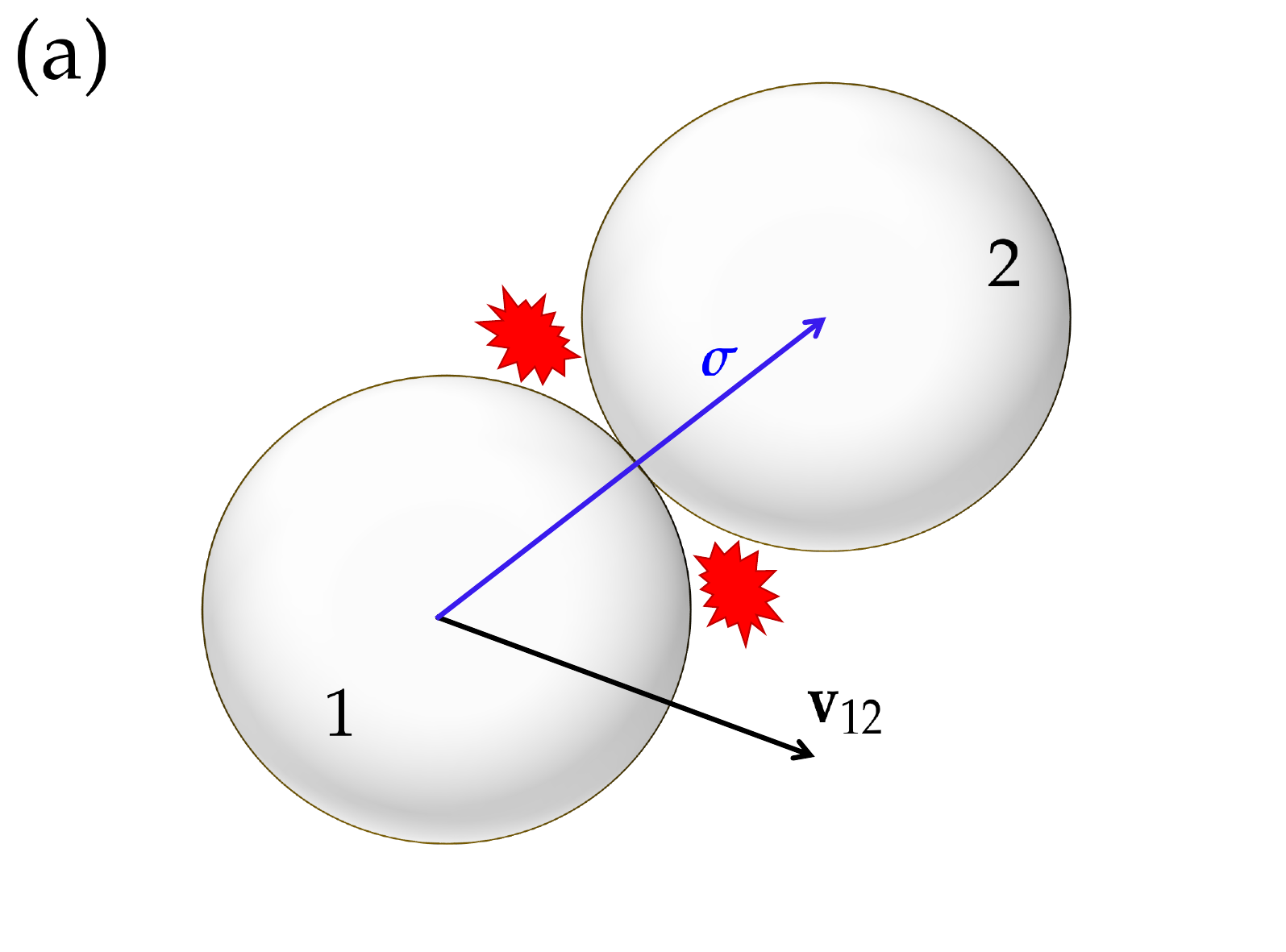}
    \includegraphics[width=0.49\textwidth]{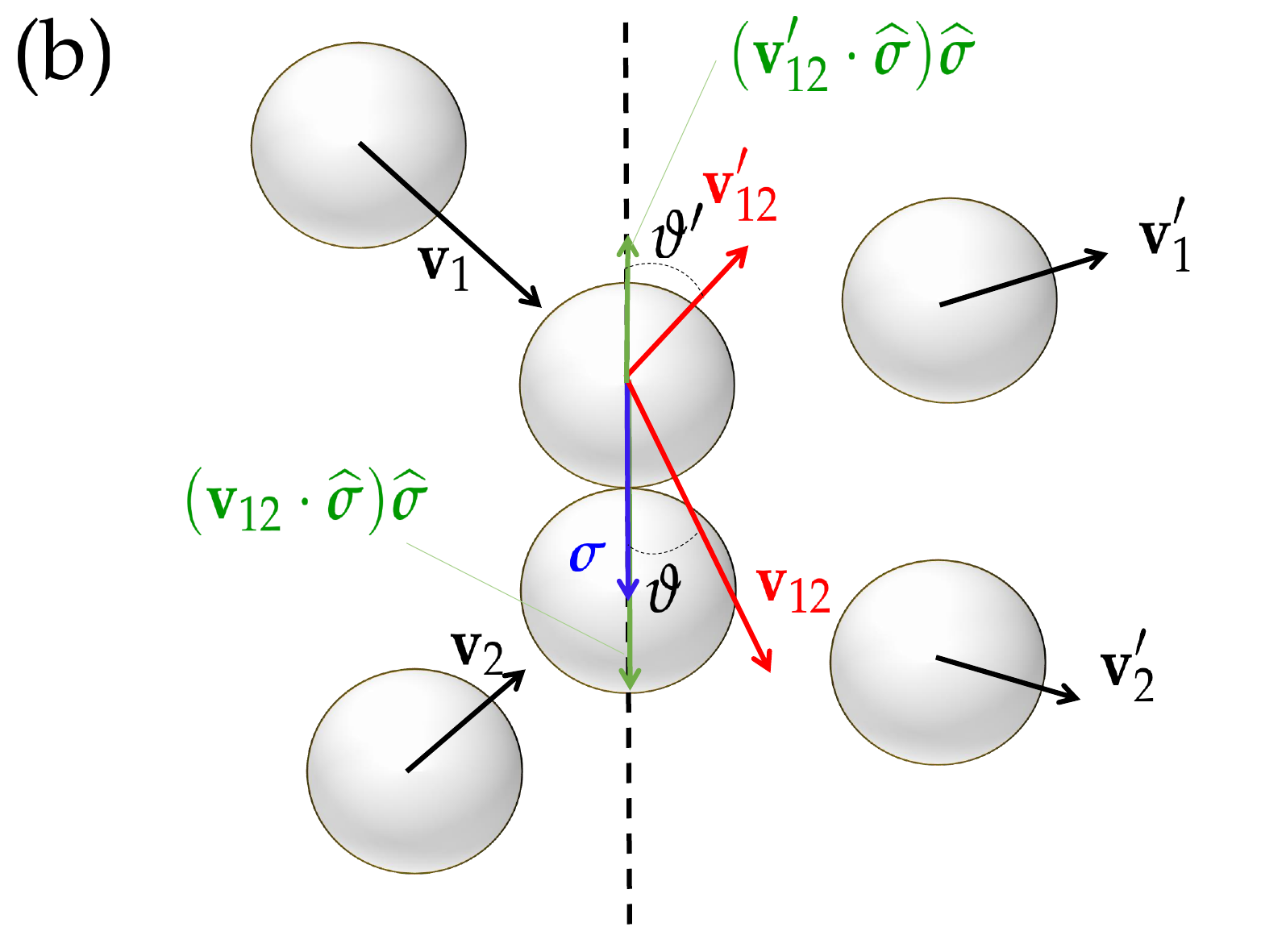}
    \caption{Sketch of (a) a collision between two hard inelastic spheres where the reddish glints represent the presence of inelasticity, and (b) the collisional process in the \acrshort{IHS} model with the notation of the direct collisional rules for the spatial and velocity quantities. Notice that, as compared with \reffig{COL_EHS}, there is no reflection any more due to generally different impact and resulting angles, $\vartheta\neq\vartheta^\prime$.}
    \labfig{COL_IHS}
\end{figure}
These collisional rules fulfill conservation of linear momentum [see \refeq{cons_lin_mom}].

However, due to the inelasticity, conservation of energy does not apply,
\begin{align}\labeq{cons_E_IHS}
   E^{\prime}_{12}-E_{12} = \frac{1}{2}m{v_1^\prime}^2+\frac{1}{2}m{v_2^\prime}^2- \frac{1}{2}m{v_1}^2+\frac{1}{2}m{v_2}^2 = -\frac{m}{4}(1-\een^2)(\vvel_{12}\cdot\ssab)^2\leq 0.
\end{align}
Notice that for elastic collisions ($\een=1$) one recovers conservation of energy. However, in case of inelasticity ($0\leq\een<1$) a loss of energy in each binary collision is ensured. Moreover, from the transformation specified in either \refeq{dir_rules_inelastic} or \refeq{inv_rules_inelastic}, one can derive that $\mathfrak{J}=\een$ and, therefore, the \index{Boltzmann collisional operator!for IHS}Boltzmann integro-differential collisional operator reads
\begin{equation}\labeq{BCO_IHS}
    J[\vvel_1|f,f] = \sigma^{\dt-1}\int\dif\vvel_2\int_{+}\dif\ssab\, 
    (\vvel_{12}\cdot\ssab)\left(\een^{-2}f_1^{\pprime}f_2^{\pprime}-f_1 f_2\right).
\end{equation}
It is important to remark that the inversion of the direct collisional rules is not possible in the case of perfectly inelastic particles ($\een=0$).

\subsection{Inelastic and rough hard spheres}\labsubsec{IRHS_KTGG}

The \acrshort{IHS} model is quite an accurate collisional model for describing granular gaseous interactions. In spite of this, its simplicity might skip some other real effects which could be crucial in the phenomenology of granular gases. The first simplest generalization that can be considered is the assumption of the presence of surface roughness in the hard \dt-spherical grains. Whereas the tangential component of the relative velocity in a binary collision in the \acrshort{IHS} model stays unaffected, this is not the case in the presence of surface roughness. Moreover, the spinning of hard particles becomes important in the kinetics in the context of this implementation. Therefore, the dynamics of particles can be split up into translational and rotational motion accounted by translational and rotational degrees of freedom, $\dt$ and $\dr$, respectively. The complexity of considering rotation in high dimensional systems, $\dt\geq 4$, and its lack of verisimilitude in those schemes invite to work just on two- and three-dimensional systems, which can be summarized as \acrshort{HD} or \acrshort{HS} systems, respectively. The surface roughness plays a dissipative role into the energy change in a binary collision, in addition to the inelasticity. This effect will be accounted for in the inelastic and rough hard-sphere (\acrshort{IRHS}) model by the definition of a coefficient of tangential restitution\index{Coefficient of tangential restitution}, $\eet$, defined by the relation
\begin{equation}\labeq{CTR}
    (\ssab\times\ggab^{\prime}) = -\eet(\ssab\times\ggab),
\end{equation}
which is assumed to be constant. Here, $\ggab$ the relative velocity of the points at contact of particles 1 and 2, defined by
\begin{equation}\labeq{ggab}
    \ggab = \vvab -\ssab\times \mathbf{S}_{12},
\end{equation}
with $\mathbf{S}_{12}=\sigma(\oo_1+\oo_2)/2$. From its definition, the range of validity of the coefficient of tangential restitution is \index{Coefficient of tangential restitution}$-1\leq \eet\leq 1$, such that $\eet=-1$ describes perfectly smooth particles and $\eet=1$ characterizes completely rough particles. Both cases preserve energy if $\een=1$, as we will show later. In fact, the conservative case with $\een=\eet=1$ is known in the literature as Pidduck's gas~\cite{P22}. Obviously, the description of the \acrshort{IHS} model developed in \refsubsec{IHS_KTGG} is recovered if $\een <1$ and $\eet=-1$.

Thus, we can consider a granular gas of identical \acrshort{HD} or \acrshort{HS} of mass $m$, diameter $\sigma$, and moment of inertia $I$, with positions $\{\rr_i\}_{i=1}^N$, and velocities $\{\vom_i\}_{i=1}^N\equiv\{ \vvel_i,\oo_i\}_{i=1}^N$. A sketch of a collision between two \acrshort{HD} and two \acrshort{HS} is shown in \reffig{COL_IRHS}. In order to derive the direct collisional rules, we impose not only conservation of linear momentum, but, additionally, conservation of angular momentum at the point of contact between the two particles, that is,
\begin{subequations}
\begin{align}\labeq{conservation}
    m\vvel_1^\prime+m\vvel_2^\prime=&m\vvel_1+m\vvel_2,\\
    I\oo_1^\prime-\frac{m}{2}\sigma(\ssab\times\vvel_1^\prime)=& I\oo_1-\frac{m}{2}\sigma(\ssab\times\vvel_1),\\
    I\oo_2^\prime+\frac{m}{2}\sigma(\ssab\times\vvel_2^\prime)=&I\oo_2+\frac{m}{2}\sigma(\ssab\times\vvel_2).
\end{align}
\end{subequations}
Let \phantomsection\label{sym:Q12}$\mathbf{Q}_{12}$ denote the impulse exerted by particle 1 on particle 2. Then, in general, the binary collisional rules for spinning spheres 1 and 2 satisfying \refeqs{conservation} are
\begin{align}\labeq{coll_rules_IRHS}
    \vvel_{1,2}^\prime= \vvel_{1,2}\mp \frac{\mathbf{Q}_{12}}{m},\qquad
    \oo_{1,2}^\prime=\oo_{1,2}-\frac{\sigma}{2I}\left(\ssab\times \mathbf{Q}_{12}\right).
\end{align}
\begin{figure}[h!]
    \centering
    \includegraphics[width=0.49\textwidth]{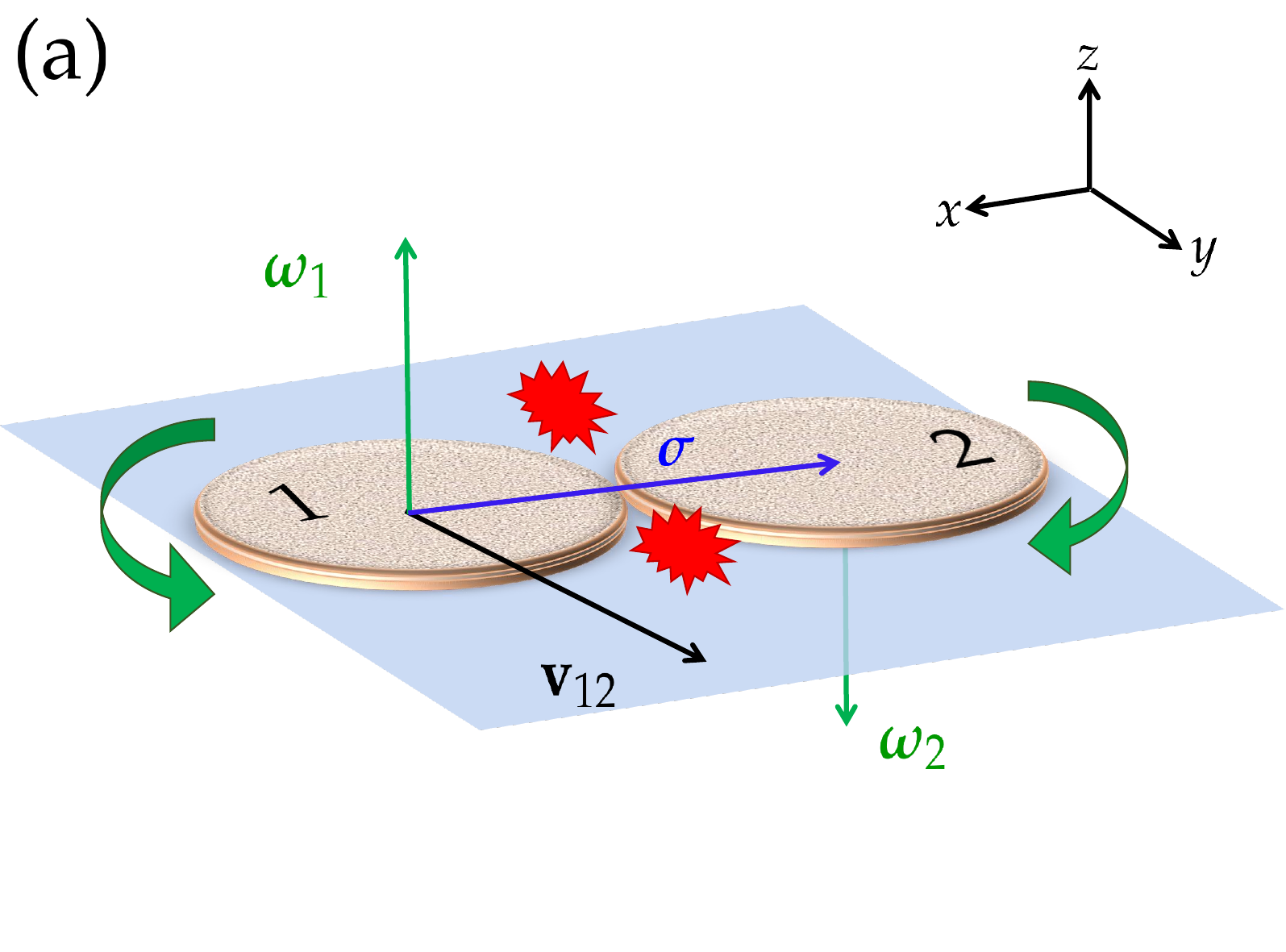}
    \includegraphics[width=0.49\textwidth]{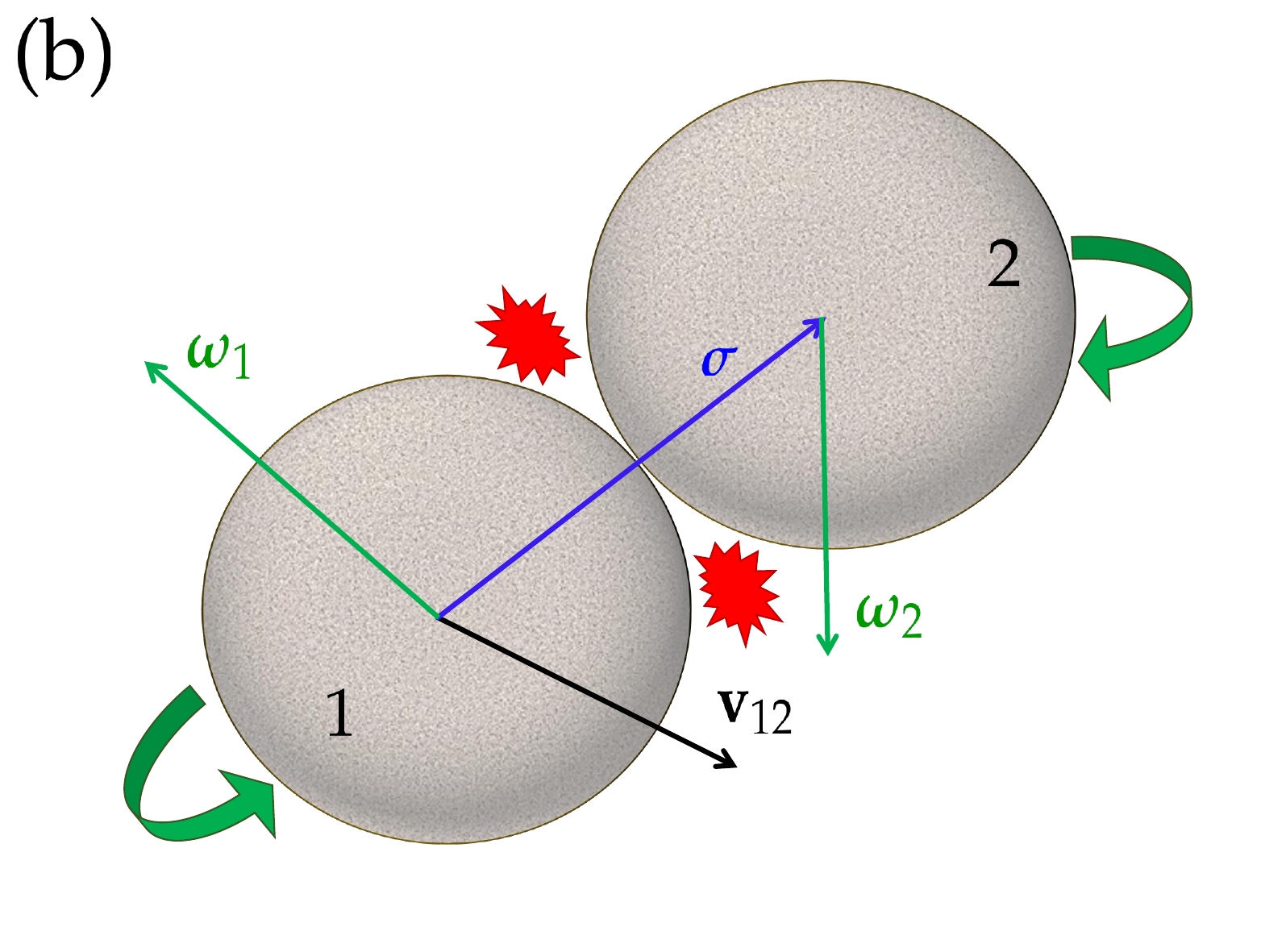}
    \caption{Representation of a collision between (a) two \acrshort{HD} and (b) two \acrshort{HS}. It can be observed that the evolution of the \acrshort{HD} is restricted to the plane and that their associated angular velocities lie along the direction orthogonal to the plane. On the other hand, there is no restriction of this kind for \acrshort{HS}, where both translational and angular velocities live in the same three-dimensional vector space.}
    \labfig{COL_IRHS}
\end{figure}
If we decompose the impulse into normal and tangential components with respect to $\ssab$, i.e., 
\begin{align}\labeq{Qab_par_perp}
    \Qqab = \Qqab^{\parallel}+\Qqab^{\perp}, \quad \Qqab^{\parallel} = (\ssab\cdot\Qqab)\ssab, \quad \Qqab^{\perp}=-\ssab\times\left(\ssab\times\Qqab\right),
\end{align}
then, from the definitions of the coefficients of restitution in \refeqs{CNR} and \eqref{eq:CTR}, one obtains
\begin{align}
    (\ssab\cdot\Qqab)=m\en (\ssab\cdot\ggab), \qquad (\ssab\times\Qqab)=m\et (\ssab\times\ggab),
\end{align}
with 
\begin{equation}\labeq{red_coef_rest}
    \en = \frac{1+\een}{2}, \qquad \et = \frac{\kappa}{1+\kappa}\frac{1+\eet}{2},
\end{equation}
$\kappa$ being the reduced moment of inertia,
\begin{equation}
    \kappa = \frac{4I}{m\sigma^2},
\end{equation}
whose value ranges from $\kappa=0$ in the case of particles with their masses concentrated at the central point, and $\kappa=1$ for \acrshort{HD} or $\kappa=2/3$ for \acrshort{HS}, in case the mass is concentrated on the surface. Therefore, the impulse is
\begin{equation}
    \frac{\Qqab}{m} = \en(\ssab\cdot\vvab)\ssab+\et\left[ \vvab-(\ssab\cdot\vvab)\ssab-(\ssab\times\mathbf{S}_{12})\right].
\end{equation}
This closes the collisional rules in \refeqs{coll_rules_IRHS}. Following similar steps, the restituting collisional rules are
\begin{align}\labeq{coll_rules_IRHS_rest}
    \vvel_{1,2}^{\pprime}= \vvel_{1,2}\mp \frac{\Qqab^-}{m},\qquad
    \oo_{1,2}^{\pprime}=\oo_{1,2}-\frac{\sigma}{2I}\left(\ssab\times \Qqab^-\right),
\end{align}
with $\Qqab^-$ defined as
\begin{equation}
    \frac{\Qqab^{-}}{m} = \frac{\en}{\een}(\ssab\cdot\vvab)\ssab+\frac{\et}{\eet}\left[ \vvab-(\ssab\cdot\vvab)\ssab-(\ssab\times\mathbf{S}_{12})\right].
\end{equation}

From all these rules, one can prove that energy is dissipated upon collisions. Considering a binary collision between particles 1 and 2, the difference between post and precollisional total kinetic energies is
\begin{align}\labeq{conservation_energy_IRHS}
    E^\prime_{12}-E_{12} =& \frac{m}{2} {v^\prime_1}^2+\frac{m}{2}{v^\prime_2}^2+\frac{I}{2}{\omega^\prime_1}^2+\frac{I}{2}{\omega^\prime_2}^2- \frac{m}{2} v_1^2-\frac{m}{2} v_2^2-\frac{I}{2} \omega_1^2-\frac{I}{2} \omega_2^2\nonumber\\
    =& -m\frac{1-\een^2}{4}\left(\ssab\cdot\vvab\right)^2-\frac{m\kappa}{1+\kappa}\frac{1-\eet^2}{4}\left[\ssab\times\left(\ssab\times\vvab+2\mathbf{S}_{12} \right)\right]^2\leq 0.
\end{align}
Notice that conservation of energy is recovered for elastic particles ($\een=1$) and for either perfectly smooth or completely rough particles ($|\eet|=1$). In any other case, energy loss is ensured at the collisional level.

Finally, once the binary collisional rules are defined, the Jacobian of the transformation $(\vvel_1,\vvel_2,\oo_1,\oo_2)\rightarrow (\vvel^\prime_1,\vvel^\prime_2,\oo^\prime_1,\oo^\prime_2)$ is $\mathfrak{J}=\een|\eet|$ for \acrshort{HD}, and $\mathfrak{J}=\een|\eet|^2$ for \acrshort{HS}, i.e.,
\begin{equation}
    \mathfrak{J}\equiv\left| \frac{\partial(\vvel^\prime_1,\vvel^\prime_2,\oo^\prime_1,\oo^\prime_2)}{\partial(\vvel_1,\vvel_2,\oo_1,\oo_2)}\right| = \een|\eet|^{2\frac{\dr}{\dt}},
\end{equation}
which is only valid specifically for \acrshort{HD} ($\dt=2$, $\dr=1$) and \acrshort{HS} ($\dt=\dr=3$) cases. The associated \index{Boltzmann collisional operator!for IRHS}Boltzmann collisional operator [Eq.~\eqref{eq:BCO}] for this particular model reads
\begin{equation}\labeq{BCO_IRHS}
    J[\vom_1|f,f] = \sigma^{\dt-1}\int\dif\vom_2\int_{+}\dif\ssab\, 
    (\vvab\cdot\ssab)\left(\frac{f_1^{\pprime}f_2^{\pprime}}{\een^{2}|\eet|^{2\frac{\dr}{\dt}}}-f_1 f_2\right).
\end{equation} 
 From the \acrshort{IHS} model, we already knew that the inversion of the direct collisional rules is not possible in the case of perfect inelastic particles ($\een=0$). In addition, this inversion is also not allowed for $\eet=0$ in the \acrshort{IRHS} model.
 
\section{Thermostatted states}\labsec{therm_models_KTGG}

In \refsec{collisional_models_KTGG}, we specified the collisional models of the gas, either molecular or granular, which is vital for the explicit form of the \index{Boltzmann collisional operator}Boltzmann collisional operator as defined in \refeq{BCO}. However, attending to the full \acrshort{BE} in \refeq{general_BE}, in order to be completed, we need to specify the generalized external forces per unit mass, $\mathbf{G}$, defined in \refsec{der_BE_KTGG} if they act on the system. As we will see in next sections, in this thesis we considered either freely evolving systems, i.e., $\mathbf{G}= \mathbf{0}$, or forced systems. In the latter case, whereas these generalized forces can be fully deterministic, such us gravity---which is widely used to study segregation in granular gases~\cite{JY02,BRM05}---along this thesis we have focused on thermostatted systems. 

First of all, let us consider Langevin-like systems. That is, we will assume that the particle dynamics of the gas is affected by the action of a thermal bath. In the most basic case, the granular gas is assumed to be immersed in another bath fluid of much smaller particles, so that the effect of the bath on particle $i$ of the granular gas can be modeled by the Langevin equation (\acrshort{LE})~\cite{L08,LG97,vK07} 
\begin{equation}\labeq{white_noise}
    \frac{\mathbf{F}_i^{\mathrm{ext}}(\rr,t)}{m}\equiv\frac{\dif\vvel_i(t)}{\dif t} = -\xi_0\vvel_i(t)+\boldsymbol{\eta}_i(t),
\end{equation}
where $\xi_0$ is a constant friction or drag coefficient\index{Drag coefficient} and $\boldsymbol{\eta}_i(t)$ is a random force, usually described as a white noise. That is, this random force is assumed to be Gaussian-distributed and $\delta$-correlated, i.e., 
\begin{align}\labeq{white_noise_corr}
    \langle \boldsymbol{\eta}_i(t)\rangle = \mathbf{0}, \quad \langle \boldsymbol{\eta}_i(t) \boldsymbol{\eta}_j(t^\prime)\rangle = \chi^2 \delta(t-t^\prime)\delta_{ij}\mathbb{1}_{\dt},
\end{align}
where $\mathbb{1}_{\dt}$ is the $\dt\times\dt$ unit matrix and $\chi^2$ represents the intensity of the thermostat. In the case of a molecular gas in a thermal bath, from the fluctuation-dissipation theorem (\acrshort{FDT})~\cite{CW51,K66} one has $\chi^2=2\xi_0^2D$ with \phantomsection\label{sym:diff_coeff}$D= T_b/m\xi_0$ being the diffusion coefficient, and $T_b$ the \index{Temperature!of the thermal bath}temperature\footnote{Throughout the thesis, \index{Temperature}temperature has units of energy, except in \refch{molecular_gases}. Then, we formally take \phantomsection\label{sym:kb}$\kb=1$, $\kb$ being Boltzmann's constant.} of the bath, here assumed to be constant. Therefore, the counterpart of the \acrshort{BE} related to the external force can be deduced from its associated Fokker--Planck equation (see \refapp{app_FPLE}), which is
\begin{equation}
    \frac{\partial}{\partial\vvel}\cdot\left(f \frac{\mathbf{F}^{\mathrm{ext}}}{m}\right) \rightarrow -\frac{\partial}{\partial\vvel}\cdot\left[\left(\xi_0\vvel+\frac{\chi^2}{2}\frac{\partial}{\partial\vvel} \right)f\right] = -\frac{\partial}{\partial\vvel}\cdot\left[\xi_0\left(\vvel+\frac{T_b}{m}\frac{\partial}{\partial\vvel} \right)f\right].
\end{equation}
Adding up this term and the \index{Boltzmann collisional operator}Boltzmann collisional operator coming from a specific collisional model to the \acrshort{BE}, we obtain the \acrshort{BE} for Brownian colliding particles. In general, the \acrshort{BE} with a nonzero stochastic external force is usually referred to as \acrshort{BFPE}.  

The exposed case is widely studied and important in the statistical-physics literature, both historically and conceptually. In this thesis, we use the knowledge from the original \acrshort{LE} to derive different \acrshort{BFPE} for other stochastic thermostats. Particularly, we have considered the generalization of the \acrshort{LE} to a nonlinear velocity-dependent drag coefficient\index{Drag coefficient!velocity-dependent} as a coarse-grained model for interactions with other gaseous particles under some conditions. This generalized thermostat, which is introduced in \refsubsec{nonlinear_KTGG}, has been applied to \acrshort{EHS} molecular gases and granular gases described by the \acrshort{IHS} model in Article 1 (\refsec{Art1}) and Article 4 (\refsec{Art4}), respectively. In addition, we also considered an energy injection to translational and rotational degrees of freedom applied to a granular gas of \acrshort{IRHS} in the absence of drag forces in Article 6 (\refsec{Art6}), this thermostat being explained in \refsubsec{splitting_KTGG}.  

\subsection{Langevin dynamics with nonlinear drag}\labsubsec{nonlinear_KTGG}

Let us consider a \emph{quasi}-Rayleigh (\acrshort{QR}) gas, that is, an ensemble of \emph{heavy} Brownian particles (of mass $m$, diameter $\sigma$, number density $n$ and velocities $\vvel$) surrounded by a dilute gas of \emph{light} particles (of mass \phantomsection\label{sym:mb}$m_b$, diameter \phantomsection\label{sym:sb}$\sigma_b$, number density \phantomsection\label{sym:nb}$n_b$, and velocities \phantomsection\label{sym:vb}$\vvel_b$) in equilibrium at \index{Temperature!of the thermal bath}temperature $T_b$, acting as a thermal bath on the heavy particles. However, light particle masses, whereas (much) smaller than the heavy ones, are not going to be completely neglected. As derived in Refs.~\cite{F00b,F07,F14}, under these conditions, a velocity-dependent nonlinear drag coefficient\index{Drag coefficient!velocity-dependent} can be obtained as an expansion in powers of the mass ratio $m_b/m$. This expansion can be derived formally from kinetic theory.

Let us assume for simplicity that both gas and bath particles are hard $\dt$-spheres such that, whereas gas-gas collisions are given from a certain model which is irrelevant for this derivation, gas-bath interactions are assumed to be elastic. Moreover, we will consider that the Brownian particles are not too far from the thermal equilibrium with the bath. This approach implies that~\cite{F00b,F07,F14},
\begin{equation}\labeq{vb_cb}
    \frac{v}{v_b}\simeq \sqrt{\frac{m_b}{m}}\frac{c}{c_b},
\end{equation}
so that expansions on $m_b/m$ are equivalent to those on $v/v_b$. Here, $\cc$ and $\cc_b$ are the reduced velocity variables,
\begin{equation}\labeq{red_velocity}
    \cc = \frac{\vvel}{v_{\mathrm{th}}},\quad \cc_b = \frac{\vvel_b}{v_{\mathrm{th},b}}
\end{equation}
with $v_{\mathrm{th}}\equiv\sqrt{2T/m}$ and $v_{\mathrm{th},b}\equiv\sqrt{2T_b/m_b}$ the thermal velocities, and, from our assumption, $T\simeq T_b$. The deterministic change of momentum due to the collisions between Brownian and bath particles will be given by the drag force, that is,
\begin{equation}\labeq{mom_drag}
    m\left(\frac{\dif \vvel}{\dif t}\right)_{\text{Brown}} = -m\xi_{\mathrm{QR}}(v)\vvel,
\end{equation}
where the velocity-dependent drag coefficient\index{Drag coefficient!velocity-dependent} $\xi_{\mathrm{QR}}(v)$ can be computed using kinetic-theory arguments. In \refapp{app_NLD} the drag coefficient is derived using arguments similar to those followed in \cite{F00b}, but for general $\dt$ dimensions, such that it reads
\begin{equation}\labeq{xi_QR_gamma}
    \xi(v) \xrightarrow{\text{QR}} \xi_0\left( 1+\gamma\frac{mv^2}{T_b}\right), \quad \gamma\equiv \frac{m_b}{2(\dt+2)m},
\end{equation}
with $\xi_0$ being a constant given by the parameters of the problem [see \refeq{xi_QR_orders_NLD}], and $\gamma$ controls the nonlinear effect, and it is expected to be small. Hence, the first correction from the \acrshort{QR} approximation to the drag coefficient corresponds to a quadratic term in the velocity modulus. 

In summary, we can consider a gas of hard spheres which, apart from their internal collisions, are in contact with a background fluid acting as a thermal bath within the conditions assumed before. The background fluid, from a coarse-grained description, will act on the gas exerting a force characterized by two components: first, a drag force \phantomsection\label{sym:Fdrag}$\mathbf{F}_{\mathrm{drag}}=-m\xi(v)\vvel$ with $\xi(v)$ given by \refeq{xi_QR_gamma}, and, second, a stochastic force with nonlinear variance $m^2\chi^2(v)$.
Then, the equation of motion of $i^{\mathrm{th}}$ gas particle is the following one
\begin{equation}\labeq{LE_NLD_KTGG_def}
    \frac{\dif\vvel_i}{\dif t} = -\xi(v)\vvel_i + \boldsymbol{\eta}_i+\mathbf{F}_{\mathrm{coll}},
\end{equation}
where $\mathbf{F}_{\mathrm{coll}}$ is the force due to collisions with other particles and $\boldsymbol{\eta}_i$ is a stochastic variable with the white-noise properties given by \refeq{white_noise_corr}\footnote{Note that the noise term in \refeq{white_noise_corr} has assumed to be white, this is reasonable due to the \emph{diagonal} and continuous drag coefficient. However, it is not true, in general, if off-diagonal drag matrices or discontinuities appear in the drag force term (see Refs.~\cite{K17,J22}).}. Moreover, from \acrshort{FDT}, we can impose that
\begin{equation}
    \chi^2(v) = \frac{2 T_b}{m}\xi(v).
\end{equation}

Therefore, the \acrshort{BFPE} for this specific case reads
\begin{equation}\labeq{BFPE_NLD}
    \frac{\partial f}{\partial t} +\vvel\cdot\nabla f - \frac{\partial}{\partial \vvel}\cdot\left[\xi(v)\left(\vvel+\frac{T_b}{m}\frac{\partial}{\partial \vvel} \right)f\right] = J[\vvel, f|f].
\end{equation}
Whereas in a molecular gas the steady \index{Temperature!molecular}temperature would coincide with \index{Temperature!of the thermal bath}$T_b$, this is not so in the granular case. In fact, dissipation of energy due to collisions between granular particles will lead to a smaller \index{Temperature!granular}temperature. As it will be described in next sections, we have implemented this background fluid interaction with a nonlinear velocity-dependent drag coefficient\index{Drag coefficient!velocity-dependent} to both molecular gases of hard spheres and the \acrshort{IHS} model for granular gases.

\subsection{Splitting thermostat}\labsubsec{splitting_KTGG}

From the \acrshort{IHS} and \acrshort{IRHS} models of interactions in granular gases, the loss of energy  has been discussed at the level of collisions. Experimentally, the process of cooling is usually very fast and it is then convenient to inject energy into the system to observe the properties of the granular fluid in \index{Nonequilibrium steady state}nonequilibrium steady states. Theoretically, this has been commonly modeled via a stochastic volume force acting on the translational degrees of freedom, i.e., a stochastic thermostat, with the properties of a Gaussian white noise. That is, $\mathbf{F}^{\mathrm{ext}}/m$ fulfills the properties in \refeq{white_noise_corr}, i.e.,
\begin{align}\labeq{ext_stoc_force}
    \langle \mathbf{F}^{\mathrm{ext}}(t)\rangle = \mathbf{0}, \quad \langle \mathbf{F}^{\mathrm{ext}}_i(t) \mathbf{F}^{\mathrm{ext}}_j(t^\prime)\rangle = m^2\chi_{t}^2 \delta_{ij}\delta(t-t^\prime)\mathbb{1}_{\dt}.
\end{align}
Nevertheless, in analogy with the generalization in \refsec{der_BE_KTGG} of the external force per unit mass $\mathbf{G}$, this thermostat can be generalized as well to act on the rotational degrees of freedom via a stochastic volume torque. That is, we will assume that, apart from the external force defined by \refeq{ext_stoc_force}, there could be an external stochastic torque per unit moment of inertia fulfilling as well the properties of a white noise, i.e.,
\begin{align}\labeq{ext_stoc_torque}
    \langle \boldsymbol{\tau}^{\mathrm{ext}}(t)\rangle = \mathbf{0}, \quad \langle \boldsymbol{\tau}^{\mathrm{ext}}_i(t) \boldsymbol{\tau}^{\mathrm{ext}}_j(t^\prime)\rangle = I^2\chi_{r}^2 \delta_{ij}\delta(t-t^\prime)\mathbb{1}_{\dr},
\end{align}
where $\mathbb{1}_{\dr}$ is the $\dr\times\dr$ unit matrix. Then, the associated \acrshort{BFPE} is
\begin{align}\labeq{FPE_ST_first}
    \frac{\partial f}{\partial t}+\vvel\cdot\nabla f -\frac{\chi_t^2}{2}\left(\frac{\partial}{\partial \vvel}\right)^2f-\frac{\chi_r^2}{2}\left(\frac{\partial}{\partial \oo}\right)^2 f= J[\vom|f ,f].
\end{align}
The action of the thermostat over the system can be controlled by the pair of noise intensities $(\chi_t^2,\chi_r^2)$. Each value of that pair will describe a specific thermostat belonging to this class. Equivalently, we can define two other quantities to define unequivocally the thermostat, which will be chosen as the \emph{total noise intensity}, $\chi^2$, and the \emph{rotational-to-total} noise intensity ratio, $\varepsilon$,
\begin{equation}\labeq{total_and_varepsilon_ST}
    \chi^2 = \chi_t^2+\frac{\dr}{\dt}\frac{I}{m}\chi_r^2, \quad \varepsilon = \frac{\dr}{\dt}\frac{I}{m}\frac{\chi_r^2}{\chi^2},
\end{equation}
such that each pair $(\chi^2,\varepsilon)$ defines a unique thermostat, with $0\leq \varepsilon\leq 1$. Obviously, the freely evolving system is recovered by taking the limit $\chi^2\to 0$, regardless of the value of $\epsilon$. Because the action of this thermostat is divided into translational and rotational energy injections, it will be called \emph{splitting thermostat} (\acrshort{ST}) throughout this thesis. In the case $\varepsilon=0$, we recover a thermostat acting only on the translational degrees of freedom, whereas $\varepsilon=1$ indicates that the action of the thermostat is concentrated on the rotational ones~\cite{CLH02}. In addition, one can define a \index{Temperature!noise}\emph{noise temperature} from dimensional analysis,
\begin{equation}\labeq{def_Twn_ST}
    T^{\mathrm{wn}} = m\left( \frac{2\sqrt{\pi}}{\Omega_{\dt}}\frac{\chi^2}{n\sigma^{\dt-1}}\right)^{2/3},
\end{equation}
where the numerical constants have been added for simplification, with $\Omega_{\dt}$ being the $\dt$-dimensional solid angle [see \refeq{solid_angle_dt}]. Then, it will be equivalent to take either the pair $(\chi^2,\varepsilon)$ or the pair $(T^{\mathrm{wn}},\varepsilon)$ to describe a \acrshort{ST}. The \acrshort{BFPE} for this description of the \acrshort{ST} reads
\begin{align}\labeq{BFPE_epsilon_ST}
    \frac{\partial f}{\partial t}+\vvel\cdot\nabla f -\frac{\nu^{\mathrm{wn}}T^{\mathrm{wn}}}{4m}\left[(1-\varepsilon)\left(\frac{\partial}{\partial \vvel}\right)^2+\varepsilon\frac{\dt m}{\dr I}\left(\frac{\partial}{\partial \oo}\right)^2\right] f= J[\vom|f ,f],
\end{align}
with
\begin{equation}\labeq{nuwn}
	\nuwn \equiv K_{\dt} n\sigma^{\dt-1}\sqrt{\frac{2T^\mathrm{wn}}{m}}, \quad K_{\dt} = \frac{\sqrt{2}\pi^{\frac{\dt-1}{2}}}{\Gamma\left(\frac{\dt}{2}\right)},
\end{equation}
where $\nuwn$ is a frequency associated with the thermostat energy injection. Mathematically speaking, one can conclude that there are bijections between the pair of variables $(\chi_t^2,\chi_r^2)$, $(\chi^2,\varepsilon)$, and $(T^{\mathrm{wn}},\varepsilon)$. Thus, their descriptions are equivalent and we use any of them as convenient. The main advantage of this \acrshort{ST} is its two-parameter description, which is used in the work described in Article 6 (\refsec{Art6}).

\section{Homogeneous states}\labsec{HS_KTGG}

Once the \acrshort{BE} is derived [see \refeq{general_BE}] and different collisional models and stochastic thermostats used in this thesis are defined in \refsec{collisional_models_KTGG} and \refsec{therm_models_KTGG}, the natural path is to look for solutions of the kinetic equation for each case. As a first step, it seems reasonable to look for solution in homogeneous systems, that is, in the absence of gradients. In that case, \refeq{general_BE} becomes
\begin{equation}\labeq{general_hom_BE}
    \frac{\partial f}{\partial t}+\frac{\partial}{\partial \vom}\cdot(f\mathbf{G}) = J[\vom|f,f],
\end{equation}
where the forms of the \index{Boltzmann collisional operator}Boltzmann collisional operator and the generalized forces are given once the system model is indicated. 

To get a deeper insight into this problem, it is important to distinguish between the molecular and granular cases.

\subsection{Homogeneous molecular gas of hard \dt-spheres}\labsubsec{HMG_KTGG}

Let us start from the simplest case of a homogeneous molecular gas of hard \dt-spheres in the absence of external deterministic forces and any other fluid or bath. Then, according to the collisional model for elastic hard \dt-spheres presented in \refsubsec{EHS_KTGG}, and its associated collisional operator defined in \refeq{coll_operator_EHS}, the respective form of the \acrshort{BE} is
\begin{equation}\labeq{HBE_free_EHS}
	\frac{\partial f(\vvel_1;t)}{\partial t} = \sigma^{\dt-1}\int\dif\vvel_2\int_{+}\dif\ssab\, 
    (\vvel_{12}\cdot\ssab)\left[f(\vvel^{\prime\prime}_1;t)f(\vvel^{\prime\prime}_2;t)-f(\vvel_1;t)f(\vvel_2;t)\right].
\end{equation}
In addition, for spatially homogeneous and isotropic systems, the \acrshort{VDF} is expected to depend only on the velocity modulus, $v$. Moreover, the \index{Temperature!molecular}temperature, which is defined as the mean kinetic energy weighted by the number of degrees of freedom, i.e.,
\begin{equation}\labeq{molecular_temp}
	\frac{\dt}{2}T(t) = \frac{1}{2}m\langle v^2\rangle(t) \equiv \frac{1}{2}\int \dif\vvel mv^2f(\vvel;t),
\end{equation} 
is preserved due to energy conservation upon collisions, i.e., because the second velocity moment, $v^2$, is a collisional invariant of the \index{Boltzmann collisional operator}Boltzmann collisional operator. This collisional model possesses three collisional invariants, $\{1,\vvel,v^2\}$, related to mass, linear momentum, and energy conservation. Hence, it is physically expected that, if the system is initially prepared with an arbitrary \acrshort{VDF}, it will evolve to equilibrium. The equilibrium \index{Velocity distribution function!equilibrium}\acrshort{VDF} must coincide with the steady-state solution of \refeq{HBE_free_EHS}, which coincides with the Maxwellian \index{Velocity distribution function!Maxwellian}\acrshort{VDF},
\begin{equation}\labeq{MDF_EHS}
 f_{\mathrm{M}}(\vvel) = n\vth^{-\dt} \pi^{-\dt/2}\exp\left(-\frac{mv^2}{2 T} \right),
\end{equation}
with $T$ preserved and therefore, coinciding with the initial \index{Temperature!molecular}temperature. This solution can be deduced from the celebrated Boltzmann's $H$-theorem (see \refsubsec{HTheorem_MG_KTGG}), which also shows the property of the Maxwellian \index{Velocity distribution function!Maxwellian}\acrshort{VDF} being an attractor of the \acrshort{BE}, not only for the homogeneous case.

In case an interaction with a thermal bath is considered in terms of a Langevin-like type equation, as it is the case in \refch{molecular_gases} with the nonlinear-drag model introduced in \refsubsec{nonlinear_KTGG}, the Fokker--Planck-like term of the resulting \acrshort{BFPE} does not violate the evolution toward equilibrium and the system will reach again the Maxwellian \index{Velocity distribution function!Maxwellian}\acrshort{VDF} with the \index{Temperature!of the thermal bath}temperature of the bath, $T_b$. That is, $v^2$ is still a collisional invariant, but \index{Temperature!molecular}temperature is not preserved until $T_b$ is reached.

A quite important difference between the original implementation of Langevin dynamics and its extension to a nonlinear drag model is the coupling of the \index{Temperature!molecular}temperature evolution rate with higher moments of the \index{Velocity distribution function!moments}time-dependent \index{Velocity distribution function!time-dependent}\acrshort{VDF}. This will be further developed in Article 1 (\refsec{Art1}), this fact being responsible for the emergence of memory effects, namely, counterintuitive phenomena due to the memory of the system about previous states or its initial preparation. In \refch{ME_KTGG}, some of these effects studied in this thesis are introduced.

\subsection{Freely evolving homogeneous granular gas: the Homogeneous Cooling State}\labsubsec{HCS_KTGG}

The introduction of the granular models associated with a loss of energy, such as the \acrshort{IHS} or \acrshort{IRHS} ones described in \refsubsec{IHS_KTGG} and \refsubsec{IRHS_KTGG}, respectively, complicates the analysis of the homogeneous \acrshort{BE}, in contrast to the apparent simplicity of the case of elastic collisions. 

Let us focus on a monocomponent homogeneous freely evolving granular gas, namely, the evolution is not influenced by any external energy injection or force. A first consequence of the granular dynamics is the energy dissipation, present in both the \acrshort{IHS} and \acrshort{IRHS} collisional models, in which the second velocity moments are not collisional invariant any more. In order to present a homogeneous reference state for granular dynamics, usually called \emph{Homogeneous Cooling State} (\acrshort{HCS}), which is governed by this energy dissipation, we will distinguish first between both granular collisional models.

\subsubsection{Homogeneous freely evolving states in the IHS model}

Let us start from the homogeneous form of the \acrshort{BE} for the \acrshort{IHS} model,
\begin{equation}\labeq{HBE_IHS}
	\frac{\partial f(\vvel_1;t)}{\partial t} = \sigma^{\dt-1}\int\dif\vvel_2\int_{+}\dif\ssab\, 
    (\vvel_{12}\cdot\ssab)\left[\een^{-2}f(\vvel^{\prime\prime}_1;t)f(\vvel^{\prime\prime}_2;t)-f(\vvel_1;t)f(\vvel_2;t)\right],
\end{equation}
describing the evolution of the one-particle \index{Velocity distribution function!one-particle}\acrshort{VDF} of a freely evolving monocomponent homogeneous granular gas of inelastic hard \dt-spheres. It is obvious that the Maxwellian \index{Velocity distribution function!Maxwellian}\acrshort{VDF} is not any more a solution of \refeq{HBE_IHS}. This fact can be considered as a first indicator of an intrinsically nonequilibrium system.

In analogy with the molecular case, one can define a \index{Temperature!granular}\emph{granular temperature} as the mean kinetic energy weighted by the degrees of freedom [see \refeq{molecular_temp}]. For this collisional model, we already know that $v^2$ is not a collisional invariant due to energy dissipation [see \refeq{cons_E_IHS}]. Then, \index{Temperature!granular}temperature is not preserved anymore in the system, and it will decay, driving the system asymptotically to a completely frozen (zero-energy) state. To infer the evolution of this \index{Temperature!granular}temperature dissipation, one can derive the associated differential equation directly from the homogeneous \acrshort{BE} related to this model. Multiplying $v^2$ in \refeq{HBE_IHS} and integrating over all velocities, we get\index{Haff's law}
\begin{equation}\labeq{diff_ev_T}
	\frac{\partial T(t)}{\partial t} = -\zeta(t)T(t),
\end{equation}
with $\zeta(t)$ being usually called the \index{Cooling rate}\index{Cooling rate!in the IHS model}\emph{cooling rate},
\begin{equation}\labeq{cool_rate_def}
	\zeta(t) \equiv -\frac{m n}{\dt T(t)}\int \dif\vvel v^2 J[\vvel|f,f] = \frac{\pi^{\frac{\dt-1}{2}}}{4\dt \Gamma\left(\frac{\dt+3}{2}\right)}(1-\een^2)\frac{mn\sigma^{\dt-1}}{T(t)}\llangle v^3_{12}\rrangle(t),
\end{equation}
where the notation $\llangle \psi(\vvel_1,\vvel_2) \rrangle$, with $\psi(\vvel_1,\vvel_2)$ being a two-body velocity related quantity, means
\begin{equation}
	\llangle \psi(\vvel_1,\vvel_2)\rrangle(t) = \frac{1}{n^2}\int \dif \vvel_1 \int\dif \vvel_2\, \psi(\vvel_1,\vvel_2) f(\vvel_1;t)f(\vvel_2;t).
\end{equation} 
Notice that for $\een<1$, $\zeta>0$ because of energy dissipation. Thus, the explicit form of \refeq{diff_ev_T} is conditioned to the time dependence of \refeq{cool_rate_def}, which is hidden in the evolution of the \acrshort{VDF}. However, it is expected that, after a first \emph{kinetic} evolution of the \acrshort{VDF}, it will admit a scaling solution of the form,\index{Velocity distribution function!reduced}
\begin{equation}\labeq{scaling_HCS}
	f_\HCS(\vvel;t) = n\vth^{-\dt}(t)\phi_\HCS(\cc),
\end{equation} 
with $\cc$ the rescaled or reduced velocity defined in \refeq{red_velocity}. When this scaling form is reached, the system evolution is just governed by the \index{Temperature!granular}temperature continuous decay. In this situation, the system is said to be at the \acrshort{HCS}, the \index{Cooling rate!in the IHS model}cooling rate at this state satisfying \phantomsection\label{sym:zetaHCS}$\zeta_\HCS(t)\propto \sqrt{T_\HCS(t)}$. In that case, the solution of the evolution equation, \refeq{diff_ev_T}, is
\begin{equation}\labeq{Haffs_law}
	T_\HCS(t) = \frac{T_{\HCS}(t_0)}{\left[1+\frac{1}{2}\zeta_\HCS(t_0)(t-t_0) \right]^2},
\end{equation}
with $t_0$ being a reference time belonging to the \acrshort{HCS}. The solution in \refeq{Haffs_law} is the so-called \index{Haff's law}Haff's law of cooling~\cite{H83}. Moreover, it is convenient to measure the evolution of the system not in the \emph{laboratory frame}, but in a \emph{collisional frame}. This alternative description is characterized by the dimensionless time scale
\begin{equation}\labeq{s_variable}
	s(t) = \frac{1}{2}\int_0^t \dif t^\prime \nu\left(t^\prime\right),
\end{equation}
$\nu(t)$ being the inverse of the mean free time, i.e.,
\begin{equation}\labeq{freq_coll}
	\nu(t) = K_{\dt} n\sigma^{\dt-1}\vth(t),
\end{equation}
with $K_{\dt}$ defined in \refeq{nuwn}. This $\nu(t)$ is actually the (local-equilibrium) collision frequency, and then, according to \refeq{s_variable}, the time scale $s(t)$ represents the (local-equilibrium) accumulated average number of collisions per particle up to time $t$. In this collisional frame, Haff's cooling law reads
\begin{equation}\labeq{Haffs_law_coll}
	T_\HCS(s) = T_\HCS(s_0) \exp[-2\zeta_\HCS^* (s-s_0)],
\end{equation}
with $\zeta_\HCS^*\equiv \zeta_\HCS(t)/\nu(t)$ the reduced cooling rate at \acrshort{HCS}, which is a constant due to the \acrshort{HCS} condition. Comparing \refeqs{Haffs_law} and \eqref{eq:Haffs_law_coll}, we see that $s$ and $t$ variables are logarithmically related in the \acrshort{HCS}.  

Furthermore, in this collisional frame, the \acrshort{BE} for an evolving reduced \index{Velocity distribution function!reduced}\acrshort{VDF} \phantomsection\label{sym:phi}$\phi(\cc;s) = n^{-1}\vth^{\dt}(t)f(\vvel;t)$ reads
\begin{equation}\labeq{red_BE_IHS_free}
	\frac{K_{\dt}}{2}\frac{\partial \phi(\cc;s)}{\partial s}+\frac{\mu_2(s)}{\dt}\frac{\partial}{\partial\cc}\cdot\left[\cc \phi(\cc;s) \right]= I[\cc|\phi,\phi],
\end{equation}
with $I[\cc|\phi,\phi]$ being a reduced collisional operator,
\begin{equation}\labeq{I_coll_def}
	I[\cc_1|\phi,\phi] \equiv \frac{\vth^{\dt-1}}{n\sigma^{\dt-1}}J[\vvel|f,f]=\int\dif\cc_2\int_{+}\dif\ssab\, (\cc_{12}\cdot\ssab)\left[\alpha^{-2}\phi_1^{\pprime}\phi_2^{\pprime}-\phi_1\phi_2\right],
\end{equation}
and $\mu_k$ being the $k^{\mathrm{th}}$ \index{Collisional moments}\index{Collisional moments!for IHS}collisional moment,
\begin{align}\labeq{coll_mom_IHS_def}
	\mu_{k}(s)\equiv -\int\dif\cc\, c^k I[\cc|\phi,\phi],
\end{align}
or, equivalently,
\begin{equation}\labeq{coll_mom_IHS_def_2}
	\mu_k(s) = -\frac{1}{2}\int\dif\cc_1\int\dif\cc_2\,\phi(\cc_1;s)\phi(\cc_2;s)\int_+\dif\ssab\, (\cc_{12}\cdot\ssab) \left(\mathfrak{B}_{12,\ssab}-1\right)\left(c_1^k+c_2^k \right),
\end{equation}
such that from the definition of \index{Cooling rate!in the IHS model}cooling rate in \refeq{cool_rate_def} one infers that $\zeta = 2\mu_2/\dt$.

Thus, the formal definition of the \acrshort{HCS} \index{Velocity distribution function!of the HCS}\acrshort{VDF} in \refeq{scaling_HCS} is equivalent to the steady-state solution of the reduced \acrshort{BE} in \refeq{red_BE_IHS_free},
\begin{equation}\labeq{eq_steady_HCS_IHS}
	\frac{\mu_2^\HCS}{\dt}\frac{\partial}{\partial\cc}\cdot\left[\cc \phi_\HCS(\cc) \right]= I[\cc|\phi_\HCS,\phi_\HCS].
\end{equation}
This \emph{steady}\index{Velocity distribution function!moments}\index{Temperature!granular}\index{Velocity distribution function!of the HCS}\index{Velocity distribution function!reduced}\footnote{The system is not really in a stationary state because its second velocity moment (granular temperature) monotonically decays due to the energy dissipation. In terms of the \acrshort{VDF}, $f$, in the \acrshort{HCS}, all the time dependence of $f_\HCS$ is controlled by the temperature decay, i.e., $\partial_t f_\HCS= (\partial_t T_\HCS)(\partial_{T_\HCS} f_\HCS)= -\zeta_\HCS T_\HCS \partial_{T_\HCS} f_\HCS= \frac{1}{2}\zeta_\HCS \partial_\vvel\cdot(\vvel f_\HCS)$, where the last equality comes from the scaling condition of the \acrshort{HCS} \acrshort{VDF} in \refeq{scaling_HCS}.  However, the reduced \acrshort{VDF} at the \acrshort{HCS}, $\phi_\HCS$, is steady, i.e., $\partial_s \phi_\HCS =0$.} condition establishes that all the time dependence of the (unscaled) \acrshort{VDF} $f_\HCS$ in the \acrshort{HCS} occurs through the continuous cooling of the \index{Temperature!granular}granular temperature, according to Haff's law. From this stationary condition, one can infer some properties of the \acrshort{HCS} \index{Velocity distribution function!of the HCS}\acrshort{VDF}, such as a hierarchy of moment\index{Velocity distribution function!moments} equations,
\begin{equation}\labeq{hierarchy_moms_IHS}
	\mu_2^{\HCS}\langle c^{2k}\rangle_\HCS = \frac{\dt}{2k}\mu^\HCS_{2k}, \quad \langle c^{p}\rangle_\HCS = \int\dif\cc\, c^p \phi_\HCS(\cc) ,
\end{equation}
coming from a multiplication of the term $c^{2k}$ in both sides of \refeq{coll_mom_IHS_def_2} and integration over the reduced velocity variable $\cc$. This hierarchy of moments\index{Velocity distribution function!moments} equation assumes in its derivation that $c^{2k}\phi_\HCS(\cc)\xrightarrow{c\to\infty}0$, which is compatible with another known feature of $\phi_\HCS$, its exponential \acrshort{HVT}, which will be commented on later.

The exact form of $\phi_\HCS$ is not known up to date, but in the last three decades there have been a vast number of works focused on the study of its properties. Only for $\een=1$ the solution is the Maxwellian \index{Velocity distribution function!Maxwellian}\acrshort{VDF}, but this is the molecular case already studied in \refsubsec{HMG_KTGG}, where cooling is eliminated, \index{Temperature!molecular}temperature is preserved, and, then, \acrshort{HCS} formally does not exist. Without the explicit form of $\phi_{\HCS}$, we cannot derive the explicit expressions of the \index{Collisional moments!for IHS}collisional moments present in the hierarchy of velocity moments\index{Velocity distribution function!moments} in \refeq{hierarchy_moms_IHS}. In order to manage these quantities, we need to make approximations for the \acrshort{HCS} \index{Velocity distribution function!of the HCS}\acrshort{VDF}. For example, let us first focus on the \index{Cooling rate}cooling rate, which is proportional to $\mu_2$. One expects that, at least for $\alpha \lesssim 1$, the \acrshort{HCS} \index{Velocity distribution function!one-particle}\acrshort{VDF} is close to the Maxwellian one in \refeq{MDF_EHS} with a time-dependent \index{Temperature!granular}granular temperature. In its reduced version, this Maxwellian \index{Velocity distribution function!Maxwellian}\acrshort{VDF} reads
\begin{equation}\labeq{phiM_def}
	\phi_{\mathrm{M}}(\cc) = \pi^{-\dt/2}e^{-c^2}.
\end{equation}
Therefore, assuming $\phi_{\HCS}\approx \phi_{\mathrm{M}}$, the cooling rate\index{Cooling rate!in the IHS model} can be explicitly computed and, under this approach, one obtains~\cite{G19}
\begin{equation}\labeq{zeta_HCS_red_MA}
	\zeta_{\HCS}^{*\mathrm{MA}} = \frac{1-\een^2}{\dt}.
\end{equation}
This approach is usually known as \emph{Maxwellian approximation} (\acrshort{MA}). However, for not quasi-elastic systems, deviations for this form of the cooling rate, as well as for the \acrshort{MA} of $\phi_{\HCS}$, are found theoretically and from computer simulations~\cite{BRC96,vNE98,MS00,HOB00,CDPT03,BP06,BP06b,SM09}. A way of obtaining information about the \acrshort{HCS} \index{Velocity distribution function!of the HCS}\acrshort{VDF} is from its velocity moments\index{Velocity distribution function!moments} $\langle c^{2k}\rangle_\HCS$ or, equivalently, from the cumulants\index{Velocity distribution function!cumulants} of the distribution. First of all, due to isotropy, one expects that $\phi_{\HCS}$ depends on $\cc$ only through its modulus. Thus, a straightforward approach consists in writing $\phi_{\HCS}$ as a series expansion of a orthogonal basis of polynomials based on the Gaussian measure. It is quite common in kinetic theory to use Sonine polynomials~\cite{CC70,FK72,BP03,G19}, which are just a special case of associated Laguerre polynomials,
\begin{equation}\labeq{Sk_sonine_def}
	S_k(c^2) \equiv L_k^{\left(\frac{\dt}{2}-1\right)}(c^2) = \sum_{j=0}^k \frac{(-1)^j\Gamma\left(\frac{\dt}{2}+k\right)}{\Gamma\left(\frac{\dt}{2}+j\right)(k-j)!j!}c^{2j},
\end{equation} 
fulfilling the orthogonal and normalization condition,
\begin{equation}
	\braket{S_{k^\prime}|S_k}\equiv \int \dif\cc\, \phi_{\mathrm{M}}(\cc)S_{k^\prime}(c^2)S_{k}(c^2) =\mathcal{N}_k \delta_{k^{\prime}k}, \quad \mathcal{N}_k\equiv \frac{\Gamma\left(\frac{\dt}{2}+k\right)}{\Gamma\left( \frac{\dt}{2}\right)k!}.
\end{equation}
Then, an arbitrary isotropic \acrshort{VDF} $\phi(\cc)$ is expressed in this expansion as
\begin{equation}\labeq{sonine_expansion_c}
	\phi(\cc) = \phi_{\mathrm{M}}(\cc)\left[1+\sum_{k\geq 2}a_k S_k(c^2) \right],\quad a_k = \frac{\braket{S_k| \phi/\phi_{\mathrm{M}}}}{\mathcal{N}_k},
\end{equation} 
the coefficients $a_k$ being the cumulants\index{Velocity distribution function!cumulants} of the distribution, where the two leading cumulants are $a_0=1$ and $a_1=0$, the latter coming from the definition of \index{Temperature!granular}temperature, \refeq{molecular_temp}, using $\langle c^2\rangle=\dt/2$. Additionally, there is a biunivocal relationship between the cumulants\index{Velocity distribution function!cumulants} and the moments\index{Velocity distribution function!moments} of the distribution,
\begin{equation}\labeq{comp_moments_ck_Max}
	\langle c^{2k}\rangle = \langle c^{2k}\rangle_{\mathrm{M}}\left[1+\sum_{j=2}^k (-1)^j\binom{k}{j}a_j \right], \quad k\geq 2,
\end{equation}
where the subscript $\mathrm{M}$ denotes that the quantity is computed using the Maxwellian \index{Velocity distribution function!Maxwellian}\acrshort{VDF}. 

The straightforward extension of the \acrshort{MA} for the \acrshort{HCS} \index{Velocity distribution function!of the HCS}\acrshort{VDF} consists in truncating the series in \refeq{sonine_expansion_c} up to a certain cumulant order. This is the basis of the \emph{Sonine approximation} (\acrshort{SA}), widely used in the literature~\cite{BRC96,BDKS98,vNE98,MS00,HOB00,BBRTvW02,CDPT03,BP06,BP06b,SM09,G19}. This method consists in the mentioned truncation plus a corresponding linearization of the \index{Collisional moments!for IHS}collisional moments\index{Velocity distribution function!moments} on the cumulants\index{Velocity distribution function!cumulants} involved in the approximation. This approach is based on the assumption that those cumulants\index{Velocity distribution function!cumulants} are increasingly small, that is, $\phi_{\HCS}$ is not very far from $\phi_{\mathrm{M}}$. Under this assumptions, mainly \phantomsection\label{sym:akHCS}$a_2^{\HCS}$ and $a_3^{\HCS}$ have been estimated theoretically by means of different linearizations and successfully compared with simulation results from either direct simulation Monte Carlo (\acrshort{DSMC}) or event-driven molecular dynamics (\acrshort{EDMD}) techniques. In the work in Article 2 (\refsec{Art2}) this problem is revisited and \acrshort{EDMD} is extended to highly inelastic systems.

However, simulation results for $a_4^\HCS$, $a_5^\HCS$, and $a_6^\HCS$ reported in Refs.~\cite{BP06,BP06b}, as well as the theoretical analysis previously developed in Ref.~\cite{HOB00}, indicate a growth of their absolute values as their order increases for highly inelastic systems, $\een\lesssim 0.6$. Then, those systems may suffer a breakdown of the Sonine expansion, which can be explained~\cite{BP06,BP06b} by the exponential \acrshort{HVT} found for the \acrshort{HCS} \index{Velocity distribution function!of the HCS}\acrshort{VDF}. This tail, predicted by kinetic theory from a standard approach in which the loss term of the collisional operator dominates over the gain term in this high-velocity limit, has the form \cite{vNE98,EP97}
\begin{equation}\labeq{HVT_IHS_KTGG}
	\phi_{\HCS}(\cc)\sim \mathcal{A} \exp\left(-\gamma_{c}^{\mathrm{IHS}}c \right),\quad \gamma_{c}^{\mathrm{IHS}} = \frac{\dt \pi^{\frac{\dt-1}{2}}}{\mu_2^{\HCS}\Gamma\left( \frac{\dt+1}{2}\right)},
\end{equation}
for $c\gg 1$. The existence of this non-Maxwellian \acrshort{HVT} has been also satisfactorily observed in computer simulations \cite{BCR99,PBF06} and in recent microgravity experiments~\cite{YSS20}. In the latter work, the \acrshort{IHS} model and the viscoelastic model are compared with the experimental outcomes, concluding a negligible influence of the impact velocity on the coefficient of restitution, thus preferring the constant coefficient of restitution model. Moreover, Haff's cooling law and the exponential form of the \acrshort{HVT} were observed. On the other hand, because of some important deviations of the cooling rate\index{Cooling rate!in the IHS model} observed experimentally with respect to the kinetic-theory predictions, the authors claimed for models that could implement surface roughness.

The \acrshort{HCS} is not only important from a descriptive point of view. This state represents the homogeneous base state in the \index{Chapman--Enskog method}\acrshort{CE} method when transport and hydrodynamics of granular gaseous rapid flows are studied. Finally, this \acrshort{HCS} is not exclusive of monodisperse systems, but it is also described within the framework of the \acrshort{IHS} model in granular mixtures in several works~\cite{GD99b,MG02,DHGD02,G19}.

\subsubsection{Homogeneous freely evolving states in the IRHS model}

Whereas the \acrshort{HCS} does not change conceptually with the introduction of the \acrshort{IRHS} model, it acquires some implications that may be important, not only from a mathematical point of view, but also phenomenologically, and could solve, for example, the numerical discrepancies observed in Ref.~\cite{YSS20}.

Let us begin, as before, with the homogeneous \acrshort{BE} for the \acrshort{IRHS} model for a monodisperse system \cite{MS19,MS19b},
\begin{equation}\labeq{HBE_free_IRHS}
	\frac{\partial f(\vom_1;t)}{\partial t} = \sigma^{\dt-1}\int\dif\vom_2\int_{+}\dif\ssab\, 
    (\vvel_{12}\cdot\ssab)\left[\frac{f(\vom^{\prime\prime}_1;t)f(\vom^{\prime\prime}_2;t)}{\een^{2}|\eet|^{2\dr/\dt}}-f(\vom_1;t)f(\vom_2;t)\right],
\end{equation}
where rotational degrees of freedom, $\dr$, are introduced coming from the angular velocity vectors, $\oo$.

In order to start the kinetic description of the system, we need to introduce again the \index{Temperature!granular}granular temperature. However, in contrast to the simple definition given by \refeq{molecular_temp}, we need to take into account that the total kinetic energy for this model is the sum of a translational and a rotational counterparts, as considered in \refeq{conservation_energy_IRHS}, i.e., for the $i^{\mathrm{th}}$ particle its total kinetic energy is
\begin{equation}\labeq{Energy_t_r}
	E^{\mathrm{total}}_{i} \equiv E_{i}^{t}+ E_i^{r}= \frac{1}{2}mv_i^2 +\frac{1}{2}I\omega_i^2.
\end{equation}
Therefore, we can define naturally two \emph{partial} temperatures, one as the average of the \index{Temperature!translational}translational kinetic energy and the other one as the average of the \index{Temperature!rotational}rotational energy, both of them weighted by their respective degrees of freedom,
\begin{equation}\labeq{tr_rot_temp}
	\frac{\dt}{2}\Ttr(t) \equiv \frac{1}{2}m\langle v^2\rangle(t) , \quad \frac{\dr}{2}\Trot(t) \equiv \frac{1}{2}I\langle \omega^2\rangle(t),
\end{equation}
where we are explicitly setting $\langle \vvel\rangle=\mathbf{0}$ without any loss of generality in homogeneous states.

Moreover, from the definition of total kinetic energy, one can define a \index{Temperature!mean granular}\emph{mean} granular temperature as the average of the total kinetic energy weighted by the total number of degrees of freedom, 
\begin{equation}\labeq{def_T_IRHS}
	\frac{\dt+\dr}{2}T(t) \equiv \frac{1}{2}m\langle v^2\rangle(t)+ \frac{1}{2}I\langle \omega^2\rangle(t) = \frac{\dt \Ttr(t)+\dr \Trot(t)}{\dt+\dr},
\end{equation}
which is identified with a weighted mean of the \index{Temperature!translational}translational and \index{Temperature!rotational}rotational partial temperatures defined in \refeq{tr_rot_temp}.

Then, from the definitions in \refeq{tr_rot_temp} and the homogeneous \acrshort{BE} for the \acrshort{IRHS} model introduced in \refeq{HBE_free_IRHS}, one can derive the evolution equations for the \index{Temperature!translational}\index{Temperature!rotational}partial temperatures,
\begin{subequations}\labeq{ev_eqs_Tt_Tr}
\begin{align}\labeq{P_Rates_t}
	\partial_t \Ttr(t) =& -\xi_t(t) \Ttr(t), \quad \xi_t(t) \equiv -\frac{m}{n\dt \Ttr(t)}\int\dif\vom\, v^2 J[\vom|f,f], \\ \labeq{P_Rates_r}
	\partial_t \Trot(t) =& -\xi_r(t) \Trot(t), \quad \xi_r(t) \equiv -\frac{I}{n\dr \Trot(t)}\int\dif\vom\, \omega^2 J[\vom|f,f],
\end{align}
\end{subequations}
with $\xi_t$ and $\xi_r$ being the translational and rotational \emph{energy production rates}, respectively, which do not represent \emph{cooling} rates necessarily. This is because, although the mean \index{Temperature!mean granular}temperature decays monotonically, the coefficients $\xi_t$ and $\xi_r$ also take into account a transfer of energy from the translational part to the rotational one, and vice versa. This implies a coupling between both differential equations. However, analogously to what happened with \refeq{diff_ev_T}, \refeqs{ev_eqs_Tt_Tr} do not form a closed set of differential equations because the energy production rates depend on the full time-dependent \index{Velocity distribution function!time-dependent}\acrshort{VDF}. Moreover, the evolution for the mean granular \index{Temperature!mean granular}temperature reads
\begin{equation}\labeq{Haffs_law_IRHS}
	\partial_t T(t) = -\zeta(t) T(t), \quad \zeta(t) \equiv \frac{\dt \xi_t(t)\Ttr(t)+\dr \xi_r(t)\Trot(t)}{(\dt+\dr)T(t)},
\end{equation}
$\zeta$ being the \index{Cooling rate!in the IRHS model}cooling rate for the \acrshort{IRHS} model. Again, this equation cannot be solved if the time dependence and the form of the \acrshort{VDF} is unknown.

\refeqs{tr_rot_temp}--\eqref{eq:Haffs_law_IRHS} reflect the dynamics inferred for the partial and mean \index{Temperature!translational}\index{Temperature!rotational}\index{Temperature!mean granular}temperatures from the introduction of the \acrshort{IRHS} model into the homogeneous \acrshort{BE}. Using the same reasoning as in the case of the \acrshort{IHS} model, one expects that, after a first rapid kinetic transient stage, all time dependence of the \acrshort{VDF} occurs through the \index{Temperature!mean granular}temperature decay equation, \refeq{Haffs_law_IRHS}. In that regime, the \acrshort{HCS} is reached and the \index{Velocity distribution function!reduced}\acrshort{VDF} fulfills the following rescaling,
\begin{equation}\labeq{f_resc_IRHS}
	f_{\HCS}(\vom;t) = n \left(\frac{2\tau^\HCS_t}{m} \right)^{-\dt/2}\left(\frac{2\tau^\HCS_r}{I} \right)^{-\dr/2}\left[T_{\HCS}(t)\right]^{-\frac{\dt+\dr}{2}}\phi_{\HCS}(\cw),
\end{equation}
with $\cw\equiv\{\cc,\ww\}$,
\begin{equation}\labeq{def_c_w}
	\cc\equiv \frac{\vvel}{\vth(t)}, \quad \ww\equiv \frac{\oo}{\oth(t)}; \qquad \vth \equiv\sqrt{\frac{2\tau_t T}{m}}, \quad \oth \equiv\sqrt{\frac{2\tau_r T}{I}}.
\end{equation}
In \refeq{f_resc_IRHS}, the \index{Temperature!ratios}temperature ratios $\tau_t \equiv \Ttr/T$ and $\tau_r\equiv \Trot/T$ acquire necessarily steady values because all time dependence is contained in $T(t)$, and $\phi_{\HCS}(\cw)$ does not evolve in time. The condition of \index{Temperature!ratios}$\tau_t$ and $\tau_r$ reaching a steady form is equivalent to the \index{Temperature!rotational-to-translational ratio}rotational-to-translational temperature ratio getting a steady value,
\begin{equation}\labeq{theta_equip_def}
	\theta \equiv \frac{\Trot}{\Ttr} \Rightarrow \tau_t = \frac{\dt+\dr}{\dt+\dr\theta}, \quad \tau_r = \frac{\dt+\dr}{\dt/\theta+\dr}.
\end{equation}
In general, from the differential equations of the \index{Temperature!translational}\index{Temperature!rotational}partial temperatures in \refeqs{ev_eqs_Tt_Tr}, the evolution of \index{Temperature!rotational-to-translational ratio}$\theta$ is given by
\begin{equation}
	\partial_t \theta = -\left(\xi_r-\xi_t\right)\theta.
\end{equation}
At the \acrshort{HCS}, the \index{Temperature!rotational-to-translational ratio}rotational-to-translational temperature ratio reaches a constant value and, therefore, \phantomsection\label{sym:xi_prod_HCS}$\xi_t^\HCS = \xi_r^\HCS = \zeta_\HCS$, where the last equality comes from the \index{Cooling rate!in the IRHS model}cooling rate definition in \refeq{Haffs_law_IRHS}. Additionally, the unique set of collisional invariants for this collisional model is $\{1,\vvel\}$ due to conservation of mass and linear momentum. However, $\oo$ it is not a collisional invariant. Thus, to complete the description, at least up to second order in velocity moments\index{Velocity distribution function!moments}, we need to derive the evolution equation for the mean angular velocity of the system, \phantomsection\label{sym:avOmega}$\bar{\mathbf{\Omega}}\equiv \langle \oo\rangle$. Then, from the homogeneous \acrshort{BE}, \refeq{HBE_free_IRHS}, one finds
\begin{equation}\labeq{ev_eq_Omega_zeta}
	\partial_t \bar{\mathbf{\Omega}}(t)= -\zeta_{\bar{\Omega}}\bar{\mathbf{\Omega}}(t), \quad \zeta_{\bar{\Omega}}\bar{\mathbf{\Omega}}(t)\equiv -\int \dif\vom \oo J[\vom|f,f].
\end{equation}
Since the \acrshort{HCS} is isotropic and there is no preferred direction. one must have \phantomsection\label{sym:omega_zeta_HCS}$\bar{\boldsymbol{\Omega}}_{\text{H}}=\mathbf{0}$.

Again, from the \acrshort{HCS} condition, $\zeta_{\HCS}(t)/\sqrt{T_\HCS(t)}\propto \text{const}$ and, therefore, the expression of \index{Haff's law}Haff's law in the laboratory time frame derived from the \acrshort{IHS} model, \refeq{Haffs_law}, is recovered, except that now the \index{Cooling rate!in the IRHS model}cooling rate is that of the \acrshort{IRHS} model. Since in the \acrshort{HCS} all the time dependence takes place through \index{Temperature!mean granular}temperature, we have,
\begin{equation}
	\frac{\partial}{\partial t} \equiv \left(\frac{\partial T_\HCS}{\partial t} \right)\frac{\partial}{\partial T_\HCS}= -\zeta_\HCS(t) T_\HCS(t)\frac{\partial}{\partial T_{\HCS}}.
\end{equation}
As a consequence, the homogeneous \acrshort{BE} for the \acrshort{IRHS} can be rewritten, still in the laboratory time frame, as~\cite{G19}
\begin{equation}\labeq{BE_IRHS_V_W}
	\frac{1}{2}\zeta_\HCS \left(\frac{\partial}{\partial\mathbf{V}}\cdot\mathbf{V}+ \frac{\partial}{\partial\oo}\cdot\oo \right)f_\HCS = J[\vom|f_\HCS,f_\HCS],
\end{equation}
where we have used the scaling form of $f_\HCS$ and the dependence of the variables in $\cw$ with $T_\HCS$ according to their definition in \refeq{def_c_w}, which are dependent on $T_\HCS$ in the \acrshort{HCS}.

Following the same steps as in the \acrshort{IHS} model, it is also convenient in the \acrshort{IRHS} model to define the collisional frame and the $s$-variable defined in \refeqs{s_variable} and~\eqref{eq:freq_coll}, but taking into account that in this model the thermal velocity is $\vth = \sqrt{2\tau_t T/m}$. Then, from similar reasoning as in the deduction of \refeq{red_BE_IHS_free} and  introducing the generalized unsteady scaling $f(\vvel;t) = n^{-1}\vth^{-\dt}(t)\oth^{-\dr}(t)\phi(\cc;s)$, the homogeneous \acrshort{BE} for a freely evolving granular gas in the \acrshort{IRHS} model for the reduced \index{Velocity distribution function!reduced}\acrshort{VDF} in the collisional frame reads
\begin{equation}\labeq{red_BE_IRHS_free}
	\frac{K_{\dt}}{2}\frac{\partial \phi(\cw;s)}{\partial s}+\frac{\mu^{(0)}_{20}(s)}{\dt}\frac{\partial}{\partial\cc}\cdot\left[\cc \phi(\cw;s) \right]+\frac{\mu^{(0)}_{02}(s)}{\dr}\frac{\partial}{\partial\ww}\cdot\left[\ww \phi(\cw;s) \right]= I[\cw|\phi,\phi],
\end{equation}
with $I[\cw|\phi,\phi]$ being a reduced collisional operator defined analogously to~\refeq{I_coll_def} for the \acrshort{IHS} case, i.e.,
\begin{equation}\labeq{I_coll_def_IRHS}
	I[\cw_1|\phi,\phi] \equiv \frac{\vth^{\dt-1}\oth^{\dr}}{n\sigma^{\dt-1}}J[\vvel|f,f]=\int\dif\cw_2\int_{+}\dif\ssab\, (\cc_{12}\cdot\ssab)\left(\frac{\phi_1^{\pprime}\phi_2^{\pprime}}{\een^2|\eet|^{2\frac{\dt}{\dr}}}-\phi_1\phi_2\right).
\end{equation}
In \refeq{red_BE_IRHS_free}, and as a generalization of \refeq{coll_mom_IHS_def_2}, $\mu^{(r)}_{pq}$ is the \index{Collisional moments!for IRHS}collisional moment of order $(p,q,r)$,
\begin{align}\labeq{coll_mom_IRHS_def}
	\mu^{(r)}_{pq}(s)\equiv -\int\dif\cw\, c^p w^q (\cc\cdot\ww)^r I[\cw|\phi,\phi]
\end{align}
or, equivalently,
\begin{align}\labeq{coll_mom_IRHS_def_2}
	\mu^{(r)}_{pq}(s) =& -\frac{1}{2}\int\dif\cw_1\int\dif\cw_2\,\phi(\cw_1;s)\phi(\cw_2;s)\int_+\dif\ssab\, (\cc_{12}\cdot\ssab)\nonumber \\
	&\times\left(\mathfrak{B}_{12,\ssab}-1\right)\left[c_1^p w_1^q(\cc_1\cdot\ww_1)^r+c_2^p w_2^q(\cc_2\cdot\ww_2)^r \right].
\end{align}
From the definition of the energy production rates in \refeqs{ev_eqs_Tt_Tr}, one infers that $\xi_t = 2\mu^{(0)}_{20}/\dt$ and $\xi_r = 2\mu^{(0)}_{02}/\dr$. From \refeq{coll_mom_IRHS_def_2}, once we know the collisional change of the velocity function $c_1^p w_1^q(\cc_1\cdot\ww_1)^r$, we can compute the $(p,q,r)$-order \index{Collisional moments!for IRHS}collisional moment. In general, given an arbitrary one-body velocity dependent function, $\psi(\cw_1)\equiv \psi_1$, even if the \acrshort{VDF} is unknown, one can formally express its corresponding collisional integral, 
\begin{align}\labeq{coll_integral_I_cal}
\mathcal{I}[\psi_1|\phi,\phi]&\equiv \int\dif\cw_1\, \psi(\cw_1)I[\cw_1|\phi,\phi]\nonumber \\
&= \int\dif\cw_1\int\dif\cw_2\int_{+}\dif\ssab\, (\cc_{12}\cdot\ssab) \phi(\cw_1)\phi(\cw_2)\left( \mathfrak{B}_{12,\ssab}-1 \right)\psi(\cw_1),
\end{align}
in terms of two-body averages. Then, the \acrshort{HCS} condition is equivalent to
\begin{equation}\labeq{HCS_cond_IRHS}
	\frac{\mu^{(0)\HCS}_{20}}{\dt}\frac{\partial}{\partial\cc}\cdot\left[\cc \phi_\HCS(\cw) \right]+\frac{\mu^{(0)\HCS}_{02}}{\dr}\frac{\partial}{\partial\ww}\cdot\left[\ww \phi_\HCS(\cw) \right]= I[\cw|\phi_\HCS,\phi_\HCS].
\end{equation}
This is the \acrshort{IRHS} counterpart of \refeq{eq_steady_HCS_IHS}.

In Refs.~\cite{MS19,MS19b}, some of the collisional integrals for the velocity moments\index{Velocity distribution function!moments} up to quadratic order are computed in terms of two-body averages, which are associated with conservation of linear momentum, the nonconservation of angular velocity encoded in the quantity $\zeta_{\bar{\Omega}}$, and the energy production rates. In \reftab{EPR_2body}, some of these quantities are summarized in their reduced-velocity version, where the quantity \phantomsection\label{sym:W12}$\mathbf{W}_{12}=(\ww_1+\ww_2)/2$ is introduced. Moreover, in the work presented in Article 5 (\refsec{Art5}) of this thesis, an extension to \index{Collisional moments!for IRHS}collisional moments of fourth order is carried out. 

\begin{table}[h!]
\caption{Collisional integrals for velocity functions up to quadratic order in terms of two-body averages. Here, $\mathcal{I}_0\equiv 2\pi^{\frac{\dt-1}{2}}/(\dt+1)$.}
\labtab{EPR_2body}
\centering
\begin{tabularx}{\textwidth}{rcl}
\hline\hline
	$\displaystyle{\psi(\cw)}$ && $\displaystyle{\mathcal{I}[\psi|\phi,\phi]/\mathcal{I}_0}$ \rule{0pt}{2.6ex}\rule[-0.9ex]{0pt}{0pt}\\
	\hline\\
	$\displaystyle{\cc_1}$ && $\displaystyle{-\left(\en+\frac{\dt-1}{2}\et\right)\llangle c_{12}\cc_{12}\rrangle+\frac{\sqrt{\pi}\Gamma\left( \frac{3+\dt}{2}\right)}{\Gamma\left( 1+\frac{\dt}{2}\right)}\et\sqrt{\frac{\theta}{\kappa}}\llangle \cc_{12}\times\mathbf{W}_{12}\rrangle}$\\
	$\displaystyle{\ww_1}$ && $\displaystyle{-\frac{\et}{\kappa}\left[3\llangle c_{12}\mathbf{W}_{12}\rrangle-\llangle c_{12}^{-1}\cc_{12}(\cc_{12}\cdot\mathbf{W}_{12})\rrangle \right]}$\\ 
	$\displaystyle{c^2_1}$ && $\displaystyle{-2\left(\en+\frac{\dt-1}{2}\et\right)\llangle c_{12}(\cc_1\cdot\cc_{12})\rrangle+\frac{1}{2}\left(\en^2+\frac{\dt-1}{2}\et^2\right)\llangle c_{12}^3\rrangle+\frac{2\et^2\theta}{\kappa}}$\\
	&& $\displaystyle{\times\left[3\llangle c_{12}W_{12}^2\rrangle-\llangle c_{12}^{-1}(\cc_{12}\cdot\mathbf{W}_{12})^2\rrangle \right]+\frac{2\sqrt{\pi}\Gamma\left(\frac{3+\dt}{2}\right)}{\Gamma\left(1+\frac{\dt}{2}\right)}\et\sqrt{\frac{\theta}{\kappa}}\llangle (\cc_{1}\times\cc_{12})\cdot\mathbf{W}_{12}\rrangle}$\\
	$\displaystyle{w^2_1}$ && $\displaystyle{-\frac{\dt-1}{2}\frac{\et^2}{\kappa\theta}\llangle c_{12}^3\rrangle-2\frac{\et^2}{\kappa^2}\left[ 3\llangle c_{12}W_{12}^2\rrangle -\llangle c_{12}^{-1}(\cc_{12}\cdot\mathbf{W}_{12})^2\rrangle\right]}$\\
	&& $\displaystyle{+2\frac{\et}{\kappa}\left[ 3\llangle c_{12}(\ww_1\cdot\mathbf{W}_{12})\rrangle -\llangle c_{12}^{-1}(\cc_{12}\cdot\mathbf{W}_{12})(\cc_{12}\cdot\ww_1)\rrangle\right]}$\\
	$\displaystyle{\frac{c^2_1+c^2_2}{2}}$ && $\displaystyle{\left[\en(1-\en)+\frac{\dt-1}{2}\et(1-\et)\right]\llangle c_{12}^3\rrangle+\frac{2\et^2\theta}{\kappa}\left[3\llangle c_{12}W_{12}^2\rrangle-\llangle c_{12}^{-1}(\cc_{12}\cdot\mathbf{W}_{12})^2\rrangle \right]}$\\
	$\displaystyle{\frac{w^2_1+w^2_2}{2}}$ && $\displaystyle{-\frac{\dt-1}{2}\frac{\et^2}{\kappa\theta}\llangle c_{12}^3\rrangle+2\frac{\et}{\kappa}\left(1-\frac{\et}{\kappa} \right)\left[ 3\llangle c_{12}W_{12}^2\rrangle-\llangle c_{12}^{-1}(\cc_{12}\cdot\mathbf{W}_{12})^2\rrangle\right]}$\\
	&&\\
	\hline\hline
\end{tabularx}
\end{table}
Whereas expressions in \reftab{EPR_2body} are exact in the context of the \acrshort{BE} and the \acrshort{IRHS} model for \acrshort{HD} ($\dt=2$, $\dr=1$) and \acrshort{HS} ($\dt=\dr=3$), it is important, from a predictive point of view, to make estimates of the involved two-body averages. In this sense, we need to propose certain forms for the \acrshort{VDF} if we want to evaluate explicitly the evolution equations and also to study the \acrshort{HCS}. As a first reasonable proposal, we can consider that translational and angular velocities are uncorrelated or that their correlation is negligible\footnote{We will see in \refch{IRHS_HS} that this is not always true.}. In this approximation, the total \acrshort{VDF} can be factorized as a product of its marginal translational and rotational \index{Velocity distribution function!marginal}\acrshort{VDF},
\begin{equation}\labeq{MA_VDF_IRHS}
	\phi(\cw)\rightarrow \phi_{\cc}(\cc)\phi_{\ww}(\ww),
\end{equation}
where $\phi_{\cc}$ and $\phi_{\ww}$ refer to the translational and rotational marginal \index{Velocity distribution function!marginal}\acrshort{VDF}, respectively, defined as
\begin{equation}\labeq{marginal_VDF}
		\phi_{\cc}(\cc)\equiv \int\dif\ww\, \phi(\cw), \quad \phi_{\ww}(\ww) \equiv \int\dif\cc\, \phi(\cw).
\end{equation}
As explained in Refs.~\cite{S18,G19,MS19,MS19b,MS21}, an apparently good approach consists in choosing $\phi_{\cc}(\cc)\rightarrow \phi_{\mathrm{M}}(\cc)$ from arguments of maximum-entropy formalism of this translational part, but not to give explicitly any form for the rotational marginal \index{Velocity distribution function!marginal}\acrshort{VDF}. This is because, up to the order of the quantities in \reftab{EPR_2body}, no powers above quadratic are observed for the angular velocity. Thus, we take
\begin{equation}\labeq{qMA_IRHS}
		\phi(\cw)\approx \phi_{\mathrm{M}}(\cc)\phi_{\ww}(\ww),
\end{equation}
as a sort of \acrshort{MA}. Thus, within this approximation, the form of the principal production rates appearing in the evolution equations of the system are summarized in \reftab{energy_production_rates}. In this table, \phantomsection\label{sym:zetaOmega_ast}$\zeta_{\bar{\Omega}}^*\equiv \zeta_{\bar{\Omega}}/\nu$, $\xi_{t,r}^*\equiv \xi_{t,r}/\nu$, and $\zeta^*\equiv\zeta/\nu$, with the collisional frequency $\nu$ defined in \refeq{freq_coll}, and the quantity \phantomsection\label{sym:X_rot}$\widetilde{X} \equiv I\bar{\Omega}^2/\dr\Trot$ is introduced.\index{Velocity distribution function!moments}

\begin{table}[h!]
\caption{Collisional production rates up to second order in velocity moments under the \acrshort{MA} in \refeq{qMA_IRHS}.}
\labtab{energy_production_rates}
\centering
\begin{tabularx}{\textwidth}{rcl}
\hline\hline
Production rate && Expression in the approximation $\phi(\cw)\rightarrow \phi_{\mathrm{M}}(\cc)\phi_{\ww}(\ww)$\rule{0pt}{2.6ex}\rule[-0.9ex]{0pt}{0pt}\\
	\hline\\	
	$\displaystyle{\zeta^*_{\bar{\Omega}}}$ && $\displaystyle{\frac{2}{\dt}\frac{1+\eet}{1+\kappa}}$\\
	$\displaystyle{\xi^*_t}$ && $\displaystyle{\frac{1-\een^2}{\dt}+\frac{2\dr\kappa(1+\eet)}{\dt^2(1+\kappa)^2}\left[\frac{(1-\eet)}{2}\left(\kappa+\theta+\widetilde{X} \right)+1-\theta-\widetilde{X}\right]}$\\
	$\displaystyle{\xi^*_r}$ && $\displaystyle{\frac{2\kappa(1+\eet)}{\dt(1+\kappa)^2}\left[ \frac{1-\eet}{2\kappa}\left(\frac{\kappa}{\theta}+1+\widetilde{X}\right)+1-\frac{1}{\theta}+\widetilde{X}\right]}$\\ 
	$\displaystyle{\zeta^*}$ && $\displaystyle{\frac{1-\een^2}{\dt+\dr\theta}+\frac{1-\eet^2}{\dt(1+\kappa)}\frac{\dr/\dt}{1+\dr\theta/\dt}\left(\kappa+\theta+\widetilde{X} \right)}$\\
	&&\\
	\hline\hline
\end{tabularx}
\end{table}

In general, the basic evolution equations in the collisional frame read
\begin{align}
	\frac{K_{\dt}}{2}\frac{\partial \ln T}{\partial s} = -\zeta^*, \quad \frac{K_{\dt}}{2}\frac{\partial \ln\theta}{\partial s} = -\xi_r^*+\xi_t^*, \quad \frac{K_{\dt}}{2}\frac{\partial \ln \widetilde{X}}{\partial_s} = -2\zeta_{\bar{\Omega}}^*+\xi_r^*. 
\end{align}
As expected, \reftab{energy_production_rates} shows that $\zeta^*\geq 0$, the equality being fulfilled only if $\een=|\eet|=1$. In addition, it can be easily proved that $2\zeta_{\bar{\Omega}}^*-\xi_r^*\geq 0$ \cite{G19,MS19b}, implying that $\widetilde{X}$ tends to 0 monotonically, so that \phantomsection\label{sym:XHCS}$\widetilde{X}^\HCS=0$. Finally, the solution for \phantomsection\label{sym:thetaHCS}$\theta^\HCS$ comes from the condition \phantomsection\label{sym:xitrHCS}$\xi_t^{*\HCS}=\xi_r^{*\HCS}$. Within the \acrshort{MA}, the \acrshort{HCS} solution for the \index{Temperature!rotational-to-translational ratio}rotational-to-translational temperature ratio is given by
\begin{equation}\labeq{expr_theta_M}
	\theta_{\mathrm{M}}^{\HCS} = \sqrt{h_{\theta}^2+\frac{\dt}{\dr}}+h_{\theta}, \quad h_{\theta} \equiv \frac{\dt(1+\kappa)^2}{2\dr\kappa(1+\eet)^2}\left[1-\een^2-\frac{1-\frac{\dr}{\dt}\kappa}{1+\kappa}(1-\eet^2) \right]+\frac{1}{2}\left(1-\frac{\dt}{\dr} \right).
\end{equation}
This expression clearly signals a nonequipartition between translational and rotational degrees of freedom. In Refs.~\cite{VSK14,G19,MS19b}, a \index{Linear stability analysis}linear stability analysis of the \acrshort{HCS} is done under \emph{homogeneous} perturbations in the context of the \acrshort{MA} given by \refeq{qMA_IRHS}.

Whereas this approach is a good reference to study the \acrshort{HCS}, correlations between translational and angular velocities are known to be present \cite{BPKZ07} and might be relevant. Then, the factorization in \refeq{MA_VDF_IRHS} for the \acrshort{HCS} \acrshort{VDF} needs to be corrected in some ranges of parameters. As occurs in the analysis of the \acrshort{IHS} model, the \acrshort{HCS} \index{Velocity distribution function!of the HCS}\acrshort{VDF} in the \acrshort{IRHS} model is non-Maxwellian. As an extension of the smooth case, applying a Grad--Sonine methodology in the \acrshort{IRHS} consists in the expansion of the \acrshort{VDF} $\phi(\cw)$ around a two-temperature Maxwellian distribution in terms of a proper set of orthogonal basis of polynomials. In this case,
\begin{subequations}\labeq{Sonine_expansion_IRHS}
\begin{align}\labeq{Sonine_expansion_IRHS_a}
	\phi(\cw) =& \phi_{\mathrm{M}}(\cw)\sum_{p=0}^{\infty}\sum_{q=0}^{\infty}\sum_{r=0}^\infty a_{pq}^{(r)} \Psi_{pq}^{(r)}(\cw), \quad \phi_{\mathrm{M}}(\cw)= \pi^{-\dt-\dr}e^{-c^2-w^2},\\
	\Psi_{pq}^{(r)}(\cw)\equiv& L_p^{\left(2r+\frac{\dt}{2}-1\right)}(c^2)L_q^{\left(2r+\frac{\dr}{2}-1\right)}(w^2)\left(c^2w^2\right)^rP_{2r}\left(\frac{|\cc\cdot\ww|}{cw}\right),
\end{align}
\end{subequations}
with $L_j^{(\ell)}(x)$ being the associated Laguerre polynomials and $P_j(x)$ being the Legendre polynomials~\cite{AS72}. Then, the orthonormalization condition for the basis $\left\{\Psi_{pq}^{(r)} \right\}_{p,q,r=1}^{\infty}$ is
\begin{equation}\labeq{Sonine_expansion_IRHS}
	\braket{\Psi_{p^\prime q^\prime}^{(r^\prime)}|\Psi_{pq}^{(r)}} = \mathcal{N}_{pq}^{(r)}\delta_{p p^\prime}\delta_{q q^\prime}\delta_{r r^\prime}, \quad \mathcal{N}_{pq}^{(r)} \equiv \frac{\Gamma\left(2r+\frac{\dt}{2}+p \right) \Gamma\left(2r+\frac{\dr}{2}+q \right)}{\Gamma\left( \frac{\dt}{2}\right)\Gamma\left( \frac{\dr}{2}\right)(4r+1)p!q!}.
\end{equation}
The coefficients $a_{pq}^{(r)}$ are the cumulants\index{Velocity distribution function!cumulants} of the \acrshort{VDF} and they are defined as
\begin{equation}
	a_{pq}^{(r)} = \frac{\langle \Psi_{pq}^{(r)}\rangle}{\mathcal{N}_{pq}^{(r)}}.
\end{equation}
The expansion in \refeq{Sonine_expansion_IRHS_a} accounts for the correlations between translational and rotational velocities. In the study of the \acrshort{HCS} \index{Velocity distribution function!of the HCS}\acrshort{VDF}, before this approach was considered standard (see Refs.~\cite{VSK14,MS23}), original works of Goldshtein and Shapiro~\cite{GS95} and Aspelmeier \emph{et al.}~\cite{AHZ01} already proposed Sonine expansions for translational and rotational velocities around $\phi_{\mathrm{M}}(\cw)$, but correlations between $\cc$ and $\ww$ were not considered. Again, a truncation of \refeq{Sonine_expansion_IRHS_a} for the first nontrivial cumulants\index{Velocity distribution function!cumulants}, together with the proper consistent linearization of the \index{Collisional moments!for IRHS}collisional moments in terms of the nonneglected cumulants\index{Velocity distribution function!cumulants}, has been the approach to study deviations from the Maxwellian in the thermal part of the \acrshort{VDF}. First, in Ref.~\cite{VSK14}, a \acrshort{SA} was performed to study the cumulants\index{Velocity distribution function!cumulants} in systems of \acrshort{HS}, namely, $a_{pq}^{(r)}$ with $p+q+2r\leq 2$ were considered nonzero, thus improving a previous approach~\cite{SKS11} where $a_{00}^{(1)}$ was not considered in the truncation for \acrshort{HS}. An extension of this study in terms of the translational and rotational degrees of freedom, valid for both \acrshort{HD} and \acrshort{HS}, is developed in Article 5 (\refsec{Art5}). Moreover, as will be summarized in Article 5 (\refsec{Art5}), the \acrshort{HVT} of the marginal \index{Velocity distribution function!marginal}\acrshort{VDF} at the \acrshort{HCS} were analyzed. 

As an important remark, the \acrshort{HCS} for both \acrshort{IHS} and \acrshort{IRHS} is not always a stable state under \emph{inhomogeneous} perturbations. The conditions under which this homogeneous state is unstable have been of interest of the granular gas community. These unstable states come in the form of \index{Instability!clustering}clustering or \index{Instability!vortices}vortices and are predicted by \index{Linear stability analysis}linear stability analysis of the \acrshort{HCS} under inhomogeneous perturbations. The inhomogeneities have been theoretically studied using the \index{Chapman--Enskog method}\acrshort{CE} method introduced in \refsec{CEM_KTGG}, from which the \index{Transport coefficient}transport coefficients to describe the granular hydrodynamics are computed. Therefore, this \index{Linear stability analysis}linear stability analysis is performed on the \index{Balance equations}balance equations of the granular gas taking advantage of the knowledge of the \index{Transport coefficient}transport coefficients. These instabilities were first predicted for inelastic systems~\cite{GZ93,BRM98,BRC99,LH99,G05,MDCPH11,MGHEH12,MZBGDH14,FH17,G19} and, in the last decade, roughness came into play for \acrshort{HS}~\cite{MDHEH13,GSK18}. In Article 7 (\refsec{Art7}) and Article 8 (\refsec{Art8}), the analysis has been extended to \acrshort{HD} and generalized the description to be dependent on the number of degrees of freedom, \dt and \dr.

\subsection{Homogeneous states for driven granular gases}\labsubsec{HDS_KTGG}

Granular gases in free evolution are very difficult to be studied experimentally. After just a few of collisions per particle, the system is almost frozen. Then, in order to study granular dynamics, usually grains are excited by some external energy injection. In addition, some granular suspensions can be found, where the granular fluid is usually assumed to be interacting with a certain background fluid in terms of Langevin-like models, in analogy to molecular gaseous models. Mathematically, this is translated into a Fokker--Planck-like term in the homogeneous \acrshort{BFPE}. In those situations, the continuous decay of the \index{Temperature!granular}temperature due to inelasticity or roughness mechanisms is counterbalanced by energy injection, either from an external source or from the interaction with a thermal bath. Therefore, the addition of these sources forces the granular gas to reach a steady state. The analysis in these systems takes advantage of quantities already computed from the purely collisional part for the different collisional models. The steady homogeneous driven states must come from the stationary form of \refeq{general_hom_BE},
\begin{equation}
    \frac{\partial}{\partial \vom}\cdot(f\mathbf{G}) = J[\vom|f,f],
\end{equation}
in which we need to specify each collisional model and thermostat.

In the case where the system is in contact with a background fluid, this is usually modeled as a Langevin-like equation for the noncollisional part of the \acrshort{BFPE} due to a coarse-grained consideration of the interaction. Furthermore, whereas a molecular fluid reaches a Maxwellian \index{Velocity distribution function!Maxwellian}\acrshort{VDF} at the \index{Temperature!of the thermal bath}temperature associated with the thermal bath, a granular gas will get a steady \index{Temperature!steady-state}temperature lower than $T_b$, due to the loss of energy upon collisions. Moreover, the stationary \index{Velocity distribution function!steady-state}\acrshort{VDF} is expected to be non-Maxwellian, but much closer to it as compared with the \acrshort{HCS} one. Those are the main conclusions obtained in the application of the \acrshort{IHS} model with Langevin-like interactions for granular suspensions \cite{CVG12,CVG13}. In this thesis, in Article 4 (\refsec{Art4}) we extended these model to the more realistic case in which the drag coefficient\index{Drag coefficient!velocity-dependent} is not constant but velocity-dependent, as described in \refsubsec{nonlinear_KTGG}.

On the other hand, the energy injection may come from an external source without any drag or thermal bath. The aim of this external driving is to compensate for the loss of energy upon collisions and then get a dynamic steady state. This driven system is generally described by a homogeneous stochastic force modeled as a white noise and characterized by a certain noise intensity. In that case, one expects that the final \index{Temperature!steady-state}temperature will be proportional to a proper power of the noise intensity and a function of the coefficients of restitution involved in the selected model. Characterization of granular gases within the \acrshort{IHS} model and driven by a stochastic force of this type has been of interest in the last few decades, were both the steady \index{Temperature!steady-state}temperature and the stationary \index{Velocity distribution function!steady-state}\acrshort{VDF} have been studied from kinetic theory and computer simulations~\cite{vNE98,MS00,CVG12}. The way of studying the non-Gaussianities of driven systems comes from the same type of mathematical tools as used in the \acrshort{HCS}. The first nontrivial cumulants\index{Velocity distribution function!cumulants} of the steady distribution are in general smaller than in the \acrshort{HCS} case. Moreover, the \acrshort{HVT} for the steady-state distribution is not exponential, but a stretched exponential, \phantomsection\label{sym:phiwn}$\phi^{\mathrm{wn}}(\cc)\sim \exp\left( -\gamma_c^{\mathrm{wn}}c^{3/2}\right)$ for $c\gg 1$, that is, between Maxwellian and \acrshort{HCS} behavior. Recently, in Ref.~\cite{CZ22} this $3/2$-exponent for the tail of the distribution has been found to be a characteristic, not only for dilute driven systems, but also for dense ones. In fact, it has been observed that, not only the driving makes the \acrshort{VDF} \emph{more Gaussian}, but also the \index{Instability}instabilities observed in the \acrshort{HCS} vanish~\cite{vNEBO97,vNBE98,vNETP99,VAZ11,GCV13,GCV13b} and the homogeneous driven states become stable under linear inhomogeneous perturbations.

The homogeneous states for rough \acrshort{HS} subjected to a stochastic force have been studied in Refs.~\cite{VS15,MS19}, where steady values of \index{Temperature!steady-state}temperature, \index{Temperature!rotational-to-translational ratio}rotational-to-translational temperature ratio, and first \index{Velocity distribution function!cumulants}cumulants were determined. In this thesis, we generalize the injection of energy using the \acrshort{ST} introduced in \refsubsec{splitting_KTGG}. The stationary and transient homogeneous states are controlled in the \acrshort{ST} by two parameters, which for convenience have been taken as $\{\Twn,\varepsilon\}$. More details about this work can be found in Article 6 (\refsec{Art6}).

\section{Nonequilibrium entropy and $H$-theorem}\labsubsec{math_entropy_KTGG}

The term \index{Nonequilibrium entropy}``entropy'' comes etymologically from the Greek word ``\textgreek{>entrop'h}'' meaning ``change''. This widely-used physical concept was introduced in the origins of classical equilibrium thermodynamics mainly by Rudolf Clausius to account for the increment of \emph{non-usable energy} or irreversibility in his studies about the power of heat and Carnot's theory~\cite{C50}. Then, the second law of thermodynamics was born and the entropy change of a system was defined as a state function whose change is given by \phantomsection\label{sym:SQ}$\dif S = T^{-1}\delta Q$, with $\delta Q$ referring to the heat transferred as if the process were reversible. Moreover, a postulate was formulated: the entropy of the universe tends to a maximum, i.e., $\dot{S}\geq 0$, where the dot notation stands for the time derivative. Next, a statistical meaning for entropy was given by L.~Boltzmann by defining entropy---in the microcanonical ensemble---as \phantomsection\label{sym:overlineW}$S = \ln \overline{W}$, with $\overline{W}$ the number of microstates compatible with a given macrostate of the system~\cite{B95}. Later, Boltzmann defined the entropy production in the context of its kinetic equation and demonstrated the existence of a quantity acting as entropy, proving the irreversibility toward the equilibrium state in this context~\cite{B72,B03}. This result was later called the $H$-theorem and it was under discussion by several scientists of the period, where some thought experiments were fundamental to understand Boltzmann kinetic theory and solve some formulated paradoxes. The implications of this $H$-theorem have been and still are immense. 

From the $H$-theorem, the concept of \index{Nonequilibrium entropy}nonequilibrium entropy started to arise. Entropy was firstly understood only from the point of view of equilibrium systems, without paying attention to nonequilibrium processes. Proper definitions have been under discussion in the community of nonequilibrium thermodynamics~\cite{JCL93} and statistical physics~\cite{BS92,MS96}. For example, in Ref.~\cite{G22} it is proposed that conceptually a \index{Nonequilibrium entropy}nonequilibrium entropy is defined from an extension of the space of equilibrium states by considering nonequilibrium states as well. Then, the function associated with the \index{Nonequilibrium entropy}nonequilibrium entropy is a state function fulfilling the second law of thermodynamics due to the internal evolution of the nonequilibrum variables, exclusively, in isolation. Of course, a \index{Nonequilibrium entropy}nonequilibrium entropy must be connected to equilibrium entropy and it must reduce to the original concept in case of equilibrium. From the original Boltzmann kinetic theory for molecular gases, we will see in \refsubsec{HTheorem_MG_KTGG} that \index{Nonequilibrium entropy}nonequilibrium entropy is related with the already mentioned $H$-quantity, but this definition represents problems in the granular gas case, as introduced in \refsubsec{HTheorem_GG_KTGG}, and developed in \refch{KLD_IHS}.

Finally, the physical interpretation of entropy and its statistical interpretation was the seed of the mathematical information theory, C.~Shannon being its main forerunner from the definition of Shannon's entropy~\cite{S48}, which was actually a mathematically founded formulation of the \phantomsection\label{sym:ShannonH}$H$ functional of Boltzmann's theory, with the aim of \emph{measuring missing information}. Information theory has been useful for several processes, not only physical ones, since its birth. Nowadays, there is a continuous feedback between physical and mathematical knowledge of entropy, which has been specially exploited by statistical physicists.

\subsection{$H$-theorem for molecular gases of hard spheres}\labsubsec{HTheorem_MG_KTGG}

Evolving toward equilibrium is a particular case of dissipation of energy or increment of entropy. Boltzmann derived a transport equation with the purpose of describing how a system out of equilibrium evolves up to it \cite{B72,B03,B86,B86b}. In this process, Boltzmann obtained a quantity which acts like an entropy, that is, it is compatible with the second law of thermodynamics, implying that the progress toward the equilibrium state is irreversible.

Let us work under the assumptions of \acrshort{BE} for an isolated molecular system, i.e., $\mathbf{F}^{\mathrm{ext}}=0$. For the sake of simplicity, a monocomponent rarefied molecular gas of elastic hard spheres is considered, that is, the collisional model described in \refsec{CEM_KTGG} will be used, so that the \acrshort{BE} reads
\begin{align}\labeq{BE_HT}
\frac{\partial}{\partial t}f+\vvel\cdot\nabla f=\sigma^{\dt-1}\int\dif\vvel_2\int_{+}\dif\ssab\, (\vvel_{12}\cdot\ssab)\left(f_1^{\prime\prime} f_2^{\prime\prime}-f_1 f_2\right).
\end{align}    
The irreversibility hidden in the \acrshort{BE} is more general than the described system, but it will be enough to capture its essence, which is contained in the Boltzmann's $H$-theorem \cite{B95,GS03}. The theorem states that the following functional $H$---originally called $E$ by Boltzmann \cite{B72,B03,B86,B86b,C98}---of the one-particle distribution function $f$,
\begin{equation}\labeq{H_functional}
    H(t) = \int \dif\rr \int\dif\vvel\, f(\rr,\vvel,t)\ln f(\rr,\vvel, t), 
\end{equation}
is a Lyapunov functional for the \acrshort{BE} or, equivalently,
\begin{equation}\labeq{S_with_H_functional}
    S(t) = -\left[H(t)-H_\eq\right] +S_\eq
\end{equation}
is the \index{Nonequilibrium entropy}nonequilibrium entropy of the system, with $H_\eq$ the value of the $H$ functional for the final equilibrium distribution and $S_\eq$ the value of the entropy at equilibrium. That is, the theorem ensures that
\begin{align}\label{eq:2ndlaw_cond}
 \dot{S}(t)\geq 0 \quad \forall t\geq 0,   
\end{align}
or, equivalently, $\dot{H}(t)\leq 0$ $\forall t\geq 0$.

In order to prove the statement, we will define the hydrodynamic quantity $\psi(\vvel) = -\ln f(\vvel)$, which, after its insertion into the \acrshort{BE} defined in \refeq{BE_HT}, and integrating over all possible velocities and positions, yields
\begin{align}\labeq{entropy_prod}
    \dot{S}(t) =& \frac{1}{4}\int \dif\rr \int\dif\vvel_1\int\dif\vvel_2 \int_+\dif\ssab\, (\vvab\cdot\ssab) \left(f_1^{\pprime}f_2^{\pprime}-f_1 f_2 \right)\ln \frac{f_1^{\pprime}f_2^{\pprime}}{f_1 f_2},
\end{align}
as the entropy production. The integrand of the entropy production is positive definite because $f\geq 0$, $(\vvab\cdot\ssab)\geq 0$, and, for positive real-valued variables $x,y$, the quantity $(x-y)\ln(x/y)\geq 0$, where the identity is fulfilled if and only if $x=y$. Thus, the time variation of the entropy fulfills condition~\eqref{eq:2ndlaw_cond} and the equality $\dot{S}=0$ is only reached if $\ln f$ is a collisional invariant. From \refsubsec{HMG_KTGG} we know that the set of collisional invariants for the molecular case are $\{1,\vvel,v^2\}$, and it can be proved that all collisional invariants must be a linear combination of those three, i.e, $A_0+\mathbf{A}_1\cdot \vvel+A_2 v^2$~\cite{GS03}, where these coefficients $A_0$, $\mathbf{A}_1$ and $A_2$ can be expressed in terms of the \index{Hydrodynamic!fields}hydrodynamic fields associated with the collisional invariants, i.e,
\begin{equation}\labeq{n_def}
    n(\rr,t) = \int \dif\vvel\, f(\rr,\vvel;t),
\end{equation}
being the number density,
\begin{equation}\labeq{u_def}
    \mathbf{u}(\rr,t) = \frac{1}{n(\rr,t)}\int\dif\vvel\, \vvel f(\rr,\vvel;t),
\end{equation}
the flow field, and $T(\rr;t)$, the \index{Temperature!field}temperature as defined in \refeq{molecular_temp}. Then, it is straightforwardly deduced that the local version of the Maxwellian \index{Velocity distribution function!Maxwellian}\acrshort{VDF} in \refeq{MDF_EHS},
\begin{equation}
	f_{\mathrm{M}}(\rr,\vvel;t) =  n(\rr,t)\left[\frac{m}{2\pi T(\rr,t)} \right]^{\dt/2}\exp\left\{-\frac{m\left[ \vvel-\mathbf{u}(\rr,t)\right]^2}{2 T(\rr,t)} \right\},
\end{equation}
is the local equilibrium solution.

After its formulation, Boltzmann got a lot of criticisms for his $H$-theorem. Some of them came from well-founded paradoxes, such as Loschmidt's, and Poincar\'e--Zermelo's ones~\cite{C98}. Loschmidt observed an apparently contradiction between the $H$-theorem and the reversibility of collisions in the molecular case, together with the time reversibility of Newton's equations of motion. Some authors proposed solutions based on the fluctuation theorem \cite{M21} or the zero-probability measure of second-law-violating initial conditions \cite{HHP87}. Additionally, Poincar\'e and Zermelo pointed out that, from the theory formulated by the former, the so-called Poincar\'e's recurrence theorem \cite{P90} and later extended by Zermelo \cite{B66}, any mechanical system with a finite number of degrees of freedom will eventually pass arbitrarily close to its initial state. This, apparently, contradicted the statement of Boltzmann's $H$-theorem. However, this paradox is solved in the thermodynamic limit and Boltzmann himself inferred a very large time---even larger than the estimated age of the Universe---of recurrence for a gas of $10^{18}$ particles in a box of $1$~$\mathrm{cm}^3$~\cite{B66,C98}. Despite the residual existence of some retractors, the $H$-theorem is fundamental in the kinetic theory of gases and has been generalized to plenty of statistical-physical systems.

\subsection{The problem of the $H$-theorem in granular gases}\labsubsec{HTheorem_GG_KTGG}

The original $H$-theorem applies for whatever molecular interaction, in which the transformation $(\vom_1^{\pprime},\vom_2^{\pprime})\to(\vom_1,\vom_2)$ is unitary, that is, the associated Jacobian $\mathfrak{J}=1$, and its differential cross section stays invariant, i.e., \phantomsection\label{sym:Bgpprime}$B(g^{\pprime}_{12},\overline{\vartheta}_1^{\pprime})=B(g_{12},\overline{\vartheta}_1)$. However, those are not the conditions of either the \acrshort{IHS} or the \acrshort{IRHS} models, which include already an irreversibility factor in their definition of the collisional rules. In those cases, it is known that the described system is always out of equilibrium and does not reach in the long-time limit an equilibrium state but the \acrshort{HCS} in the free-cooling homogeneous case. It is straightforward to show that the Maxwell--Boltzmann \index{Velocity distribution function!Maxwellian}\acrshort{VDF} is not a solution of the inelastic \acrshort{BE} characterized by the collisional operator in \refeq{BCO_IHS} or the inelastic and rough version in \refeq{BCO_IRHS}.

Let us now change a few aspects of the paradigm. We consider grains instead of molecules in homogeneous states and replace the word ``equilibrium'' by \acrshort{HCS}. Then, the relevant question is, is there any Lyapunov functional analogue to the $H$ quantity that ensures the evolution of the granular gas toward the \acrshort{HCS}? Whereas the arguments of \refsubsec{HTheorem_MG_KTGG} are very simple, there is still no answer---either positive or negative---to that question. However, Garc\'ia de Soria \emph{et al.} proposed in Ref.~\cite{GMMMRT15} that the proper \index{Nonequilibrium entropy}nonequilibrium entropy in the case of inelastic granular gases described by the \acrshort{IHS} model would not be related with Shannon's entropy, but with the \emph{relative entropy} or \emph{Kullback--Leibler divergence} (\acrshort{KLD}) of the \acrshort{VDF} with respect to the \acrshort{HCS}. In general, the \acrshort{KLD} of a normalized probability distribution, $f(x)$, with respect to a \index{Velocity distribution function!reference}reference one, \phantomsection\label{sym:fref}$f_{\mathrm{ref}}(x)$, is defined by
\begin{equation}\labeq{KLD_def}
	\KLD(f|f_{\mathrm{ref}}) = \int \dif x\, f(x) \ln \frac{f(x)}{f_{\mathrm{ref}}(x)}.
\end{equation}
The arguments in Ref.~\cite{GMMMRT15} were developed at the level of kinetic Kac's equation, proving the validity only for quasielastic systems. Moreover, they also tested the hypothesis with computer simulations. However, in the works in Article 2 (\refsec{Art2}) and Article 3 (\refsec{Art3}) appearing in \refch{KLD_IHS} of this thesis, we extend and focus the discussion at the level of the \acrshort{BE}, studying the problem for \acrshort{HD} and \acrshort{HS} for all values of the coefficient of restitution. Furthermore, some arguments based on a toy model are made.

Moreover, in this thesis the \acrshort{KLD} as a \index{Nonequilibrium entropy}nonequilibrium entropy has been used also in Article 1 (\refsec{Art1}) and Article 6 (\refsec{Art6}) to study the \acrshort{ME} in molecular and granular gases, respectively. Additionally, we exploit the \acrshort{KLD} not only as a \index{Nonequilibrium entropy}nonequilibrium entropy, but also as a functional quantifying how much a given probability distribution diverges from a reference one, taking advantage of its meaning from information theory. Concretely, it is useful in Article 5 (\refsec{Art5}) to measure the goodness of the \acrshort{SA} in the \acrshort{HCS} for granular gases of inelastic and rough \acrshort{HD} and \acrshort{HS}, as well as in Article 8 (\refsec{Art8}) to detect the appearance of \index{Instability!clustering}clustering in \acrshort{EDMD} simulations.

\section{The Chapman--Enskog method}\labsec{CEM_KTGG}

In order to solve the \acrshort{BE} for molecular realistic nonequilibrium situations or for granular gaseous nonhomogeneous flows, perturbative \index{Velocity distribution function!cumulants}methods\footnote{In homogeneous states we have also introduced the perturbative method coming from the Grad--Sonine expansions explained in \refsubsec{HCS_KTGG} for granular gases, where the perturbative parameters were the cumulants of the \acrshort{VDF}.} become fundamental. For that, we need to define a proper perturbative parameter. In order to infer it, we need to investigate the order of magnitude of typical quantities which can control the evolution of $f$ under different stages. To simplify the scheme, we will ignore external forces without any loss of generalization because the perturbation method will be essentially performed to the \index{Boltzmann collisional operator}Boltzmann collisional operator. Let us denote by \phantomsection\label{sym:tau0}$\tau_0$ a typical time scale, by \phantomsection\label{sym:L}$L$ a typical length scale of the problem, and by $\vth$ the typical velocity scale. Then, the orders of magnitude of terms conforming the \acrshort{BE} of a monocomponent isolated system are
\begin{align}
    \frac{\partial f}{\partial t} =  \mathcal{O}\left(\tau_0^{-1}f\right), \quad \vvel\cdot\nabla f = \mathcal{O}\left(\vth L^{-1}f\right), \quad J[\vom|f,f] = \mathcal{O}\left(n\sigma^{\dt-1} \vth f\right),
\end{align}
where we can relate $n\sigma^{\dt-1}$ to the mean free path for hard $\dt$-spheres,
\begin{equation}\labeq{mean_free_path}
    \ell = \frac{1}{K_{\dt} n\sigma^{\dt-1}}.
\end{equation}
In addition, one can define the mean free time as $\tau=\ell/\vth$. Then, two dimensionless numbers naturally arise. The first one, related to the length scales, is the Knudsen number, \phantomsection\label{sym:Kn}\Kn, defined as
\begin{equation}
    \Kn = \frac{\ell}{L}.
\end{equation}
The second one is $\tau/\tau_0$, which can be related to \Kn via the Strouhal number \phantomsection\label{sym:Sr}$\mathrm{Sr}$, common in the context of continuum theory, as~\cite{C88}
\begin{align}
    \frac{\tau}{\tau_0} = \mathrm{Sr}\,\Kn, \quad \mathrm{Sr}= \frac{L}{\tau_0\vth}.
\end{align}
As a first approximation, one can take $\mathrm{Sr}\simeq 1$, so that time and length scales are comparable, and one could compare the orders of magnitude in both sides of the \acrshort{BE} by means of \Kn. Then, if $\Kn \ll 1$, the mean free path is much smaller than the characteristic length of the system. We will work under this assumption.

\subsection{A brief description of the Hilbert expansion}\labsubsec{HExp_KTGG}

The first attempt to solve the \acrshort{BE} through an expansion in terms of a bookkeeping parameter \phantomsection\label{sym:epsilon_book}$\epsilon \sim \Kn$, assumed to be small, was described by Hilbert~\cite{C88,T08}, who proposed the following series expansion of the \acrshort{VDF}
\begin{equation}\labeq{Hilbert_expansion}
    f=f^{(0)}+\epsilon f^{(1)}+\epsilon^2 f^{(2)}+\cdots.
\end{equation}
In addition, from dimensional and scale analysis, he introduced another $\epsilon$ into the left hand-side of the \acrshort{BE}, that is,
\begin{equation}
    \epsilon\left(\frac{\partial f}{\partial t}+\vvel\cdot\nabla f \right) = J[\vom|f,f].
\end{equation}
Thus, inserting Eq.~\eqref{eq:Hilbert_expansion} into the collisional operator, we can expand the latter as follows,
\begin{subequations}\labeq{J_expansion}
    \begin{align}
    J =&J^{(0)}+\epsilon J^{(1)}+\epsilon^2 J^{(2)}+\cdots, \\
    J^{(0)} =& J[\vom|f^{(0)},f^{(0)}], \quad J^{(n)} = \sum_{k=0}^{n} J[\vom|f^{(k)},f^{(n-k)}].
    \end{align}
\end{subequations}
This scheme leads to the following recurrence problem by equating powers of $\epsilon$,
\begin{align}
    J^{(0)}=0, \quad \frac{\partial f^{(n-1)}}{\partial t}+\vvel\cdot\nabla f^{(n-1)} = J^{(n)},\quad (n\geq 1).
\end{align}
Note that, in the elastic case, the condition $J^{(0)}=0$ ensures that $f_0$ is the Maxwellian \index{Velocity distribution function!Maxwellian}\acrshort{VDF}. However, this expansion faces some mathematical problems as described in \cite{C88}. Some of these are solved by using the \acrshort{CE} expansion.

\subsection{Chapman--Enskog expansion}\labsubsec{CE_Exp_KTGG}

The \index{Chapman--Enskog method}\acrshort{CE} expansion \cite{CC70,C88} is based on a perturbative scheme, not only for the \acrshort{VDF} (as Hilbert's), but also on the derivative operators. This is related to the assumption that the spatio-temporal dependence of the one-particle \index{Velocity distribution function!one-particle}\acrshort{VDF} is through a functional dependence on the \index{Hydrodynamic!fields}hydrodynamic fields of the theory. Let us briefly develop this, first for a monocomponent isolated molecular gas, and then extended to granular gases. 

In nonhomogeneous flows of a molecular gas, one can define the \index{Hydrodynamic!length scale}\emph{hydrodynamic} length scale, \phantomsection\label{sym:h}$h$, which is the typical distance along which the \acrshort{VDF} changes notably. This \index{Hydrodynamic!length scale}hydrodynamic scale is assumed to be of the order of the typical length of the system, $L$. Apart from the lengths $L$, $\ell$, and $h$, a fourth length scale is the particle diameter $\sigma$, which, in the context of the \acrshort{BE} is $\sigma\ll \ell$. Therefore, one can define two other time scales associated with their respective length scales: the duration of a collision \phantomsection\label{sym:tauc}$\tau_c=\sigma/\vth$, which is assumed to be instantaneous in the \acrshort{BE} framework, i.e., $\tau_c\ll\tau = \ell/\vth$, and the \index{Hydrodynamic!time}hydrodynamic time \phantomsection\label{sym:tauh}$\tau_h= h/\vth$. Then, under the assumption of small \Kn, i.e., $\Kn\ll 1$, we have $\tau\ll\tau_h$.

Immersed in this description, one can decompose the evolution of the \acrshort{VDF} into two different evolution stages. First, for times of the order of $\tau$, we have the \emph{kinetic regime}, in which the \acrshort{VDF} relaxes rapidly toward \emph{local equilibrium}. The latter is achieved for times $\tau\ll t\ll \tau_h$, where, given a small region around a space point $\rr$, the \acrshort{VDF} is close to the Maxwellian\index{Velocity distribution function!Maxwellian} one with the values of the proper collisional-invariant \index{Hydrodynamic!fields}hydrodynamic fields at that point. Finally, for times of the order of $\tau_h$, the \acrshort{VDF} evolves slowly up to \emph{global equilibrium} reached at $t\gg\tau_h$. This last evolution stage is dominated by hydrodynamics and is then called \index{Hydrodynamic!regime}\emph{hydrodynamic regime}.

This two-stage description for the \acrshort{VDF} can be extended to granular gaseous flows and has been conveniently adapted to freely evolving inelastic and rough \acrshort{HD} and \acrshort{HS} in Article 7 (\refsec{Art7}). A granular gas in free evolution does not reach equilibrium any more, either locally or globally. Then, the homogeneous base state for this method in the granular context is the \acrshort{HCS}. That is, in the kinetic stage, the \acrshort{VDF} is expected to reach the local version of the \acrshort{HCS} \index{Velocity distribution function!of the HCS}\acrshort{VDF}. In the \index{Hydrodynamic!regime}hydrodynamic stage, the system then relaxes slowly toward the \acrshort{HCS}. According to this picture, one should choose properly the \index{Hydrodynamic!fields}hydrodynamic fields that take place in this description. Straightforwardly from the two collisional-invariant quantities $\{1,\vvel\}$, it is clear that the number density,
\begin{equation}\labeq{n_def_field}
    n(\rr,t) = \int \dif\vom\, f(\vom,\rr;t),
\end{equation}
and the flow field,
\begin{equation}\labeq{u_def_field}
    \mathbf{u}(\rr,t) = \frac{1}{n(\rr,t)}\int\dif\vom\, \vvel f(\vom,\rr;t),
\end{equation}
are fundamental fields in the \index{Hydrodynamic!description}hydrodynamic description of the system\footnote{Notice that \refeqs{n_def_field} and~\eqref{eq:u_def_field} extend \refeqs{n_def} and~\eqref{eq:u_def}, respectively, to the cases where the angular velocities are relevant.}. On the other hand, the \index{Temperature!granular}\index{Temperature!mean granular}granular temperature $T(\rr;t)$ field is not preserved. It is given by
\begin{equation}\labeq{T_CE_IRHS}
	T(\rr;t) = \frac{1}{(\dt+\dr) n(\rr,t)} \int \dif\vom\, \left(mV^2+I\omega^2\right)f(\vom,\rr;t),
\end{equation}
with $\mathbf{V} = \vvel -\mathbf{u}$ the \index{Peculiar velocity}peculiar velocity\footnote{Notice that in \refeq{def_T_IRHS}, being applicable to homogeneous and isotropic states, we took $\mathbf{u}=0$ without loss of generality.}. With this definition of the \index{Temperature!field}temperature field, \refeq{T_CE_IRHS}, we are extending the analysis for the most general case, the \acrshort{IRHS} model, in which surface roughness is accounted for. In spite of the cooling, the mean granular \index{Temperature!mean granular}temperature is assumed to be a good slow hydrodynamic quantity \cite{DB11,G19} for the \acrshort{CE} description\footnote{Alternative expansions around $1-\een$ for nearly elastic \acrshort{HD} or \acrshort{HS} based on the analysis of molecular gases and not considering the \acrshort{HCS} as the base state were done in, e.g., Refs.~\cite{LSJC84,JR85,JM89,J98,GS95,SG98}, but their results did not reproduce satisfactorily well the \index{Transport coefficient}transport coefficient as reported later for \acrshort{CE} method (see, for example, Ref.~\cite{BDKS98,G19}).}. Arguments in favor of this hypothesis come from kinetic model and computer simulation results for the \acrshort{IHS} model \cite{BDKS98,BC01,BD05,SA05,MSG05,MSG07,GSM07,G19}. This reasoning leads to propose the following functional dependence for the \acrshort{VDF} in the \index{Hydrodynamic!regime}hydrodynamic regime, 
\begin{equation}\labeq{CE_hypothesis}
    f(\vom,\rr;t) = f[\vom,\rr|n(t),\mathbf{u}(t),T(t)].
\end{equation}
That is, in order to know the value of the complete one-particle \index{Velocity distribution function!one-particle}\acrshort{VDF} $f$ at a certain instant $t$, it is enough to know the value of the \index{Hydrodynamic!fields}hydrodynamic fields $n$, $\mathbf{u}$ and $T$ at this same time. However, to know the \acrshort{VDF} at the point $\rr$, it is not enough to know the \index{Hydrodynamic!fields}hydrodynamic fields at that point but also, at least, within a neighborhood of the point. This is equivalent to know the gradients of the \index{Hydrodynamic!fields}\index{Hydrodynamic!fields}hydrodynamic fields and, assuming they are small, perform an expansion in those gradients. Hence, not only a Hilbert-like expansion is considered for $f$ [see \refeq{Hilbert_expansion}], but also
 \begin{subequations}\labeq{ders_expansion}
 \begin{align}\labeq{exp_mat_time_der}
    \mathcal{D}_t =& \mathcal{D}_t^{(0)}+\epsilon \mathcal{D}_t^{(1)}+ \epsilon^2 \mathcal{D}_t^{(2)}+\cdots, \\
    \nabla^k \rightarrow & \epsilon^k \nabla^k,
\end{align}
\end{subequations}
where
\begin{equation}\labeq{mat_time_der}
\mathcal{D}_t\equiv\partial_t+\mathbf{u}\cdot\nabla
\end{equation}
is the \index{Material time-derivative}material time-derivative. The functional dependence of $f$ on the \index{Hydrodynamic!fields}\index{Hydrodynamic!fields}hydrodynamic fields established in \refeq{CE_hypothesis} induces the following important property,
\begin{equation}\labeq{prop_Dt_f}
    \mathcal{D}_t f^{(k)} =\mathcal{D}_t n\frac{\partial f^{(k)}}{\partial n}+ \mathcal{D}_t \mathbf{u}\cdot\frac{\partial f^{(k)}}{\partial \mathbf{u}}+\mathcal{D}_t \frac{\partial f^{(k)}}{\partial T},
\end{equation}
which must be applied to any other hydrodynamic quantity $\psi(\vom,\rr;t)$,
\begin{equation}\labeq{prop_Dt_psi}
    \mathcal{D}_t \psi =\mathcal{D}_t n \frac{\partial \psi}{\partial n}+ \mathcal{D}_t\mathbf{u}\cdot\frac{\partial \psi}{\partial \mathbf{u}}+\mathcal{D}_tT\frac{\partial \psi}{\partial T}.
\end{equation}
Here, $\mathcal{D}_t n$, $\mathcal{D}_t\mathbf{u}$, and $\mathcal{D}_tT$ are given by their respective associated \index{Balance equations}balance equations, where the most significant difference with respect to the molecular case is the appearance of the \index{Cooling rate}cooling rate due to the \index{Temperature!granular}\index{Temperature!mean granular}temperature dissipation (see \refapp{app_balance_equations}). Putting together the expansions in \refeqs{Hilbert_expansion},~\eqref{eq:J_expansion},~\eqref{eq:ders_expansion},~\eqref{eq:prop_Dt_f}, and~\eqref{eq:prop_Dt_psi}, a hierarchy of equations appear for the different orders of truncation. The zeroth-order equations are compatible with the local version of the \acrshort{HCS}, while the first order refers to the \acrshort{NSF} transport equations. Comparing \index{Balance equations}balance equations with the \acrshort{NSF} and using phenomenological forms for the fluxes, one can compute the \acrshort{NSF} \index{Transport coefficient}transport coefficients. This methodology has been used in Article 7 (\refsec{Art7}) to compute these coefficients for dilute granular gases of inelastic and rough \acrshort{HD} and \acrshort{HS}.

\chapter{Memory effects}
\labch{ME_KTGG}
\vspace*{-2cm}
\begin{minipage}[t]{.6\textwidth}
\vspace{0pt}
\begin{center}
\includegraphics[width=\linewidth]{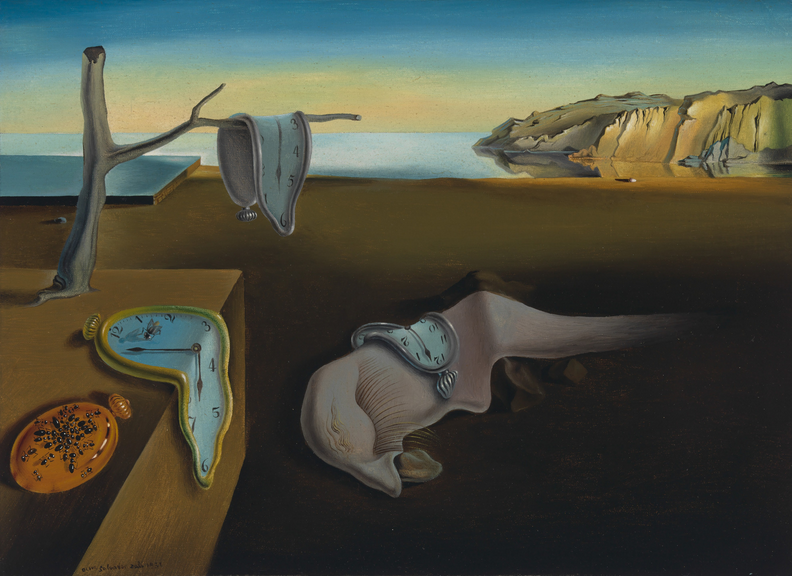}
\\
{\tiny{\textcopyright\, 2023 Salvador Dal\'i, Gala-Salvador Dal\'i
Foundation / Artists Rights Society (ARS), New York}}\\
{Salvador Dal\'i, \emph{La persistance de la m\'emoire} (1931). \href{https://www.moma.org/collection/works/79018}{The Museum of Modern Art, New York}}
\end{center}
\end{minipage}

\vspace*{0.5cm}
\section{Introduction}
For human beings, the ability to remember facts, experiences, dates, images, events and many concepts is quite an important fact that is certainly present during our lifetime. However, the memory softens over time, as Dal\'i's limp watches do. The passage of time wrecks havoc not only on our external appearance, but also on our mind. And this aging cannot be avoided either for humans or for everything in the universe. Nevertheless, we do not only store information in our minds to avoid its loss. Since the very early times of humanity, our ancestors transmitted their art and discoveries to future generations in many different ways: old paintings and constructions, lately writings, then photographs and movies, and nowadays, we also sophisticatedly save data in magnetic materials. We are indeed obsessed about saving information with the aim (sometimes) of progressing in knowledge. But this is not something just human. The universe is actually sharing with us different events of its past, accessible by local and present observations, like recently discovered gravitational waves or the useful cosmic microwave background radiation. Even more, nature is full of these examples. Biologists can study the aging of a tree by just observing the different rings in a transverse section of its trunk. Moreover, the geological history of Earth can be studied by different patterns that are not yet erased. Then, we are surrounded by signs of our natural past, and their identification and interpretation are of huge importance to give an answer to the big question of where we come from. However, as occurs with human minds, the effect of time makes these elements and signs to deteriorate and, at the end, they will completely perish.

We can infer from this argumentation that memory is lost when balance is reached. Therefore, in words of Keim \emph{et al.} in the review paper~\cite{KPZSN19}: \emph{Memory engages us in a study that targets phenomena related to transient or far-from-equilibrium behavior.} And this is the key of memory, at least, from a statistical physics point of view. A system at equilibrium satisfies a Markovian-process description, which by definition is memoryless, namely, it does not preserve---and then, erases---any sort of remembrance about its past. Roughly speaking, this is because the probability of the next state does only depend on the current state of the system. Then, when a system equilibrates, one cannot infer anything about its past or initial preparation. Hence, a memory effect is a phenomenon which emerges during a nonequilibrium stage of a physical system due to any reminiscence of its previous states. In physics, there is a huge number of memory effects found in different kind of systems. Maybe one of the most studied examples is the hysteresis loops in magnetic materials~\cite{RCW76,B98}\footnote{Whereas it was discovered in magnetic systems, it is not restricted to them. See, for example, Refs.~\cite{D14,PSF03}.}. Other examples are echoes~\cite{H50,KPZSN19}, which are characterized by a time reversal path or return of some signal, in analogy to the popular echo of sound; associative memory coming from collective behavior in a complex system and used, for example, in network theory~\cite{H82}; some systems might exhibit memory about the initial conditions in their dynamics, like in air bubble formation~\cite{SKZN09}; among others.

Let us come back to the context of the thesis. Kinetic theory of gases, by definition, describes the evolution of a gaseous system toward equilibrium. Then, this stage of relaxation is a nonequilibrium process and memory effects might arise. Furthermore, inelastic (or rough) interactions make granular gases, by definition, to be out of equilibrium. On the other hand, whereas the classical example of freely evolving elastic hard \dt-spheres is too simple and \emph{very Markovian} to show up any sort of phenomenon of this type, the action of thermostats might induce memory at the level of statistical correlations. In this thesis, we have studied two types of memory effects: the \acrshort{KE} and the \acrshort{ME}, which will be introduced in the next two sections.

\section{Kovacs effect}\labsec{ME_KE}

The \acrshort{KE} is no more than an echo response due to the inner variables of a system, when it is tried to be suddenly equilibrated, attending just to the macrostate. Then, the system is observed first to recede from this reference state, and then relax toward it, guided by some of its macrostate variables. 

\begin{figure}[h]
\centering
\includegraphics[width = 0.98\textwidth]{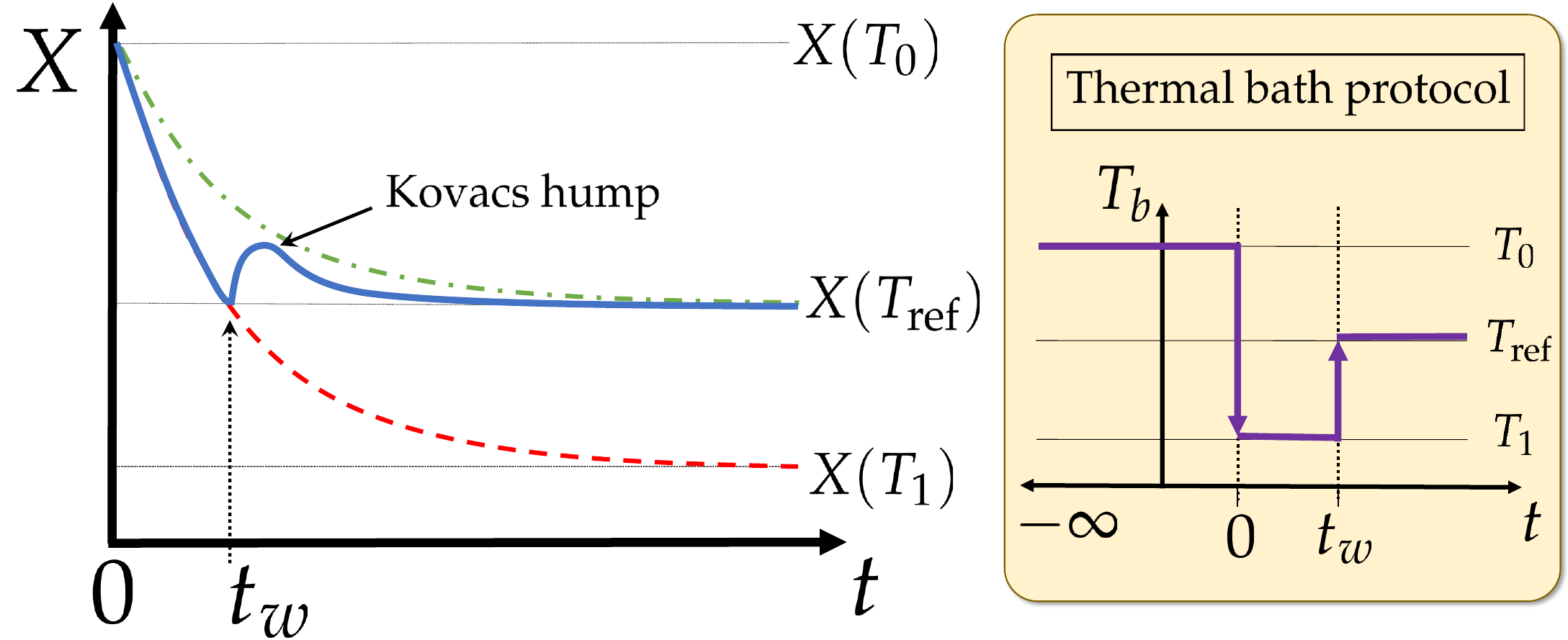}
\caption{Sketch of a general Kovacs-like protocol on a macroscopic variable $X$, with $T_0>T_{\mathrm{ref}}>T_1$. On the left-hand side of the figure there is a graph of the evolution of $X$ during the Kovacs-like experiment, complemented by a sketch of the protocol for the thermal bath \index{Temperature!of the thermal bath}temperature, $T_b$, on the right-hand side. The blue thick continuous curve corresponds to the real time evolution of the $X$ macroscopic variable. The red dashed curve (-- --) refers to the relaxation of $X$ from $T_0$ to $T_1$, while the green dashed-dotted (--\,$\cdot$\,--) line depicts the direct evolution of $X$ from $T_0$ to $T_{\mathrm{ref}}$.}
\labfig{Kovacs_protocol}
\end{figure}

This effect is called after A.~J.~Kovacs due to his discovery of the phenomenon in amorphous polymers, as reported in Refs.~\cite{K63,KSF63}. Kovacs and collaborators first studied the relaxation of a sample of polyvinil acetate in an isobaric cooling process from an initial \index{Temperature!of the thermal bath}temperature, $T_0$, to a lower reference one, $T_{\mathrm{ref}}$. Then, the system was let to equilibrate at a certain volume $V_{\mathrm{eq}}(T_{\mathrm{ref}})$, and this value was annotated. After this first measurement, the original experiment consisted in two main stages. First, the system at the same initial temperature was quenched to a lower \index{Temperature!of the thermal bath}temperature compared to the previous reference one, $T_0>T_{\mathrm{ref}}>T_1$, letting the system now to get the equilibrium volume, $V_{\mathrm{eq}}(T_{\mathrm{ref}})$, at a certain \emph{waiting time}, $t_w$. At this time, they heated the sample up to $T_{\mathrm{ref}}$ immediately, from a sudden change in the thermal bath \index{Temperature!of the thermal bath}temperature which was in contact to the sample. The system is, at this instant, characterized by the \index{Temperature}temperature $T_{\mathrm{ref}}$ and the volume value at equilibrium, $V_{\mathrm{eq}}(T_{\mathrm{ref}})$. Despite the thermodynamic variables being at the equilibrium values, the volume of the sample was observed to first increase and, then, to evolve back to $V_{\mathrm{eq}}(T_{\mathrm{ref}})$ in an isothermal process. Attending to the evolution of the volume, a hump appeared at $t>t_{w}$. Kovacs \emph{et al.} explained this anomaly in Ref.~\cite{KAHR79} by the action of inner variables of the system in a nonequilibrium process. Therefore, whereas its macrostate corresponded to the equilibrium variables, the system was not at the equilibrium state. The system response showed in fact a reminiscence of its previous states.

In general, we can define a Kovacs-like protocol for a macroscopic variable $X$ playing the role of the volume in the original experiment for a system in contact with a thermal bath. Again, a reference \index{Temperature!of the thermal bath}temperature is fixed, and the value of this variable at equilibrium or steady thermalized system is known, i.e., $X(T_{\mathrm{ref}})$. Then, the system from an \index{Temperature!of the thermal bath}initial temperature $T_0$ is cooled down to $T_1$, such that, $T_0>T_{\mathrm{ref}}>T_1$. At the waiting time $t_w$, the system is at $X(T_{\mathrm{ref}})$ and, then, the bath temperature\index{Temperature!of the thermal bath} is suddenly changed to $T_{\mathrm{ref}}$. If the system is at a nonequilibrium state, it is expected to exhibit a nonmonotonic behavior in terms of a Kovacs hump. A sketch of this protocol is represented in \reffig{Kovacs_protocol}. This is the direct generalization of the Kovacs experiment. However, one could also consider a heating process rather than the original cooling one. A system exhibiting the \acrshort{KE} is expected to be dynamically described by a system of coupled differential equations,
\begin{align}\labeq{memory_descr}
	\left\{ \begin{array}{l}
	\dot{X} = F_X(X,Y_1,Y_2,\dots), \\
	\dot{Y_1} = F_{Y_1}(X,Y_1,Y_2,\dots), \\
	\dot{Y_2} = F_{Y_2}(X,Y_1,Y_2,\dots), \\
	\vdots
	\end{array}\right.
\end{align}
with $\{Y_n\}$ being the set of inner variables acting on the evolution of $X$, and $F_X$ and $\{F_{Y_n}\}$ being the corresponding functions for the rates of change of the macrostate guiding variable and inner variables, respectively. Furthermore, it is common to choose $X$ as a nonequilibrium \index{Temperature!nonequilibrium}temperature instead of the volume (see, for example, Refs.~\cite{PB10,PT14,PSP21,MLLVT21}). 

This \acrshort{KE} has been observed in different systems, many of them characterized by slow glassy dynamics~\cite{H94,BB02,AM04,CLL04,MS04,ALN06,MARBS20,LLZDL21}, studied in the context of linear response theory~\cite{PB10,RP14}, in active matter~\cite{KSI17}, observed theoretically in driven granular gases of hard~\cite{PT14} and viscoelastic particles~\cite{MLLVT21}, and also in molecular gases under nonlinear drag~\cite{PSP21}, among other systems. In the latter three works, the guiding quantity is the \index{Temperature!granular}\index{Temperature!molecular}temperature---either granular or molecular---and its evolution equation exhibits a coupling with higher moments of the \index{Velocity distribution function!moments}\acrshort{VDF} playing the role of inner variables of the system. In the case of inelastic granular gases, the \index{Temperature!granular}granular temperature evolution depends explicitly on the whole \acrshort{VDF} throughout the cooling rate expression, i.e., $Y_{n-1} = a_n$ with $n\geq 2$. On the other hand, in the work of molecular gases, the velocity dependence of the drag coefficient (see \refsubsec{nonlinear_KTGG}) implies an explicit coupling of the \index{Temperature}temperature evolution with the fourth cumulant, $Y_1 = a_2$, of the \acrshort{VDF}. However, in the latter example, the evolution of $a_2$ of course depends explicitly on higher-order cumulants. 

Furthermore, we study the Kovacs humps mechanism in an inelastic granular gas described by the \acrshort{IHS} model and interacting with a thermal bath with a nonlinear drag introduced in Article 4 (\refsec{Art4}). As we will see, the \index{Temperature!granular}granular temperature dependence is more complex due to the combination of the latter two couplings.

\section{Mpemba effect}\labsec{ME_ME}

The \acrshort{ME} is usually formulated as a counterintuitive phenomenon in which, given two samples (generally of a fluid) at different \index{Temperature}temperatures, the initially hotter one may arrive more quickly to equilibrium than the other colder sample. We will see throughout this section that more interpretations of the \acrshort{ME} are possible.

The effect was originally studied in water. In the 1960s, it was observed that hot water may freeze down faster than initially colder samples, as reported by the person that gave it the name to the \acrshort{ME}, Erasto B.~Mpemba. At that time, E.~B.~Mpemba (1950 -- 2020) was a Tanzanian high-school student, who was helped by D.~Osborne (1932 -- 2014), publishing the finding in Refs.~\cite{MO69,O79}. It can be said that the discovery of the effect by Mpemba was a stroke of serendipity. In 1963, Mpemba had to make ice cream with his schoolmates. The method they applied consisted in boiling a mixture of milk and sugar and, after a first cooling at room temperature, the mixture was put inside a refrigerator. However, one day, due to the high demand of the refrigerator usage, Mpemba decided to enter directly the boiling milk inside the refrigerator---skipping the second step of the recipe---at the same time a schoolmate entered his tempered sample. An hour later, Mpemba was astonished to observe that his sample had already frozen down, whereas his schoolmate's had not. Mpemba asked his physics teacher about that, but the teacher did not contemplate that Mpemba's observation could be possible, and argued that Mpemba was certainly confused. Mpemba first believed his teacher, but then, ice-cream sellers agreed with him. Young Erasto insisted in his findings, whereas his thoughts were disregarded at the high school until Dr. Denis Osborne appeared. Osborne visited the school to give a talk about ``Physics and national development'', and during the question turn Mpemba asked about the effect he observed. Dr. Osborne, instead of ignoring him, was curious about this observation for which he had not an answer, and asked a technician to reproduce it. The results were in favor of Mpemba's findings and they were later reported in Ref.~\cite{MO69}, where experimental results of the time a sample of water took to start freezing against the initial \index{Temperature}temperature were presented. It is reproduced in \reffig{mpemba_1969}. It is curious that, despite his discovery, Mpemba never studied physics but devoted himself to the preservation of the wildlife in a Tanzanian natural reserve.

\begin{figure}[ht]
\centering
\includegraphics[width=0.4\textwidth]{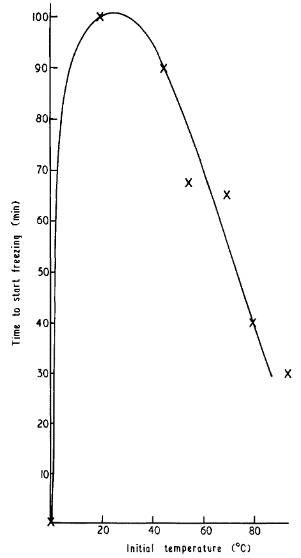}
\caption{E.~B.~Mpemba and D.~G.~Osborne, Cool?, \emph{Phys. Educ.} \textbf{4} pp.~172-175 (1969), DOI:~\href{www.doi.org/10.1088/0031-9120/4/3/312}{10.1088/0031-9120/4/3/312}. \textcopyright IOP Publishing. Reproduced with permission. All rights reserved. Graph of the fitting curve for several experimental points reproducing the time a sample of liquid water started to freeze up in minutes, with respect to the initial \index{Temperature}temperature of the cooling process. The results come from the original and very first experiments of Mpemba and Osborne.}
\labfig{mpemba_1969}
\end{figure}

One can say that the \acrshort{ME} is an example of Stigler's law of eponymy~\cite{S80}, since Mpemba was not actually the first one in observing the effect, but classical authors like Aristotle~\cite{aristotle_works_1931}, F.~Bacon~\cite{Bacon1620}, or R.~Descartes~\cite{Descartes1637} commented it before. Moreover, the \acrshort{ME} was common in the popular belief of cold countries, whereas scientifically it was hidden under Newton's law of cooling~\cite{Newton1701} which, by definition, denies the existence of this phenomenon. Although the \acrshort{ME} in water is not a very recent issue in science, it is still unsolved~\cite{E18}. In fact, the scientific community is divided into supporters, who observe the effect experimentally and propose different explanations~\cite{K69,F71,D71,F74,G74,W77,F79,K80,H81,WOB88,A95,K96,M96,J06,ERS08,K09,VM10,B11,VM12,BT12,ZHMZZZJS14,VK15,S15,JG15,IC16,GLH19,BKC21,THZLZ23} and detractors, who are skeptical and either consider that the experimental methods used were flawed or give negative results~\cite{B06b,BL16,BH20,ES21}.

Whereas the discussion in water is still alive, the \acrshort{ME} went beyond this substance and has been studied in a vast variety of complex physical systems, like carbon nanotube resonators~\cite{GLCG11}, clathrate hydrates~\cite{AKKL16}, Ising models~\cite{GMLMS21,TYR21,VD21}, spin glasses~\cite{Betal19}, non-Markovian mean-field systems~\cite{YH20,YH22}, active systems~\cite{SL22}, quantum systems~\cite{CLL21,CTH23}, ideal gases~\cite{ZMHM22}, nonequilibrium systems with Markovian dynamics~\cite{LR17,KRHV19,CKB21,BGM21}\footnote{At the beginning of this chapter it was said that to observe memory effects we cannot have Markovian processes. Whereas the dynamics are Markovian, in these models the systems are initially very far from equilibrium and suffer a first nonequilibrium relaxation in usually very asymmetric potential landscapes, playing the role of the necessary \emph{non-Markovianity}.}, models based on Landau's theory of phase transition~\cite{HR22}, and colloidal experiments~\cite{KB20,KCB22}, among others. In 2017, Lasanta \emph{et al.} described the \acrshort{ME} in granular gases in Ref.~\cite{LVPS17} for homogeneous states, both in free evolution and driven by a stochastic thermostat, with the binary collisions being described by the \acrshort{IHS} model. Afterwards, it has been observed in other collisional models, like the \acrshort{IRHS} one~\cite{TLLVPS19}, Maxwell models~\cite{BPRR20}, or viscoelastic particles~\cite{MLLVT21}; also in anisotropically driven granular gases~\cite{BPR21,BPR22}, inertial suspensions~\cite{THS21,T21}---where the \acrshort{ME} has been described also by the viscosity of the system as the guiding macrostate quantity\cite{THS21}---, or binary mixtures~\cite{GG21}. Furthermore, Santos and Prados described the conditions for the observation of the \acrshort{ME} in a molecular gas in contact with a background fluid~\cite{SP20}, where the interaction between the gas and the background fluid was described by the coarse-grained model explained in \refsubsec{nonlinear_KTGG}. Later, in Ref.~\cite{PSP21}, a glassy behavior and its implication on the \acrshort{ME} in the same system was detailed.

Therefore, the \acrshort{ME} is observed in several nonequilibrium processes. However, the nature of this effect is not unique and, unlike the \acrshort{KE}, it is not described by a given \emph{a priori} protocol. There are mainly two different ways to look for \acrshort{ME} in a complex system. The first one is due to an anomalous relaxation of the samples, and the second one consists in considering that the effect is immersed in a phase transition, like occurs in models based on Landau's theory. In the terminology of water studies, the first methodology corresponds to an anomalous cooling, whereas the second one refers to differences between both samples in the freezing process. In this thesis, following the granular gas methodologies, we considered the anomalous relaxation picture. Then, we could categorize the \acrshort{ME} as a memory effect due to a dependence of the dynamics on the initial conditions or initial preparation.

Furthermore, it is worth mentioning that the effect has been originally studied in terms of the temperature associated with the physical system. However, in other situations the definition of \index{Temperature}temperature is meaningless, and then the relaxation of its probability distribution through a \index{Nonequilibrium entropy}nonequilibrium entropy functional is used as the guiding quantity. We will refer from now on as \emph{thermal} \acrshort{ME} (\acrshort{TME}) or \emph{entropic} \acrshort{ME} (\acrshort{EME}) if the description is the former or the latter, respectively. Interestingly, these two descriptions can be equivalent or not, depending on the studied system. On the one hand, the \acrshort{TME} states that the \index{Temperature}temperature of the hotter system arrives earlier to its steady-state value as compared with an initially colder one, according to the definition given at the very beginning of this section. On the other hand, the \acrshort{EME} corresponds to the effect associated with the possibility that a sample starting from an initial state further from equilibrium (or steadiness), as compared to other sample of the same system, evolves earlier toward it, as measured by some entropy functional. The latter quantity would play the role of a \index{Nonequilibrium entropy}nonequilibrium entropy, typically with a negative sign to study a monotonic decreasing quantity, instead of an increasing one. Intuitively, if the relaxation times of the higher moments of the distribution are comparable to, or greater than, the typical time corresponding to the hotter sample overpassing the relaxation of the colder sample, then the entropic description may yield results different from the thermal one. By contrast, if the relaxation of those higher moments is much more rapid than the \index{Temperature}temperature relaxation (so that, the probability distribution is fully described by the instantaneous temperature), both descriptions are then expected to be equivalent. Thus, one must be cautious in the choice of the framework, in order to interpret properly the results. 

Another point to highlight is how to predict theoretically the \acrshort{ME}. Let us focus  on the anomalous cooling context and kinetic-theory based systems, where one looks for a dynamical description like \refeq{memory_descr}, with $X$ being either the \index{Temperature}temperature in a \acrshort{TME} or the \index{Nonequilibrium entropy}nonequilibrium entropy in an \acrshort{EME} description. In the context of the \acrshort{TME}, we will work firstly under the assumption that the \index{Temperature!nonequilibrium}temperature curve does not \index{Temperature!overshoot}overshoot the steady value, i.e., \phantomsection\label{sym:Tst}$T(t)-T^{\mathrm{st}}\geq 0$ $\forall t\geq t_0$, $T^{\mathrm{st}}=\lim_{t\rightarrow \infty} T(t)$ being the steady-state \index{Temperature!steady-state}temperature, and $t_0$ the initial time. Within this analysis, the appearance of the \acrshort{TME} is used to be linked with a single crossing of the two \index{Temperature!nonequilibrium}temperature curves, usually at short times. Then, a linear analysis of the relaxation with respect to the initial conditions applies, see Refs.~\cite{LVPS17,SP20,BPRR20,GG21,BPR21,MLLVT21,BPR22}. In these works, the authors can predict analytically the crossing time, i.e., the moment when both curves intersect. Nevertheless, the effect is not generally just restricted to short times, as observed in Ref.~\cite{TLLVPS19}. Therefore, in general, the \acrshort{TME} must be detected if an odd number of crossing times appear, something that can be extrapolated to the \acrshort{EME}. If there is no crossing, or there are an even number of them, the initially hotter (or further from steadiness) sample will in the end take longer to reach the final state than the initially colder sample. 

Although it may seem that those are the only possibilities for observing \acrshort{ME}, this is not the case for \acrshort{TME}. The reason resides in the initial assumption $T(t)-T^{\mathrm{st}}\geq 0$ for all finite $t\geq t_0$. If this condition is not fulfilled, even appearing a crossing (or an odd number of them), the initially hotter sample could overshot the value $T^{\mathrm{st}}$ in such a way that it will ultimately arrive the last one to the final steady state. Let us see the problem the other way around. One can face the case of two samples initially at different temperature, whose associated evolution curves do not intersect, but the initially colder system \index{Temperature!overshoot}overshoots the $T^{\mathrm{st}}$ value and finally reaches the asymptotic value later than the other sample. This is an example of \acrshort{TME} without any crossing point, and we called it \emph{overshoot} \acrshort{ME} (\acrshort{OME}) in Ref.~\cite{MSP22}, which recalls the supercooling mechanism proposed for water~\cite{A95}. As a matter of fact, this cannot occur in the entropic interpretation because of a positive semidefinite \emph{thermal distance} for the difference $T-T^{\mathrm{st}}$ would constrain us to work just on the number-of-crossing detection framework, capturing the \acrshort{OME} to this classical interpretation.

The last but not least remark is that the \acrshort{ME} experiment can also be generalized to a heating process in lieu of the cooling one. This is usually referred to as \emph{inverse} \acrshort{ME} (\acrshort{IME}) and it is only meaningful in the context of \acrshort{TME}. If a system admits \acrshort{IME}, then one could find an initially colder sample that warms up more quickly than a hotter one. Again, a thermal distance for $T-T^{\mathrm{st}}$ already considers these cases with the same detection mechanism based on the crossing points. On the contrary, the \acrshort{TME} appearing in the classical cooling experiment is sometimes called \emph{direct} \acrshort{ME} (\acrshort{DME}). See \reftab{phenom_ME} for a visual characterization of all the explained phenomenology.

In Article 1 (\refsec{Art1}), we study the \acrshort{ME} phenomenology in the model proposed in Ref.~\cite{SP20} of a molecular gas under nonlinear drag. Differences between \acrshort{TME} and \acrshort{EME} are described, as well as the introduction of the \acrshort{OME}. Moreover, in Article 4 (\refsec{Art4}), the \acrshort{TME} is studied for a granular gas also under nonlinear drag. In the kinetic-theory based systems, it is difficult to create a protocol to generate the proper initial states of the dynamics needed for the emergence of the effect. The initialization protocol is something fundamental if the effect is wanted to be replicated in real experiments. Takada \emph{et al.} proposed in Ref.~\cite{THS21} a protocol based on sheared states. In Article 6 (\refsec{Art6}), we develop a protocol to generate the proper initial conditions, but in homogeneous states for a granular gas based on the \acrshort{IRHS} model with the \acrshort{ST}.
\newpage
\begin{landscape}
\begin{table}[t]
\caption{Classification of sketches for the different scenarios that can be found in a dynamical study of the \acrshort{ME}. The figure is divided into the \acrshort{TME} and the \acrshort{EME} interpretations, the latter being guided by the quantity $\KLD$ [according to its use in Article 1 (\refsec{Art1})]. The \acrshort{TME} is described in the \index{Temperature!picture}temperature picture, where the effect is studied by means of the \index{Temperature!nonequilibrium}nonequilibrium temperature $T$, and the thermal-distance picture, where $\mathfrak{D}$ is the chosen quantity [according to the notation of the quantity used in Article 6 (\refsec{Art6}), defined such that $\mathfrak{D}=0$ if and only if $T=T^{\mathrm{st}}$]. Both descriptions are applied to the the \acrshort{DME} and the \acrshort{IME}, in their standard form or via the \acrshort{OME}. The red curves refer to the initially hotter system, whereas the blue ones to the initially colder one. In addition, thick continuous lines correspond to the system initially further from the steady state, whereas the dashed (-- --) ones represent the initially closest. Notice that in the \acrshort{EME} there is neither red nor blue lines because temperature could be meaningless. The green points appearing in the different intersections are crossing points. Finally, notice that the different scenarios are simplified to a single crossing, which is equivalent to find an odd number of them in the \acrshort{EME} or in the thermal-distance picture of the \acrshort{TME}.}
\labtab{phenom_ME}
\begin{tabular}{c||c|c||c|c||c}
\hline\hline
&\multicolumn{4}{c}{\acrshort{TME}}&\multicolumn{1}{c}{\acrshort{EME}}
\\
\cline{2-6}
 &\multicolumn{2}{c}{standard \acrshort{TME}}&\multicolumn{2}{c}{\acrshort{OME}}&\multicolumn{1}{c}{\acrshort{EME}}\\
 \cline{2-6}
 &Temperature&Thermal distance&Temperature&Thermal distance&Entropy\\
\cline{2-6}
\begin{tabular}{c}\acrshort{DME}\end{tabular}&
\begin{tabular}{c}\\ \includegraphics[clip,width=\ancho\columnwidth]{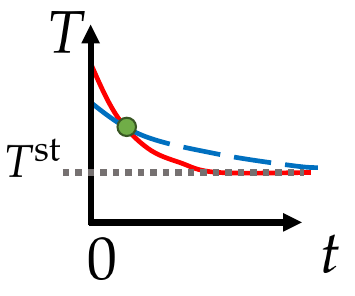}\end{tabular}&
\begin{tabular}{c}\\ \includegraphics[clip,width=\ancho\columnwidth]{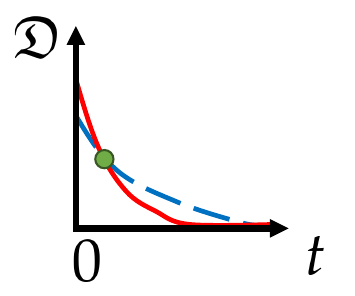}\end{tabular}&
\begin{tabular}{c}\\ \includegraphics[clip,width=\ancho\columnwidth]{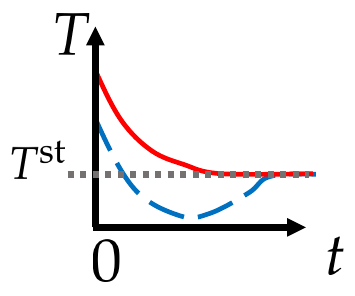}\end{tabular}&
\begin{tabular}{c}\\ \includegraphics[clip,width=\ancho\columnwidth]{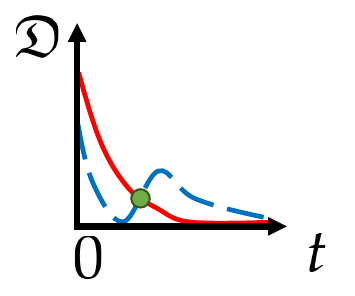}\end{tabular}&
\multirow{2}{*}{\begin{tabular}{c}\\ \includegraphics[clip,width=\ancho\columnwidth]{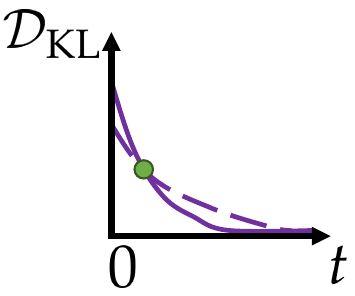}\end{tabular}}\\
\cline{1-5}
\begin{tabular}{c}\acrshort{IME}\end{tabular}&
\begin{tabular}{c}\\ \includegraphics[clip,width=\ancho\columnwidth]{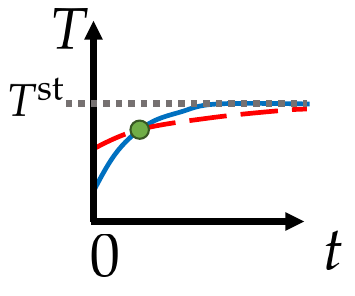}\end{tabular}&
\begin{tabular}{c}\\ \includegraphics[clip,width=\ancho\columnwidth]{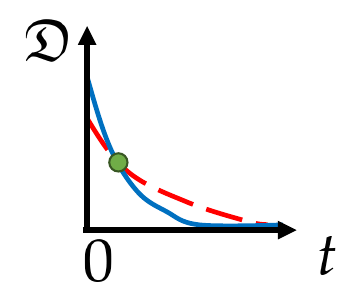}\end{tabular}&
\begin{tabular}{c}\\ \includegraphics[clip,width=\ancho\columnwidth]{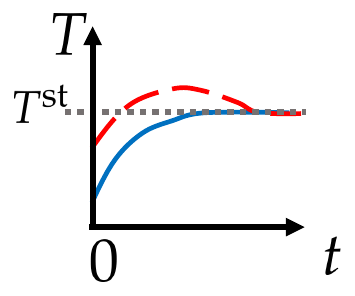}\end{tabular}&
\begin{tabular}{c}\\ \includegraphics[clip,width=\ancho\columnwidth]{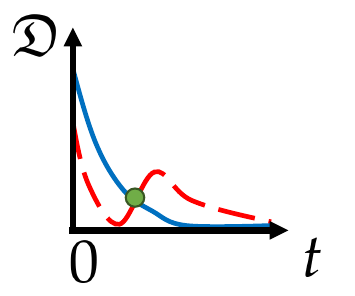}\end{tabular}&\\
\hline\hline
\end{tabular}
\end{table}
\end{landscape}

\chapter{Simulation Methods}
\labch{SM}
\vspace*{-2cm}
\begin{minipage}[t]{.6\textwidth}
\vspace{0pt}
\begin{center}
\includegraphics[width=0.7\linewidth]{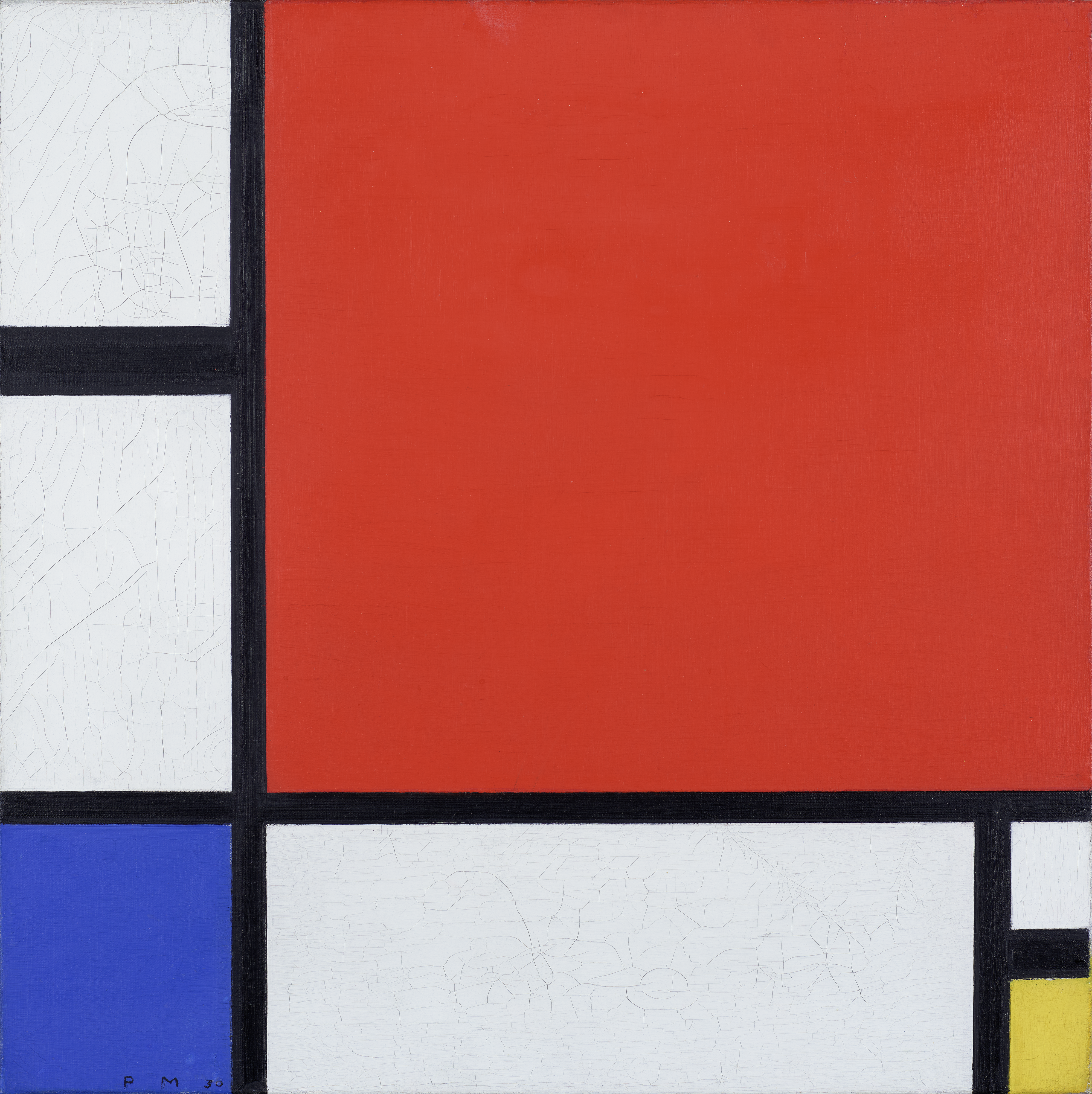}
\\
Piet Mondrian, \emph{Komposition mit Rot, Blau und Gelb} (1930). \href{https://collection.kunsthaus.ch/en/collection/item/2455/}{Kunsthaus Z\"urich}.
\end{center}
\end{minipage}

\vspace*{0.5cm}
\section{Introduction}

In order to check the validity of the theoretical results derived in this thesis, it is important to compare them with computer simulations as a first \emph{quasi-experimental} approach. In the granular matter world, different computational schemes have been developed. In this thesis, only methods applicable to low-density states have been considered, with \acrshort{DSMC} and \acrshort{EDMD} methods being the most common ones. Whereas the former is stochastic, the latter is deterministic, and both give us different useful pieces of information to compare with the theoretical predictions and approaches. Although molecular dynamics (\acrshort{MD}) schemes are older than \acrshort{DSMC}, the latter is firstly introduced here due to its \emph{simplicity}.

\section{Direct Simulation Monte Carlo method}\labsec{DSMC}

The \acrshort{DSMC} method consists in a probabilistic computational method to simulate a physical rarefied gaseous system. This method was developed sixty years ago by G.~A.~Bird, who firstly proposed it in Ref.~\cite{B63} as an alternative to \acrshort{MD} simulations, which tend to be computationally harder and slower at low densities. Despite its lack of attention during the first few years, it became later a fundamental simulation method in the rarefied gas dynamics community. The original algorithms~\cite{B94,B13} have been extended and continuously improved to an ample variety of systems, being not only important in molecular gases but also in dilute granular gases~\cite{PS05}, atmospherical systems~\cite{AG00}, aerospace problems~\cite{SC13}, astrophysical flows~\cite{W14}, etc. 

 The method uses Monte Carlo techniques to determine the time-dependent \acrshort{VDF} from \emph{quasi-particles} or \emph{simulators}, each one being a \emph{probability quantum} in the phase space, and representing a great number of real particles. Formally speaking, each simulator is a representative state of a stochastic variable. From now on, we will abuse the language by terming particles to these simulators. The statistical base of the method is closely related to the \acrshort{BE}, even though it is formally considered as a numerical resolution of it~\cite{W92,S19}. That is, the particle positions and velocities are random variables following a stochastic process based on the \acrshort{BE}, where the Monte Carlo-like scheme is immersed in the collisional part of the process using an acceptance-rejection Metropolis-like criterion~\cite{MRRTT53} adapted to a proper scattering process~\cite{B94,B13}. Unless otherwise indicated, only hard \dt-sphere interactions are considered.

 Basically, the method assumes that the evolution of the \acrshort{VDF} can be separated into a collisional and free-streaming part, as considered in the deduction of the BE [see Eq.~\eqref{eq:dfdt_str_coll}]. As every spatio-temporal numerical finite-difference method, time and space are discretized. Space is usually divided into cells of typical length smaller than the mean free path. However, in this thesis only homogeneous schemes for \acrshort{DSMC} have been considered and, therefore, positions play no role anymore in the method. Moreover, in homogeneous flows, the free streaming part is only characterized by the action of the external forces. In this thesis, external actions on the system have only being considered through thermostats, as explained in \refsec{therm_models_KTGG}.

 Let us assume a \acrshort{DSMC} scheme for a homogeneous gas of hard $\dt$-spheres simulated by $N$ particles with velocities $\{ \vom_i(t)\}_{i=1}^{N}$ at a certain instant $t$. The discrete \acrshort{VDF} for the numerical scheme is given by,
\begin{equation}\label{eq:discrete_VDF}
     n^{-1} f(\vom;t) = \frac{1}{N}\sum_{i=1}^{N}\delta(\vom_i(t)-\vom).
\end{equation}

At the very beginning, it is necessary to initialize the set of velocities in accordance with a certain probability distribution. According to \refeq{discrete_VDF}, the time evolution of the discrete \acrshort{VDF} is driven through the change of the velocities of the particles. After this first initialization process, velocities are updated from $t$ to  \phantomsection\label{sym:timestep}$t+\Delta t$, where the time step $\Delta t$ is much smaller than the mean free time. This is done by splitting the algorithm into the two different stages mentioned before.

In the collisional stage, the method approximates the change of the \acrshort{VDF} due to an estimation of collisional encounters. First, a number \phantomsection\label{sym:floor_op}$\lfloor \frac{1}{2}N\varpi_{\max}\Delta t\rfloor$ of pairs~\cite{B94,MS00,PS05,SM09,B13} are randomly and equiprobably chosen, with \phantomsection\label{sym:varpimax}$\varpi_{\max}$ being an upper bound estimate for the one-particle collision rate, such that the ignored decimals in the before rounding off are saved for the next iterative step. Given a chosen pair $ij$, a collision is accepted with probability \phantomsection\label{sym:varpi_ij}$\varpi_{ij}/\varpi_{\max}$, where $\varpi_{ij}=\Omega_{\dt}n\sigma^{\dt-1}\left|\vvel_{ij}\cdot\ssab_{ij}\right|$ is the estimator of the collision rate for the specific pair, $\vvel_{ij}\equiv \vvel_i-\vvel_j$ being the relative velocity of the particles involved in the chosen pair and $\ssab_{ij}$ being a random vector drawn from a uniform probability distribution in the unit $\dt$-sphere. If the collision is accepted for such a pair, the velocities of the particles are updated according to the collisional rules corresponding to the chosen model, $\vom_{i,j}(t+\Delta t)=\mathfrak{B}_{ij,\ssab_{ij}}\vom_{i,j}(t)$. 

Whereas $\varpi_{\max}$ is usually chosen as a constant, this is not the case for freely evolving granular gases~\cite{PS05,SM09}, where this quantity must be updated due to the continuous cooling of the system. This can be done in two ways: first, since $\varpi_{\max}$ is usually related to a maximum velocity proportional to $\vth(t)$, one can rescale $\varpi_{\max}$ according to the instantaneous value of $\vth(t)$, which is related to the \index{Temperature}temperature; another way is to rescale all velocities periodically\footnote{This method used to work better than the first one. This occurs because, if we simply correct $\varpi_{\max}$, one might get very small values in large runs due to the continuous cooling, not only in the value of the estimated frequency, but, more importantly, in all velocity vectors. Then, numerical errors due to the computational management of float numbers likely appear.} to a fixed value of the mean \index{Temperature}temperature, saving the multiplicative factors to reconstruct the real instantaneous temperatures of the system. Notice that the latter method does not affect the evolution of the \emph{rescaled} \acrshort{VDF}, but only the temperature quantities. In order to be more accurate, not only in the freely evolving case, but in every considered system, $\varpi_{\max}$ should be related to the real maximum velocity of the system. However, this consideration has a computational cost in the algorithm.

Afterwards, in the free-streaming stage, if there is an external force, velocities are updated according to its action. In this thesis, only thermostats have been considered and, therefore, all the velocities at $t+\Delta t$ are modified according to the associated Langevin-like stochastic differential equation (\acrshort{SDE}). A general \acrshort{DSMC} algorithm for our homogeneous (granular) gas can be read in \refalg{DSMC_main}. 

\begin{algorithm}[H]
\begin{algorithmic}[1]
\State Initialize the $N$ velocity vectors $\{\vom_i\}_{i=1}^N$. Set to zero the number of collisions {\verb+number_colls+} and the cumulative decimals of sampled collisions {\verb+dcolls+}. Set $t\leftarrow 0$.
\State If the predefined condition for the velocity rescaling is satisfied, rescale the velocities.
\State Get $\varpi_{\max}$.
\State Estimate the number of colliding particles: {\verb+ncolls+}$\leftarrow \lfloor\frac{1}{2}N\varpi_{\max}\Delta t\rfloor+${\verb+dcolls+}. Save the decimals not considered in the before estimation: {\verb+dcolls+}$\leftarrow \frac{1}{2}N\varpi_{\max}\Delta t-${\verb+ncolls+}.
\State Get the sample of colliding pairs randomly: {\verb+i1+}$\,\leftarrow\,${\verb+int_rand+}$(N$,{\verb+ncolls+}), {\verb+i2+}$\,\leftarrow(i$+{\verb+int_rand+}$(N-1$,{\verb+ncolls+}$)+1)$\% $N$.
With {\verb+int_rand+}$(N$,{\verb+ncolls+}) get a vector of {\verb+ncolls+} components whose elements are integer random number up to $N$.
\State Set $k\leftarrow 0$.
\State \textbf{for} $0\leq k<${\verb+ncolls+} \textbf{do}:
 \State Set $\{i,j\}\leftarrow$ $\{$\texttt{i1}[$k$],\,\texttt{i2}[$k$]$\}$.
 \State Get the unit intercerter vector $\ssab_{ij}$ from a uniform random distribution.
 \State Apply the acceptance-rejection algorithm. If the collision is accepted, update the velocities of the colliding pair according to the binary collisional rules: $\vom_{i,j}\leftarrow \mathfrak{B}_{ij,\ssab_{ij}}\vom_{i,j}$.
\State \textbf{end for}
\State Increase the number of collisions by the number of accepted collisions {\verb+icolls+}: {\verb+number_colls+}$\leftarrow${\verb+icolls+}.
\State If a thermostat is considered, update the velocities by solving numerically the associated Langevin-like \acrshort{SDE} between $t$ and $t+\Delta t$.
\State Update $t\leftarrow t+\Delta t$.
\State If the final condition is reached, this is the end of the program. Otherwise, go back to step 2.
\end{algorithmic}
\caption{General algorithm for \acrshort{DSMC}.}\labalg{DSMC_main}
\end{algorithm} 


\section{Event Driven Molecular Dynamics}\labsec{EDMD}

Whereas the \acrshort{DSMC} method is a stochastic algorithm based on the assumptions underlying the \acrshort{BE}, the \acrshort{MD} technique solves numerically the equations describing the dynamics of the constituents of a physical system. 

If we have a classical, many-body mechanical system, \acrshort{MD} solves Newton's equations of motion for the particles in order to elaborate a computational experiment similar to a real one. In fact, the idea behind a \acrshort{MD} algorithm is very close to the idea of an experiment: we first prepare the system (sample) and we measure numerically certain quantities during the process. The measurements are subdued to statistical errors that are avoided as much as possible by replicating its realization several times. The observables computed from a \acrshort{MD} simulation must be functions of the positions and momenta of the particles. 

Newton's equations of motion are numerically solved in this technique, which depends on the interaction potential of the system constituents. This is usually named as \emph{force-based} \acrshort{MD} algorithms. For integrating the equation of motion, it is necessary to discretize time in units of a time step, $\Delta t$. The system advances in multiples of this time step and the integration is reduced to a discretization algorithm for solving differential equations. Additionally to this scheme, we need to define the proper boundary conditions according to the experiment we want to computationally reproduce.

However, in this thesis only hard interactions are considered, such that the potential is singular and, therefore, no interparticle force is actually considered in this case. Instead, hard particles interact via collisions, so that, if no collision occurs, the particles evolve in uniform rectilinear motion in the absence of external forces or accelerate in case they are moving under a certain force field. Let us assume that we use a \emph{time-driven} molecular dynamics (\acrshort{TDMD}) algorithm for simulating $N$ hard \dt-spheres under the absence of external forces. Then, after each time step $\Delta t$, we must update positions of the hard particles according to their constant velocity and rectilinear motion and check whether there is any overlap. In case there are overlaps, we need to go back to the time when the collision occurs and update the velocities according to the collision rules and, then, continue with the simulation. The idea of this kind of algorithm is simple and is an extension of a typical force-based \acrshort{MD} simulation. Nevertheless, this approach has mainly two disadvantages. First, we need to define a very small---compared with the mean free time---$\Delta t$ in order not to miss any collision. The other one is the need of performing a $\mathcal{O}(N^2)$ checking of the overlaps for each time step, which is computationally very inefficient, and even more if $\Delta t$ is restricted to be small.

Hence, a way of solving those drawbacks is to impose the progress of the system by just considering the times where an event occurs, this events commonly being the particle-particle collisions and particle-boundary interactions. This type of algorithms is called \emph{event-driven} molecular dynamics (\acrshort{EDMD}), and its challenge is to efficiently list and update the events characterizing the system improving the complexity associated with an ordinary time-driven algorithm. The very first \acrshort{EDMD} algorithms were proposed in the pioneering works of Alder and Wainwright \cite{AW57,AW59}. In this thesis, in Article 2 (\refsec{Art2}) and Article 3 (\refsec{Art3}), the \acrshort{EDMD} simulations were performed by using the software DynamO~\cite{BSL11}. However, in the other works described in Article 1 (\refsec{Art1}), Article 4 (\refsec{Art4}), Article 5 (\refsec{Art5}), Article 6 (\refsec{Art6}), and Article 8 (\refsec{Art8}), a hand-made \acrshort{EDMD} Python-based program created from scratch was used~\cite{M23_github}. This latter program is sequential, since it was proved to be fast enough acting on the considered problems, but parallel implementations could speed up the algorithms~\cite{ML04}. Depending on the problem, a vast number methods to improve and accelerate the \acrshort{EDMD} has been proposed, see for example Refs.~\cite{I99,I16,I17,IK15,KE19} .A brief description of a general \acrshort{EDMD} program and specific algorithms used in our own are introduced in what follows.

\subsection{A general EDMD algorithm}\labsubsec{general_EDMD}

In principle, an \acrshort{EDMD} algorithm is based on the steps presented in \refalg{EDMD_basic}. Here, it is assumed the absence of external forces.\footnote{If there is any deterministic external force field affecting the particles dynamics, particle positions should be updated according to the equations of motion.}

\begin{algorithm}[H]
\begin{algorithmic}[1]
\State Initialize the particle positions $\{\rr_i\}_{i=1}^N$ and velocities $\{\vom_i\}_{i=1}^N$.
\State Initialize the ordered event list by determining their occurring times.
\State Extract the earliest valid event, occurring at time $t^*$.
\State Update the positions of the particles according to a free streaming, i.e., $\rr_k\leftarrow \rr_k + \vvel_k\left(t^*-t\right)$, $\forall k \in\{1,\dots,N\}$.
\State If the earliest event is a particle-particle collision between particles $i$ and $j$, update the velocities according to the binary collisional rules, $\vom_{i,j}\leftarrow \mathfrak{B}_{ij,\ssab_{ij}}\vom_{i,j}$. If not, in case of a particle-boundary interaction, the position and velocity of the involved particle must change according to the boundary conditions.
\State Compute the earliest future events for the particles involved in the previous event and update the list.
\State Update the system time $t\leftarrow t^*$.
\State Check the end-of-program condition. If it is not fulfilled, return to step 3.
\end{algorithmic}
 \caption{Basic \acrshort{EDMD} loop.}\labalg{EDMD_basic}
\end{algorithm}

\subsection{Initialization}\labsubsec{init_EDMD}

The first part of an \acrshort{EDMD} algorithm corresponds to initialize the positions and velocities of the particles. As in the \acrshort{DSMC} algorithm we need to determine the initial velocities of the particles according to a certain \acrshort{VDF}, managing the discrete \acrshort{VDF} as defined in Eq.~\eqref{eq:discrete_VDF}.

With respect to the positions, they are described as points in a discrete finite volume (box) of the $\dt$-dimensional space. In analogy to the velocities, they must be initially organized according to a certain initial distribution. The initial arrangement of the particles depends on the problem to be studied, from ordered states to completely randomized distributions. In the original algorithm used in this thesis, the implemented way of initializing has been hybrid. We divided the simulated box into identical square or cubic cells in two and three dimensional setups, respectively. The idea is to choose a fixed number of particles per cell and to assign random positions inside a cell avoiding overlaps. If the number of particles per cell is \emph{a priori} established, the number of cells is constrained to the number density of the problem, $n\sigma^{\dt}$. The definition of these cells is very useful to speed up the event detection and scheduling, as will be explained later. In \refalg{CA2D}, the algorithm for this hybrid arrangement is presented in a rectangular box, fixing one particle per cell for a monodisperse system of hard disks of diameter $\sigma$.

\begin{algorithm}[H]
\begin{algorithmic}[1]
\Procedure{CellArrangement2D}{$r_x$, $r_y$, $\sigma$, $N^{\mathrm{cell}}_x$, $N^{\mathrm{cell}}_y$, $L^{\mathrm{cell}}_x$, $L^{\mathrm{cell}}_y$} \Comment{$(r_x[i],r_y[i])$ is the position of the $i^{\mathrm{th}}$ particle.}
\State \textbf{for} $0\leq k<N^{\mathrm{cell}}_x$ \textbf{do}:
\State \textbf{for} $0\leq \ell<N^{\mathrm{cell}}_y$ \textbf{do}:
\State $r_x[\ell+k N^{\mathrm{cell}}_y]\leftarrow \text{randf}(k L^{\mathrm{cell}}_x+\sigma/2,(k+1) L^{\mathrm{cell}}_x-\sigma/2)$ \Comment{$\text{randf}(a,b)$ gives a uniform random number in $(a,b)$.}
\State $r_y[\ell+k N^{\mathrm{cell}}_y]\leftarrow \text{randf}(\ell L^{\mathrm{cell}}_y+\sigma/2,(\ell+1) L^{\mathrm{cell}}_y-\sigma/2)$ 
\State \textbf{end for}
\State \textbf{end for}
\State \textbf{return} $r_x$, $r_y$
\EndProcedure
\end{algorithmic}
\caption{Function for the hybrid arrangement of particle positions in rectangular cells of volume $L^{\mathrm{cell}}_x\times L^{\mathrm{cell}}_y$, in a box with $N^{\mathrm{cell}}_x\times N^{\mathrm{cell}}_y$ cells, for one particle per cell.}\labalg{CA2D}
\end{algorithm}

\subsection{Event detection}\labsubsec{detect_EDMD}

The event detection refers to the part of the \acrshort{EDMD} program that predicts new events. In fact, in the steps 2 and 6 of \refalg{EDMD_basic}, new events are detected and their associated time is computed. There are, essentially, two types of events, as introduced before: a pairwise particle collision and a boundary interaction. 

First, to predict a particle-particle collision, we need to solve, for each particle $i$, the equation $r_{ij}(t^*)=\sigma_{ij}$ $\forall j\neq i$, with $r_{ij}\equiv|\rr_i-\rr_j|$ and $\sigma_{ij}\equiv(\sigma_i+\sigma_j)/2$ being the mean diameter. Therefore, given an instant $t$, the collision condition for a pair of particles with positions $\{\rr_i,\rr_j\}$ and translational velocities $\{\vvel_i,\vvel_j\}$ can be rewritten, in the absence of an external deterministic force field, as
\begin{equation}
	\left|\rr_i(t)+\vvel_i(t)(t^*-t)- \rr_j(t)-\vvel_j(t)(t^*-t)\right| = \sigma_{ij}.
\end{equation}
After squaring, this condition can be reduced to the following second degree equation,
\begin{equation}
	v_{ij}^2(t^*-t)^2+2\rr_{ij}\cdot\vvel_{ij}(t^*-t)+r_{ij}^2-\sigma_{ij}^2=0.
\end{equation}
A collision will actually occur if $\rr_{ij}\cdot\vvel_{ij}<0$ (particles are approaching) and $\left(\rr_{ij}\cdot\vvel_{ij}\right)^2-v_{ij}^2\left(r_{ij}^2-\sigma_{ij}^2\right)>0$ (physical solution). Assuming there is no overlap between particles, the solution that accounts for a future collision between particles $i$ and $j$ is
\begin{equation}
 t^* = t- \frac{\rr_{ij}\cdot\vvel_{ij}}{v_{ij}^2}-\sqrt{\left(\frac{\rr_{ij}\cdot\vvel_{ij}}{v_{ij}^2}\right)^2-\frac{r_{ij}^2-\sigma_{ij}^2}{v_{ij}^2}}.
\end{equation}
However, it is discussed in Ref.~\cite{PS05} that the latter equation could lead to some numerical errors, which are mostly solved if the implemented (equivalent) equation is the following one,
\begin{equation}\labeq{t^*_sol_alg}
 t^* = t+ \left(r_{ij}^2-\sigma_{ij}^2\right)\left[-\vvel_{ij}\cdot\rr_{ij}+\sqrt{(\vvel_{ij}\cdot\rr_{ij})^2-v_{ij}^2\left(r_{ij}^2-\sigma_{ij}^2\right)} \right]^{-1}.
\end{equation}
Moreover, to avoid forbidden states, the algorithm introduced in Ref.~\cite{BSFP14} is executed for particle-particle collision detection. Given the state of particles $i$ and $j$ at time $t$, this algorithm works as written in \refalg{Stable_PPC}. Notice that in the step 1 of this algorithm we are asking for the necessary condition for a collision to occur, in step 2 we are detecting an overlap, and in step 3 we are checking for the other condition to ensure the collision.

\begin{algorithm}[H]
\begin{algorithmic}[1]
\State If $\rr_{ij}\cdot\vvel_{ij}\geq 0$ then return $\infty$.
\State If $r_{ij}^2-\sigma_{ij}\leq 0$ (overlapping particles) then return $t$.
\State If $\left(\rr_{ij}\cdot\vvel_{ij}\right)^2-v_{ij}^2\left(r_{ij}^2-\sigma_{ij}^2\right)\leq 0$ then return $\infty$.
\State Else, return $t^*$ as expressed in \refeq{t^*_sol_alg}.
\end{algorithmic}
\caption{Stable algorithm for predicting pairwise particle collisions proposed in Ref.~\cite{BSFP14}.}\labalg{Stable_PPC}
\end{algorithm}

To detect a boundary event, we need to compute the time the particle takes to arrive to a certain border. Let us imagine a three-dimensional problem, such that the position of particle $i$ is given by the vector $\rr_i = (r_x,r_y,r_z)$ inside of the rectangular box $[0,L_x]\times [0,L_y]\times [0,L_z]$, and with translational velocity $\vvel_i = (v_x,v_y,v_z)$. To compute the time to the closest boundary event, we need to determine first each individual time to arrive to a boundary. For example, the \refalg{EDMD_BC_x} summarizes how to predict boundary events in the direction of the $x$ axis.

\begin{algorithm}[H]
\begin{algorithmic}[1]
\State If $v_x>0$, then $t_x\leftarrow (L_x-r_x-\sigma_i/2)/v_x$.
\State If $v_x<0$, then $t_x\leftarrow (-r_x+\sigma_i/2)/v_x$.
\State Else (if $v_x=0$), then $t_x\leftarrow \infty$.
\end{algorithmic}
\caption{Prediction of a boundary event time at the $x$ axis.}\labalg{EDMD_BC_x}
\end{algorithm} 

\begin{figure}[ht]
	 \centering
	 \includegraphics[width=0.45\textwidth]{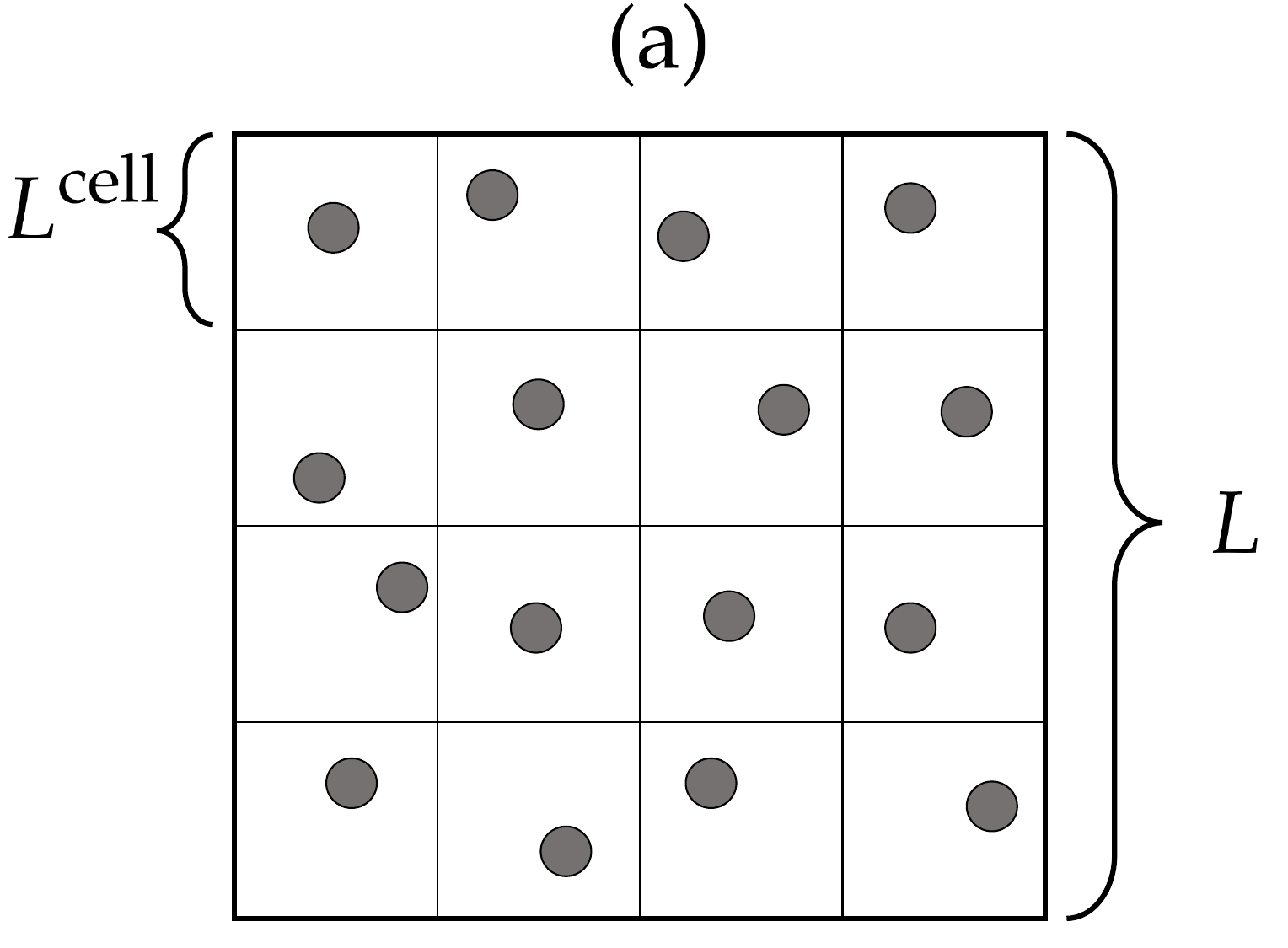}
	 \includegraphics[width=0.45\textwidth]{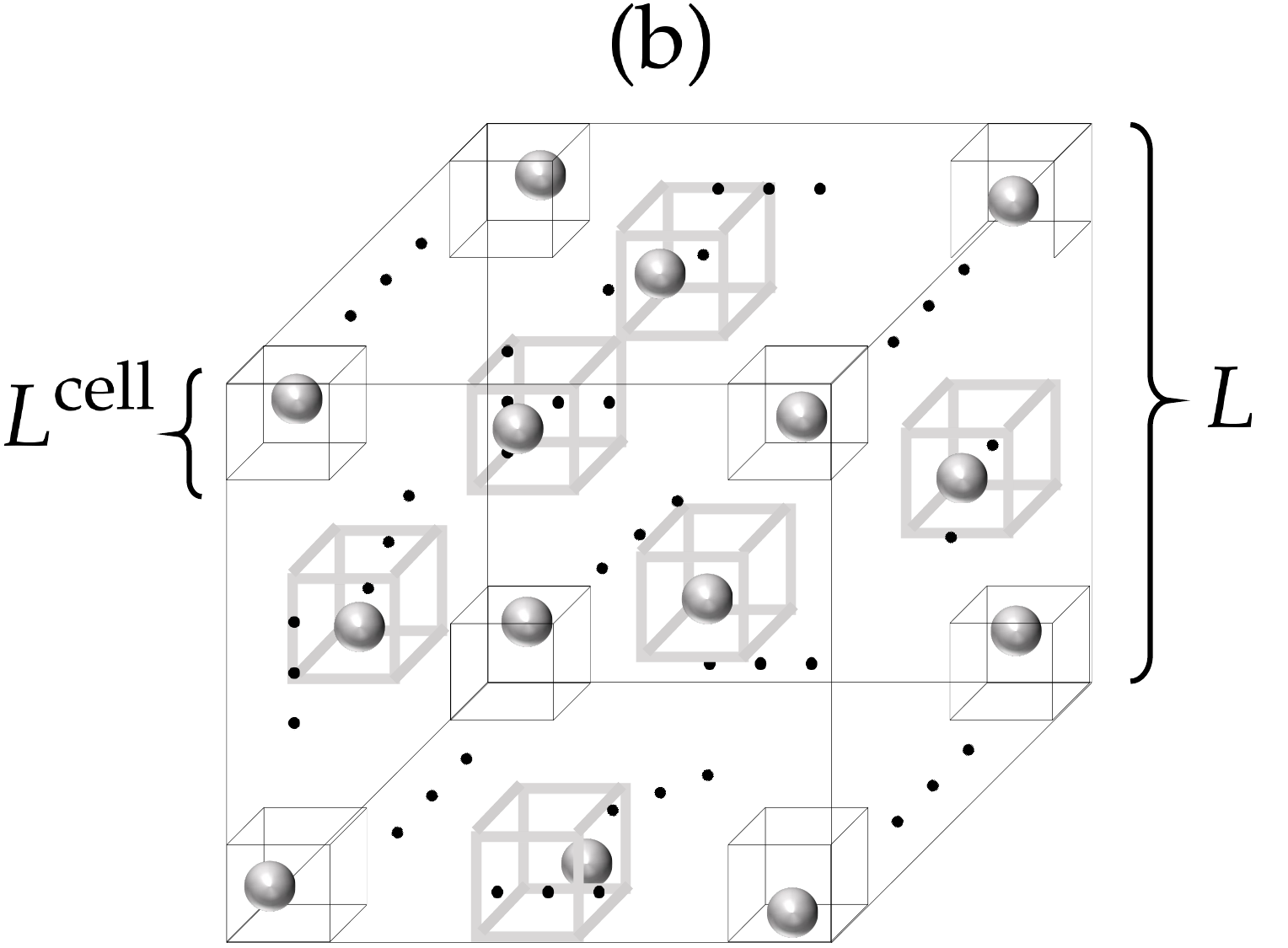}
	 \caption{Sketch of initialization cells for (a) squared and (b) cubic, with one particle per cell arrangements. In the three-dimensional cubic case not all cells are displayed to ensure a proper visualization.}
	 \label{fig:init_cells}
\end{figure}

Whereas the procedure for prediction of boundary events has a linear complexity $\mathcal{O}(N)$, the detection of pairwise collisions increases quadratically with respect to the number of particles, $\mathcal{O}(N^2)$. In order to speed up the algorithm for event detection, we took advantage of the cell division of the simulation box already introduced in \refsubsec{init_EDMD}. Then, given a particle in a certain cell, we will only look for events in the closest neighbor cells. Therefore, fixing a particle, their possible collisions are searched inside the proper cell and within its next neighboring ones. Moreover, the transfer of a particle from one cell to another is considered now as a boundary event, whereas boundary conditions are only applied to those cells in contact with the borders of the simulation box. In order to look for efficiency, it is essential to choose an appropriate number of particles per cell. A suitable value is the result of a competence between the number of transferring cell events per particle and the frequency of collision, both being highly dependent on the number density of the system. A very high number of particles per cell might cause an excessively large number of collisions within a cell, approaching the original quadratic complexity. On the other hand, a very small number of particles per cell determines a transfer-dominating algorithm, thus delaying the advance of the program. According to the discussion in Ref.~\cite{ML04}, a low-density system, which is the case considered throughout this thesis, has an optimal number of particles per cell of order unit, reaching the unity equality in two-dimensional systems. Therefore, in the developed \acrshort{EDMD} program, we have fixed a rate of one particle per cell, which is imposed since initialization (see \refalg{CA2D}), as depicted in Fig.~\ref{fig:init_cells}. We assumed squared or cubic boxes in the systems studied in this thesis, i.e., $L=L_i$ $\forall i\in\{1,\dots,\dt\}$. However, other more complex arrangements could be considered for specific problems with a certain geometry.

\subsection{Event scheduling} \labsubsec{event_scheduling_EDMD}

One of the challenges of the \acrshort{EDMD} since its beginning was to choose a proper scheduling algorithm, not only for initializing but also for sorting the events.

The initialization of the event list was already discussed in \refsubsec{detect_EDMD}, where a \emph{cell-list} method, widely accepted in literature~\cite{ML04,R04,BSL11}, was explained. It improves the initialization of the event list from being quadratically complex in the number of particles to be linear. This method is proved to be very efficient in the case of monodisperse hard \dt-spheres, in contrast to other algorithms as \emph{nearest-neighbor-list} (\acrshort{NNL}) methods proposed in case of high polydispersity and nonspherical particles~\cite{DTS05,dM10}. In these \acrshort{NNL} methods, instead of considering cells, each particle identifies its closest neighbors in a certain range of action to look for events there. Nevertheless, since polydisperse or nonspherical particles are not cases explored in this thesis, these \acrshort{NNL} methods have not been implemented. 

The way of sorting the events and maintaining the event list updated represents the core of the computational cost in an \acrshort{EDMD} program. An appropriate manner would consist in using a proper data structure for the event list which would allow for an efficient sorting and updating. In the pioneering works of Rapaport~\cite{R80}, binary-tree structures were proposed and, later, complete binary trees found to be optimal~\cite{MC95}. Therefore, in our Python-based program, this complete binary tree has been implemented by using a heap queue (priority queue)~\cite{SW11}, specifically from the Python library {\verb+heapq+}~\cite{webPython}.

In fact, we proceed by defining a heap queue for each particle and the closest event feeds the main event heap queue. These sublists of events are not saved during all the program since they are only defined during the initialization or updating of the main event priority queue. When the closest future event is extracted from the main list, it is first checked to be valid, that is, we do examine if the particles involved have not suffered any other event since its prediction and if that time is actually referring to a future (and not past) event that could be duplicated. For that, we identify the state of a particle, not only by their positions and velocities, but also by their current cell, the times it has suffered a collision, or the total times it has changed from one box cell to another, as proposed in Ref.~\cite{ML04}. Hence, the event array is made of the time of occurrence, the particles taking part in it, its type, and the particle states at its prediction, in order to guarantee a proper event validation. If the event results to be invalid, the event list is updated by searching the closest future events for the particles involved. In case the event extracted is concluded to be valid, the particles are upgraded to their new states and the main event list is again updated by inserting new future closest events of the particles involved. The routine for the event processing during the main loop, taking place after initialization and during the final condition is not reached, is summarized in \refalg{EDMD_event_loop}.

\begin{algorithm}[H]
\begin{algorithmic}[1]
\State Extract the next event from the priority queue.
\State Check whether the event is valid or not. If it is valid, upgrade the time to the event time and the particles states according to the event type.
\State Predict new future closest events for the particles involved in the extracted event.
\State Update the event heap queue by scheduling these new events.
\State Check the stop condition. If it is not fulfilled return to 1.
\end{algorithmic}
\caption{Event processing loop in an \acrshort{EDMD} algorithm.}\labalg{EDMD_event_loop}
\end{algorithm}

\subsection{Last remarks}\labsubsec{last_remarks_EDMD}

The \acrshort{EDMD} main algorithm is almost finished after the description of the last processes. However, some problems might appear during the running of the program associated with numerical errors. First of all, although we have considered only low-density systems, the probability of an overlap increases for moderately ones. The algorithm introduced in Ref.~\cite{BSFP14} is very efficient to avoid those problems, but in case they were present, it is important to perform periodically two processes: to save the state of the system and to check the presence of overlaps. Whereas the former is memory consuming, the latter, as already discussed, is time consuming. In case an overlap is detected and persistently present, we can turn back the simulation to a previous state again to correct the numerical error. The overlap checking was one of the reasons of choosing \acrshort{EDMD} over \acrshort{TDMD} algorithms. Nevertheless, their detection in \acrshort{EDMD} is not necessary to be as recurrent as it must be in \acrshort{TDMD}.

Moreover, the granular gaseous case is quite special if we compare with the \acrshort{EHS} model for molecular gases. In freely evolving granular gaseous systems, their velocities decrease continuously according to the energy loss in the collisional process and reflected in \index{Haff's law}Haff's law of cooling. Therefore, computationally speaking, the managing of these small quantities can lead to strong numerical errors and a deceleration of the program running until an almost stopping of the system. This issue is solved by a periodical rescaling of the particle velocities, as proposed for example in Ref.~\cite{L01,FR13}, and already considered in the \acrshort{DSMC} algorithm. This rescaling allows for the long time study of a granular gas.

Finally, it is worth mentioning the possibility for a homogeneous granular gas to spontaneously face  \index{Instability!clustering}cluster or \index{Instability!vortex}vortex instabilities, as already introduced in \refsubsec{HCS_KTGG}. The rescaling permits the long-time observation of this phenomenon. Moreover, an important and exclusive issue in a granular gas \acrshort{MD} simulation is the inelastic \index{Inelastic collapse}collapse associated with clustering formation, in which a chain of sticked-together particles enters in an infinite entangled state that makes the \acrshort{EDMD} algorithm to fall down in an infinite number of collisional events in finite time, thus stopping the progress of the program \cite{TP19}. This granular phenomenon was first studied in one-dimensional problems and subsequently in two- and three-dimensional setups for the \acrshort{IHS} model. If we want to study large-scale simulations with clustering \index{Instability!clustering}instability, we must avoid the program to stop at this \index{Inelastic collapse}inelastic collapse. This issue is solved in an \acrshort{EDMD} algorithm with some techniques proposed in the literature~\cite{MY94,PS05}, like the introduction of the \emph{TC model}~\cite{LM98} or a slight modification of the collisional rule~\cite{DB97,G97}. We chose the first method to be implemented in our program.

\subsection{EDMD with Langevin-like dynamics}\labsubsec{LD_EDMD}

Up to now, we just only considered gaseous systems whose particle evolution is only submitted to a collisional process. However, particles can be subjected to other interactions due to their presence in a certain medium or the application of external force fields. In the latter case, e.g., considering the action of gravity, the modification of the classical program resides only in the free evolution equation of motion, where the associated acceleration due to the action of the force must be properly integrated. However, a nontrivial situation comes in the case of the action of stochastic forces, which transform the deterministic equations of motion into Langevin-type equations for particle velocities. These associated \acrshort{SDE} produce erratic Brownian-like motions, i.e., the Langevin-type differential equation for velocity induces as well a \acrshort{SDE} for the particle positions. This becomes a problem for the event prediction process because of the random trajectories of the particles. Hence, the implementation of stochastic forces, no matter if they come from possibly external sources or due to the action of a background fluid, is not an easy task.

We address this issue using the \emph{Approximate Green Function} (\acrshort{AGF}) algorithm proposed in Refs.~\cite{SVdM07,S12}. To expose the derivation of the crucial equations of the problem, we will start from the original case of the \acrshort{LE} with constant friction coefficient. That is, the time evolution of the translational velocities is described by Eq.~\eqref{eq:white_noise}. Moreover, taking into account the time evolution of the particle positions, we have the following system of \acrshort{SDE},
\begin{subequations}\labeq{Langevin_SDE_LE}
\begin{align}
	\frac{\dif \vvel_i(t)}{\dif t} =& -\xi_0\vvel_i(t)+v_b\xi_0^{1/2}\bar{\boldsymbol{\eta}}_i(t), \\
	\frac{\dif \rr_i(t)}{\dif t} =& \vvel_i(t), \medspace \forall i \in\{1,\dots,N\},
\end{align}
\end{subequations}
where we used the \acrshort{FDT}, with $v_b$ being the thermal velocity associated with the bath \index{Temperature!of the thermal bath}temperature, and each $\overline{\boldsymbol{\eta}}_i$ being a Gaussian random vector representing the effect of a zero-mean and delta-correlated white noise with unit variance,
\begin{equation}\labeq{white_noise_eta_bar}
	\langle \bar{\boldsymbol{\eta}}_i(t)\rangle = \mathbf{0}, \qquad \langle \bar{\boldsymbol{\eta}}_i(t)\bar{\boldsymbol{\eta}}_j(t^\prime)\rangle = \mathbb{1}_{\dt}\delta_{ij}\delta(t-t^\prime).
\end{equation}
Let us integrate the system of \acrshort{SDE} from $t$ to $t+\Delta t$. Hence~\cite{S12,AT17},
\begin{subequations}
\begin{align}\label{eq:v_and_r_LE}
    \vvel_i(t+\Delta t)=& \vvel_i(t)e^{-\xi_0 \Delta t}+v_b\xi_0^{1/2}\int_0^{\Delta t}\dif t^\prime\, e^{\xi_0(t^\prime-\Delta t)}\bar{\boldsymbol{\eta}}_i(t^\prime), \\
    \rr_i(t+\Delta t)=& \rr_i(t)+\xi_0^{-1}\vvel_i(t)\left(1-e^{-\xi_0 \Delta t}\right)+v_b\xi_0^{1/2}\int_0^{\Delta t}\dif t^\prime\,\int_0^{t^\prime} \dif t^{\prime\prime}\, e^{\xi_0(t^{\prime\prime}-t^\prime)}\bar{\boldsymbol{\eta}}_i(t^{\prime\prime}).
\end{align}
\end{subequations} 
In these equations, two random variables appear,
\begin{align}\labeq{W_rand_def_Lang_LE}
    \boldsymbol{\mathcal{W}}_i = \int_0^{\Delta t}\dif t^\prime\, e^{\xi_0(t^\prime-\Delta t)}\bar{\boldsymbol{\eta}}_i(t^\prime), \quad
    \bar{\boldsymbol{\mathcal{W}}}_i = \int_0^{\Delta t}\dif t^\prime\,\int_0^{t^\prime} \dif t^{\prime\prime}\, e^{\xi_0(t^{\prime\prime}-t^\prime)}\bar{\boldsymbol{\eta}}_i(t^{\prime\prime}),
\end{align}
which are Gaussian random variables with the following properties
\begin{align}\labeq{W_rand_LE}
    \mathbfcal{W} = \sqrt{\langle {\mathcal{W}}^2\rangle}\boldsymbol{\mathcal{Y}}, \quad
    \bar{\mathbfcal{W}} = \frac{\langle \mathbfcal{W}\cdot \bar{\mathbfcal{W}}\rangle}{\sqrt{\langle {\mathcal{W}^2}\rangle}}\mathbfcal{Y}+\sqrt{\langle \bar{\mathcal{W}}^2\rangle-\frac{\langle \mathbfcal{W}\cdot \bar{\mathbfcal{W}}\rangle^2}{\langle {\mathcal{W}}^2\rangle}}\bar{\mathbfcal{Y}},
\end{align}
where $\boldsymbol{\mathcal{Y}},\bar{\boldsymbol{\mathcal{Y}}}$ are drawn from Gaussian probability distributions, i.e.,
\begin{align}\labeq{gaussian_Y_and_Ybar}
	P(\boldsymbol{\mathcal{Y}}) = \left(2\pi\right)^{-\dt/2}e^{-\mathcal{Y}^2/2}, \quad P(\bar{\boldsymbol{\mathcal{Y}}}) = \left(2\pi\right)^{-\dt/2}e^{-\bar{\mathcal{Y}}^2/2},
\end{align}
and the expected quantities in \refeq{W_rand_LE} in the case of the \acrshort{LE} read
\begin{subequations}\labeq{W_values_LE}
\begin{align}
	\langle \mathcal{W}^2\rangle =& \frac{1}{2}\xi_0^{-1}\left(1-e^{-2\xi_0\Delta t}\right), \\
    \langle \boldsymbol{\mathcal{W}}\cdot\bar{\boldsymbol{\mathcal{W}}}\rangle =& \frac{1}{2}\xi_0^{-2}\left(1-e^{-\xi_0\Delta t}\right)^2, \\ 
    \langle \bar{\mathcal{W}}^2\rangle =& \frac{1}{2}\xi_0^{-3}\left(2\xi_0\Delta t-3-e^{-2\xi_0\Delta t}+4e^{-\xi_0\Delta t}\right).
\end{align}
\end{subequations}
Thus, the solutions of the equations of motion during the free-streaming stage at $t+\Delta t$ in \refeq{v_and_r_LE}
\begin{subequations}\labeq{v_and_r_LE_2}
\begin{align}
    \vvel_i(t+\Delta t)=& \vvel_i(t)e^{-\xi_0 \Delta t}+\frac{v_b}{\sqrt{2}}\sqrt{1-e^{-2\xi_0\Delta t}}\mathbfcal{Y}_i, \\
    \rr_i(t+\Delta t)=& \rr_i(t)+\xi_0^{-1}\vvel_i(t)\left(1-e^{-\xi_0 \Delta t}\right)+\frac{v_b}{\sqrt{2}}\xi_0^{-1}\left(1-e^{-\xi_0\Delta t}\right)^{3/2}\left(1+e^{-\xi_0\Delta t}\right)^{-1/2}\mathbfcal{Y}_i\nonumber\\
    &+v_b\xi_0^{-1}\sqrt{\xi_0\Delta t-1+e^{-2\xi_0\Delta t}\left(\frac{3-e^{-\xi_0\Delta t}}{1+e^{-\xi_0\Delta t}}\right)}\bar{\mathbfcal{Y}}_i, \quad \forall i \in\{1,\dots,N\}.
\end{align}
\end{subequations}
Equations~\eqref{eq:v_and_r_LE_2} solve the dynamics between $t$ and $t+\Delta t$ in free evolution due to the action of the thermal bath inducing the \acrshort{LE}. From them, the \acrshort{AGF} algorithm splits up the dynamics in two stages. In the first one, starting from the system at $t$, $\{\rr_i(t),\vvel_i(t)\}$, and according to Eqs.~\eqref{eq:v_and_r_LE_2}, one computes the \emph{putative} positions and velocities at $t+\Delta t$, \phantomsection\label{sym:rput}$\{\rr_i^{\mathrm{put}}, \vvel_i^{\mathrm{put}} \}$. Thus, one can define \emph{fictive} velocities as $\vvel_i^{\mathrm{fic}}= [\rr_i^{\mathrm{put}}-\rr_i(t)]/\Delta t$. In the second stage, we run a classical \acrshort{EDMD} algorithm for events taking place only in the time interval $(t,t+\Delta t]$. In this latter stage, including the event-list prediction, putative positions and fictive velocities are used as the proper variables, i.e., the states are described by $\{\rr_i^{\mathrm{put}},\vvel_i^{\mathrm{fic}}\}$. All valid events within this interval are performed. Then, the states of the particles are updated: the states at $t+\Delta t$ for particles that did not suffer any collision in the interval are given by $\{\rr_i^{\mathrm{put}}, \vvel_i^{\mathrm{put}} \}$; otherwise, their states will be updated according to a two-step numerical scheme, where positions are modified using the resulting fictive velocities after the collision rules are applied. The resulting velocities are coming from the application of the collision rules to the putative velocities using the position unit vectors computed in the event prediction. That is, in the case of a collision between particles $i$ and $j$ at \phantomsection\label{sym:tcoll}$t_{\mathrm{coll}}\in (t,t+\Delta t]$, the latter steps follow,
\begin{subequations}\labeq{rv_final_Lan_EDMD}
\begin{align}
	\rr_i(t+\Delta t) \approx& \rr_i(t)+\vvel_i^{\mathrm{fic}}\Delta t_{\mathrm{coll}}+ {\vvel_i^{\mathrm{fic}}}^\prime(\Delta t-\Delta t_{\mathrm{coll}}),\labeq{r_final_Lan_EDMD} \\
	\vvel_i(t+\Delta t)\approx& \mathfrak{B}^{ij,\mathrm{put}}_{\ssab_{ij}^{\mathrm{fic}}}\vvel_i^{\mathrm{put}},\labeq{v_final_Lan_EDMD}
\end{align}
\end{subequations}
where $\Delta t_{\mathrm{coll}}=t_{\mathrm{coll}}-t$, $\ssab_{ij}^{\mathrm{fic}}= [\rr_{ij}(t)+\vvel_{ij}^{\mathrm{fic}}\Delta t_{\mathrm{coll}}]/|\rr_{ij}(t)+\vvel_{ij}^{\mathrm{fic}}\Delta t_{\mathrm{coll}}|$ is a fictive unit intercenter vector at the time of collision, ${\vvel_i^{\mathrm{fic}}}^\prime=\mathfrak{B}^{ij,\mathrm{fic}}_{\ssab_{ij}^{\mathrm{fic}}}\vvel_i^{\mathrm{fic}}$ is the postcollisional fictive velocity, and the operators $\mathfrak{B}^{\mathrm{fic}}_{ij,\ssab}$ and $\mathfrak{B}^{\mathrm{put}}_{ij,\ssab}$ refer to the postcollisional operators for a pair $ij$ with intercenter unit vector $\ssab$ and using the fictive and putative velocities in the collisional rules, respectively. 

To ensure the convergence of the method, the time step must be chosen such that $v_b\sqrt{\xi_0^{-1}\Delta t}\ll \sigma$, that is, the mean free displacement in $\Delta t$ is much smaller than the diameter, which is the typical length of the interaction, as explained in Refs.~\cite{SVdM07,S12}.

Whereas this is the original algorithm for a gas of hard spheres with Langevin dynamics, it can be generalized to an arbitrary bath interaction. Let us assume that the particle velocities follow a Langevin-like \acrshort{SDE}, so that their solutions at $t+\Delta t$ can be expressed as functions of their states at $t$ (explicit scheme), the noise parameters (noise intensity, constants related to the drag coefficient, etc.) \phantomsection\label{sym:bargamma}$\bar{\gamma}$, the time-step $\Delta t$, and the corresponding random variables driven from a Gaussian distribution in Eq.~\eqref{eq:gaussian_Y_and_Ybar}, 
\begin{subequations}
	\begin{align}\labeq{vels_pos_FrakF}
		\vom_i(t+\Delta t) \approxeq & \mathfrak{F}_1[\vom_i(t),\bar{\gamma},\Delta t,\mathbfcal{Y}_i], \\
		\rr_i(t+\Delta t) \approxeq & \mathfrak{F}_2[\rr_i(t),\vvel_i(t),\bar{\gamma},\Delta t,\mathbfcal{Y}_i,\bar{\mathbfcal{W}}_i].
	\end{align}
\end{subequations}
Notice that we are considering the usual and simple case that the rotational degrees of freedom, while possibly present, are not explicitly coupled to the positions $\rr_i(t)$. In case of nonanalytic solutions like \refeqs{rv_final_Lan_EDMD}, the numerically alternative used is the Euler-Maruyama method for \acrshort{SDE}, which can be improved by using Runge-Kutta techniques~\cite{KP92}. The concrete schemes for the Langevin dynamics with nonlinear drag (see \refsubsec{nonlinear_KTGG}) and the \acrshort{ST} (see \refsubsec{splitting_KTGG}) are explicitly derived in \refapp{app_AGF}. Hence, for an arbitrary stochastic thermostat or bath interaction, the algorithm for a collision event is given by \refalg{AGF_collision} and the whole \acrshort{AGF} algorithm loop can be read in \refalg{AGF+EDMD}.

\begin{algorithm}[H]
\begin{algorithmic}[1]
    \State Get the collisional time interval $\Delta t_{\mathrm{coll}}$, the total one $\Delta t$, the fictive velocities of the colliding particles $\vom^{\mathrm{fic}}_{i,j}$ and their positions at current time $t$, $\rr_{i,j}$.
    \State Compute the intercenter vector at contact time $\ssab_{ij}^{\mathrm{fic}}\leftarrow ( \rr_{ij}+\vvel_{ij}^{\mathrm{fic}}\Delta t_{\mathrm{coll}})/|\rr_{ij}+\vvel_{ij}^{\mathrm{fic}}\Delta t_{\mathrm{coll}}|$.
    \State Determine the postcollisional fictive velocities according to the collisional rules, i.e., ${{\vom}^{\mathrm{fic}}_{i,j}}^\prime\leftarrow \mathfrak{B}^{\mathrm{fic}}_{ij,\ssab_{ij}^{\mathrm{fic}}}\vom_{i,j}^{\mathrm{fic}}$.
    \State Calculate the final positions for the colliding particles according to \refeqs{rv_final_Lan_EDMD}: $\rr_{i,j}\leftarrow\rr_{i,j}+\vvel_{i,j}^{\mathrm{fic}} \Delta t_{\mathrm{coll}}+{\vvel_{i,j}^{\mathrm{fic}}}^\prime(\Delta t-\Delta t_{\mathrm{coll}})$, taking into account boundary conditions.
    \State Compute the resulting velocities of the colliding particles: $\vom_{i,j}\leftarrow \mathfrak{B}^{\mathrm{put}}_{ij,\ssab_{ij}^{\mathrm{fic}}} \vom_{i,j}^{\mathrm{put}}$. 
    \State Return $\rr_{i,j}$ and $\vom_{i,j}$.
  \end{algorithmic}
 \caption{Collisional routine in an \acrshort{AGF} algorithm.}\labalg{AGF_collision}
\end{algorithm}

\begin{algorithm}[H]
\begin{algorithmic}[1]
    \State Get the current state of the particles, $\{ \rr_i, \vom_i\}_{i=1}^N$ and current time $t$.
    \State Compute the putative positions and velocities as the solutions of the associated Langevin-like \acrshort{SDE} after a $\Delta t$, i.e., $ \vom_i^{\mathrm{put}}\leftarrow  \mathfrak{F}_1[\vom_i,\bar{\gamma},\Delta t,\mathbfcal{Y}_i]$ and $\rr_i^{\mathrm{put}}\leftarrow  \mathfrak{F}_2[\rr_i,\vvel_i(t),\bar{\gamma},\Delta t,\bar{\mathbfcal{Y}}_i]$.
    \State Get the fictive velocities as $\vvel_i^{\mathrm{fic}}\leftarrow \left[\rr_i-\rr_i^{\mathrm{put}}\right]/\Delta t$. Considering the case that noise is not coupled to angular displacements, fictive angular velocities (if defined) are set to coincide with putative angular velocities.
    \State Create the event list with the fictive states, $\{ \rr_i^{\mathrm{fic}}, \vom_i^{\mathrm{fic}}\}_{i=1}^N$.
    \State Run the \acrshort{EDMD} algorithm (see, for example, \refalg{EDMD_basic}) between $t$ and $t+\Delta t$, using \refalg{AGF_collision} in case of collision events, and save the indices of colliding particles in a list, $\{i_1,\dots,i_{n_\mathrm{cp}}\}$, with $n_\mathrm{cp}$ being the total number of colliding particles.
    \State For noncolliding particles, update positions and velocities, i.e., $\rr_i\leftarrow \rr_i^{\mathrm{put}}$ and $\vom_i\leftarrow \vom_i^{\mathrm{put}}$, $\forall i\notin \{i_1,\dots,i_{n_\mathrm{cp}}\}$.
    \State Update time, $t\leftarrow t+\Delta t$.
    \State Check the end-of-program condition. If it is not fulfilled, go back to 1.
  \end{algorithmic}
 \caption{Main \acrshort{AGF} loop.}\labalg{AGF+EDMD}
\end{algorithm}

This algorithm is used in the works exposed in Article 1 (\refsec{Art1}), Article 4 (\refsec{Art4}), and Article 6 (\refsec{Art6}) to reproduce the action of the nonlinear drag Langevin-like dynamics introduced in \refsubsec{nonlinear_KTGG}, and the \acrshort{ST} described in \refsubsec{splitting_KTGG}.

\addpart{Molecular gases}

\chapter{Thermal and entropic Mpemba effects in molecular gases under nonlinear drag}
\labch{molecular_gases}

\section{Summary}

The \acrshort{ME} is studied in detail in a dilute and homogeneous molecular gas made of hard \dt-spherical identical particles of mass $m$ and diameter $\sigma$, and surrounded by an interstitial fluid at equilibrium, at temperature\index{Temperature!of the thermal bath} $T_b$. The interaction of the background fluid with the molecular gas is assumed to be described by a Langevin-like relation, as described in the thermostatted model presented in \refsubsec{nonlinear_KTGG}. That is, the interaction is assumed to be split into a drag force, with a coefficient assumed to depend quadratically on the velocity modulus, and a stochastic force, whose noise intensity is related to the drag coefficient\index{Drag coefficient!velocity-dependent} via a fluctuation-dissipation relation. The dynamics of the system is studied from a kinetic-theory point of view, whose associated homogeneous \acrshort{BFPE} corresponds to \refeq{BFPE_NLD} in the absence of gradients. Moreover, it was already known that, in this model, the \acrshort{TME} is present~\cite{SP20,PSP21}, also in the collisionless case~\cite{PSP21}, and had been previously studied in terms of crossing times of the temperature evolution curves for the samples involved in the effect (see \refsec{ME_ME}). Furthermore, as already introduced in \refsec{ME_ME}, in previous literature about the \acrshort{ME}, both the \acrshort{TME} and the \acrshort{EME} descriptions have been developed. Usually, the former description has been applied to kinetic-theory-based systems~\cite{LVPS17,TLLVPS19,BPRR20,SP20,GG21,BPR21,PSP21,MLLVT21,THS21,BPR22}, whereas the second one has been employed in stochastic-process descriptions of physical systems~\cite{LR17,KRHV19,KB20,CKB21,BGM21,CLL21,LLHRW22,SL22,YH22}. However, the emergence of the \acrshort{ME}  has typically been interpreted with a similar argumentation from both descriptions.

The main goal of this chapter is to stress the conceptual characterization of the \acrshort{ME} from the thermal and entropic interpretations and expose their differences and common characteristics in the context of the introduced molecular gaseous system. This discussion is based on the whole phenomenology arising in the \acrshort{ME}. For this objective, we use the \acrshort{BFPE} to introduce the kinetic-theory description of the system. For the thermal description, the evolution equations of the first three moments of the time-dependent \index{Velocity distribution function!time-dependent}\index{Velocity distribution function!moments}\acrshort{VDF} are derived as a truncation of a full hierarchy of moment equations from a Sonine expansion of the isotropic \acrshort{VDF}. To work in dimensionless terms, we define a time and \index{Temperature!of the thermal bath}temperature reference scale, \phantomsection\label{sym:nub}$\nu_b^{-1}$ and $T_b$, respectively. This hierarchy is truncated in two different ways, that is, two \acrshort{SA} were implemented. In the first one, termed in Article 1 (\refsec{Art1}) as \emph{basic} \acrshort{SA} (\acrshort{BSA}), the system of differential equations is just made for the evolution of the \index{Temperature!nonequilibrium}temperature, $T$, and the fourth cumulant, $a_2$. The \acrshort{BSA} is the approach indeed previously considered in Ref.~\cite{SP20}. In this \acrshort{BSA}, higher-order cumulants are neglected, as well as nonlinear terms of $a_2$. This is compared with an \emph{extended} \acrshort{SA} (\acrshort{ESA}), which considers the evolution equation of the sixth cumulant, $a_3$, and neglects $a_n$ $\forall n\geq 4$, as well as the nonlinear terms of $a_2$ and $a_3$. These systems of coupled differential equations are built from Eqs.~(17) and (19) of Article 1 (\refsec{Art1}). Then, if we compare the dynamic description of the molecular gas with that introduced in \refeqs{memory_descr} of \refch{ME_KTGG}, the inner variables acting on the thermal evolution are the cumulants of the \index{Velocity distribution function!cumulants}\acrshort{VDF}, especially $a_2$, which appears explicitly in the equation of the slope of the temperature, $\dot{T}$. In fact, the \acrshort{ESA} is mainly introduced to improve the theoretical approach, as compared with simulation results, already observed in Ref.~\cite{PSP21}. In Ref.~\cite{SP20}, the explicit coupling of $a_2$ with the evolution equation of $T/T_b$ was already derived, inferring the emergence of the \acrshort{TME} from it for short times of the evolution. Under this approach, a linearization of the evolution equations around the initial conditions applies, this being called in Article 1 (\refsec{Art1}) as linearized \acrshort{BSA} (\acrshort{LBSA}). Then, one can even derive algebraically the crossing time, denoted as \phantomsection\label{sym:t_theta}$t_\theta$ and defined in Eq.~(26) of Article 1  (\refsec{Art1})\footnote{Note that, in Article 1 (\refsec{Art1}), $\theta$ denotes the ratio $T/T_b$.}. This short evolution is guaranteed when the system is not subjected to extreme either quenching or heating~\cite{PSP21}.

Once the thermal description is introduced, we characterize entropically the system by its associated \index{Nonequilibrium entropy}nonequilibrium entropy. The functional used for that description is the \acrshort{KLD} with the Maxwellian \index{Velocity distribution function!Maxwellian}\acrshort{VDF} at \index{Temperature!of the thermal bath}$T_b$---the equilibrium \index{Velocity distribution function!equilibrium}\acrshort{VDF}---as the reference \acrshort{VDF}. For this case, $\KLD(\cdot|f^{\mathrm{eq}}_{\mathrm{M}})$ is equivalent, up to a constant, to the $H$-functional, $H(\cdot)$, which was already introduced as the \index{Nonequilibrium entropy}nonequilibrium entropy for molecular gases in \refsubsec{HTheorem_MG_KTGG}. Then, the discussion about the \acrshort{EME} is based on the decomposition of the \index{Nonequilibrium entropy}nonequilibrium entropy into a kinetic and a local-equilibrium counterparts, i.e., \phantomsection\label{sym:kin_LE}$\KLD(f(t)|f^{\mathrm{eq}}_{\mathrm{M}})=\mathcal{D}^{\mathrm{kin}}(t)+\mathcal{D}^{\mathrm{LE}}(T(t))$ [see Eq.~(11) of Article 1 (\refsec{Art1})]. The kinetic part accounts for the \emph{distance}\footnote{Note that the $\KLD$ is not, strictly speaking, a distance function. In general, it is not symmetric and it does not fulfill the triangle inequality~\cite{KL51}.} between the time-dependent \index{Velocity distribution function!time-dependent}\acrshort{VDF} and the local-equilibrium \index{Velocity distribution function!local-equilibrium}\acrshort{VDF}, which coincides with the Maxwellian \index{Velocity distribution function!Maxwellian}\acrshort{VDF} at the instantaneous \index{Temperature!nonequilibrium}temperature, $T(t)$ [see Eqs.~(10) and (12a) of Article 1 (\refsec{Art1})]. On the other hand, the local-equilibrium counterpart measures how far the local-equilibrium \index{Velocity distribution function!local-equilibrium}\acrshort{VDF} is from the equilibrium \index{Velocity distribution function!equilibrium}\acrshort{VDF}, which results to be a convex and positive function of $T(t)/T_b$ [see Eq.~(12b) of Article 1 (\refsec{Art1})]. A schematic representation of this two-stage description is sketched in \reffig{EME_MSP22}. Therefore, if the \acrshort{ME} arises at large times, that is, in the \index{Hydrodynamic!stage}hydrodynamic stage, when the instantaneous \acrshort{VDF} coincides with the local-equilibrium one, then $\mathcal{D}^{\mathrm{kin}}\approx 0$, whereas $\KLD(t)\approx \mathcal{D}^{\mathrm{LE}}(T(t))$. Hence, \acrshort{TME} and \acrshort{EME} become approximately equivalent during that stage. However, it was observed that \acrshort{TME} tends to arise during the first part of the evolution~\cite{SP20}, becoming $\mathcal{D}^{\mathrm{kin}}$ essential in this discussion and breaking down a general equivalence between the thermal and entropic descriptions. In fact, the \acrshort{VDF} at this kinetic regime is in a completely nonequilibrium state and therefore, memory effects, like \acrshort{ME}, are expected to occur before the \index{Velocity distribution function!local-equilibrium}\acrshort{VDF} acquires the local-equilibrium form. To simplify the entropic analysis, we derive a \emph{renormalized} \acrshort{SA} [based on the toy model derived in Article 2 (\refsec{Art2})] for $\mathcal{D}^{\mathrm{kin}}(t)$. In this approach, the \index{Velocity distribution function!Gamma}\acrshort{VDF} is assumed to be compatible with a Gamma distribution along the system evolution, resulting that $\mathcal{D}^{\mathrm{kin}}(t)\propto a_2^2(t)$ [see Eq.~(24) of Article 1 (\refsec{Art1})]. This Gamma approximation is introduced because of two facts: first, it is a two-parameter exponential probability distribution than can be made to accommodate any other exponential-like one (for example, it can be reduced straightforwardly to the Maxwellian one); second, the initial state of our computer simulated systems are generated from this distribution.\index{Velocity distribution function!local-equilibrium} \index{Velocity distribution function!kinetic}

\begin{figure}[ht]
\centering
\includegraphics[width=0.85\textwidth]{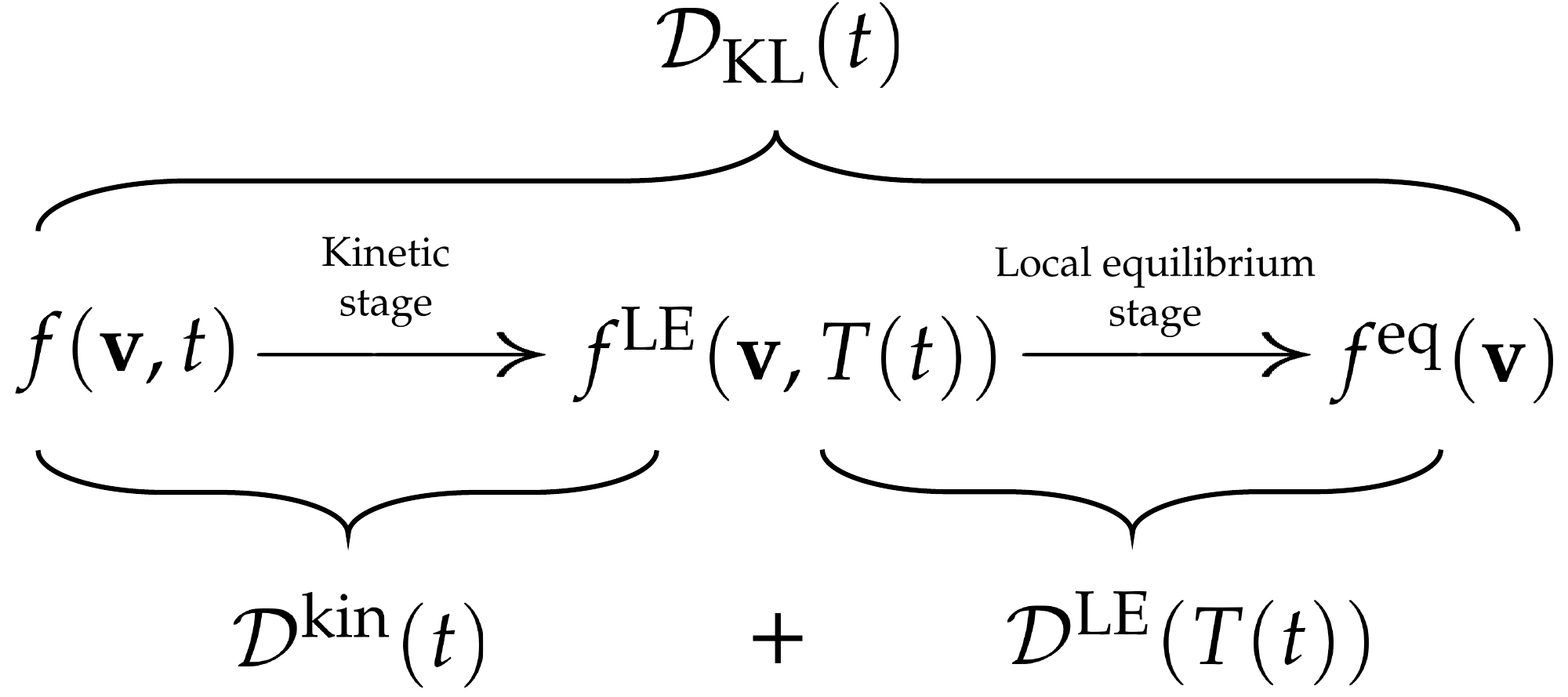}
\caption{Sketch of the evolution of the \acrshort{VDF}, $f$. The decomposition of the \acrshort{KLD} into the kinetic and local-equilibrium counterparts is represented. First, the evolution of the \acrshort{VDF} is involved into a kinetic stage, in which the \acrshort{VDF} stays completely at a nonequilibrium regime. Then, $f$ enters in local equilibrium ($f^{\mathrm{LE}}$) and the \acrshort{VDF} is fully described by its instantaneous \index{Temperature!nonequilibrium}temperature, $T(t)$, before the relaxation toward equilibrium is completed.}
\labfig{EME_MSP22}
\end{figure}

Once the mathematical descriptions of the \acrshort{TME} and \acrshort{EME}, respectively guided by the nonequilibrium \index{Temperature!nonequilibrium}temperature and entropy, are introduced, the possible scenarios in which the \acrshort{ME} can arise are discussed. Let us imagine two samples, A and B, the former referring to that sample whose initial temperature is further from the equilibrium value. As a first approach, we assume that $T(t)$ does not cross its steady equilibrium value, $T_b$, during the evolution. If the \index{Temperature!nonequilibrium}temperature curve of system A arrives earlier to equilibrium than B, \acrshort{TME} will be present. According to the description in \refsec{ME_ME}, if the initial conditions fulfill \phantomsection\label{sym:TA_TB}$T_A^0/T_b>T_B^0/T_b>1$ (cooling process) then, if a \acrshort{TME} emerges, it is said to be a \acrshort{DME}. On the other hand, if $T_A^0/T_b<T_B^0/T_b<1$, a \acrshort{TME} is classified as an \acrshort{IME}. In the former case, we need that $\dot{T}_A(0)/T_b<\dot{T}_B(0)/T_b$, which is found to be equivalent to \phantomsection\label{sym:a2A_a2B}$a_{2A}^0>a_{2B}^0$, in the \acrshort{LBSA}, and the other way around in the latter preparation. Moreover, under the \acrshort{LBSA} scheme, this necessary condition for the \acrshort{TME} is completed by imposing $t_\theta>0$ to its expression, finally resulting in Eqs.~(27) and (28) of Article 1 (\refsec{Art1}). On the other hand, if the initially further from equilibrium---as accounted for by the full \acrshort{VDF} via the \acrshort{KLD}---relaxes earlier to the equilibrium distribution, the \acrshort{EME} is said to be present.

Under those considerations, the \acrshort{ME} is characterized by the appearance of a crossing between the evolution curves at either $t_\theta$ or \phantomsection\label{sym:t_KLD}$t_{\mathcal{D}}$, in the \acrshort{TME} or the \acrshort{EME} contexts, respectively. As explained in \refsec{ME_ME}, an even number of crossings indicates the existence of the \acrshort{ME}, as well. However, as we are working in the rapid kinetic stage, more than one crossing is quite unlikely. The possible phenomenology is then summarized into 8 possible events, they being referred to in Article 1 (\refsec{Art1}) as
\begin{itemize}
\item \underline{ET1:} \acrshort{EME} and \acrshort{DME} are present and $t_{\mathcal{D}}<t_{\theta}$.
\item \underline{TE1:} \acrshort{EME} and \acrshort{DME} are present and $t_{\mathcal{D}}>t_{\theta}$.
\item \underline{ET2:} \acrshort{EME} and \acrshort{IME} are present and $t_{\mathcal{D}}<t_{\theta}$.
\item \underline{TE2:} \acrshort{EME} and \acrshort{IME} are present and $t_{\mathcal{D}}>t_{\theta}$.
\item \underline{T1:} only \acrshort{DME} (and not \acrshort{EME}) is present.
\item \underline{T2:} only \acrshort{IME} (and not \acrshort{EME}) is present.
\item \underline{E1:} only \acrshort{EME} (and not \acrshort{DME}) is present in a cooling process.
\item \underline{E2:} only \acrshort{EME} (and not \acrshort{IME}) is present in a heating process.
\end{itemize} 
The comparisons between $t_\theta$ and $t_{\mathcal{D}}$ are coming from the result $\mathcal{D}^{\mathrm{kin}}(t)\propto a_2^2(t)$, as derived in the toy model approach [see Eq.~(24) of Article 1 (\refsec{Art1})]. Thus, whereas one could think about the case in which both \acrshort{EME} and \acrshort{TME} are present and $t_{\mathcal{D}}\approx t_\theta$, this is concluded to be just present in the \index{Hydrodynamic!stage}hydrodynamic stage and, then, discarded from our kinetic-stage analysis. The conditions under which these events arise are classified in Table~I of Article 1 (\refsec{Art1}).

After the identification of the phenomenology in the absence of \index{Temperature!overshoot}overshoot, we compute numerically the phase diagrams of the different latter events. This can be observed in Figure~4 of Article 1 (\refsec{Art1}). Some illustrative examples, whose details are exposed in Table~II of Article 1 (\refsec{Art1}), are shown from numerical solutions of the \acrshort{LBSA} in Figures~5--8 of Article 1 (\refsec{Art1}).  

Furthermore, we consider the case in which the \index{Temperature!nonequilibrium}temperature evolution of (at least) one of the samples \index{Temperature!overshoot}overshoots the equilibrium value. Then, we characterize a new event compatible with the \acrshort{TME}, such that $T_A$ will arrive earlier to \index{Temperature!of the thermal bath}$T_b$ as compared with $T_B$, but in the absence of crossovers (or with an odd number of them). One expects that system B \index{Temperature!overshoot}overshoots \index{Temperature!of the thermal bath}$T_b$ in such a way that $T_A$ can overtake the evolution of $T_B$ toward $T_b$. This is called the \acrshort{OME}, already introduced in \refsec{ME_ME}, and it is expected to occur in states initially close to the equilibrium \index{Temperature!of the thermal bath}temperature, $T_b$, for this system. To analyze the \acrshort{OME}, a \acrshort{LBSA} is built from initial conditions at $|T(0)-T_b|\ll 1$ [see Eqs.~(30) of Article 1 (\refsec{Art1})]. Then, the crossover time of $T(t)$ through $T_b$ is computed, \phantomsection\label{sym:t_O}$t_O$, together with the necessary conditions for the \index{Temperature!overshoot}overshoot to appear [see Eq.~(33) of Article 1 (\refsec{Art1})]. In contrast to what happened in the standard version of the \acrshort{TME}, one must look for an initially much faster decay (growth) of system B than A in a cooling (heating) experiment. Moreover, whereas there is no crossing between the thermal curves, we realize that there is, in fact, a crossover between \phantomsection\label{sym:KLD_LE_A_B}$\mathcal{D}^{\mathrm{LE}}_A(t)$ and $\mathcal{D}^{\mathrm{LE}}_B(t)$ at \label{sym:t_KLD_LE}$t= t_{\mathcal{D}^{\mathrm{LE}}}$. The latter crossing time is predicted from \acrshort{LBSA}, and the conditions for the initial preparations to observe the \acrshort{OME} are also derived [see Eqs.~(35)--(37) of Article 1 (\refsec{Art1})]. To derive the latter necessary conditions, we impose that $ t_{\mathcal{D}^{\mathrm{LE}}}>0$ and $t_\theta <0$ (no crossing between $T_A$ and $T_B$). A comparison between the standard \acrshort{TME} and the \acrshort{OME} in this system is summarized in \reftab{TME_and_OME}. We then conclude that $\mathcal{D}^{\mathrm{LE}}$, being a convex and positive function of $T(t)/T_b$, which, in addition, $\mathcal{D}^{\mathrm{LE}}=0$ if and only if \index{Temperature!of the thermal bath}$T(t)=T_b$, describes both the standard \acrshort{TME} (\acrshort{DME} and \acrshort{IME}) and the \acrshort{OME} in the crossing-time characterization of the \acrshort{TME}. That is, the $\mathcal{D}^{\mathrm{LE}}$ acts as the thermal distance introduced in \refsec{ME_ME}. Afterwards, some numerical examples from the \acrshort{LBSA} for the \acrshort{OME} were given in Figures~9 and 10 of Article 1 (\refsec{Art1}).

\begin{table}[t]
\caption{Summary of crossing times $t_X$ (with $X=\theta,\mathcal{D}^{\mathrm{LE}}$) and the necessary conditions derived from \acrshort{LBSA} for the emergence of the standard \acrshort{TME} and the \acrshort{OME}, in the case of a molecular gas of hard \dt-spheres in contact with a background fluid with a nonlinear drag. Here, $R^0 \equiv (T_A^0-T_B^0)/T_b(a_{2A}^0-a_{2B}^0)$, $R_+^0 \equiv (T_A^0+T_B^0+2T_b)/T_b(a_{2A}^0+a_{2B}^0)$, $R^0_{\max}\equiv A_{12}/A_{11}$, $A_{ij}$ and $\lambda_{+,-}$ are coming from the \acrshort{LBSA} developed in Appendix~C of Article 1 (\refsec{Art1}). Overlined quantities refer to their values at $T^0=T_b$.}
\labtab{TME_and_OME}
\centering
\begin{tabularx}{\textwidth}{lcc}
\hline\hline
	 & Standard \acrshort{TME} & \acrshort{OME} \\
	\hline\\
	$t_X$ & ${\displaystyle{t_\theta =\frac{1}{\lambda_+-\lambda_{-}}\ln\left(1+\frac{A_{11}^{-1}}{R^0_{\max}/R^0-1}\right)}}$  & ${\displaystyle{t_{\mathcal{D}^{\mathrm{LE}}} =\frac{1}{\overline{\lambda}_+-\overline{\lambda}_{-}}\ln\left(1+\frac{\overline{A}_{11}^{-1}}{\overline{R}^0_{\max}/R_+^0-1}\right)}}$ \\
	If~\dots & ${\displaystyle{0<R^0<R^0_{\max}}}$ & ${\displaystyle{(0<R_+^0<R^0_{\max}) \wedge (R^0<0 \lor R^0>\overline{R}^0_{\max})}}$ \\
	&&\\
	\hline\hline
\end{tabularx}
\end{table}

Finally, we compare the theoretical predictions with \acrshort{DSMC} and \acrshort{EDMD} computer simulations from hand-made computer programs~\cite{M23_github}. Here, it is necessary to implement the Langevin dynamics of the system according to the Fokker--Planck term of the \acrshort{BFPE} in the free-streaming stages of both algorithms, as indicated in \refch{SM}. Concretely, we use the \acrshort{AGF} algorithm reported on Refs.~\cite{SVdM07,S12} adapted to the \acrshort{LE} of this problem (see \refapp{app_FPLE}), which is detailed both in Article 1 (\refsec{Art1}) and in \refapp{app_AGF}. Representative cases for the different predicted events are shown. It is observed that the simulation outcomes reproduce the predictions. Moreover, numerical solutions coming from \acrshort{BSA} and \acrshort{ESA} are compared with the simulation data, concluding a quantitatively better agreement with the \acrshort{ESA}.

\section{Article 1}\labsec{Art1}
\underline{\textbf{Title:}} Thermal versus entropic Mpemba effect in molecular gases with nonlinear drag\\
\underline{\textbf{Authors:}} Alberto Meg\'ias$^1$, Andr\'es Santos$^{1,2}$, and Antonio Prados$^3$\\
\noindent \underline{\textbf{Affiliations:}}\\
$^1$ Departamento de Física, Universidad de Extremadura, E-06006 Badajoz,
Spain\\
$^2$  Instituto de Computación Científica Avanzada
(ICCAEx), Universidad de Extremadura, E-06006 Badajoz, Spain\\
$^3$ F\'isica Te\'orica, Universidad de Sevilla, Apartado de Correos 1065, E-41080 Sevilla, Spain\\ \vspace{-0.6cm}
\begin{flushleft}
\begin{minipage}{0.5\textwidth}\raggedright
\underline{\textbf{Journal:}} Physical Review E\vspace{0.2cm}

\underline{\textbf{Volume:}} 105\vspace{0.2cm}

\underline{\textbf{Pages:}} 054140\vspace{0.2cm}

\underline{\textbf{Year:}} 2022\vspace{0.2cm}

\underline{\textbf{DOI:}} \href{https://doi.org/10.1103/PhysRevE.105.054140}{10.1103/PhysRevE.105.054140}
\end{minipage}
\begin{minipage}{0.49\textwidth}\raggedleft
\includegraphics[width=0.49\textwidth]{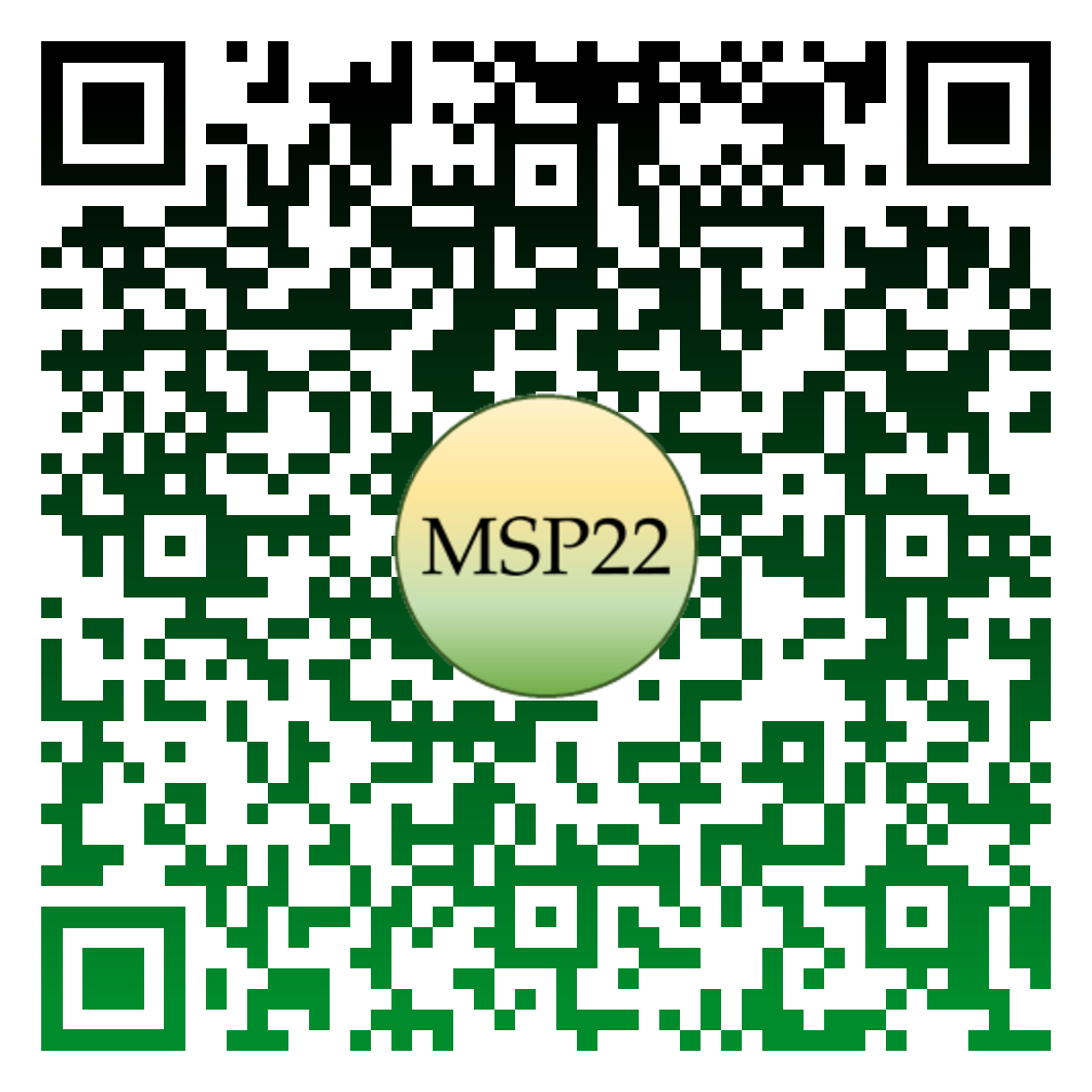}
\end{minipage}
\end{flushleft}
Copy of the preprint of the work: ``Alberto Meg\'ias, Andr\'es Santos, and Antonio Prados, 'Thermal versus entropic Mpemba effect in molecular gases with nonlinear drag', \emph{Physical Review E} \textbf{105}, 054140 (2022) \href{https://doi.org/10.1103/PhysRevE.105.054140}{https://doi.org/10.1103/PhysRevE.105.054140}.''

\includepdf[clip,
    trim=4mm 4.5mm 10mm 1.5mm,pages=-,
   scale=0.8,pagecommand={},offset=8mm -6mm,
   frame]{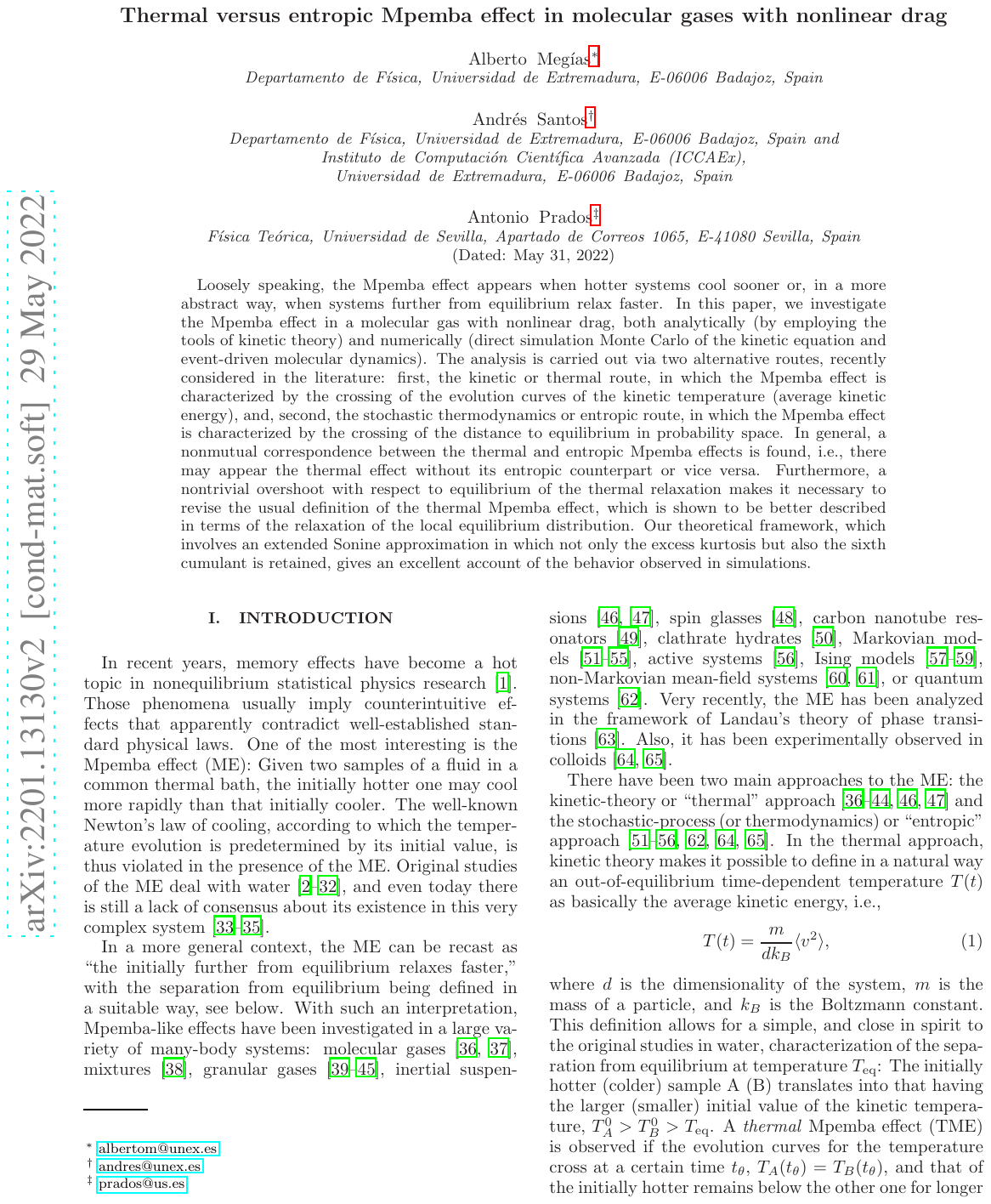}

\addpart{Granular gases of inelastic and smooth particles}

\chapter{Nonequilibrium entropy of inelastic hard $\dt$-spheres}
\labch{KLD_IHS}

\section{Summary}

In this chapter, the \index{Nonequilibrium entropy}nonequilibrium entropy of a homogeneous freely evolving dilute granular gas made of hard \dt-spheres, described by the \acrshort{IHS} collisional model, is studied. Concretely, the particles are considered to be identical, with diameter $\sigma$, mass $m$, and coefficient of normal restitution\index{Coefficient of normal restitution} $\alpha$, assumed to be constant. Their evolution is just subjected to their mutual collisions, and neither external force fields nor spatial gradients are considered. The system is then described by the homogeneous \acrshort{BE} written in \refeq{HBE_IHS}.

The aim of this chapter is to address the problem of the existence of an $H$-theorem associated with the homogeneous and inelastic \acrshort{BE}. We recall that an inelastic granular gas reaches, during its time evolution, a scaling solution, as already introduced in \refsubsec{HCS_KTGG}, which corresponds to the \acrshort{HCS}. Then, one expects that if the system remains in homogeneous states, the \acrshort{HCS} solution might be an attractor of the evolution of the system.  Whereas this argument seems to be reasonable, it has not yet been mathematically proven. In fact, inelasticity breaks down the symmetry underlying in the collisional rules, violating the original hypotheses, on which the original proof of the molecular $H$-theorem is sustained (see \refsubsec{HTheorem_GG_KTGG}). In this chapter, we try to give arguments in favor of an $H$-theorem for homogeneous and inelastic granular gases based on the choice of the \acrshort{KLD} functional (identifying the reference \acrshort{VDF} as the \acrshort{HCS} one) as the possible Lyapunov functional that would solve the problem. This conjecture has been previously formulated in Refs.~\cite{BPV13,GMMMRT15,PP17}, but it is still not proved or disproved.

In Article 2 (\refsec{Art2}), the homogeneous states of the granular gas of inelastic \dt-spheres is studied from a Sonine expansion of the \acrshort{VDF} both in transient states and in the \acrshort{HCS}. To develop the evolution equations, we truncate the Sonine expansion up to the sixth cumulant $a_3$, and the considered slopes are linearized with respect to $a_2$ and $a_3$. Then, the system of evolution equations for the granular gas is formed by \index{Haff's law}Haff's law with the \index{Cooling rate!in the IHS model}cooling rate approximated within this \acrshort{SA}, and the differential equations of $a_2$ and $a_3$. Moreover, the proper linearizations of the equations for $\dot{a}_2$ and $\dot{a}_3$ are derived in such a way that their steady states coincide with the best approximate \acrshort{HCS} values reported on Ref.~\cite{SM09}. In fact, this \acrshort{SA} allows us to solve explicitly the system of differential equations for $\dot{a}_2$ and $\dot{a}_3$, see Eqs.~(25)--(29) of Article 2 (\refsec{Art2}). These theoretical approaches are tested with \acrshort{EDMD} simulation results from the DynamO software~\cite{BSL11} for a wide range of values of $\alpha$, covering elastic, quasielastic, intermediately inelastic, and highly inelastic systems. The \acrshort{HCS} values for the fourth and sixth cumulants are compared and concluded to be in good agreement with already reported \acrshort{DSMC} values in Ref.~\cite{SM09}. Moreover, the evolution equations for $a_2$ and $a_3$ under this \acrshort{SA} are compared for a wide variety of initial conditions, getting quite a good quantitative agreement for the evolution of $a_2$ and a more qualitatively agreement for the evolution of $a_3$, which improves for initial conditions not far from a Maxwellian distribution.

Once the homogeneous states of the system are studied, the quest of a proper \index{Nonequilibrium entropy}nonequilibrium entropy is discussed. First of all, the original $H$ functional [see \refeq{H_functional}] presents certain problems that preclude its candidature to be the \index{Nonequilibrium entropy}nonequilibrium entropy of the system. First of all, as already mentioned, the proof of the original $H$-theorem does not fit any more. Moreover, it presents the so-called measure problem, as introduced in Ref.~\cite{MT11}. That is, the $H$-functional is not invariant under nonunitary transformations. In this system, there exists a natural scaling which is not unitary, $\vvel\rightarrow \cc$. In fact, it is proved that Shannon's measure in the $\vvel$ space coincides with the $H$ functional for the rescaled \acrshort{VDF}, $\phi(\cc)$, called \phantomsection\label{sym:Hast}$H^*$ in Article 2 [see Eq.~(34) of Article 2 (\refsec{Art2})], plus an additional term, which increases with time. This $H^*$, which could be thought as a possibly candidate, still presents the measure problem. This is solved by the \acrshort{KLD}. Then, admitting that a relative measure of entropy as the \acrshort{KLD} is required, we need to choose the proper reference \acrshort{VDF}, $f_{\mathrm{ref}}$. We compare two possible choices. First, if this reference \acrshort{VDF} is chosen as the Maxwellian distribution, $f_{\mathrm{ref}}=f_{\mathrm{M}}$, then, the time dependent \acrshort{KLD} will coincide with $H^*$ plus a numerical constant [see Eq.~(37) of Article 2 (\refsec{Art2})]. Moreover, the other choice is the well-based conjecture $f_{\mathrm{ref}} = f_{\HCS}$. These two choices are analyzed via a numerical Sonine-approximated scheme [see Eqs.~(39) and (40) of Article 2 (\refsec{Art2})] and \acrshort{EDMD} simulations for a wide variety of initial conditions and both \acrshort{HD} ($\dt=2$) and \acrshort{HS} ($\dt=3$) systems, as can be observed in Figures~6--9 of Article 2 (\refsec{Art2}). Here, we do observe cases that reject the former choice, whereas all of them are in agreement with the conjecture. Finally, we develop a \emph{toy model} based on a \acrshort{SA} of the \acrshort{KLD} up to $a_2$. That approximate scheme, together with the evolution equation of the fourth cumulant, indicates a monotonic decay of $\KLD(f|f_{\HCS})$. This latter argument is in favor of the conjecture of $S=-\KLD(f|f_{\HCS})$ being a proper \index{Nonequilibrium entropy}nonequilibrium entropy for this system. At the end of Article 2 (\refsec{Art2}), the quantity $\KLD(f_{\HCS}|f_{\mathrm{M}})$ is computed numerically and from \acrshort{EDMD} simulation results, which indicates how far one \acrshort{VDF} is from the other.

In addition, the toy model derived in Article 2 (\refsec{Art2}) has been extended to an arbitrary reference \acrshort{VDF} in Article 3 (\refsec{Art3}). Here, in Article 3 (\refsec{Art3}), we find that, under this approach, there are counterexamples for the choice $\KLD(f|f_{\mathrm{M}})$ as a possible \index{Nonequilibrium entropy}nonequilibrium entropy, in quite good agreement with new \acrshort{EDMD} simulation results for \acrshort{HS}. Moreover, these additional simulations are still in accordance with the conjecture $S = -\KLD(f|f_{\HCS})$ [see Figure~1 of Article 3 (\refsec{Art3})]. Finally, in Article 3 (\refsec{Art3}), a Maxwell-demon-like experiment in \acrshort{EDMD} simulations was carried out, based on Orban and Bellemans' pioneering works~\cite{OB67,OB69} for the \acrshort{EHS} model, and Aharony's work~\cite{A71}, including a source of irreversibility in the particle dynamics. We let the system evolve up to a certain \emph{waiting} number of collisions per particle, \phantomsection\label{sym:sw}$s_w$, and, then, all the velocities are inverted. The evolution of the system is guided by $\KLD(f|f_{\mathrm{M}})$ in the elastic case, and by $\KLD(f|f_{\HCS})$ for $\alpha<1$. In the elastic case ($\alpha=1$), original results of Refs.~\cite{OB67,OB69} were observed, with the system entering in an anti-kinetic stage, trying to restore its initial state, and this stage being symmetric with respect to the previous evolution. After this returning regime, the system again evolves up to equilibrium. Notice that, in the elastic case, $H$ and $\KLD(f|f_{\mathrm{M}})$ coincide up to a numerical constant. Moreover, for $\alpha<1$, this anti-kinetic stage is almost completely suppressed. Only a small hump in the evolution of the \acrshort{KLD} is observed in the inelastic case. This is a consequence of the irreversibility introduced into the granular dynamics by the \acrshort{IHS} model [see Eqs.~(8) and (9) of Article 3 (\refsec{Art3})]. The results are quite similar to Aharony's observations~\cite{A71}. These humps are indeed observed to decrease as inelasticity increases [see Figure~2 of Article 3 (\refsec{Art3})].
 
\newpage

\section{Article 2}\labsec{Art2}
\underline{\textbf{Title:}} Kullback--Leibler Divergence of a Freely Cooling Granular Gas \\
\underline{\textbf{Authors:}} Alberto Meg\'ias$^1$ and Andr\'es Santos$^{1,2}$\\
\noindent \underline{\textbf{Affiliations:}}\\
$^1$ Departamento de Física, Universidad de Extremadura, E-06006 Badajoz,
Spain\\
$^2$  Instituto de Computación Científica Avanzada
(ICCAEx), Universidad de Extremadura, E-06006 Badajoz, Spain\\\vspace{-0.6cm}
\begin{flushleft}
\begin{minipage}{0.5\textwidth}\raggedright
\underline{\textbf{Journal:}} Entropy\vspace{0.2cm}

\underline{\textbf{Volume:}} 22\vspace{0.2cm}

\underline{\textbf{Pages:}} 1308\vspace{0.2cm}

\underline{\textbf{Year:}} 2020\vspace{0.2cm}

\underline{\textbf{DOI:}} \href{https://doi.org/10.3390/e22111308}{10.3390/e22111308}
\end{minipage}
\begin{minipage}{0.49\textwidth}\raggedleft
\includegraphics[width=0.49\textwidth]{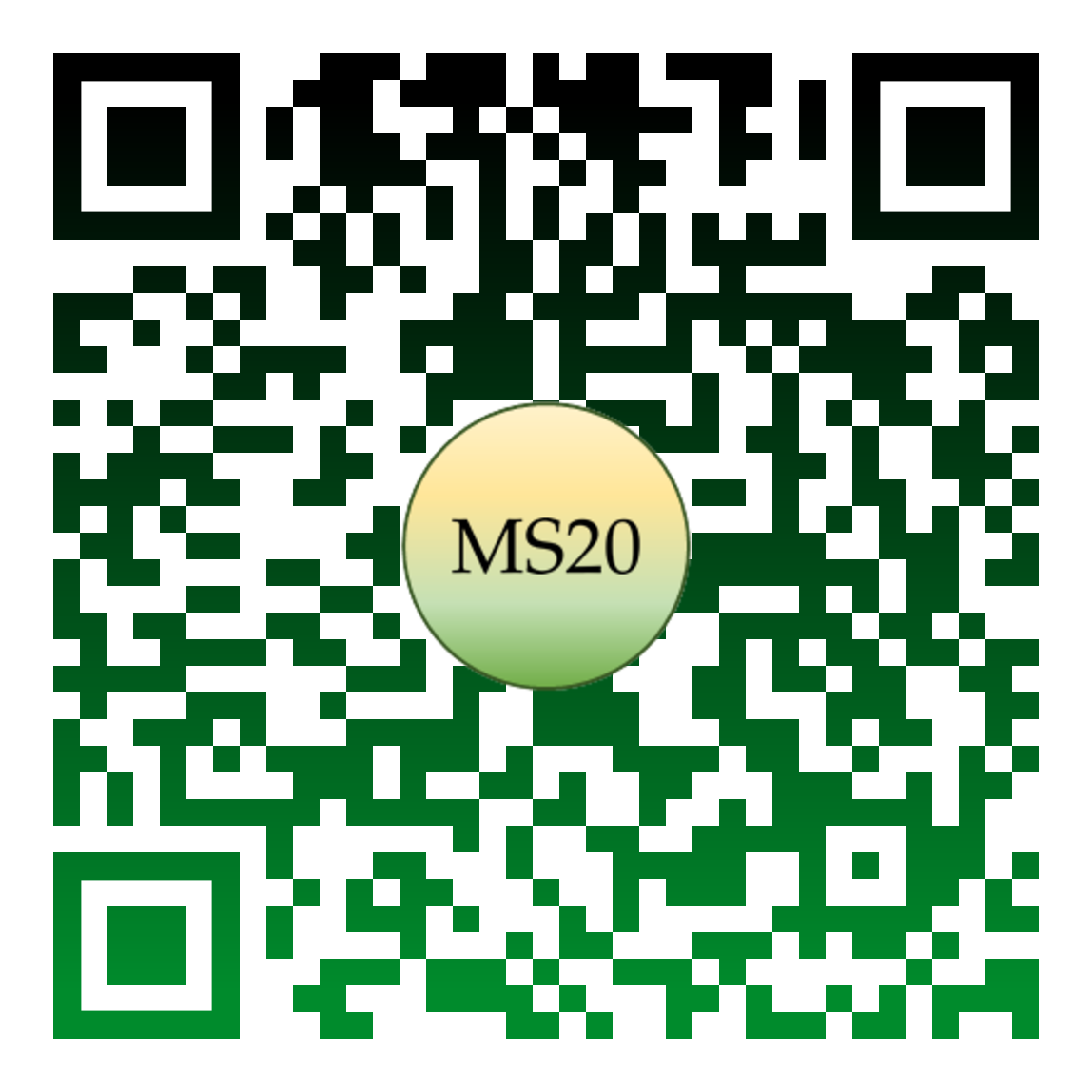}
\end{minipage}
\end{flushleft}
Copy of the preprint of the work: ``Alberto Meg\'ias, and Andr\'es Santos, 'Kullback--Leibler Divergence of a Freely Cooling Granular Gas', \emph{Entropy} \textbf{22} 11, 1308 (2020) \href{https://doi.org/10.3390/e22111308}{https://doi.org/10.3390/e22111308}.''

\includepdf[clip,
    trim=4mm 4.5mm 10mm 1.5mm,pages=-,
   scale=0.8,pagecommand={},offset=8mm -6mm,
   frame]{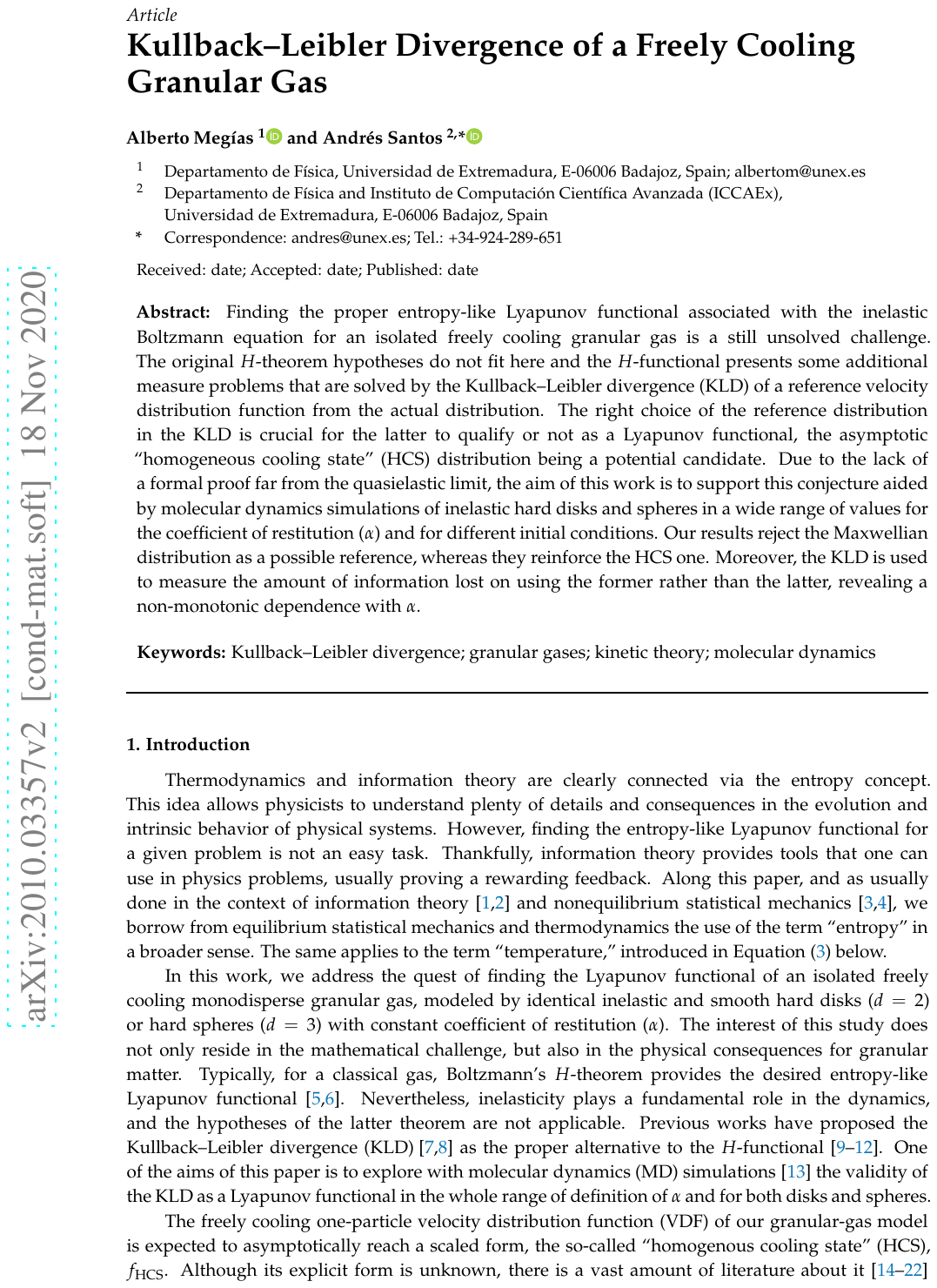}

\section{Article 3}\labsec{Art3}

\underline{\textbf{Title:}} Relative entropy of freely cooling granular gases. A molecular dynamics study \\
\underline{\textbf{Authors:}} Alberto Meg\'ias$^1$ and Andr\'es Santos$^{1,2}$\\
\noindent \underline{\textbf{Affiliations:}}\\
$^1$ Departamento de Física, Universidad de Extremadura, E-06006 Badajoz,
Spain\\
$^2$  Instituto de Computación Científica Avanzada
(ICCAEx), Universidad de Extremadura, E-06006 Badajoz, Spain\\\vspace{-0.6cm}
\begin{flushleft}
\begin{minipage}{0.5\textwidth}\raggedright
\underline{\textbf{Journal:}} EPJ Web of Conferences\vspace{0.2cm}

\underline{\textbf{Volume:}} 246\vspace{0.2cm}

\underline{\textbf{Pages:}} 04006\vspace{0.2cm}

\underline{\textbf{Year:}} 2021\vspace{0.2cm}

\underline{\textbf{DOI:}} \href{https://doi.org/10.1051/epjconf/202124904006}{10.1051/epjconf/202124904006}
\end{minipage}
\begin{minipage}{0.49\textwidth}\raggedleft
\includegraphics[width=0.49\textwidth]{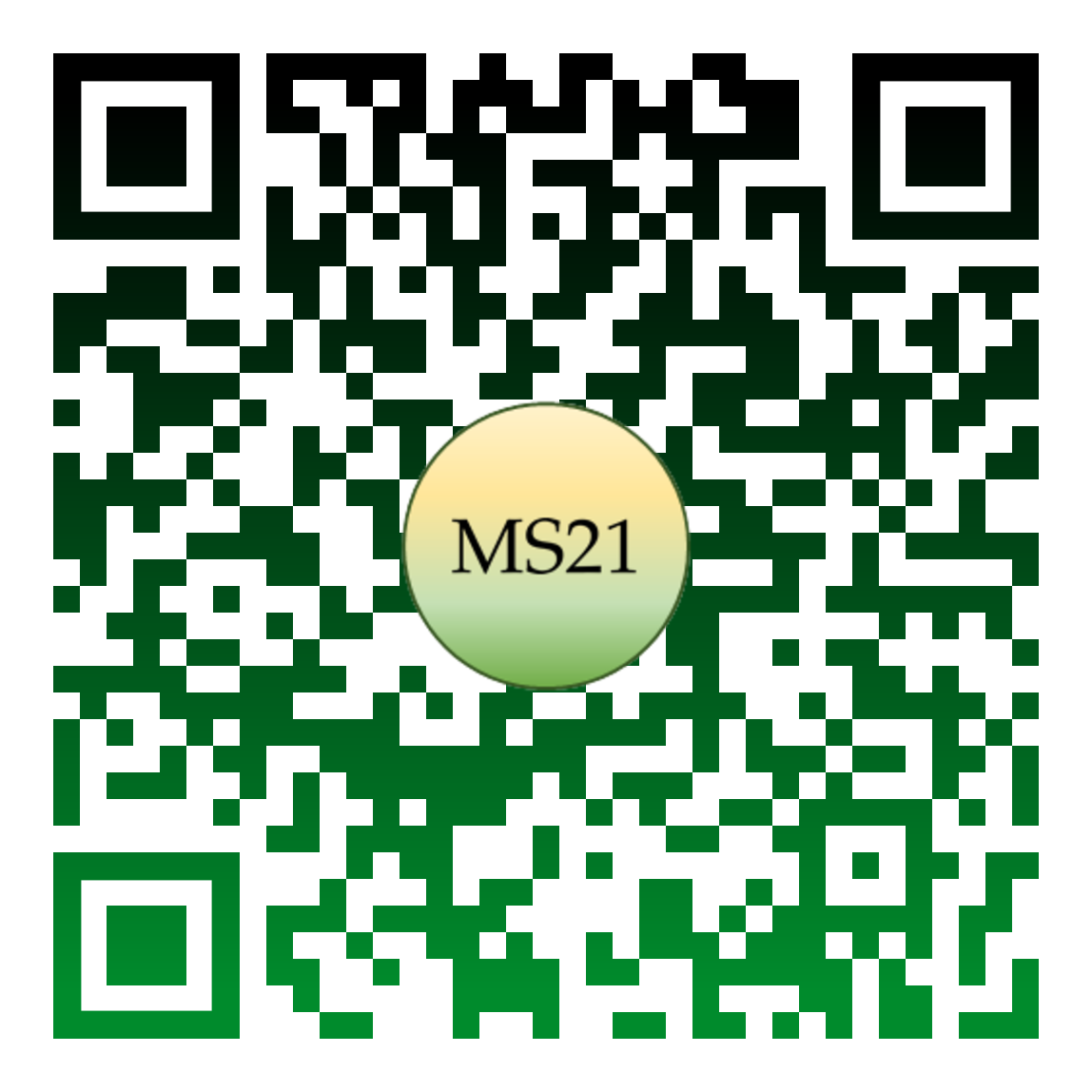}
\end{minipage}
\end{flushleft}
Copy of the preprint of the work: ``Alberto Meg\'ias, and Andr\'es Santos, 'Relative entropy of freely cooling granular gases. A molecular dynamics study', \emph{EPJ Web of Conferences} \textbf{249}, 04006 (2021) \href{https://doi.org/10.1051/epjconf/202124904006}{https://doi.org/10.1051/epjconf/202124904006}.''

\includepdf[clip,
    trim=4mm 4.5mm 10mm 1.5mm,pages=-,
   scale=0.8,pagecommand={},offset=8mm -6mm,
   frame]{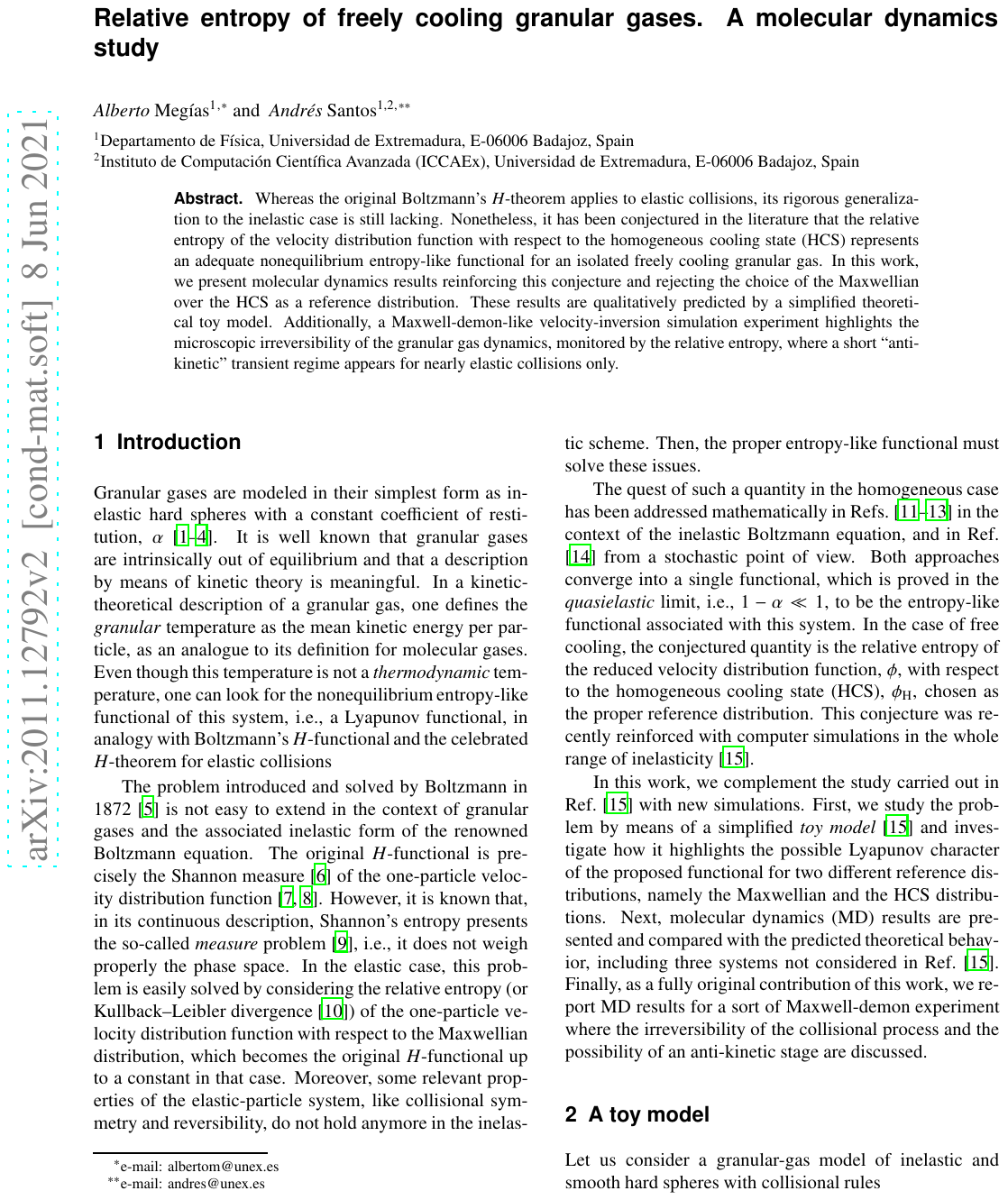}
\chapter{Inelastic hard \dt-spheres under nonlinear drag}
\labch{IHS_ND}

\section{Summary}

The homogeneous states of a dilute granular gas made of inelastic and smooth identical \dt-spheres of diameter $\sigma$, mass $m$, and coefficient of normal restitution\index{Coefficient of normal restitution} $\een$ (assumed to be constant), and surrounded by a background fluid, equilibrated at \index{Temperature!of the thermal bath}temperature $T_b$, are studied. The binary collisions between the granular particles are described by the \acrshort{IHS} model, whereas the interaction between the gas and its surrounding fluid is approximated via the coarse-grain model described in \refsubsec{nonlinear_KTGG}. Granular particles are subjected to drag and stochastic forces, and assumed to follow, in free flow, the \acrshort{LE} written in \refeq{LE_NLD_KTGG_def}. Then, the evolution of its one-body \acrshort{VDF} is described by the \acrshort{BFPE} written in \refeq{BFPE_NLD}, particularized to the context of the \acrshort{IHS} collisional model.

The main objective of this chapter is to provide a detailed description of the evolution and steady-state properties of this homogeneous system. For that, we first formally derive, from the \acrshort{BFPE}, the evolution equation of the granular \index{Temperature!granular}temperature [see Eq.~(8) of Article 4 (\refsec{Art4})]. The granular temperature slope is explicitly coupled to the fourth cumulant of the \acrshort{VDF}, $a_2$, as occurs in Article 1 (\refsec{Art1}) in the case of molecular gases. Additionally to the latter term, the \index{Cooling rate!in the IHS model}cooling rate appears, which is a function of the full \acrshort{VDF}. Hence, the cooling term breaks down the molecular equilibrium solution, but the system still admits a steady state. To work in dimensionless terms, we define the same time and \index{Temperature!of the thermal bath}temperature reference scales as in Article 1 (\refsec{Art1}), \phantomsection\label{sym:taub}$\tau_b$ and $T_b$, respectively. In order to complete the evolution description of the \index{Temperature!granular}granular temperature, we introduce a Sonine expansion for both the time-dependent and the steady \acrshort{VDF}. Therefore, from the \acrshort{BFPE}, an infinite hierarchy of moment equations arise [see Eq.~(15) of Article 4 (\refsec{Art4})], quite similar to that of Article 1 (\refsec{Art1}), but with the additional cooling term coupling. 

Since the derived infinite set of coupled nonlinear differential equations is impossible to be solved, either analytically or numerically, we introduce two approximate schemes for both evolution and steady states. First, we consider a \acrshort{MA} to approximate the \acrshort{VDF}, neglecting the differential equations of all the nontrivial cumulants of the \acrshort{VDF}, as well as their role in the granular \index{Temperature!granular}temperature evolution. Therefore, the \index{Temperature!nonequilibrium}temperature evolves with a self-consistent closed differential equation, from which we infer the \acrshort{MA} of the steady-state value of the \index{Temperature!steady-state}temperature, $T^{\mathrm{st}}/T_b$, this being the physical root of a fourth-degree polynomial [see Eq.~(18) of Article 4 (\refsec{Art4})]. Whereas this is the simplest approach, the already known evolution schemes in the molecular case~\cite{SP20,PSP21,MSP22} [see Article 1 (\refsec{Art1})] indicates that the evolution of the cumulants is essential to infer a good relaxation of the \index{Temperature!nonequilibrium}temperature. Moreover, the breakdown of the equilibrium solution due to the cooling is a signature that the steady-state \acrshort{VDF} deviates from the Maxwellian distribution. Therefore, the second approximation that we introduce consists in a \acrshort{SA} characterized by a truncation and linearization up to the fourth cumulant, $a_2$ [see Eqs.~(19)--(21) of Article 4 (\refsec{Art4})], referred to as \emph{first} \acrshort{SA} (\acrshort{FSA}) throughout Article 4 (\refsec{Art4}). From the latter scheme, we also obtain an approximation for the stationary values of the granular \index{Temperature!granular}temperature, as well as an expression for the steady-state values of the fourth cumulant, \phantomsection\label{sym:a2st}$a_2^{\mathrm{st}}$ [see Eqs.~(22)--(24) of Article 4 (\refsec{Art4})].

After the derivation of the theoretical predictions, we compare the values of $T^{\mathrm{st}}/T_b$ from the \acrshort{MA} and the \acrshort{FSA}. The results, from both approaches, are quite similar, but they present small discrepancies. These differences increase with the value of the nonlinear parameter $\gamma$ [see \refeq{xi_QR_gamma}], which controls the nonlinear part of the drag coefficient\index{Drag coefficient!velocity-dependent} [see Figure~2 of Article 4 (\refsec{Art4})]. All the stationary values of the granular \index{Temperature!steady-state}\index{Temperature!granular}temperature are below the bath \index{Temperature!of the thermal bath}temperature, i.e., $T^{\mathrm{st}}/T_b<1$, for every value of the parameters in the inelastic case, $\een<1$. This is a consequence of the influence of the granular cooling. Moreover, it is predicted that the excess kurtosis is positive in most cases and, only in a small region of the parameter space, $a_2^{\mathrm{st}}$ is negative for \acrshort{HD} and \acrshort{HS} [see Figure~3 of Article 4 (\refsec{Art4})]. In fact, there exists a critical value of the nonlinear parameter, \phantomsection\label{sym:gammac_IHS}$\gamma_c=1/3(\dt+2)$, such that for $\gamma>\gamma_c$, $a_2^{\mathrm{st}}$ is always positive for whatever inelastic value of the coefficient of normal restitution \index{Coefficient of normal restitution}($\een<1$). 

To complete the theoretical prediction, we compare the \acrshort{FSA} steady-state predictions with several well-known limiting cases. First, in the absence of drag (but in the presence of the stochastic force), the granular gas would be driven by just a stochastic thermostat, and the corresponding steady conditions are recovered. Then, if no interstitial fluid is contemplated, the \acrshort{HCS} conditions are retaken (see \refsubsec{HCS_KTGG}). Moreover, in the case of the presence of a interstitial fluid, but with a linear drag coefficient\index{Drag coefficient}, i.e., $\gamma\to 0$, results of Refs.~\cite{CVG12,CVG13} are recovered. Also, in the limit where the collision frequency $\nu_b$ is negligible as compared with the linear part of the drag coefficient\index{Drag coefficient!velocity-dependent} (i.e., the gas is assumed to be collisionless), equilibrium is restored because inelasticity does not play any role.

As a test for the theoretical predictions, we show \acrshort{DSMC} and \acrshort{EDMD} outcomes from hand-made computer programs~\cite{M23_github}. In the \acrshort{EDMD} algorithm, the \acrshort{AGF} algorithm is used to reproduce the Langevin-like dynamics in the free-streaming stage of the evolution. We test the steady and evolution states for different values of the coefficient of normal restitution\index{Coefficient of normal restitution}, $\een$, and the nonlinear parameter, $\gamma$. The \acrshort{FSA} is quantitatively better than the \acrshort{MA}, with the $a_2^{\mathrm{st}}$ predictions underestimating the simulation values for very inelastic particles and for $\gamma\sim 0.1$, which is the limiting regime of the \acrshort{QR} approximation [see \refeq{xi_QR_gamma}]. 

The second aim of the chapter is to study the emergence of some memory effects in this system. The \acrshort{FSA} indicates that the evolution of $T(t)/T_b$ is coupled to the excess kurtosis, recalling the scheme presented in \refeqs{memory_descr}. Thus, we expect that, as occurs with the molecular case and other driven granular systems, memory effects, such as \acrshort{ME} and \acrshort{KE}, must be present in this gaseous system. First, we show a pair of examples where the \acrshort{TME} appears. One of them corresponds to the \acrshort{DME} and another to the \acrshort{IME} [see Figure~6 of Article 4 (\refsec{Art4})]. In these cases, \acrshort{FSA} theoretical curves and simulation results, from \acrshort{DSMC} and \acrshort{EDMD} algorithms, are in good agreement. Afterwards, the appearance of Kovacs-like humps is discussed. Foremost, we consider the system to be immersed in a Kovacs-like protocol, as introduced in \refsec{ME_KE}, with the granular \index{Temperature!granular}temperature playing the role of the guiding macrostate variable. Then, in the \acrshort{FSA}, Eq.~(29) of Article 4 (\refsec{Art4}) establishes the form of the temperature slope at the waiting time, \phantomsection\label{sym:t_K_ast}$t^*_{\mathrm{K}}$, whose sign is determined by the difference $a_2^{\mathrm{st}}-a_2(t^*_{\mathrm{K}})$. Thus, one observes an upward hump if $a_2^{\mathrm{st}}>a_2(t^*_{\mathrm{K}})$ and a downward hump if $a_2^{\mathrm{st}}<a_2(t^*_{\mathrm{K}})$. This is confirmed by means of \acrshort{DSMC} and \acrshort{EDMD} simulation outcomes, as reproduced in Figure~7 of Article 4 (\refsec{Art4}).

\newpage 

\section{Article 4}\labsec{Art4}

\underline{\textbf{Title:}} Kinetic Theory and Memory Effects of Homogeneous Inelastic Granular Gases under Nonlinear Drag \\
\underline{\textbf{Authors:}} Alberto Meg\'ias$^1$ and Andr\'es Santos$^{1,2}$\\
\noindent \underline{\textbf{Affiliations:}}\\
$^1$ Departamento de F\'isica, Universidad de Extremadura, E-06006 Badajoz,
Spain\\
$^2$  Instituto de Computaci\'on Cient\'ifica Avanzada
(ICCAEx), Universidad de Extremadura, E-06006 Badajoz, Spain\\\vspace{-0.6cm}
\begin{flushleft}
\begin{minipage}{0.5\textwidth}\raggedright
\underline{\textbf{Journal:}} Entropy\vspace{0.2cm}

\underline{\textbf{Volume:}} 24\vspace{0.2cm}

\underline{\textbf{Pages:}} 1436\vspace{0.2cm}

\underline{\textbf{Year:}} 2022\vspace{0.2cm}

\underline{\textbf{DOI:}} \href{https://doi.org/10.3390/e24101436}{10.3390/e24101436}
\end{minipage}
\begin{minipage}{0.49\textwidth}\raggedleft
\includegraphics[width=0.49\textwidth]{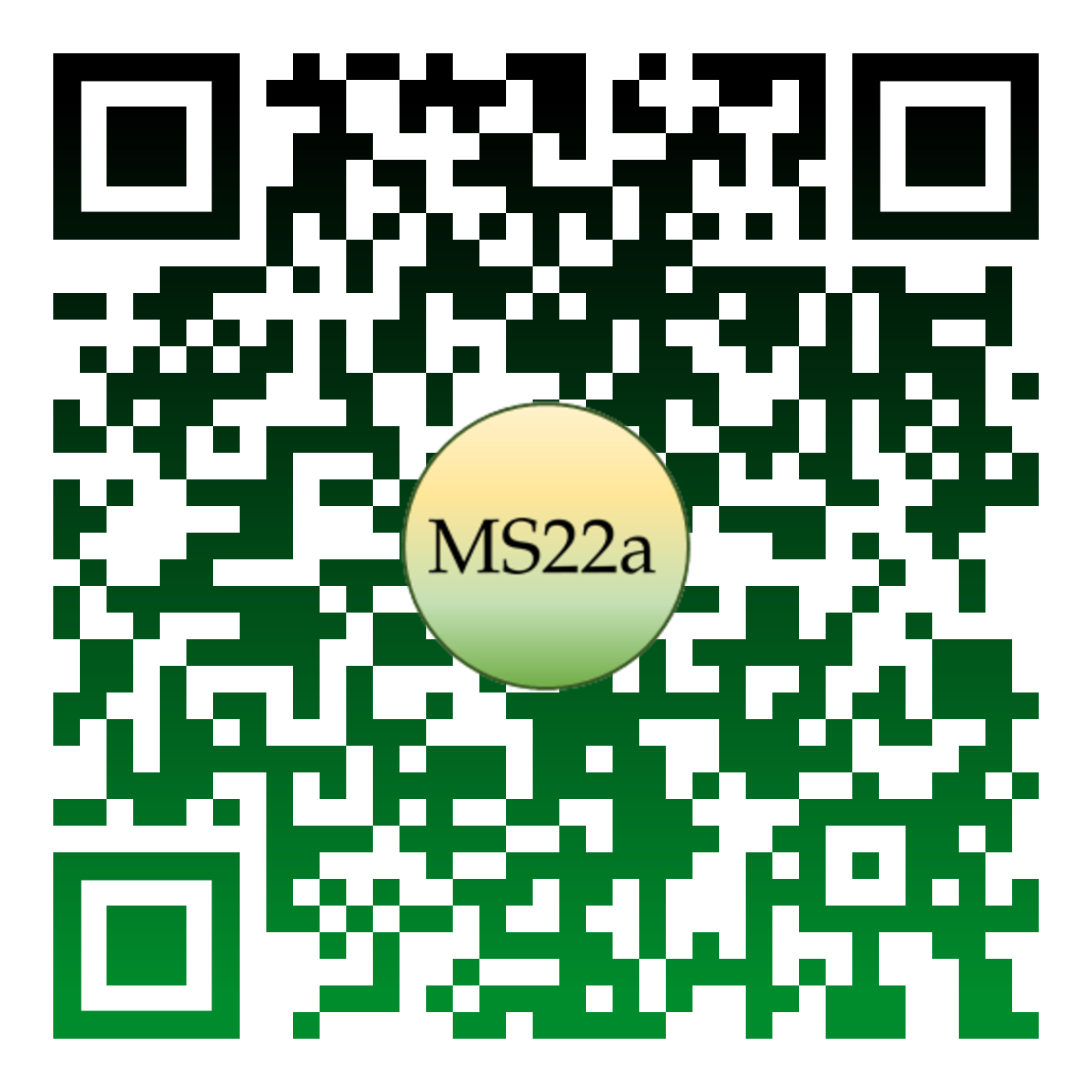}
\end{minipage}
\end{flushleft}
Copy of the preprint of the work: ``Alberto Meg\'ias, and Andr\'es Santos. 'Kinetic Theory and Memory Effects of Homogeneous Inelastic Granular Gases under Nonlinear Drag', \emph{Entropy} \textbf{24} 10, 1436 (2022) \href{https://doi.org/10.3390/e24101436}{https://doi.org/10.3390/e24101436}.''

\includepdf[clip,
    trim=4mm 4.5mm 10mm 1.5mm,pages=-,
   scale=0.8,pagecommand={},offset=8mm -6mm,
   frame]{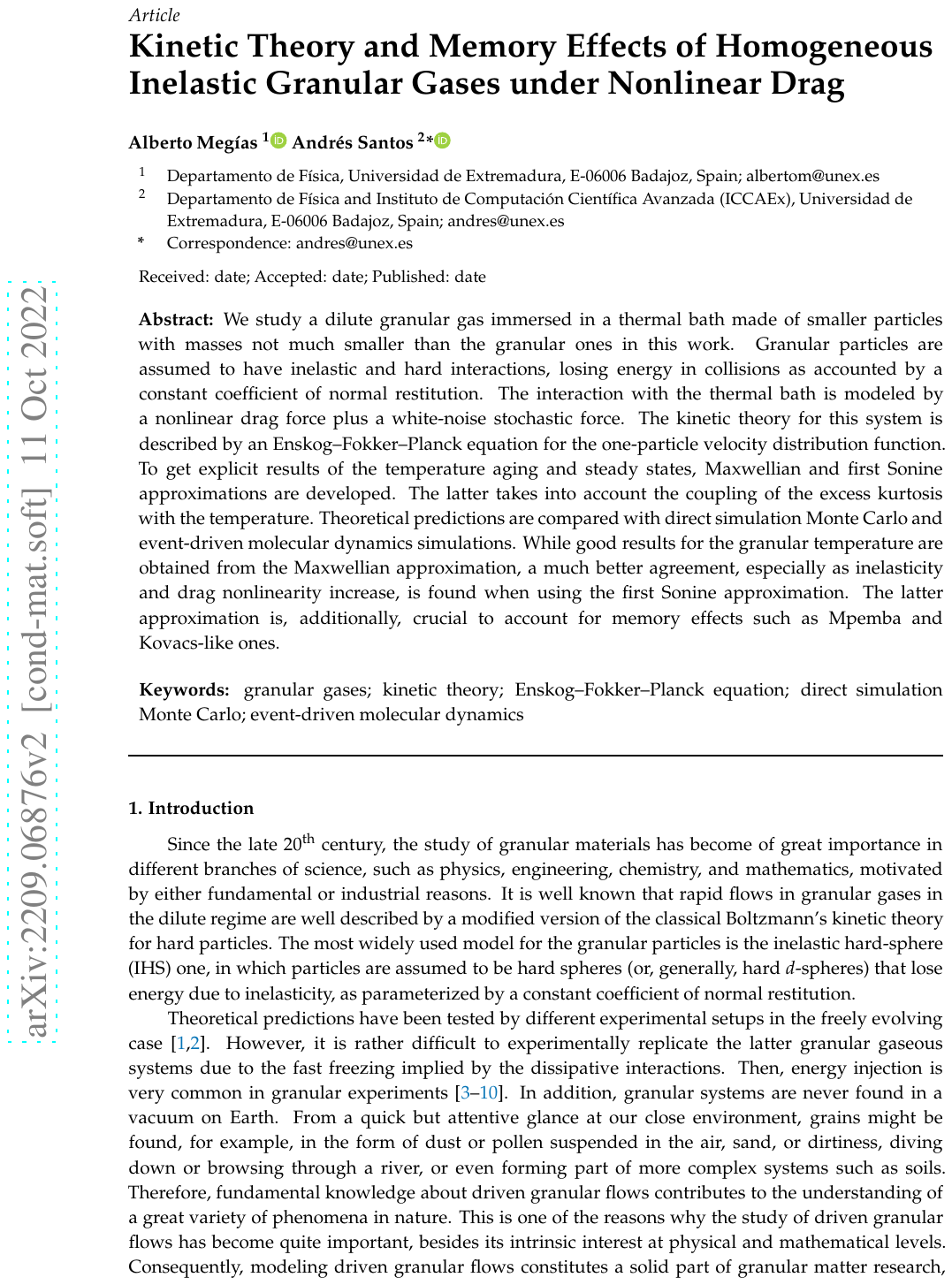}

\addpart{Granular gases of inelastic and rough particles}

\chapter{Homogeneous states in the IRHS model}
\labch{IRHS_HS}

\section{Summary}

In this chapter, the homogeneous states of a freely evolving dilute and monodisperse granular gas made of inelastic and rough \acrshort{HD} ($\dt=2,\,\dr=1$) or \acrshort{HS} ($\dt=3,\,\dr=3$) are studied. The granular particle are characterized by a common diameter ($\sigma$), mass ($m$), reduced moment of inertia ($\kappa$), their binary collisions being described by the \acrshort{IRHS} model. That is, the inelasticity of the system is parameterized by a coefficient of normal restitution\index{Coefficient of normal restitution}, $\een$, and the effect of the surface roughness in a binary collision is accounted for by a coefficient of tangential restitution\index{Coefficient of tangential restitution}, $\eet$, both considered to be constant. The evolution of the granular gas one-body \acrshort{VDF} is characterized by its associated \acrshort{BE}, written in \refeq{HBE_free_IRHS}.

It is widely known that the \acrshort{HCS} solution is not strictly compatible with the Maxwellian \acrshort{VDF} (see \refsubsec{HCS_KTGG}). Then, the main aim of this chapter is to characterize the non-Gaussianities of the \acrshort{HCS} \acrshort{VDF} by means of its first nontrivial cumulants from a \acrshort{SA}, and the \acrshort{HVT} of some marginal \acrshort{VDF} at the \acrshort{HCS}, which are studied from an asymptotic analysis of the \acrshort{BE}. This study is performed in a common framework for \acrshort{HD} and \acrshort{HS} systems in terms of the translational and rotational degrees of freedom, $\dt$ and $\dr$, respectively.

To address this problem, first the homogeneous states are formally introduced and described in the context of the homogeneous \acrshort{BE} for the \acrshort{IRHS} model, in Article 5 (\refsec{Art5}). Moreover, we assume, from isotropy arguments, that the instantaneous and \acrshort{HCS} \acrshort{VDF} can be written in terms of a Sonine expansion [see \refeqs{Sonine_expansion_IRHS}], and the associated infinite hierarchy of moment equations is derived. In order to obtain explicit results, we introduce certain approximations. First, from a \acrshort{MA}, we obtain a closed evolution equation for the rotational-to-translational \index{Temperature!rotational-to-translational ratio}temperature ratio, $\theta$, as well as an approach to its \acrshort{HCS} values. Afterwards, we construct a \acrshort{SA} based on the truncation of the full Sonine expansion up to the first nontrivial cumulants of the \acrshort{VDF}, that is, those Sonine coefficients $a_{pq}^{(r)}$ such that $p+q+2r\geq 3$ are neglected. In other words, the nonneglected cumulants are \phantomsection\label{sym:apq}$a_{20}\equiv a_{20}^{(0)}$, $a_{02}\equiv a_{02}^{(0)}$, $a_{11}\equiv a_{11}^{(0)}$, and $a_{00}^{(1)}$, the latter being meaningless in the \acrshort{HD} case. This \acrshort{SA} allows us to get a closed set of differential equations for the time-dependent quantities $\{\theta,a_{20},a_{02},a_{11}\}$ for the \acrshort{HD} case and $\{\theta,a_{20},a_{02},a_{11},a_{00}^{(1)}\}$ for the \acrshort{HS} case, from a proper linearization of the relevant \index{Collisional moments!for IRHS}collisional moments with respect to the considered cumulants [see Table~I of Article 5 (\refsec{Art5})]. Therefore, from the steady solutions of the latter differential equations, we obtain the \acrshort{HCS} values of the involved quantities\footnote{To localize the regions of the parameter space where the \acrshort{SA} is expected to work worse, we derive, in the Supplemental Material, a \acrshort{KLD}-like functional from an approximation similar to the toy model derived in \refch{KLD_IHS}.}, recovering the already known \acrshort{HS} results of Ref.~\cite{VSK14}. Even more, the choice of the proper linearized scheme is complemented by a linear stability analysis under homogeneous perturbations of the transient equations, which appears in Appendix~A of Article 5 (\refsec{Art5}).

To complete the theoretical description of the \acrshort{HCS}, we compute the \acrshort{HVT} for the translational, \phantomsection\label{sym:phi_marg_HCS}$\phi_{\cc}^{\HCS}(\cc)$, rotational, $\phi_{\ww}^{\HCS}(\ww)$, and the joint marginal \acrshort{VDF} $\phi^{\HCS}_{cw}(c^2w^2)$ [see their definitions in \refeq{marginal_VDF} and Eq.~(5.1) of Article 5 (\refsec{Art5})]. The translational marginal \acrshort{VDF} admits the same exponential form for the \acrshort{HVT} that the smooth result, $\phi^{\HCS}_{\cc}\sim e^{-\gamma_c c}$, with $\gamma_c\rightarrow \gamma_c^{\mathrm{IRHS}} = \dt\pi^{\frac{\dt-1}{2}}/\mu_{20}^{\HCS}\Gamma\left(\frac{\dt+1}{2}\right)$, where the expressions of $\gamma_c^{\mathrm{IHS}}$ [see \refeq{HVT_IHS_KTGG}] and \phantomsection\label{sym:gammacIRHS}$\gamma_c^{\mathrm{IRHS}}$ are formally the same, except for the generalization $\mu_2^{\HCS}\to \mu_{20}^{\HCS}$. Moreover, the rotational marginal \acrshort{VDF} adopts an algebraic or scale-free form, \phantomsection\label{sym:gammaw}$\phi^{\HCS}_{\ww}\sim w^{-\gamma_w}$, implying the divergence of the Sonine coefficients $a_{0q}^{\HCS}$ if $2q\geq \gamma_w-1$. Finally, we obtain for the \acrshort{HVT} of the marginal \acrshort{VDF} of the variable $c^2w^2$ an algebraic form as well, \phantomsection\label{sym:gammacw}$\phi^{\HCS}_{cw}\sim (c^2w^2)^{-\gamma_{cw}}$. The latter result determines the divergence of the cumulants of the form $a_{jj}^{\HCS}$ if $j\geq \gamma_{cw}-1$.

Furthermore, we compare the theoretical predictions with \acrshort{DSMC} and \acrshort{EDMD} simulation results, coming from custom-built computer programs~\cite{M23_github} that mimic homogeneous systems of uniform \acrshort{HD} ($\kappa=1/2$) for a wide variety of values of the coefficients of restitution. We obtain the \acrshort{HCS} results for the rotational-to-translational \index{Temperature!rotational-to-translational ratio}temperature ratio and the cumulants of the \acrshort{SA}, i.e., $\theta^{\HCS}$, $a_{20}^{\HCS}$, $a_{02}^{\HCS}$ and $a_{11}^{\HCS}$ with good agreement, in general. Moreover, the \acrshort{HVT} are compared with the theoretical predictions, obtaining good accordance between the numerical histograms and the scaling theoretical forms. In fact, there is a good quantitative agreement of the simulation outcomes with the theoretical predicted equation for $\gamma_w$, but a more qualitative matching in the case of $\gamma_c$ and $\gamma_{cw}$ [see Fig.~9 of Article 5 (\refsec{Art5})].

\newpage

\section{Article 5}\labsec{Art5}

\underline{\textbf{Title:}} Translational and rotational non-Gaussianities in homogeneous freely evolving granular gases \\
\underline{\textbf{Authors:}} Alberto Meg\'ias$^1$ and Andr\'es Santos$^{1,2}$\\
\noindent \underline{\textbf{Affiliations:}}\\
$^1$ Departamento de F\'isica, Universidad de Extremadura, E-06006 Badajoz,
Spain\\
$^2$  Instituto de Computaci\'on Cient\'ifica Avanzada
(ICCAEx), Universidad de Extremadura, E-06006 Badajoz, Spain\\\vspace{-0.6cm}
\begin{flushleft}
\begin{minipage}{0.5\textwidth}\raggedright
\underline{\textbf{Journal:}} Physical Review E\vspace{0.2cm}

\underline{\textbf{Volume:}} 108\vspace{0.2cm}

\underline{\textbf{Pages:}} 014902\vspace{0.2cm}

\underline{\textbf{Year:}} 2023\vspace{0.2cm}

\underline{\textbf{DOI:}} \href{https://doi.org/10.1103/PhysRevE.108.014902}{10.1103/PhysRevE.108.014902}
\end{minipage}
\begin{minipage}{0.49\textwidth}\raggedleft
\includegraphics[width=0.49\textwidth]{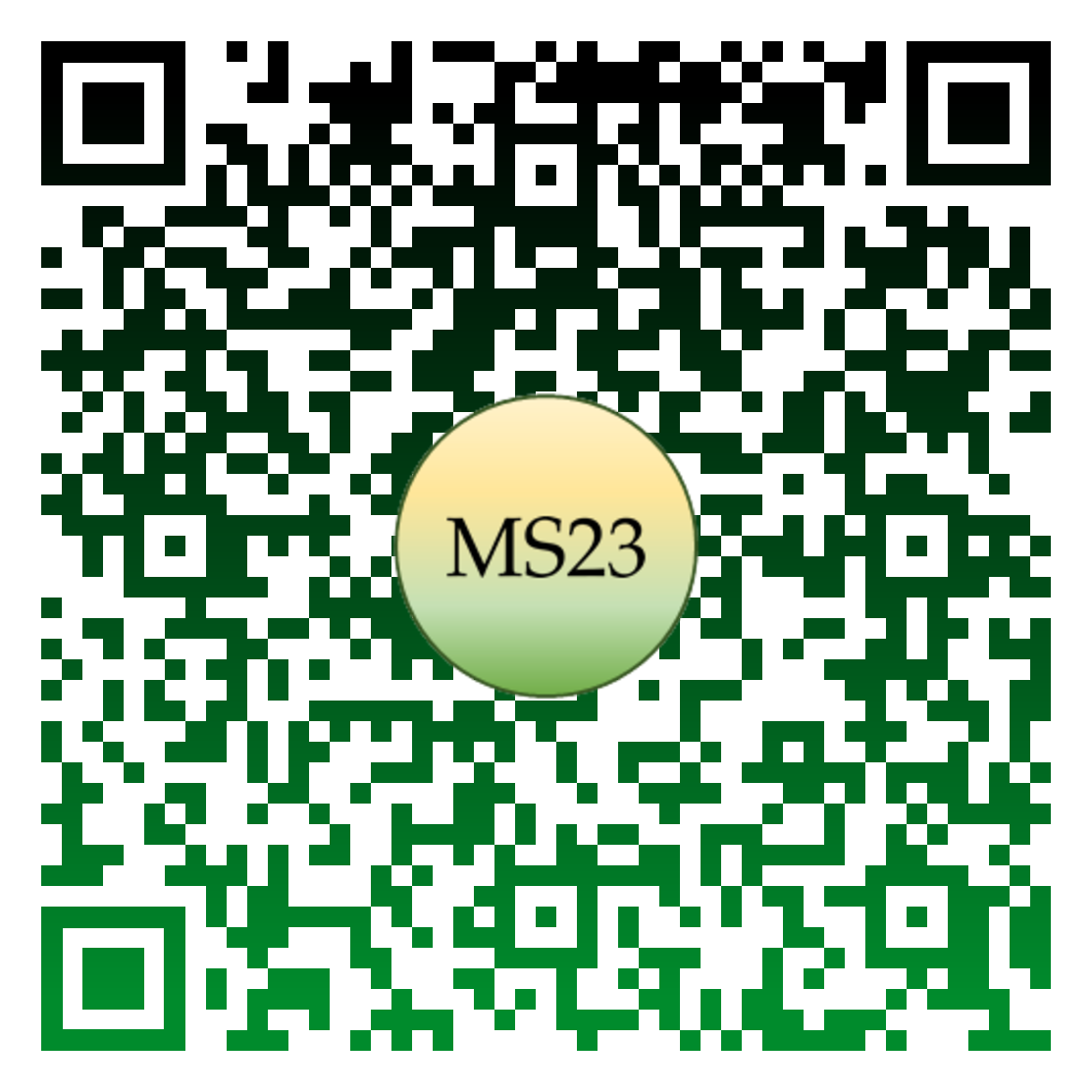}
\end{minipage}
\end{flushleft}
Copy of the preprint of the work: ``Alberto Meg\'ias, and Andr\'es Santos. 'Translational and rotational non-Gaussianities in homogeneous freely evolving granular gases', \emph{Physical Review E} \textbf{108}, 014902 (2023) \href{https://doi.org/10.1103/PhysRevE.108.014902}{https://doi.org/10.1103/PhysRevE.108.014902}.''

\includepdf[clip,
    trim=4mm 4.5mm 10mm 1.5mm,pages=-,
   scale=0.8,pagecommand={},offset=8mm -6mm,
   frame]{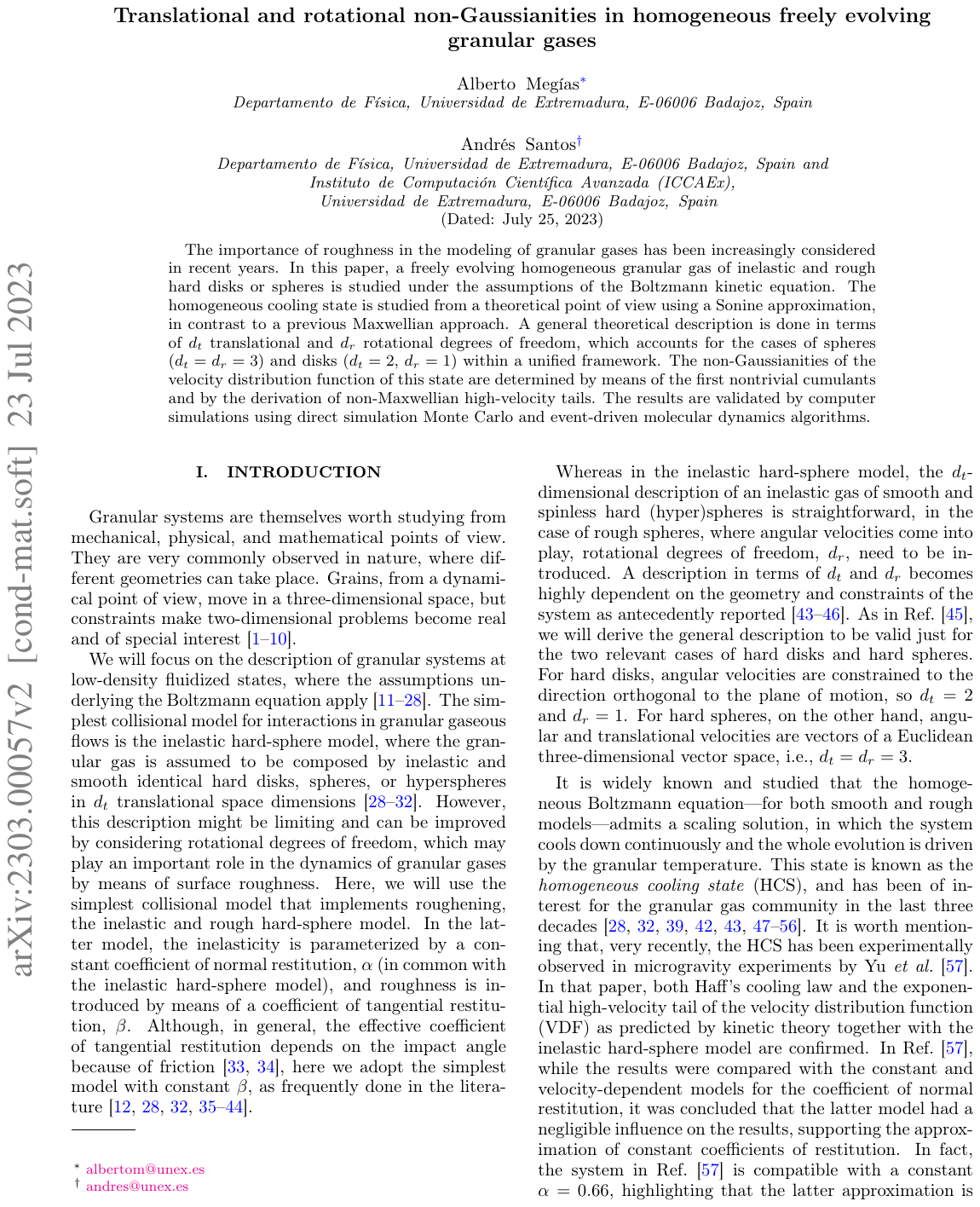}
\chapter{Mpemba-like effect in the IRHS model}
\labch{IRHS_ME}

\section{Summary}

In this chapter, the emergence of the \acrshort{ME} in a homogeneous and dilute granular gas of inelastic and rough \acrshort{HD} driven by the \acrshort{ST} is studied. The granular gas is assumed to be made of identical disks of diameter $\sigma$, mass $m$, and reduced moment of inertia $\kappa$. The binary collisions between granular particles are described by the \acrshort{IRHS} model, where the inelasticity is accounted for by a coefficient of normal restitution\index{Coefficient of normal restitution}, $\een$, and the surface roughness by a coefficient of tangential restitution\index{Coefficient of tangential restitution}, $\eet$, both assumed to be constant. The \acrshort{ST} models an injection of energy into the granular gaseous flow via its translational and rotational degrees of freedom, by means of a stochastic force and a stochastic torque, both with the properties of a white noise, and with noise intensities $\chi^2_t$ and $\chi_r^2$, respectively. A particular case of the thermostat can be determined by a noise \index{Temperature!noise}temperature, $\Twn$, and a rotational-to-total relative noise intensity, $\varepsilon$, which is equivalent to give the values of the noise intensities, as introduced in \refsubsec{splitting_KTGG}. This system is studied in the context of the homogeneous version of the \acrshort{BFPE} appearing in \refeq{BFPE_epsilon_ST} [see Eq.~(11) of Article 6 (\refsec{Art6})].

The main aim of this chapter is to design, in this system, a preparation protocol to generate the proper initial states of the samples involved in a Mpemba-like effect experiment. In fact, from previous works of the same collisional model, but with the external energy injection supplying just to the translational degrees of freedom ($\varepsilon = 0$), we know that \acrshort{ME} appears~\cite{TLLVPS19}. However, it does not seem obvious how to initially prepare the system. In Article 6 (\refsec{Art6}), after the introduction of the model, we characterize the homogeneous states of the system. Here, a noise-induced time and temperature scales of the problem appear, these being $1/\nuwn$ and $\Twn$, respectively. In order to do that, the evolution equation of the mean granular \index{Temperature!mean granular}temperature, $T/\Twn$, and the rotational-to-translational \index{Temperature!rotational-to-translational ratio}temperature ratio, $\theta$, are derived from the \acrshort{BFPE} [see Eqs.~(23) of Article 6 (\refsec{Art6})]. However, the system of differential equations formed by $\dot{T}/\Twn$ and $\dot{\theta}$ is not closed, since the energy production rates or, equivalently, the \index{Collisional moments!for IRHS}collisional moments $\mu_{20}$ and $\mu_{02}$, depend on the whole \acrshort{VDF}. Thus, we need to introduce some approximations to give quantitative predictions of the system. From previous studies of the homogeneous states of a granular gas under the \acrshort{IRHS} model, but restricted to $\varepsilon=0$, the steady-state values of the cumulants are not too large~\cite{VS15} as compared with the freely evolving case~\cite{VSK14,MS23}. Then, if the initial conditions are not too far from a Maxwellian distribution, a \acrshort{MA} for the time-dependent and steady \acrshort{VDF} seems to be plausible, at least a first approach to the problem. This is found to be a good approach in accordance with \acrshort{DSMC} and \acrshort{EDMD}--with the \acrshort{AGF} algorithm applied to the \acrshort{ST}--computer simulation results from hand-made programs~\cite{M23_github}. We reproduce sets of uniform hard disks, that is, with $\kappa=1/2$, and for different values of the coefficients of restitution, $\een$ and $\eet$, and of the parameter $\varepsilon$. From them, the \acrshort{MA} is tested both for steady-state values\index{Temperature!steady-state}\index{Temperature!rotational-to-translational ratio}, \phantomsection\label{sym:thetast}$(T^{\mathrm{st}}/\Twn, \theta^{\mathrm{st}})$, and for transient stages, $(T(t)/T^{\mathrm{st}}, \theta^{\mathrm{st}})$, as can be observed in Figures~5 and 6 of Article 6 (\refsec{Art6}). From the steady states of the system, it is important to highlight that $\theta^{\mathrm{st}}$ is an increasing function of $\varepsilon$. In addition, it is found from the theoretical prediction of the slope of $\dot{T}/T^{\mathrm{st}}\equiv \Phi$, and confirmed by the latter simulation results, that for every pair of initial conditions \phantomsection\label{sym:theta0}$(T^0/T^{\mathrm{st}}, \theta^0)$, there exists a critical value of the rotational-to-total noise intensity ratio, \phantomsection\label{sym:epsi_cr}$\varepsilon_{\mathrm{cr}}(T^0/T^{\mathrm{st}}, \theta^0)$, such that, for every $\varepsilon>\varepsilon_{\mathrm{cr}}$, the mean granular \index{Temperature!mean granular}\index{Temperature!overshoot}temperature overshoots, at a certain time, its future steady-state value. Then, we numerically compute the value of $\varepsilon_{\mathrm{cr}}$ with respect to different initial conditions, and different values of the coefficients of restitution, for uniform \acrshort{HD}, as showed in Figure~7 of Article 6 (\refsec{Art6}).

Once the homogeneous states have been theoretically described, and the approximated scheme is validated by means of computer simulation results, we study the conditions under which the standard \acrshort{DME} shows up in this system. Let us first imagine that, given a reference \acrshort{ST}, \phantomsection\label{sym:epsi_ref}$(\Twn_{\mathrm{ref}},\varepsilon_{\mathrm{ref}})$, the value of the noise intensity ratio, $\varepsilon_{\mathrm{ref}}$, is below the critical value for every initial condition. That is, we impose that there is no \index{Temperature!overshoot}overshoot of any thermal curve. Then, for two different systems, A and B, the former,  assumed to be initially hotter than the latter, will evolve up to the same steady state. The appearance of the \acrshort{DME} is detected if there exists a crossover between the two temperature curves at a certain crossing time. Therefore, we study the necessary conditions for the \acrshort{DME} to appear in terms of the mean temperature slopes. That is, we impose \phantomsection\label{sym:Phi_slope}$\Phi(T^0_A/T^{\mathrm{st}}_{\mathrm{ref}},\theta^0_A)< \Phi(T^0_B/T^{\mathrm{st}}_{\mathrm{ref}},\theta^0_B)$. From the \acrshort{MA}, it is deduced that $\theta^0_A<\theta^0_B$ is needed, which was already known from Ref.~\cite{TLLVPS19} in the particular case of $\varepsilon=0$. Then, to enhance (and ensure) the effect, the condition should be exaggerated to $\theta^0_A\ll\theta^0_B$.

On the other hand, we also analyze the \acrshort{DME} via the \acrshort{OME}. Then, we impose that, at least, the initially colder system must suffer an \index{Temperature!overshoot}overshoot with respect to its steady value, i.e., $\varepsilon_{\mathrm{ref}}>\varepsilon_{\mathrm{cr}}(T_B^0/T^{\mathrm{st}}_{\mathrm{ref}},\theta_B^0)$. Following the results in Article 1 (\refsec{Art1}), we study the \acrshort{OME} by means of a functional, $\mathfrak{D}$, which is a convex and positive function of $T(t)/T^{\mathrm{st}}$, recalling the expression of the local-equilibrium functional introduced in Article 1 (\refsec{Art1}). Then, to detect the direct \acrshort{OME}, we need a crossover between \phantomsection\label{sym:DfrakA_B}$\mathfrak{D}_A$ and $\mathfrak{D}_B$ at a certain crossing time, with system B \index{Temperature!overshoot}overshooting the steady state, and without any intersection (or an odd number of them) between the mean granular \index{Temperature!mean granular}temperatures of A and B. The necessary condition for the initial slopes is the opposite to that of the standard version of the \acrshort{DME}, that is, $\Phi(T^0_A/T^{\mathrm{st}}_{\mathrm{ref}},\theta^0_A)>\Phi(T^0_B/T^{\mathrm{st}}_{\mathrm{ref}},\theta^0_B)$, which translates into \phantomsection\label{sym:theta0A_B}$\theta^0_A> \theta^0_B$. However, according to the theoretical description of the thermal slope $\Phi$, we prefer for $\theta^0_A$ not to be excessively large, in order to avoid an extremely slow relaxation of system $A$ that could preclude the emergence of the effect.

After the latter analysis, we derive preparation protocols for the initial states in a Mpemba-like cooling experiment, which allows us to observe the standard \acrshort{DME} and the \acrshort{OME} in the described system. These initial states are based on the steady states of the system. Those protocols are described in Section~3.3 of Article 6 (\refsec{Art6}), and sketched in Figures~8 and 9 of Article (\refsec{Art6}). Finally, \acrshort{DSMC} and \acrshort{EDMD} computer simulation results validate the theoretical protocols, as can be observed in Figures~11 and 12 of Article 6 (\refsec{Art6}).

\section{Article 6}\labsec{Art6}

\underline{\textbf{Title:}} Mpemba-like effect protocol for granular gases of inelastic and rough hard disks \\
\underline{\textbf{Authors:}} Alberto Meg\'ias$^1$ and Andr\'es Santos$^{1,2}$\\
\noindent \underline{\textbf{Affiliations:}}\\
$^1$ Departamento de F\'isica, Universidad de Extremadura, E-06006 Badajoz,
Spain\\
$^2$ Instituto de Computaci\'on Cient\'ifica Avanzada
(ICCAEx), Universidad de Extremadura, E-06006 Badajoz, Spain\\\vspace{-0.6cm}
\begin{flushleft}
\begin{minipage}{0.5\textwidth}\raggedright
\underline{\textbf{Journal:}} Frontiers in Physics\vspace{0.2cm}

\underline{\textbf{Volume:}} 10\vspace{0.2cm}

\underline{\textbf{Pages:}} 971671\vspace{0.2cm}

\underline{\textbf{Year:}} 2022\vspace{0.2cm}

\underline{\textbf{DOI:}} \href{https://doi.org/10.3389/fphy.2022.971671}{10.1051/epjconf/202124904006}
\end{minipage}
\begin{minipage}{0.49\textwidth}\raggedleft
\includegraphics[width=0.49\textwidth]{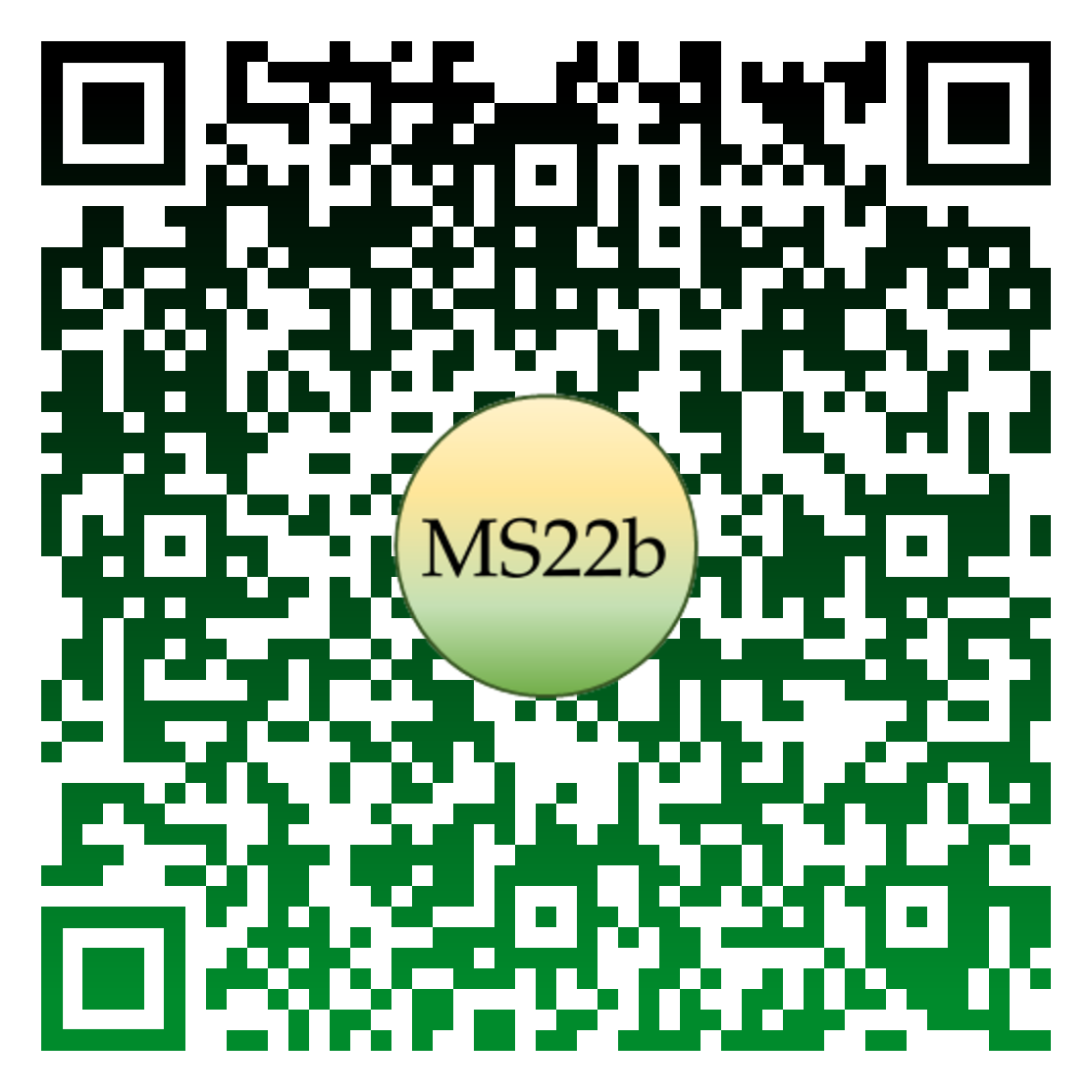}
\end{minipage}
\end{flushleft}
Copy of the preprint of the work: ``Alberto Meg\'ias, and Andr\'es Santos, 'Mpemba-like effect protocol for granular gases of inelastic and rough hard disks', \emph{Frontiers in Physics} \textbf{10}, 971671 (2022) \href{https://doi.org/10.3389/fphy.2022.971671}{ https://doi.org/10.3389/fphy.2022.971671}.''

\includepdf[clip,
    trim=4mm 4.5mm 10mm 1.5mm,pages=-,
   scale=0.8,pagecommand={},offset=8mm -6mm,
   frame]{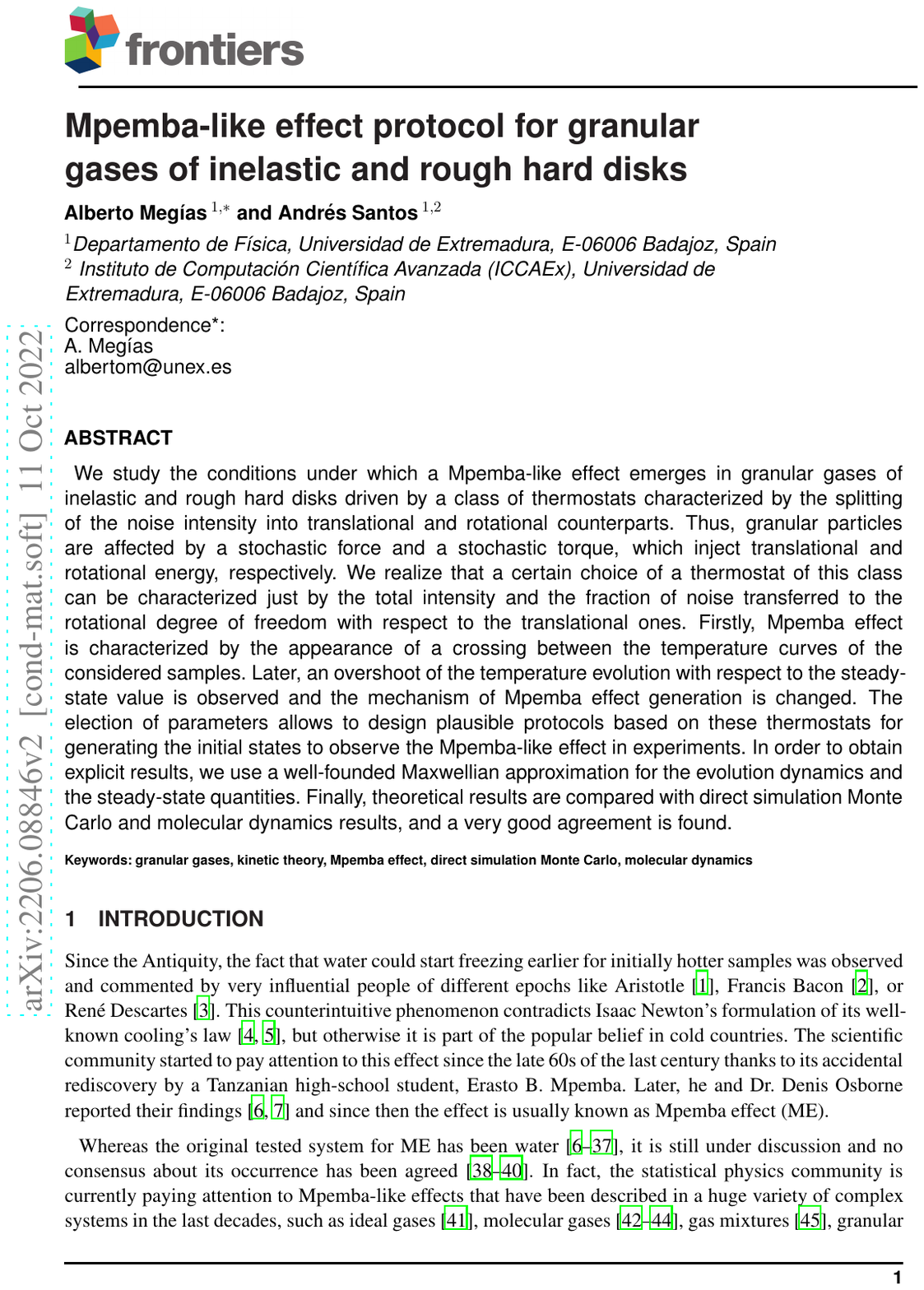}

\chapter{Nonhomogeneous states in the IRHS model}
\labch{IRHS_IS}

\section{Summary}

The transport properties of a dilute granular gas of freely evolving inelastic and rough identical \acrshort{HD} or \acrshort{HS}, under a common mathematical framework, in terms of the translational and rotational degrees of freedom, $\dt$ and $\dr$, respectively, is studied. The granular particles are set to be identical with diameter $\sigma$, mass $m$, and reduced moment of inertia $\kappa$. Their collisional rules are characterized by the \acrshort{IRHS} model, where the inelasticity of the particles is accounted for by a coefficient of normal restitution\index{Coefficient of normal restitution}, $\een$, and a coefficient of tangential restitution\index{Coefficient of tangential restitution}, $\eet$. Both of them are assumed to be velocity independent, as described by the \acrshort{IRHS} model. 

In this chapter, the main objective consists in solving the inhomogeneous \acrshort{BE} for this freely evolving dilute granular gaseous system from a \index{Chapman--Enskog method}\acrshort{CE} expansion up to the \acrshort{NSF} level. This leads to the computation of the \acrshort{NSF} \index{Transport coefficient}transport coefficient, which is detailed in Article 7 (\refsec{Art7}). Furthermore, this solution allows one to determine the stability of the homogeneous solution under linear inhomogeneous perturbations, such an analysis being carried out in Article 8 (\refsec{Art8}). First, the macroscopic hydrodynamic description of the system is given by the set of \index{Hydrodynamic!fields}hydrodynamic fields $\{n,\mathbf{u},T\}$, as already discussed in \refsubsec{CE_Exp_KTGG}, with $T$ being the mean granular \index{Temperature!mean granular}temperature. Then, we first derive the hydrodynamic balance equations associated with those fields. More details about their derivation can be found in \refapp{app_balance_equations}.

Before entering into more details about the inhomogeneous description of the system, the homogeneous \acrshort{BE} must be first characterized. The homogeneous state for this system has been already studied, first introduced in \refsubsec{HCS_KTGG} and, then, its non-Gaussianities being studied in detail in Article 5 (\refsec{Art5}). A granular gas in free evolution and in the absence of spatial gradients is known to reach a scaling solution for its \acrshort{VDF} at the so-called \acrshort{HCS}, where all the time dependence of the system is driven through the mean granular \index{Temperature!mean granular}temperature continuous decay. As a first approach to the problem, despite the non-Gaussianities of the \acrshort{HCS} \acrshort{VDF}, it seems to be reasonable to work under the \acrshort{MA} proposed in \refeq{qMA_IRHS}, as a first approach, which has been already used in Ref.~\cite{KSG14}. That is, the \acrshort{HCS} \acrshort{VDF} is assumed to decompose into a product of its translational and rotational marginal \acrshort{VDF}, the former being approximated by a Maxwellian \acrshort{VDF} at the instantaneous translational granular \index{Temperature!translational}temperature, whereas the latter does not need to be specified.

After the description of the \acrshort{HCS}, we look for a normal solution to the \acrshort{BE} described by the \acrshort{IRHS} collisional model and in the absence of external forces, by means of the \index{Chapman--Enskog method}\acrshort{CE} expansion around the local version of the \acrshort{HCS} solution, which coincides with the zeroth order contribution to the \acrshort{VDF}, $f^{(0)}$. As introduced in \refsubsec{CE_Exp_KTGG}, the \acrshort{VDF} is proposed to have a functional dependence on the set of the \index{Hydrodynamic!fields}hydrodynamic fields, see \refeq{CE_hypothesis}. Therefore, to characterize the \acrshort{VDF}, it suffices to know the instantaneous values of those fields, as well as their spatial gradients. In addition, all quantities different from the \index{Hydrodynamic!fields}hydrodynamic fields would be characterized similarly, and expanded in terms of gradients of the \index{Hydrodynamic!fields}hydrodynamic fields. This occurs with the \index{Cooling rate!in the IRHS model}cooling rate $\zeta$ and the fluxes appearing in the balance equations: the heat flux \phantomsection\label{sym:qflux}$\mathbf{q}$ and the pressure tensor \phantomsection\label{sym:Ptensor}$\mathsf{P}$. The first-order contribution to the \acrshort{VDF}, $f^{(1)}$, is a function of gradients of the \index{Hydrodynamic!fields}hydrodynamic fields [see Eq.~(29) of Article 7 (\refsec{Art7})], whose coefficients \phantomsection\label{sym:ABCE_quant}$\boldsymbol{\mathcal{A}}$, $\boldsymbol{\mathcal{B}}$, $\mathcal{C}_{ij}$, and $\mathcal{E}$ are the solutions to certain linear integral equations [see Eqs.~(30) of Article 7 (\refsec{Art7})]. To complete the description, we need to introduce the first-order contributions of the \index{Cooling rate!for IRHS}cooling rate, due to a dependence on the \index{Transport coefficient!of the cooling rate}velocity divergence [see Eq.~(32) of Article 7 (\refsec{Art7})], and the fluxes via their phenomenological constitutive equations [see Eq.~(34) of Article 7 (\refsec{Art7})]. From the constitutive relation of the pressure tensor, the \index{Transport coefficient!shear viscosity}shear and \index{Transport coefficient!bulk viscosity}bulk viscosities appear, \phantomsection\label{sym:viscosities}$\eta$ and $\eta_b$, respectively. Moreover, the proposed form for the heat flux assumes a dependence on the gradients of the mean granular \index{Temperature!mean granular}temperature and the number density. The former is accompanied by the thermal conductivity \phantomsection\label{sym:lambda}$\lambda$, whereas the latter arises from a nonzero \index{Transport coefficient!Dufour-like coefficient}Dufour-like coefficient \phantomsection\label{sym:Dufour}$\mu$, which is observed to be present in granular gaseous flows~\cite{BC01,BP03,GTSH12,KSG14,GSK18,G19}. These two latter coefficients have a translational and a rotational contribution to their complete expression. Then, our main interest resides in the computation of those coefficients, which is straightforward from the knowledge of the coefficients of $f^{(1)}$. They are derived from the simplest Sonine-like approximation, permitting to solve the corresponding linear integral equations [see Eqs.~(40) of Article 7 (\refsec{Art7})]. Thus, we obtain explicit expressions in terms of $\dt$ and $\dr$---only valid for \acrshort{HD} ($\dt=2\dr=1$) and \acrshort{HS} ($\dt=\dr=3$)---which are summarized in Table~I of Article 7 (\refsec{Art7}). Even more, we obtain the expressions of such coefficients for different limiting situations, the smooth limit ($\eet=-1$), the quasismooth limit ($\eet\to -1$), and Pidduck's gas limit ($\een=\eet=1$), which are summed up in Table~II of Article 7 (\refsec{Art7}).

Once the transport coefficients are computed, in Article 8 (\refsec{Art8}), we generalize the \index{Linear stability analysis}linear stability analysis of Ref.~\cite{GSK18} to the introduced unified scheme for \acrshort{HD} and \acrshort{HS} in terms of the translational and rotational degrees of freedom, allowing us to obtain novel results for the \acrshort{HD} case. From previous studies, the \acrshort{HCS} is unstable by means of the emergence of either clusters or vortices, both predicted in the smooth~\cite{GZ93,BRM98,BRC99,LH99,G05,MDCPH11,MGHEH12,MZBGDH14,FH17,G19} case and for rough \acrshort{HS}~\cite{MDHEH13,GSK18}. We deduce the presence of these two types of \index{Instability}instabilities by means of \index{Critical wave number}critical wave numbers, \phantomsection\label{sym:ins_modes}$k_\perp$ and $k_{\parallel}$, associated with the divergence of the transverse and longitudinal modes, respectively. Then, in the regions where $k_{\parallel}>k_\perp$, the onset of the instability is due to the longitudinal modes, which give rise to the appearance of clusters \index{Instability!clustering}(clustering instability). On the other hand, if $k_\perp>k_{\parallel}$, the instability is due to the transverse modes and it is then dominated by the emergence of vortices \index{Instability!vortices}(vortex instability). Then, although our analysis of the \acrshort{BE} assumes an infinite system, in the real (and \acrshort{MD}-computational) world, the granular gaseous systems are formed by a finite number of particles and confined within a certain finite volume. From those \index{Critical wave number}critical wave numbers, we compute the \index{Critical length}critical length of the system, \phantomsection\label{sym:Lc}$L_c$, such that if the characteristic length of the system is larger than $L_c$, then the \acrshort{HCS} becomes unstable. The theoretical computations show a divergence of $k_{\parallel}$ in a certain region of the parameter space, always present for \acrshort{HD}, and only in some cases for \acrshort{HS}. Hence, according to this description, one would expect that cluster \index{Instability!clustering}instability is always present and hydrodynamics would break down within that region of the parameter space [see Figures~3 and 5 of Article 8 (\refsec{Art8})]. This result seems to be too devastating. In order to check whether the prediction is reliable or not, we performed \acrshort{EDMD} computer simulations from a hand-made program~\cite{M23_github}.

The objective of these computer simulations is to check whether this highly unstable region (\acrshort{HUR}) is present or not for freely cooling \acrshort{HD} systems. We elaborate a protocol to detect the clustering \index{Instability!clustering}instability by means of the \acrshort{KLD} of the particle spatial distribution compared with a Poisson distribution, as a reference for a homogeneous distribution. According to this protocol, a spontaneous and noticeable---compared with the statistical error---increment of the \acrshort{KLD} is associated with the formation of a granular cluster. Moreover, we compare time-dependent simulation values of the rotational-to-translational \index{Temperature!rotational-to-translational ratio}temperature ratio $\theta$ and first nontrivial cumulants of the \acrshort{VDF}, $a_{20}$, $a_{02}$, $a_{11}$, with their \acrshort{HCS} values derived from the \acrshort{SA}. Large deviations from these values indicate the possible presence of an instability. We compare first two different systems in a point inside the predicted \acrshort{HUR} ($\een=0.2$, $\eet=0.25$, $\kappa=1/2$), and we add a third one in a point outside this region ($\een=0.7$, $\eet=0.25$, $\kappa=1/2$). Whereas in the former point, the two compared systems are expected to be unstable, we do observe \index{Instability}instabilities just in one of them. On the other hand, in the point of the parameter space outside the \acrshort{HUR}, we obtain an agreement with the theoretical prediction [see Figures~8--11 of Article 8 (\refsec{Art8})].

\section{Article 7}\labsec{Art7}
\underline{\textbf{Title:}} Hydrodynamics of granular gases of inelastic and rough hard disks or spheres. I. Transport coefficients\\
\underline{\textbf{Authors:}} Alberto Meg\'ias$^1$ and Andr\'es Santos$^{1,2}$\\
\noindent \underline{\textbf{Affiliations:}}\\
$^1$ Departamento de Física, Universidad de Extremadura, E-06006 Badajoz,
Spain\\
$^2$ Instituto de Computación Científica Avanzada
(ICCAEx), Universidad de Extremadura, E-06006 Badajoz, Spain\\ \vspace{-0.6cm}
\begin{flushleft}
\begin{minipage}{0.5\textwidth}\raggedright
\underline{\textbf{Journal:}} Physical Review E\vspace{0.2cm}

\underline{\textbf{Volume:}} 104\vspace{0.2cm}

\underline{\textbf{Pages:}} 034901\vspace{0.2cm}

\underline{\textbf{Year:}} 2021\vspace{0.2cm}

\underline{\textbf{DOI:}} \href{https://doi.org/10.1103/PhysRevE.104.034901}{10.1103/PhysRevE.104.034901}
\end{minipage}
\begin{minipage}{0.49\textwidth}\raggedleft
\includegraphics[width=0.49\textwidth]{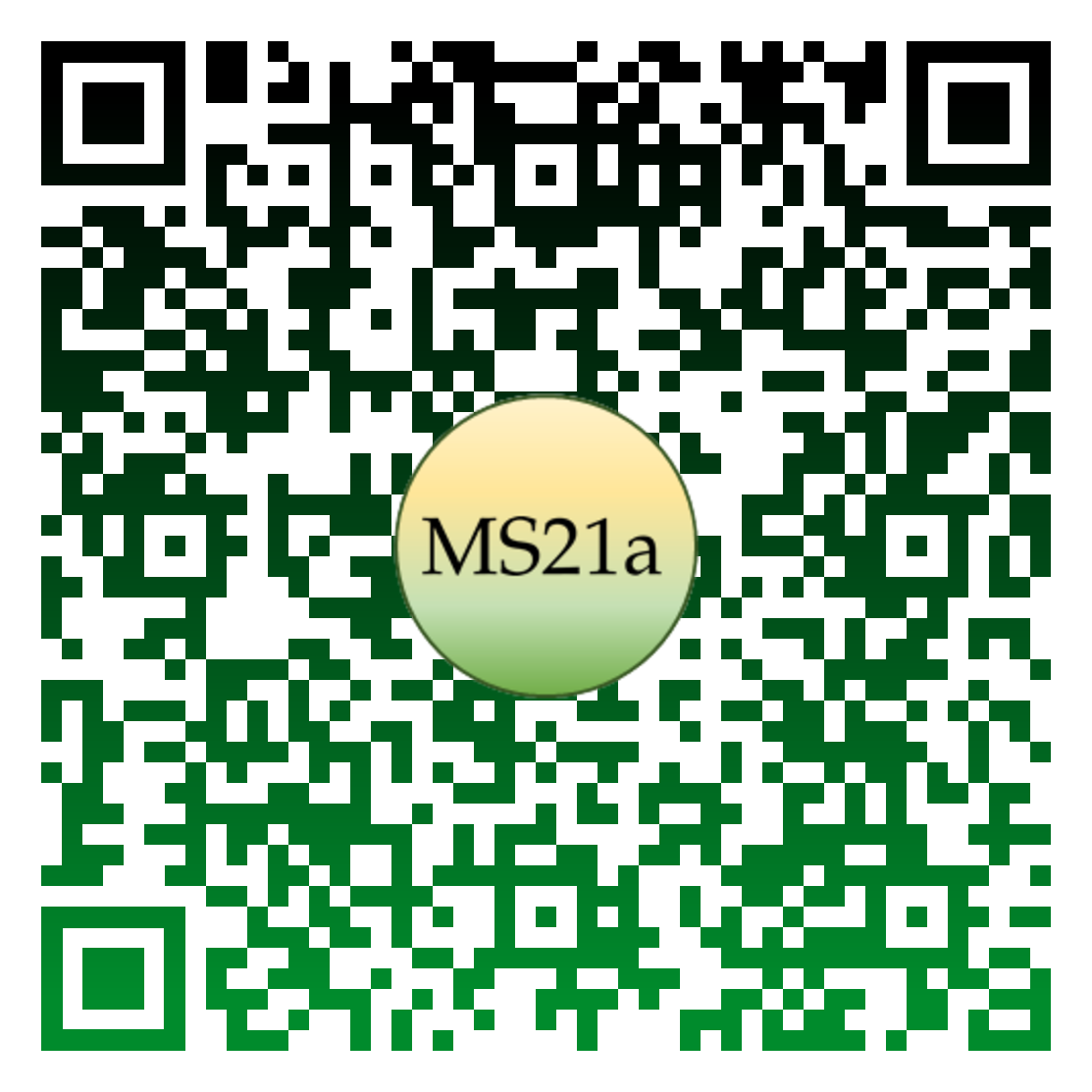}
\end{minipage}
\end{flushleft}
Copy of the preprint of the work: ``Alberto Meg\'ias, and Andr\'es Santos, 'Hydrodynamics of granular gases of inelastic and rough hard disks or spheres. I. Transport coefficients', \emph{Physical Review E} \textbf{104}, 034901 (2021) \href{https://doi.org/10.1103/PhysRevE.104.034901}{https://doi.org/10.1103/PhysRevE.104.034901}.''

\includepdf[clip,
    trim=4mm 4.5mm 10mm 1.5mm,pages=-,
   scale=0.8,pagecommand={},offset=8mm -6mm,
   frame]{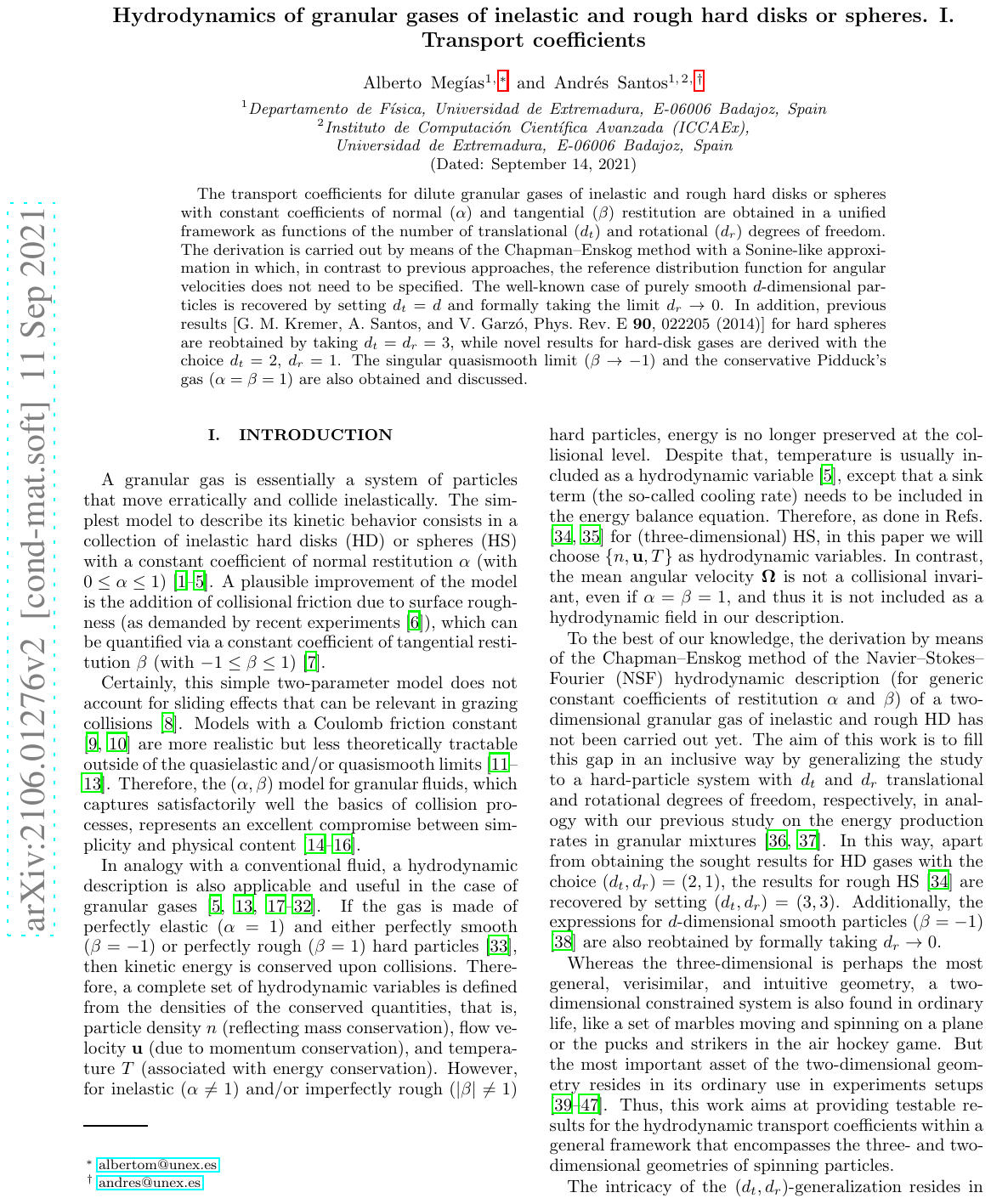}

\section{Article 8}\labsec{Art8}

\underline{\textbf{Title:}} Hydrodynamics of granular gases of inelastic and rough hard disks or spheres. II. Stability analysis\\
\underline{\textbf{Authors:}} Alberto Meg\'ias$^1$ and Andr\'es Santos$^{1,2}$\\
\noindent \underline{\textbf{Affiliations:}}\\
$^1$ Departamento de Física, Universidad de Extremadura, E-06006 Badajoz,
Spain\\
$^2$ Instituto de Computación Científica Avanzada
(ICCAEx), Universidad de Extremadura, E-06006 Badajoz, Spain\\ \vspace{-0.6cm}
\begin{flushleft}
\begin{minipage}{0.5\textwidth}\raggedright
\underline{\textbf{Journal:}} Physical Review E\vspace{0.2cm}

\underline{\textbf{Volume:}} 104\vspace{0.2cm}

\underline{\textbf{Pages:}} 034902\vspace{0.2cm}

\underline{\textbf{Year:}} 2021\vspace{0.2cm}

\underline{\textbf{DOI:}} \href{https://doi.org/10.1103/PhysRevE.104.034902}{10.1103/PhysRevE.104.034902}
\end{minipage}
\begin{minipage}{0.49\textwidth}\raggedleft
\includegraphics[width=0.49\textwidth]{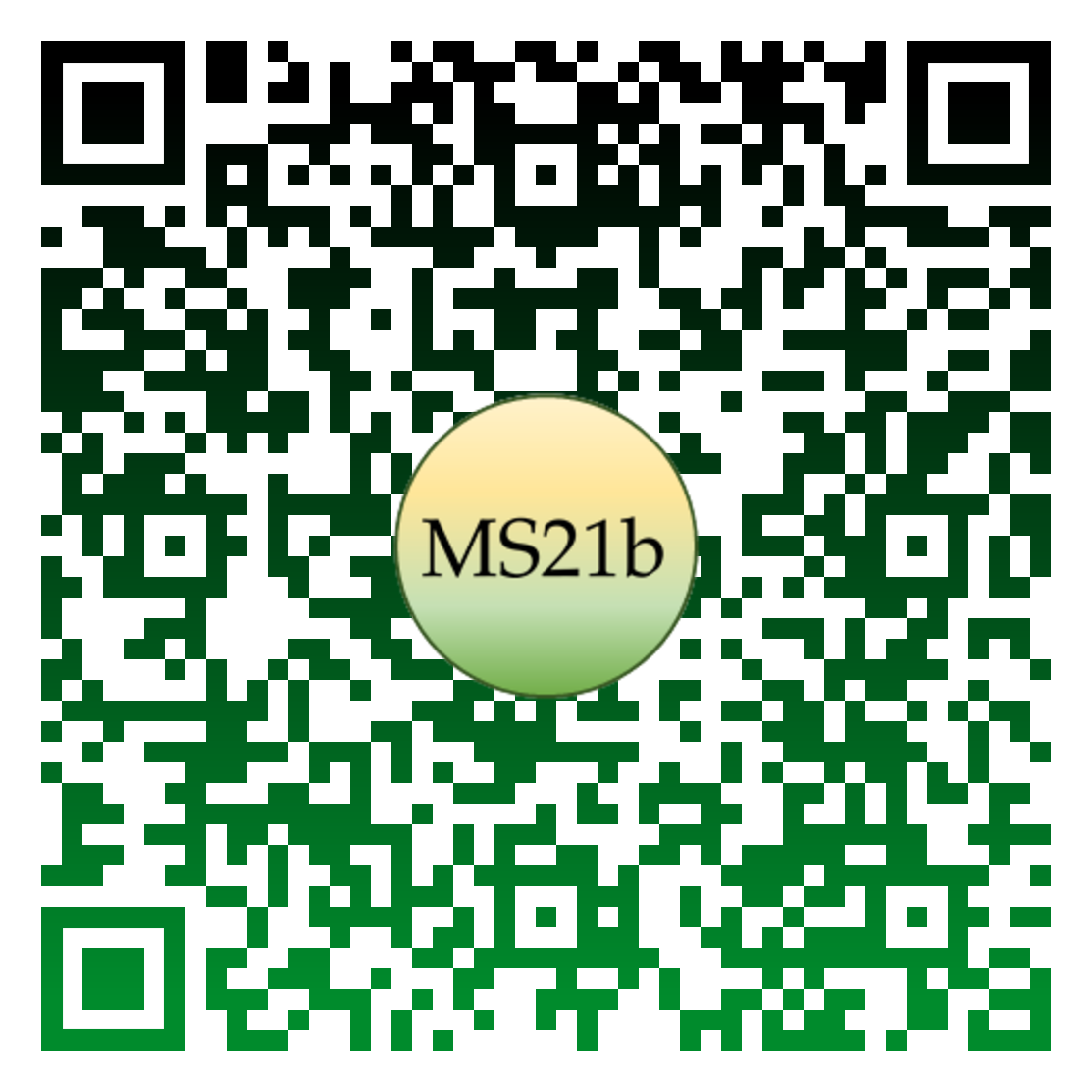}
\end{minipage}
\end{flushleft}
Copy of the preprint of the work: ``Alberto Meg\'ias, and Andr\'es Santos, 'Hydrodynamics of granular gases of inelastic and rough hard disks or spheres. II. Stability analysis', \emph{Physical Review E} \textbf{104}, 034902 (2021) \href{https://doi.org/10.1103/PhysRevE.104.034902}{https://doi.org/10.1103/PhysRevE.104.034902}.''

\includepdf[clip,
    trim=4mm 4.5mm 10mm 1.5mm,pages=-,
   scale=0.8,pagecommand={},offset=8mm -6mm,
   frame]{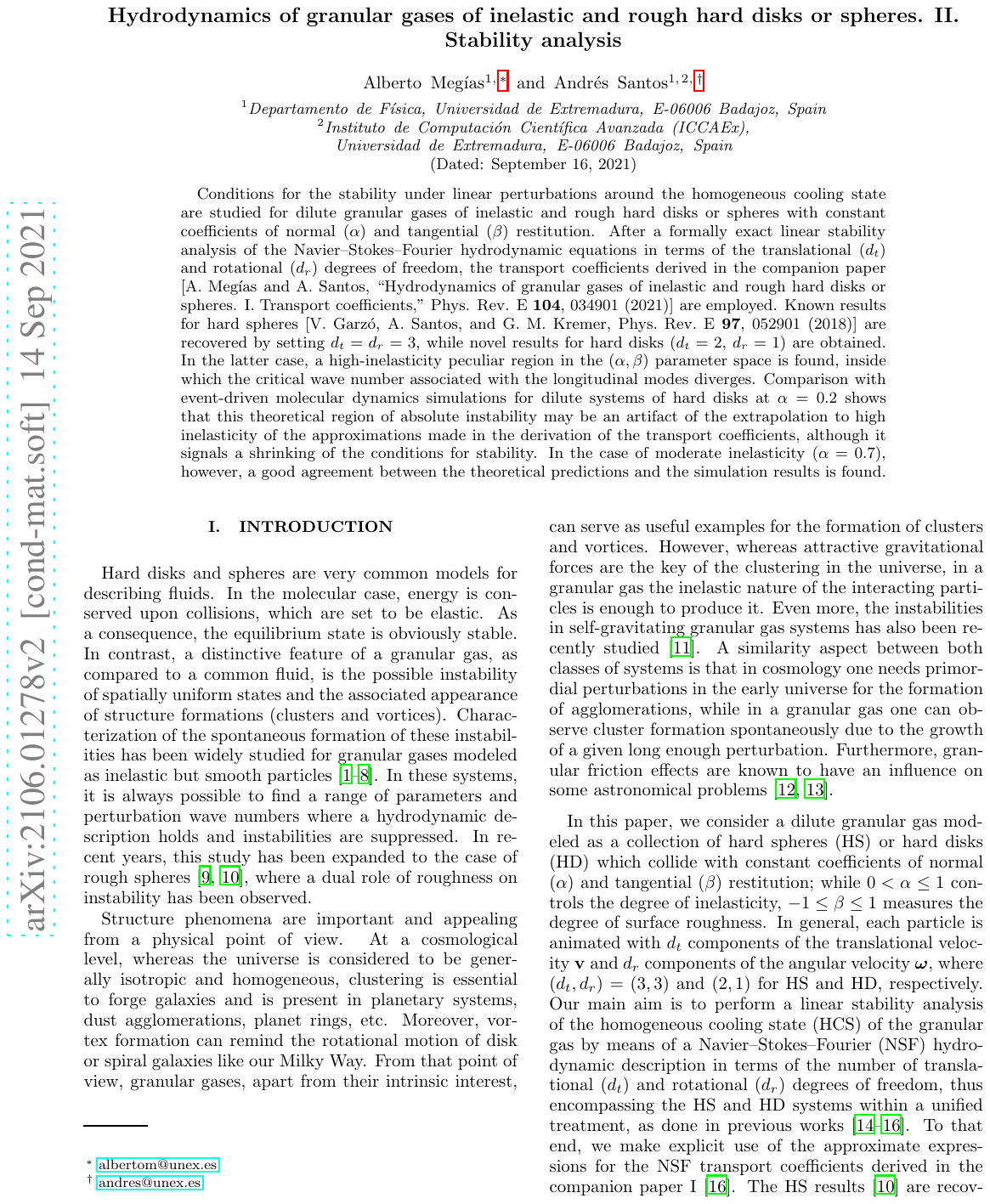}

\addpart{Results and Conclusions}

\chapter{Summary of Results and Discussion}
\labch{R_D}

The aim of this chapter is to explain and highlight the different results derived from this thesis, as well as comment on their implications in the considered system and introduce possible improvements of the analyses carried out. This discussion is classified into the different systems considered throughout this thesis.

\section{Molecular gases}

We have studied the homogeneous states of a dilute molecular gas of identical elastic hard \dt-spheres in contact with a background fluid at \index{Temperature!of the thermal bath}temperature $T_b$, which is acting on the molecular gas as a thermal bath. The nonequilibrium states of this system have been studied by means of the corresponding \acrshort{BFPE}. The evolution of the one-particle \acrshort{VDF} is due to two main events. On the one hand, the \acrshort{VDF} evolves due to binary collisions between the gas particles, which is accounted for by the \index{Boltzmann collisional operator!for EHS}Boltzmann collisional operator, in this case, described by the \acrshort{EHS} model. On the other hand, the \acrshort{VDF} changes due to the interaction between the gas and the surrounding fluid. This latter evolution is modeled by two force terms: a viscous drag force, whose associated drag coefficient is assumed to be velocity dependent, together with a stochastic force with nonlinear variance. The drag coefficient and the noise intensity are related via the \acrshort{FDT}.

The goal of this thesis, with respect to this system, resides in the analysis of the \acrshort{ME}, which is already known to emerge in this kinetic description~\cite{SP20,PSP21}. In fact, we have distinguished between a thermal description, \acrshort{TME}, and an entropic one, \acrshort{EME}. The former is guided by the kinetic nonequilibrium \index{Temperature!nonequilibrium}temperature of the system, whereas the latter has been described by the \index{Nonequilibrium entropy}nonequilibrium entropy by means of the \acrshort{KLD} of the one-body \acrshort{VDF} with respect to the equilibrium one, the Maxwellian \acrshort{VDF} at $T_b$. 

Regarding thermal description, we firstly derived the evolution equation of the \index{Temperature}temperature, coming from the \acrshort{BFPE}, and we already knew that it must be directly coupled to the excess kurtosis of the \acrshort{VDF} from the results of Ref.~\cite{SP20}. This is an indicator of the presence of memory effects due to the dependence of the nonequilibrium states of a macroscopic variable, the \index{Temperature}temperature, on an inner variable of the system, the excess kurtosis. Then, to complete the description, we have derived the infinite hierarchy of moment or, equivalently, cumulant equations. It was observed that the evolution equation of the $k^{\mathrm{th}}$ cumulant, $a_k$, is related to the equation of $a_{k-1}$ and $a_{k+1}$, for $k\geq 2$, implying that all the equations of the hierarchy are coupled. Then, in order to manage the system of differential equations, we introduced some approximations, based on a Sonine expansion of the \acrshort{VDF}, to truncate the infinite system of coupled differential equations. The first approach is referred as the \acrshort{BSA} in Article 1 (\refsec{Art1}) and it consisted in the truncation of the description up to $a_2$, considering $a_k$ with $k\geq 3$ to be negligible, and properly linearizing the differential equations with respect to the excess kurtosis, $a_2$. Moreover, we improved this approach by a more \emph{sophisticated} one, called the \acrshort{ESA} in Article 1 (\refsec{Art1}), where the Sonine coefficient that determines the truncation limit is $a_3$, plus a proper linearization of the equations with respect to $a_2$ and $a_3$. Moreover, in Ref.~\cite{SP20}, it was observed that the \acrshort{TME} occurs at the first stages of the evolution---if there is no an exaggerated quenching or heating~\cite{PSP21}---implying that a solution of the \acrshort{BSA} in terms of a linearization of the evolution equations around the initial conditions applies. The latter approach was referred to as \acrshort{LBSA} in Article 1 (\refsec{Art1}) and it allowed for an analytical solution of the system. Furthermore, with respect to the evolution of $\KLD(f(t)|f_{\mathrm{M}}^{\mathrm{eq}})$, we demonstrated that it could be divided into kinetic and local-equilibrium counterparts (see \reffig{EME_MSP22}). The former absorbs the evolution of the \acrshort{VDF} with respect to the local-equilibrium \acrshort{VDF}, which was approximated by a \acrshort{SA}, obtaining that $\mathcal{D}_{\mathrm{kin}}\propto a_2^2$. On the other hand, the latter describes the distance of the local-equilibrium distribution from the equilibrium \acrshort{VDF}, which was found to be a function only of $T(t)/T_b$. Due to the rapid relaxation of the system---from initial conditions not excessively far from the equilibrium \index{Temperature!of the thermal bath}temperature---we concluded that the kinetic part of the \acrshort{KLD} evolution determines the appearance or not of the \acrshort{EME}.

In order to study the \acrshort{ME}, we have defined two arbitrary samples of our molecular fluid, A and B, the latter with a \index{Temperature!of the thermal bath}temperature further from $T_b$. The \acrshort{TME} was said to occur if \index{Temperature!nonequilibrium}$T_A$ relaxes toward $T_b$ earlier than $T_B$. On the other hand, the \acrshort{EME} was established to emerge if a system with an initially greater \acrshort{KLD}\footnote{Whereas for systems close to the equilibrium \acrshort{VDF}, ${\KLD}_A^0>{\KLD}_B^0$, this might not be the case if system B is initialized to a \acrshort{VDF} much further from the Maxwellian one than A.} arrives earlier to equilibrium. To gain some insight, we firstly restricted the \acrshort{TME} to cases in the absence of any \index{Temperature!overshoot}overshoot of the equilibrium \index{Temperature!of the thermal bath}temperature, $T_b$. Thus, from heuristic arguments, we have identified different events related to the \acrshort{TME} and \acrshort{EME}, as Table~I of Article 1 (\refsec{Art1}) summarizes. Under these considerations, the \acrshort{TME} (\acrshort{EME}) was characterized by the emergence of a crossover---or an odd number of them---between the thermal (entropic) curves at a certain crossing time, $t_\theta$ ($t_{\mathcal{D}}$). From the \acrshort{LBSA}, we have predicted the regions of the initial-temperature conditions, where these different phenomena arise, for representative choices of the initial values of the excess kurtosis, as pictured in Figure~4 of Article 1 (\refsec{Art1}). Afterwards, we have observed that a type of \acrshort{TME} may arise without a crossover (or with an even number of crossovers) between $T_A(t)$ and $T_B(t)$. This may occur due to an \index{Temperature!overshoot}overshoot of (at least) system B through the equilibrium \index{Temperature!of the thermal bath}temperature, $T_b$. If the \index{Temperature!overshoot}overshoot lasts enough, system A could take advantage and relax to equilibrium sooner than B. This phenomenon has been termed as the \acrshort{OME}, which is meaningless in the \acrshort{EME} description, due to the monotonic decay of the \acrshort{KLD}. However, as an important result, the local-equilibrium counterpart of the \acrshort{KLD}---which is indeed only a function of $T(t)/T_b$---absorbs both the standard version of the \acrshort{TME} and the \acrshort{OME} via the single (or odd number of) crossover of the $\mathcal{D}^{\mathrm{LE}}$ curves, acting as the thermal distance introduced in \refsec{ME_ME}.

In order to check the validity of our heuristic argumentation and the approximations, we have compared the theoretical results with \acrshort{DSMC} and \acrshort{EDMD} computer simulation outcomes. The excellent agreement between the simulation results and the theoretical predictions have confirmed the great phenomenological variety observed in this system. These results open the way to different manners to look for the emergence of \acrshort{ME} in complex systems.
 
\section{Granular gases of inelastic particles}

The very first and simplest description of the interaction of particles in a granular gas is the \acrshort{IHS} model~\cite{BP04,G19}. In addition, granular matter, under rapid-flow conditions, can be described by the Boltzmann kinetic equation, whose collisional operator must be adapted to the choice of the collisional model. In the inelastic and smooth case for hard \dt-spheres, the properties of the granular gaseous flows have been widely studied in the last three decades, some examples being Refs.~\cite{C90,M93,GS95,GZB97,vNE98,SG98,D00,D01,GD02,G03,BP03,BP04,GNB05,SGNT06,VSG10,DB11,GS11,GMT13,G19}.
In this thesis, we have got immersed more in some properties of this model for the monodisperse case. First, to the freely evolving system and, then, to the case of a granular gas interacting with an interstitial fluid.

\subsection{Freely evolving granular gases of inelastic particles}

Whereas new and interesting phenomena arise in an inelastic granular gas, as compared with the widely studied case of molecular gaseous flows, most of the tools derived from the latter have been applied to the former. The simplest studied states of the system are the homogeneous ones, as in molecular gases. It is widely known and studied that the homogeneous and inelastic \acrshort{BE} is compatible with a limiting scaling solution, called the \acrshort{HCS} (see \refsubsec{HCS_KTGG}). This state has been observed experimentally~\cite{YSS20} and turns out to be unstable under certain conditions, as derived from linear stability analyses~\cite{GZ93,M93,MY96,BDKS98,BRM98,LH99,SMM00,G05,G19}. However, the $H$-theorem, which is one of the most important results of the Boltzmann kinetic theory applied to molecular fluids, proving the continuous evolution of the system toward equilibrium via its \index{Nonequilibrium entropy}nonequilibrium entropy, has not been extended rigorously to the inelastic case, mathematically speaking. References~\cite{BPV13,GMMMRT15} address this problem and conjecture, from different arguments, that, given a granular gas described by the \acrshort{IHS} model, the quantitity $\KLD(f|\lim_{t\to \infty} f(t))$ may act as the \index{Nonequilibrium entropy}nonequilibrium entropy of the system. In a driven granular gaseous system, $\lim_{t\to \infty} f(t)$ coincides clearly with the steady-state \acrshort{VDF}, whereas, in the freely evolving case, this limit is more subtle and admitted to be the \acrshort{HCS} \acrshort{VDF}. However, whereas, in Ref.~\cite{BPV13} the conjecture is presented and some arguments are given in its favor, for the freely evolving case, Ref.~\cite{GMMMRT15} gave positive arguments for the conjecture in the context of Kac's master equation~\cite{K56}. Moreover, in both works some computer simulations, for systems of inelastic \acrshort{HD} ($\dt=2$) and \acrshort{HS} ($\dt=3$), also supported the hypothesis.

In this thesis, we have studied this problem in Article 2 (\refsec{Art2}) and Article 3 (\refsec{Art3}). To gain some insight into the issues making this $H$-theorem problem to be hard be formally solved, we have firstly studied the homogeneous states of the \acrshort{BE} under standard Sonine-like approximations in Article 2 (\refsec{Art2}). We have derived the evolution equations for the second and third Sonine coefficients, $a_2$ and $a_3$, under a truncation and proper linearization of the moment evolution equations. The way of linearizing the rates of change of those two cumulants was compatible with the asymptotic \acrshort{HCS} values discussed in Ref.~\cite{SM09}. The approximate forms of the evolution equations were supported by comparison with \acrshort{EDMD} computer simulation results for transient and steady states, thus extending and improving the analysis and results of Ref.~\cite{HOB00}. 

Once the homogeneous evolution and the \acrshort{HCS} \acrshort{VDF} have been approximately characterized, the \acrshort{KLD} is introduced as a proper functional to describe the \index{Nonequilibrium entropy}nonequilibrium entropy of the system. It solves certain issues that Shannon's $H$ functional presents, associated with problems of measure~\cite{MT11}. Then, we have focused the problem on the choice of the proper reference \acrshort{VDF}. We have compared the choices $\KLD(f|f_{\mathrm{M}})$ and $\KLD(f|f_{\HCS})$ from numerical approximate schemes and computer simulations, for a wide variety of initial conditions, observing monotonic evolutions for the latter and not for the former. These results support the original conjecture and discard the choice of $\KLD(f|f_{\mathrm{M}})$. Moreover, to elucidate why the monotonic parts of the evolution appear in the \acrshort{HCS} choice and not in the Maxwellian one, we have derived a toy model for the \acrshort{KLD}, which is based on a \acrshort{SA}. Introducing the evolution equations derived under our initial approach into this toy model, $\KLD(f|f_{\HCS})$ resulted to be a Lyapunov functional, proved only under this approach. On the other hand, in Article 3 (\refsec{Art3}), this has been further analyzed and the toy model also explains why, for certain initial conditions, the nonmonotonic behaviors of the $\KLD(f|f_{\mathrm{M}})$ appear.

Furthermore, in Article 2 (\refsec{Art2}), we have used the mathematical meaning of the \acrshort{KLD} in the context of information theory to study how far the \acrshort{HCS} \acrshort{VDF} is from the Maxwellian one. This relation has been found to be nonmonotonic with respect to the coefficient of normal restitution\index{Coefficient of normal restitution}, $\een$, for the least inelastic values. More specifically, within approximately the interval $0.7 \lesssim \een \lesssim 1$, we cannot obtain univocally the coefficient of restitution from $\KLD(f_{\HCS}|f_{\mathrm{M}})$. This has been inferred from numerical schemes and \acrshort{EDMD} results, as showed in Figure~10 of Article 2 (\refsec{Art2}). For the sake of completeness, we have carried out \acrshort{EDMD} simulations for the inelastic generalization of the original velocity-inversion computer experiments of Orban and Bellemans~\cite{OB67,OB69} by a sort of a Maxwell-demon-based mechanism. The evolution of the \acrshort{VDF} is guided by $\KLD(f(t)|f_{\HCS})$, such that in the elastic case ($\een=1$) it was set to be $\KLD(f(t)|f_{\mathrm{M}})$. We have recovered, for the elastic case, an anti-kinetic stage in the evolution, symmetric to the previous kinetic evolution up to the waiting time, $t_w$, when this velocity inversion was carried out. However, in granular gases of inelastic particles, $\een<1$, this anti-kinetic stage is found to be almost fully suppressed as a result of the breakdown of the \emph{time} symmetry of the binary collisional rules. This suppression was found to be present even in the quasi-elastic limit.

We expect that these results could encourage a formal proof for the conjecture of the \index{Nonequilibrium entropy}nonequilibrium entropy of the granular gas of inelastic hard \dt-spheres. Moreover, we hope the analysis derived could be useful to other nonequilibrium systems in statistical physics contexts.

\subsection{Granular gases of inelastic particles under nonlinear drag}

After the analysis of the freely evolving case, we took a step further in the study of granular gases under the \acrshort{IHS} model. Whereas, in the outer universe, grains, like dust, evolve freely, on planet Earth, they are not in pure vacuum, except if we force them to be isolated in a laboratory setup. Thus, a granular gas can be found surrounded by other fluids, this system being usually called a \emph{granular suspension}. It is assumed that the background fluid acts as a thermal bath at a certain \index{Temperature!of the thermal bath}temperature, $T_b$. The interaction between the granular phase and the surrounding fluid has been usually modeled in the literature by a coarse-grained description. This consists in the assumption of the action of a linear drag force, plus a stochastic force with the properties of a white noise. The noise intensity and the drag coefficient values are usually constrained to obey a fluctuation-dissipation relation. Then, in the dilute regime, this system can be studied in terms of a \acrshort{BFPE}, where the Fokker--Planck part of the equation is related to the associated \acrshort{LE} for the velocities. This kind of description can be found in, for example, Refs.~\cite{CVG12,CVG13,GKG20,THS21}. In this thesis, we have studied the monodisperse case and have improved the latter Langevin-like model from the inelastic extension of the molecular system introduced in Article 1 (\refsec{Art1}). That is, we have assumed that the drag force is nonlinear, i.e., the drag coefficient is, in general, a function of the velocity modulus. The simplest nonlinear term has a quadratic dependence on $v$, which can be derived from a \acrshort{QR} approximation, as detailed in \refapp{app_NLD}. The effect of the nonlinear term is assumed to be small and is controlled by a constant coefficient, $\gamma$, which is defined through the ratio between the mass of the grains and the mass of the background fluid particles from the \acrshort{QR} approach [see \refeq{gamma_NLD}].

In Article 4 (\refsec{Art4}), our analysis of the latter system has focused on its homogeneous states. After the identification of the corresponding \acrshort{BFPE}, we have derived formally the evolution equation of the granular \index{Temperature!granular}temperature and a full hierarchy of moment equations. Attending to the structure of the differential equation of the granular \index{Temperature!granular}temperature, a two-source dependence from the cumulants of the instantaneous \acrshort{VDF} has been identified. The first dependence is explicit on $a_2$ due to the coupling of the quadratic nonlinear term of the drag coefficient, as already observed in the molecular model of Article 1 (\refsec{Art1}). On the other hand, the granular inelastic interactions made the energy to be not conserved upon collisions, this energy dissipation being accounted for by the \index{Cooling rate!in the IHS model}cooling rate. This quantity is a function of all the moments of the \acrshort{VDF}, thus acting as a the second coupling source. All this complex dependence has been studied in the context of a Sonine expansion of the \acrshort{VDF} and, in order to get a closed and finite system of differential equations for a manageable mathematical description, we have truncated the expansion in two manners. First, the simplest approach consisted in the assumption of the time-dependent and steady-state \acrshort{VDF} to be a Maxwellian distribution. However, from this approach, the couplings to the cumulants of the \acrshort{VDF} are completely lost. Then, we introduced a \acrshort{FSA} to study the transient and steady states of the system in terms of the granular \index{Temperature!granular}temperature, $T$, and the excess kurtosis, $a_2$. This approach consists in the truncation of the Sonine expansion up to its second coefficient plus a proper linearization of the dynamical set of equation with respect to $a_2$. For the sake of completeness, we have checked that the stationary values of those quantities coincide, under certain limits, with already known expressions of different granular systems. Once the theoretical bases have been established, the predictions have been compared with \acrshort{DSMC} and \acrshort{EDMD}, obtaining quite a good agreement with the \acrshort{FSA}.

From the \acrshort{FSA}, we have observed a nontrivial relaxation toward the steady state. As already studied in Article 1 (\refsec{Art1}) in the molecular case, this is due to the explicit dependence of the rate of change of the granular \index{Temperature!granular}temperature on the instantaneous value of $a_2$. However, in the granular case, this gets more involved due to the presence of the \index{Cooling rate!in the IHS model}cooling rate. The dependence of the granular temperature evolution on the inner variables of the system, during nonequilibrium stages, is proper of systems that exhibit memory effects (see \refch{ME_KTGG}). Then, using the knowledge of Refs.~\cite{SP20,PSP21,MSP22}, and as an application of the theoretical scheme, we have predicted the presence of the \acrshort{ME} and the \acrshort{KE}. It must be highlighted that the coupling of $a_2$ with the evolution of the temperature dominates the description of the memory effects. Moreover, from the difference $a_2(t)-a_2^{\mathrm{st}}$, the sign of the humps appearing in a Kovacs-like protocol can be deduced [see Eq.~(29) of Article 4 (\refsec{Art4})]. In fact, the theoretical predictions of the latter two effects agree with \acrshort{DSMC} and \acrshort{EDMD} computer simulation results. These results open the path to study more complex granular gaseous suspensions flows from a more realistic description.

\section{Granular gases of inelastic and rough particles}

Although the \acrshort{IHS} model is very simple, it has an important predictive power, at least, qualitatively or semi-quantitatively. Recent experiments~\cite{YSS20}, in fact, favor this model against velocity-dependent coefficients of normal restitution, such as in the viscoelastic model. However, the numerical discrepancies between experimental results and kinetic theory predictions are expected to be solved by the introduction of rotational degrees of freedom into the system dynamics, as the authors of Ref.~\cite{YSS20} claim. The simplest introduction of this effect is done by means of the \acrshort{IRHS} model, where energy is lost upon collision not only by the inelasticity of the particles, but also due to their surface roughness, which is characterized by a coefficient of tangential restitution\index{Coefficient of tangential restitution}, $\eet$, assumed to be constant. This model has also been studied before in the literature, see, for example, Refs.~\cite{BP04,BPKZ07,KBPZ09,VSK14,S18,G19,MS19,MS19b}. In this thesis, we have studied deeply the homogeneous and inhomogeneous states of its free-evolution version, as well as its homogeneous states and the presence of memory effects in driven setups.

\subsection{Freely evolving granular gases of inelastic and rough particles}

A homogeneous granular gas in free evolution, described by the \acrshort{IRHS}, admits a scaling limiting form for its \acrshort{VDF}, as occurs in the \acrshort{IHS} model. When this happens, the system is said to be at the \acrshort{HCS}, and all the time evolution of the system is absorbed by the temperature evolution. In general, the homogeneous transient and \acrshort{HCS} states of a freely evolving granular gases have been previously studied. First, in terms of a \acrshort{MA} (see, for example, Refs.~\cite{AHZ01,Z06,S18,MS19,MS19b}), where rotations and translations are assumed to be uncorrelated at the level of the one-particle \acrshort{VDF}. However, this was found to be false, as observed in, for example, Refs.~\cite{BPKZ07,KBPZ09}. Hence, the \acrshort{HCS} has been observed to be non-Maxwellian, similarly as occurs in the inelastic and smooth case. Approaches to their deviations from the Maxwellian \acrshort{VDF} had been studied only in the context of \acrshort{HS}, and only attending to the first nontrivial cumulants by means a Sonine expansion of the \acrshort{HCS} \acrshort{VDF}~\cite{SKS11,SK12,VSK14,VSK14b}. Moreover, the transport properties and the instability of the \acrshort{HCS} had only been studied in systems of \acrshort{HS}~\cite{KSG14,GSK18}. 

In this thesis, the goals related to this topic have been mainly four, and they have been restricted to the monodisperse case. First, we have generalized the whole description related to this system in terms of the translational and rotational degrees of freedom of the problem, $\dt$ and $\dr$, respectively, with the aim of being descriptive for both \acrshort{HD} ($\dt=2$, $\dr=1$) and \acrshort{HS} ($\dt=3$, $\dr=3$) systems. For that, steps similar to those carried out in Refs.~\cite{MS19,MS19b} have been introduced. The second goal consisted in analyzing the non-Gaussianities of the \acrshort{HCS} \acrshort{VDF}, within this generalized context, embedding the results for \acrshort{HD} and \acrshort{HS} in a single description. The studied non-Gaussian properties have not been just the first nontrivial Sonine coefficients, but also the \acrshort{HVT} of the most important marginal \acrshort{VDF} at the \acrshort{HCS}. This is summarized in Article 5 (\refsec{Art5}). On the other hand, the transport properties of a granular gas of inelastic and rough \acrshort{HD} or \acrshort{HS} have been investigated by means of the \acrshort{NSF} transport coefficients. These quantities have been derived from the \index{Chapman--Enskog method}\acrshort{CE} method applied to this context, as described in Article 7 (\refsec{Art7}). Finally, the last aim corresponds to the study of the instabilities of the \acrshort{HCS} from inhomogeneous and linear perturbations, as an application of the knowledge of the transport coefficients [see Article 8 (\refsec{Art8})].

With respect to the description of the \acrshort{HCS}, we have studied a monodisperse dilute granular gas of inelastic and rough \acrshort{HD} or \acrshort{HS}, in the context of the homogeneous \acrshort{BE}, for the \acrshort{IRHS} model. To study, in general, the homogeneous transient and \acrshort{HCS} states of the system, we have firstly introduced a Sonine expansion of the \acrshort{VDF}. Then, we have truncated the expansion up to the first nontrivial Sonine coefficients. In the \acrshort{HD}, these are $a_{20}$, $a_{02}$, and $a_{11}$. However, in the \acrshort{HS}, we had one extra coefficient, $a_{00}^{(1)}$. Thus, the transient states of the system under this approach, see Article 5 (\refsec{Art5}), were guided by the evolution of the latter Sonine coefficients plus the \index{Temperature!rotational-to-translational ratio}rotational-to-translational temperature ratio, $\theta$. Afterwards, we have derived the \index{Collisional moments!for IRHS}collisional moments involved in this scheme in terms of two-body averages. These expressions are exact in the context of the \acrshort{BE}. Then, by introducing the mentioned truncation of the \acrshort{VDF}, we have obtained certain approximate values of the two-body averages and, thus, of the \index{Collisional moments!for IRHS}collisional moments. All these quantities have been then expressed in terms of the dynamic quantities $\theta$, $a_{20}$, $a_{02}$, $a_{11}$ [and $a_{00}^{(1)}$ for \acrshort{HS}], and as functions of the coefficients of restitution, $\een$ and $\eet$, the degrees of freedom, $\dt$ and $\dr$, and the mechanical properties of the system, i.e., the diameter, $\sigma$, the mass, $m$, and the reduced moment of inertia, $\kappa$. In other words, the evolution of the system toward the \acrshort{HCS} has been described by a closed set of differential equations in terms of the parameters of the problem. It is worth highlighting that the implemented \acrshort{SA} scheme has been validated by a homogeneous linear stability analysis, as explained in Appendix~A of Article 5 (\refsec{Art5}).

From the development of the Sonine description, we have obtained the \acrshort{HCS} values of the dynamic quantities. The results for \acrshort{HS} reported in Ref.~\cite{VSK14} were again obtained, as well as novel results for the \acrshort{HD} case. Furthermore, we have observed very large values of $a_{02}$, indicating a possible breakdown of the \acrshort{SA}. In order to localize the regions of the parameter space where this approach is expected to worsen, we have developed a scheme in terms of an entropy-like functional derived from a toy-model-like approach of $\KLD(\phi_\HCS|\phi_{\mathrm{M}})$. These regions have been identified to appear as lobes emerging, paradoxically, from the elastic and smooth case ($\een=-\eet=1$), as can be observed in Figure~4 of Article 5 (\refsec{Art5}). These regions are dominated by the values of $a_{02}$, as expected. As a matter of fact, generally speaking, \acrshort{HD} systems present slightly higher deviations from the Maxwellian than \acrshort{HS} ones. Furthermore, we have determined the \acrshort{HVT} of the marginal \acrshort{VDF}, $\phi_{\cc}$, $\phi_{\ww}$, and $\phi_{cw}$, at the \acrshort{HCS}. The asymptotic treatment of this problem has been founded from the prevalence of the loss term of the \index{Boltzmann collisional operator!for IRHS}Boltzmann collisional operator over the gain one (those terms have been introduced in \refsec{der_BE_KTGG}). This is a standard approach already used in the \acrshort{IHS} model~\cite{EP97,vNE98}. In the case of $\phi_{\cc}$, the \acrshort{IHS} exponential form of the tail has been recovered, i.e., $\phi^{\HCS}_{\cc}\sim e^{-\gamma_c^{\mathrm{IRHS}}c}$. In fact, $\gamma_c^{\mathrm{IHS}}\to\gamma_c^{\mathrm{IRHS}}$ if $\mu_2^\HCS\to \mu_{20}^\HCS$. Moreover, for the rotational marginal \acrshort{VDF}, an algebraic decay has been predicted after certain approximations, i.e., $\phi^{\HCS}_{\ww}\sim w^{-\gamma_w}$. Finally, for the joint distribution of the product $c^2w^2$, an algebraic decay has also been obtained, $\phi^{\HCS}_{cw}\sim (c^2w^2)^{-\gamma_{cw}}$. The latter two resulting \acrshort{HVT} imply the divergence of cumulants of the forms $a_{0k}$ if $2k\geq \gamma_w-1$ and $a_{kk}$ if $k\geq \gamma_{cw}-1$. These consequences are another indicator of a possible breakdown of the \acrshort{SA}.

Finally, we have performed \acrshort{DSMC} and \acrshort{EDMD} computer simulations for \acrshort{HD} systems in the absence of instabilities. We have firstly compared the simulation and theoretical outcomes for the quantities $\theta^{\HCS}$, $a_{20}^{\HCS}$, $a_{02}^{\HCS}$, and $a_{11}^{\HCS}$, concluding that the predictions from the \acrshort{HCS} reproduce qualitatively well the outcomes from the \acrshort{SA}, even in the regions of larger values, as observed in Figures~5 and 6 of Article 5 (\refsec{Art5}). Furthermore, the averaged histograms for the marginal \acrshort{VDF} at the \acrshort{HCS} seem to fulfill the predicted \acrshort{HVT}, as shown in Figures~7 and 8 of Article 5 (\refsec{Art5}). In fact, the derived form of the rotational exponent, $\gamma_w$, has been very well reproduced by the simulation results. However, in the case of $\gamma_c$ and $\gamma_{cw}$, their values agree with simulation in a more qualitative way [see Figure~9 of Article 5 (\refsec{Art5})].

Once the reference homogeneous state, the \acrshort{HCS}, has been characterized, we have analyzed the influence of roughness in the transport properties of monodisperse granular gases. We have worked under the assumptions underlying the \index{Chapman--Enskog method}\acrshort{CE} method, as described in detail in \refsubsec{CE_Exp_KTGG}. The fields considered for the hydrodynamic description, in this case, are the number density $n$, the flow field $\mathbf{u}$, and the mean granular temperature $T$. The main difference with the smooth case is the influence of the rotational degrees of freedom, which translates, for example, translated in two contributions of $T$ by means of the translational partial temperature, $\Ttr$, and the rotational one, $\Trot$. Then, we have applied a \acrshort{CE} expansion and retained terms up to the \acrshort{NSF} order. This unified analysis is applicable to both \acrshort{HD} ($\dt=2$, $\dr=1$) and \acrshort{HS} ($\dt=3$, $\dr=3$) systems. With respect to the proper application of the \index{Chapman--Enskog method}\acrshort{CE} method, the zeroth order \acrshort{VDF}, $f^{(0)}$, is compatible with a local version of the \acrshort{HCS} \acrshort{VDF}. Although we knew that the \acrshort{HCS} is non-Maxwellian, we have followed the steps in Ref.~\cite{KSG14}, in which, as a first reasonable insight to the problem, a \acrshort{MA} is used. In the final computations of the \acrshort{NSF} transport coefficients, which were described by the integration of certain quantities with respect to the zeroth-order \acrshort{VDF}, the thermal part of the distribution, which was expected to be well described by a Maxwellian \acrshort{VDF}, was expected to dominate. The same occurs with the inelastic and smooth case, but it is worth mentioning that corrections by means of the introduction of the Sonine coefficients improve the results~\cite{BC01,GSM07,G19}. Nonetheless, we have worked within the \acrshort{MA} for $f^{(0)}$.

The derivation of the \acrshort{NSF} transport coefficients has been divided into two parts. First, the attainment of the balance equations directly from the \acrshort{BE}, represented in Eqs.~(8) of Article 7 (\refsec{Art7}) and detailed in \refapp{app_balance_equations}. The main difference with respect to the molecular case has been the contribution of the \index{Cooling rate!in the IRHS model}cooling rate, whereas, with respect to the smooth case, it has been the appearance of a contribution of the rotational degrees of freedom to the heat flux, $\mathbf{q}_r$, as well as to the \index{Cooling rate!in the IRHS model}cooling rate via the energy production rates [see Eqs.~(9) of Article 7 (\refsec{Art7})]. Therefore, comparing the first-order \acrshort{VDF} of the \index{Chapman--Enskog method}\acrshort{CE} expansion, $f^{(1)}$, with the balance equations and the phenomenological form of the fluxes [see Eqs.~(34) of Article 7 (\refsec{Art7})], we have obtained explicit forms for the transport coefficients as functions of the parameters of the system. These computations were performed under a sort of \acrshort{SA} for the coefficients of $f^{(1)}$, as detailed in Eqs.~(40) of Article 7 (\refsec{Art7}). The \acrshort{NSF} transport coefficients are, using the notation of Article 7 (\refsec{Art7}), $\{\eta,\eta_b,\lambda,\mu,\xi\}$, namely, the \index{Transport coefficient!shear viscosity}shear viscosity, $\eta$, the \index{Transport coefficient!bulk viscosity}bulk viscosity, $\eta_b$, the \index{Transport coefficient!thermal conductivity}thermal conductivity, $\lambda$, the \index{Transport coefficient!Dufour-like coefficient}Dufour-like coefficient, $\mu$, and the \index{Transport coefficient!of the cooling rate}velocity-divergence transport coefficient, $\xi$, which is associated with the first-order correction to the cooling rate.\footnote{Do not confuse this $\xi$, which corresponds to the contribution of the cooling rate to the velocity divergence, i.e., $\zeta^{(1)}=-\xi\nabla\cdot\mathbf{u}$, with the notation of the friction coefficient for the \acrshort{LE} in other contexts.}

We have explicitly reproduced the theoretical values of the transport coefficients, in Figures~3-6 of Article 7 (\refsec{Art7}), for uniform ($\kappa=1/2$) \acrshort{HD} systems and with a mass distribution concentrated on the outer surface ($\kappa = 1$). Moreover, we have done the same for \acrshort{HS} granular gases, again with a uniform mass distribution ($\kappa = 2/5$) and with a mass distribution concentrated on the outer surface ($\kappa = 2/3$). The results for \acrshort{HS} are in agreement with the ones previously reported on Ref.~\cite{KSG14}. All the relevant theoretical expressions for the transport coefficients have been summarized in Table~I of Article 7 (\refsec{Art7}). For the sake of completeness, we have checked that results for smooth particles are recovered, and they have also been computed in the quasismooth limit ($\eet\to -1$) and in the case of a Pidduck's gas ($\een=\eet=1$). Apart from the new dependences of $\eet$ and $\kappa$, as compared with the smooth case, we can extract some important remarks. For example, the Dufour-like coefficient, $\mu$, characteristic of granular gases, becomes more relevant in the rough case than in the perfectly smooth one. Moreover, the effect of the rotational degrees of freedom make the bulk viscosity, $\eta_b$, to be nonzero, contrarily to the dilute smooth case. In fact, in arbitrarily small presence of roughness, $\eta_b$ is still nonzero, as showed by the quasismooth limit. This recalls some arguments about the origin of the bulk viscosity in molecular fluids due to the rotational (and vibrational) molecular motions~\cite{T81,DG02}. Finally, in the quasielastic ($1-\een\ll 1$) smooth case, the velocity-divergence transport coefficient, $\xi$, admits small negative values, which is identically null in the dilute regime in the smooth limit.

As a first application of the transport properties of a granular gas of inelastic and rough \acrshort{HD} or \acrshort{HS}, we have studied the stability of the \acrshort{HCS} under linear inhomogeneous perturbations, within the generalized degrees-of-freedom scheme, as reported on Article 8 (\refsec{Art8}). We have first slightly perturbed the \index{Hydrodynamic!fields}hydrodynamic fields for small gradients around the \acrshort{HCS}. After the proper linearization of the \acrshort{NSF} equations with respect to the perturbations, and transforming the time-evolving equations from spatial gradients to wave vectors via Fourier analysis, we have obtained the \index{Hydrodynamic!modes}hydrodynamic modes as functions of the parameters of the system. These modes can decompose into a $(\dt-1)$-degenerated shear mode [see Eq.~(13) of Article 8 (\refsec{Art8})], $\varpi_{\perp}$, and three other transverse modes, $\varpi_{\parallel,i}$ with $i=1,2,3$, which are the solutions to the cubic equation written in Eq.~(18) of Article 8 (\refsec{Art8}). Their dispersion relations show the existence of critical value of the wave number, from which some of the modes become positive [see Figures~1 and 2 of Article 8 (\refsec{Art8})]. This implies that the system becomes unstable under perturbations related to values of  the wave number $k$ greater than this critical one, which has been already observed to occur in the smooth case~\cite{GZ93,BRM98,BRC99,LH99,G05,MDCPH11,MGHEH12,MZBGDH14,FH17,G19}. In fact, the instabilities occur due to the positiveness of the shear and one of the longitudinal modes, but the critical value of $k$, below which the mode becomes negative, is different for both type of modes. In fact, we have computed the explicit forms of these two \index{Critical wave number}critical wave numbers, $k_{\perp}$ and $k_{\parallel}$ [see Eqs.~(14) and (21) of Article 8 (\refsec{Art8})], as a generalization of the \acrshort{HS} results reported in Refs.~\cite{MDHEH13,GSK18}. From them, we can deduce a \index{Critical length}critical length, $L_c$, of a real (or \acrshort{MD}-computational) system, such that the \acrshort{HCS} is stable only for characteristic lengths $L<L_c$. 

The most outstanding result of this analysis is the observation of the \acrshort{HUR}, in which our approximate theory predicts that hydrodynamics breaks down for certain values of the coefficients of restitution due to the spontaneous emergence of granular clusters. This occurs due to a divergence and complex-valued regime of $k_{\parallel}$, as a consequence of a negative-valued difference between the (reduced) thermal conductivity and the (reduced) Dufour-like coefficients in the equation of this \index{Critical wave number}critical wave number [see Eq.~(21) of Article 8 (\refsec{Art8})]. The origin lies on the large values that the Dufour-like transport coefficient admits, as compared with the smooth case [see Figure~6 of Article 7 (\refsec{Art7})]. Whereas this region is slightly observed in \acrshort{HS}---just for a small region of values of the reduced moment of inertia---it is always present in \acrshort{HD} systems [see Figure~3 of Article 8 (\refsec{Art8})]. Due to the importance of this result, we have performed \acrshort{EDMD} simulations to check it. In order to detect the possible appearance of clustering, we have developed a decision scheme based on the \acrshort{KLD} of the position distribution inside the simulation box, with respect to a reference binomial homogeneous distribution, as well as a comparison of the long-time values of the quantities $\theta$, $a_{20}$, $a_{02}$, and $a_{11}$ with respect to the \acrshort{HCS} predicted values within the methods of Article 5 (\refsec{Art5}). The \acrshort{EDMD} outcomes have shown that, outside the \acrshort{HUR}, theoretical predictions about the stability are apparently valid. However, inside the \acrshort{HUR}, we have observed a stable system in a computer realization up to 1000 of collisions per particle. The simulations results indicate that this region might be \emph{very} unstable, instead of totally unstable. Thus, the theoretical prediction might be an artifact of the approximations in the evaluation of the transport coefficients. In addition, we have observed an interesting qualitative connection between the spatially defined \acrshort{KLD} in systems at \acrshort{HCS} and the values of the \index{Hydrodynamic!modes}hydrodynamic modes, as reported in Figure~7 of Article 8 (\refsec{Art8}).

We expect this work could motivate further research in this topic, especially in the nature of the \acrshort{HUR}, from theory and computer simulations. In addition, the determination of the transport coefficients by means of computer simulations would establish the goodness of the approximations. Moreover, we hope that the simplicity of the \acrshort{HD} systems could encourage the comparison of these results with real experiments. Finally, it would be interesting to investigate how the non-Gaussianities of the \acrshort{HCS} of this freely evolving gas of rough particles, studied in \refch{IRHS_HS}, affect to the final results of the transport properties.

\subsection{Driven granular gases of inelastic and rough particles}

A very interesting problem in the study of granular fluids is the consideration of driven states, both in theory~\cite{WM96,W96,vNE98,CCG00,MS00,BT02a,GM02,EB02c,HCZHL05,vMR06,GMT12,CVG12,MGT12,CVG13,KG13,G19,KG20} and experiments~\cite{GSVP11a,GSVP11b,MSGE22,CZ22}. Moreover, driven granular gases of inelastic and rough hard particles have been also considered in the context of kinetic theory~\cite{CLH02,VS15,MS19,TLLVPS19,GG20}. In this thesis, we have focused on the homogeneous states of a monodisperse and dilute granular gas driven by the \acrshort{ST}. That is, we have assumed that the energy injection is performed via a stochastic force and a stochastic torque to the translational and rotational degrees of freedom, respectively. The \acrshort{ST} can be characterized by two quantities, a reference temperature ($\Twn$) associated with the noise intensity and the rotational-to-total noise intensity ratio ($\varepsilon$), the latter accounting for the part of the energy injected directly to the rotational degrees of freedom. This thermostat includes as particular cases the system already proposed in Ref.~\cite{CLH02} for purely rotationally driven states ($\varepsilon=1$), the one in Refs.~\cite{VS15,MS19,TLLVPS19} for purely translational driven systems ($\varepsilon=0$), as well as all their intermediate cases ($0<\varepsilon<1$). In fact, our goal has been the extension of the study of the \acrshort{TME} in homogeneous and driven granular gases of rough \acrshort{HD}, which began in Ref.~\cite{TLLVPS19}, with the aim of proposing a preparation protocol for the observation of the effect. Recall that in \refch{ME_KTGG} we already mentioned that, whereas the \acrshort{KE} is associated with a given protocol, this is not the case, in general, with the \acrshort{ME}. Therefore, the absence of a predetermined preparation protocol makes more difficult its experimental reproducibility.

In Article 6 (\refsec{Art6}), we have described this system by means of its corresponding homogeneous \acrshort{BFPE} [see Eq.~(3) of Article 6 (\refsec{Art6})]. From it, we have derived the thermal evolution of the system for the mean granular temperature, $T$, and the rotational-to-translational temperature ratio, $\theta$ [see Eqs.~(23) of Article 6 (\refsec{Art6})]. As usual, to manage analytically the equations, we have introduced approximations. From a previous work in the $\varepsilon=0$ case~\cite{VS15}, we already know that the homogeneous steady states are also not too far from the two-temperature Maxwellian \acrshort{VDF}. Therefore, if the initial conditions are not too far from it, a \acrshort{MA} seems to apply at all stages of the evolution. Under this approach, Eqs.~(23) of Article 6 (\refsec{Art6}) for $\dot{T}\equiv \Phi(T/T^{\mathrm{st}},\theta)$ and $\dot{\theta}$, which are exact in the context of the \acrshort{BFPE}, can be written as Eqs.~(30) of Article 6 (\refsec{Art6}), the latter consisting in a closed set of two coupled differential equations. From them, we have numerically inferred the evolution of the system and also we have derived explicit expressions for the steady-state values of $T^{\mathrm{st}}/\Twn$ and $\theta^{\mathrm{st}}$ [see Eqs.~(26) of Article 6 (\refsec{Art6})]. In fact, we have tested these approximate transient and steady states by means of \acrshort{DSMC} and \acrshort{EDMD} simulations, obtaining an excellent agreement, as can be observed in Figures~5 and 6 of Article 6 (\refsec{Art6}). It is worth stressing that $\theta^{\mathrm{st}}$ is always an increasing function of $\varepsilon$.

Once the theoretical scheme for transient and steady states was developed and checked by computer simulations, we focused on the description of the \acrshort{TME} in cooling experiments, i.e., the \acrshort{DME}. First, we have studied the effect from the mean granular temperature as the guiding variable. That is, assuming two systems, A and B, such that the former is initially hotter than the latter, and both of them hotter than the steady state, the \acrshort{DME} occurs if A relaxes toward the steady state earlier than B. In the first instance, we have described the \acrshort{DME} in its standard form, that is, assuming that there is no \index{Temperature!overshoot}overshoot of the temperature through the stationary value during its evolution. We realized that, for each initial condition, and fixed values of the coefficients of restitution and the reduced moment of inertia, there exists a critical value of the rotational-to-total noise ratio, $\varepsilon_{\mathrm{cr}}(T_0/T^{\mathrm{st}},\theta_0)$, then such that if $\varepsilon> \varepsilon_{\mathrm{cr}}(T_0/T^{\mathrm{st}},\theta_0)$, the temperature will surpass its steady-state value at finite time. For a fixed value of $\theta_0$, this $\varepsilon_{\mathrm{cr}}$ is an increasing function of $T_0/T^{\mathrm{st}}$ [see Figure~7 of Article 6 (\refsec{Art6})]. Therefore, we have inferred that the necessary condition is that the initial slopes of the evolution must fulfill $\Phi_{0,A}<\Phi_{0,B}$, which is equivalent to the condition $\theta_{0,A}<\theta_{0,B}$ for the initial values of $\theta$. However, this is, of course, not sufficient, and we need to study numerically which initial conditions ensure the \acrshort{DME} emergence. Moreover, this is expected to be better satisfied for $\theta_{0,A}\ll\theta_{0,B}$.

Furthermore, we have also considered the presence of the \acrshort{OME} version of the \acrshort{DME}, assuming that $T$ \index{Temperature!overshoot}overshoots its steady-state value. Then, the necessary condition for this effect is the opposite to the previous one. We have imposed that $\Phi_{0,A}>\Phi_{0,B}$ because we need for system B to  exhibit a sufficiently strong \index{Temperature!overshoot}overshoot effect, which is equivalent to $\theta_{0,A}>\theta_{0,B}$. However, the extreme condition $\theta_{0,A}\gg\theta_{0,B}$ might not be valid because, even if we want for system B to surpass enough system A, very high initial values of the temperature slope $\Phi_{0,A}$ could prevent system A from arriving earlier because of a very slow relaxation. Whereas the standard version of the \acrshort{DME} is described by a crossing (or an odd number of them) of the corresponding thermal curves, the \acrshort{OME} has no crossing (or an even number of them) in the standard thermal picture. Therefore, we have proposed a function of $T/T^{\mathrm{st}}$, similar to the local-equilibrium \acrshort{KLD} functional derived in Article 1 (\refsec{Art1}), and denoted as $\mathfrak{D}$ [see Eq.~(32) for Article 6 (\refsec{Art6})], which plays the role of the thermal distance introduced in \refsec{ME_ME}. That is, the properties of $\mathfrak{D}$ implies that both the standard \acrshort{TME} and the \acrshort{OME} are described by a crossover (or an odd number of them) of the evolution curves $\mathfrak{D}_A(t)$ and $\mathfrak{D}_B(t)$.

From the identification of the necessary conditions for the initial values of the variables of the system, we have elaborated a preparation protocol for observing the \acrshort{DME} in this type of setup. This preparation protocol is based on the steady states of the described system, in which we have properly used the property of $\theta^{\mathrm{st}}$ to be an increasing value of $\varepsilon$, as well as the characterization of the absence or presence of \index{Temperature!overshoot}overshoot via $\varepsilon_{\mathrm{cr}}$. Owing to these results, for the observation of the standard \acrshort{DME}, we have proposed the initial values $\theta_{0,A}=\theta^{\mathrm{st}}(\varepsilon=0)$ and $\theta_{0,B}=\theta^{\mathrm{st}}(\varepsilon=1)$, in order to ensure the necessary condition $\theta_{0,A}<\theta_{0,B}$. Afterwards, we must choose the proper values of $\Twn_A$ and $\Twn_B$, such that the associated steady states ($T_{0,A}$ and $T_{0,B}$) verify the conditions $T_{0,A}/\Twn_{\mathrm{ref}}>T_{0,B}/\Twn_{\mathrm{ref}}>T^{\mathrm{st}}_{\mathrm{ref}}/\Twn_{\mathrm{ref}}$, with $\Twn_{\mathrm{ref}}$ and $T^{\mathrm{st}}_{\mathrm{ref}}$ being the noise and stationary temperature at the reference or final thermostat. In addition, we must choose $\varepsilon_{\mathrm{ref}}< \min\{\varepsilon_{\mathrm{cr}}(T_{0,A}/T^{\mathrm{st}}_{\mathrm{ref}}, \theta_{0,A}), \varepsilon_{\mathrm{cr}}(T_{0,B}/T^{\mathrm{st}}_{\mathrm{ref}} , \theta_{0,B})\}$ to prevent \index{Temperature!overshoot}overshoots. Finally, we let both samples evolve. 

The proposed protocol to observe the \acrshort{OME} has been practically identical, except for two main points. Firstly, the choice of the initial conditions of $\theta$ is exchanged, i.e., $\theta_{0,A}=\theta^{\mathrm{st}}(\varepsilon=1)$ and $\theta_{0,B}=\theta^{\mathrm{st}}(\varepsilon=0)$. This has been imposed to ensure the necessary condition for the \acrshort{OME}, $\theta_{0,A}>\theta_{0,B}$. Secondly, the reference value of the noise ratio has set to be $\varepsilon_{\mathrm{ref}} > \varepsilon_{\mathrm{ref}}(T_{0,B}/T^{\mathrm{st}}_{\mathrm{ref}} , \theta_{0,B})$ to force the \index{Temperature!overshoot}overshoot in system B. 

These two protocols have been summarized in Figures~8 and 9 of Article 6 (\refsec{Art6}). Moreover, in Figure~10 of Article 6 (\refsec{Art6}), we have computed numerically a phase diagram in which the presence or absence of the standard \acrshort{DME} and the \acrshort{OME} is settled.

Finally, we have prepared \acrshort{DSMC} and \acrshort{EDMD} computer simulations following the protocols proposed, and for initial conditions inside the standard \acrshort{DME} and \acrshort{OME} regions of the mentioned phase diagrams. As shown in Figures~11 and 12 of Article 6 (\refsec{Art6}), there is quite an excellence agreement between simulation results and theoretical predictions.

We hope this work could encourage the experimental reproducibility of this system, motivated by an apparent simplicity of the two-dimensional setup, as well as further investigation in \acrshort{ME} protocols. In fact, it would be interesting to compute the non-Gaussianities of these homogeneous states to reinforce the \acrshort{MA} used in this work, which has been concluded to be in good accordance with the computer simulations.

\chapter{Conclusions and Outlook}
\labch{C_O}

The aim of this chapter is to expose the main concluding remarks that must be highlighted from the results and procedures developed in this thesis, as well as planned or possible future work continuing this research line.

\section{Conclusions}

The core of the thesis has been structured in three main blocks: one dedicated to molecular gases and the other two centered on granular gases, first the \acrshort{IHS} model and then the \acrshort{IRHS} one. In the three blocks, dynamical properties of homogeneous states of freely evolving and driven setups have been studied, as well as the emergence of memory effects. Remarkably, in the last block, the influence of roughness on the transport properties and the stability of the \acrshort{HCS} have been detailed. All these analyses were performed in the context of the \acrshort{BE}, which includes Fokker--Planck-like terms in the cases of thermostatted systems. In what follows, the main conclusions for each block of thesis will be specified. 

\subsection{Molecular gases}

We have described the relaxation of the homogeneous states of a dilute molecular gas of identical \acrshort{HS} immersed in an interstitial fluid acting as a thermal bath at constant \index{Temperature!of the thermal bath}temperature $T_b$ and interacting with it via a random force and a nonlinear drag force. This has been developed in the context of the corresponding \acrshort{BFPE}.

\subsubsection{Article 1}

\begin{itemize}

\item The phenomenology of the \acrshort{ME} emergence in the latter system has been described in much detail. In addition, a conceptual and mathematical differentiation between the thermal and entropic descriptions has been developed.

\item We have observed a novel \acrshort{TME} scenario based on an \index{Temperature!overshoot}overshoot through the equilibrium \index{Temperature!of the thermal bath}temperature of one of the samples in a \acrshort{ME} experiment, referred to as the \acrshort{OME}. In a description based on the nonequilibrium (time-dependent) \index{Temperature!nonequilibrium}temperature, $T$, there is no crossing between the thermal curves. To solve this issue, we have proposed the study of the \acrshort{TME} through a thermal distance---coinciding with the local-equilibrium \acrshort{KLD}---which detects either the usual forms of \acrshort{TME} and the \acrshort{OME} by means of the crossing-point-based arguments.

\item Finally, \acrshort{DSMC} simulations agree with the approximations of the theory, reinforcing the hypotheses of the developed mathematical approach. Moreover, \acrshort{EDMD} simulation outcomes show quite a good accordance with \acrshort{DSMC} results and the theory, indicating a validation of the description with respect to realistic (computer) experiments.
\end{itemize}

\subsection{Inelastic granular gases}

We have addressed two problems in granular gases described by the \acrshort{IHS} model. The first one, in Article 2 (\refsec{Art2}) and Article 3 (\refsec{Art3}), has been the absence of a formal proof for a granular version of the $H$-theorem in freely evolving granular gases. On the other hand, in Article 4 (\refsec{Art4}), we have developed the Boltzmann kinetic theory and an analysis of some memory effects, which are present in a dilute and homogeneous granular gas of inelastic and hard \dt-spheres immersed in an interstitial fluid, whose interaction with the granular fluid is characterized via the same model as in the molecular gaseous system studied in Article 1 (\refsec{Art1}). 

\subsubsection{Articles 2 and 3}

\begin{itemize}

\item In Article 2 (\refsec{Art2}), we have proposed an approximate theoretical scheme based on a \acrshort{SA}, in which we have recovered previous values for the cumulants at the \acrshort{HCS}, and the evolution of $T$, $a_2$ and $a_3$ from arbitrary initial conditions toward the \acrshort{HCS}.

\item Moreover, in Article 2 (\refsec{Art2}), the problems of the Boltzmann's $H$ quantity that prevent it from being a proper \index{Nonequilibrium entropy}nonequilibrium entropy functional have been identified. Fundamentally, they are based on the noninvariance of $H$ under nonunitary transformations and they are solved by the introduction of the \acrshort{KLD}. Then, the problem has been reduced to find the proper reference \acrshort{VDF} (if it exists) that could allow this quantity to behave as a \index{Nonequilibrium entropy}nonequilibrium entropy.

\item In Articles 2 (\refsec{Art2}) and 3 (\refsec{Art3}), a numerical scheme and a toy model to study the evolution toward the \acrshort{HCS} of the \acrshort{KLD} with an arbitrary reference \acrshort{VDF} has been derived. In fact, it has been proved under the \acrshort{SA} theoretical scheme that the \acrshort{HCS} \acrshort{VDF} could solve the problem, as it has been conjectured by other authors~\cite{BPV13,GMMMRT15,PP17}. Moreover, this theoretical approach indicates that the Maxwellian \acrshort{VDF} is not a proper reference, in which case nonmonotonic behaviors are predicted.

\item The before theoretical analysis is supported, in Articles 2 (\refsec{Art2}) and 3 (\refsec{Art3}), by \acrshort{DSMC} and \acrshort{EDMD} simulations from a wide range of values of the coefficient of normal restitution\index{Coefficient of normal restitution}, $\een$, and different initial \acrshort{VDF}.

\item Furthermore, the proper meaning of the \acrshort{KLD} functional has indicated the deviations of the \acrshort{HCS} \acrshort{VDF} from a Maxwellian one, in Article 2 (\refsec{Art2}), revealing a nonmonotonic dependence with $\een$.

\item A velocity-inversion experiment is developed in Article 3 (\refsec{Art3}) for granular gases described by the \acrshort{IHS} model. It has revealed a strong suppression of the antikinetic stages observed in the elastic case due to the role of inelasticity at the collisional level, which is present even in the quasi-elastic case.
\end{itemize}

\subsubsection{Article 4}

\begin{itemize}
\item The \index{Nonequilibrium steady state}nonequilibrium time-dependent and steady states of a granular gas of inelastic hard \dt-spheres under nonlinear drag have been described in Article 4 (\refsec{Art4}). The theoretical analysis, under a \acrshort{SA}, reveals that this system is, at its steady state, further from equilibrium as compared with the linear drag case. This is deduced from the steady-state values of the quantities $T/T_b-1$ and $a_2$, which are, in general, further from the equilibrium values as the nonlinear control quantity, $\gamma$, increases. In fact, a critical value of the nonlinear quantity has found beyond which the steady-state \acrshort{VDF} is leptokurtic for every value of the coefficient of normal restitution\index{Coefficient of normal restitution}. Those results have been confirmed by \acrshort{DSMC} and \acrshort{EDMD} simulation results.

\item This system has shown the emergence of the Mpemba and Kovacs-like memory effects. The characterization of the evolution of the system toward the steady state from the \acrshort{SA} has allowed us to describe them. In the case of the \acrshort{KE}, we have identified the conditions related to the direction of the Kovacs humps which depends on the sign of the difference $a_2^{\mathrm{st}}-a_2(t^*_K)$. Concretely, upward humps correspond with the fulfillment of the condition $a_2(t^*_K ) < a_2^{\mathrm{st}}$ and a downward hump with $a_2(t^*_K ) > a_2^{\mathrm{st}}$. Outcomes from \acrshort{DSMC} and \acrshort{EDMD} simulations are in quite a good agreement with the theoretical predictions, supporting the findings.
\end{itemize}

\subsection{Inelastic and rough granular gases}

We have studied how the implementation of surface roughness via the \acrshort{IRHS} model affects the granular gaseous dynamics. We have developed the description of the homogeneous states for the case of free evolution in Article 5 (\refsec{Art5}) and for the driven case with the \acrshort{ST} in Article 6 (\refsec{Art6}). Moreover, the transport properties of the freely evolving granular gas and the stability of the \acrshort{HCS} have been analyzed in Articles 7 (\refsec{Art7}) and Article 8 (\refsec{Art8}), respectively.

\subsubsection{Article 5}

\begin{itemize}

\item We have characterized the non-Gaussianities of the \acrshort{HCS} of a dilute granular gas described by the \acrshort{IRHS} model via the first-order nontrivial cumulants and the \acrshort{HVT} of the main marginal \acrshort{VDF} for \acrshort{HD} and \acrshort{HS}.

\item The analysis of the cumulants has been carried out by a \acrshort{SA}. The theoretical results have shown values for \acrshort{HD} further from equilibrium than those for \acrshort{HS}, in general, and some values $a_{02}^{\HCS},\,a_{11}^{\HCS}\sim \mathcal{O}(1)$. This could indicate a possible breakdown of the \acrshort{SA}. 

\item The theoretical predictions of the cumulants for \acrshort{HD} are accurate as compared with \acrshort{DSMC} and \acrshort{EDMD} results for small values of these quantities and agree, in a more qualitative way, with the cases of larger predicted values, where the \acrshort{SA} is expected to worsen.

\item We have extended the well-known overpopulated \acrshort{HVT} of the \acrshort{HCS} \acrshort{VDF} to the rough case. The \acrshort{HVT} form of the translational marginal \acrshort{VDF} coincides with the \acrshort{IHS} case, as expected, which fulfills a exponential decay, i.e., $\phi_{\cc}^{\HCS}\sim e^{-\gamma_c c}$, but with the exponent modified by the implementation of the surface roughness of the particles. Moreover, we have made predictions, under certain approximations, of the \acrshort{HVT} for $\phi_{\ww}^{\HCS}$ and $\phi_{cw}^{\HCS}$, in which scale-free forms have been derived, i.e., $\phi_{\ww}^{\HCS}\sim w^{-\gamma_w}$ and $\phi_{cw}^{\HCS}\sim (cw)^{-\gamma_{cw}}$. The latter results indicate the existence of divergences for an infinite set of higher-order cumulants related essentially to the rotational degrees of freedom.

\item The latter scaling forms have been observed in \acrshort{DSMC} and \acrshort{EDMD} simulations for \acrshort{HD}. A quite good accurate prediction for the theoretical value of $\gamma_w$ has been observed as compared with simulations, whereas for $\gamma_c$ and $\gamma_{cw}$ the agreement is more qualitative.
\end{itemize}

\subsubsection{Article 6}

\begin{itemize}
\item The homogeneous states of a dilute granular gas of inelastic and rough \acrshort{HD} under the action of the \acrshort{ST} have been described from a kinetic-theory perspective and under a \acrshort{MA} for the time-dependent and the \index{Nonequilibrium steady state}nonequilibrium steady-state \acrshort{VDF}.

\item The steady-state value of the granular mean \index{Temperature!mean granular}temperature, $T^{\mathrm{st}}$, is proportional to the noise \index{Temperature!noise}temperature, $\Twn$, and depends, in a nonmonotonic way, on the rotational-to-total noise ratio, $\varepsilon$. On the other hand, the steady-state value of the rotational-to-translational \index{Temperature!rotational-to-translational ratio}temperature ratio, $\theta^{\mathrm{st}}$, follows a monotonic dependence on $\varepsilon$.

\item The \acrshort{DME} condition has been extended for the \acrshort{TME} (without \index{Temperature!overshoot}overshoot) of Ref.~\cite{TLLVPS19} (which was based on $\varepsilon=0$) to all possible values of $\varepsilon$. The condition of the standard form of the \acrshort{DME} reveals that a system with initially more translational energy than rotational could relax earlier to the steady state. On the contrary, a system with much more initial rotational energy than translational could evolve more slowly to the steady state as compared with the latter case. The main reason is that translational energy favors collisions which produce the cooling toward the steady-state temperature.

\item We have detailed the emergence of the \acrshort{OME} and derived its necessary condition for the initial slopes of a pair of samples in a \acrshort{ME} experiment. In addition, the \index{Temperature!overshoot}overshoot phenomenon of a thermal curve has been demonstrated to emerge for values of $\varepsilon$ greater than a critical value, $\varepsilon_{\mathrm{cr}}$, for each initial condition. The detection of the \acrshort{OME} has been guided by a thermal distance recalling the one derived in Article 1 (\refsec{Art1}).

\item From the knowledge of the dependence of \index{Nonequilibrium steady state}nonequilibrium steady states of the system on the parameters of the \acrshort{ST}, $(\Twn,\varepsilon)$, as well as the developed necessary conditions, we have elaborated a protocol to detect the \acrshort{DME} in the standard and \acrshort{OME} forms. In those protocols, using that the necessary conditions for the initial slopes depend on very disparate values of $\theta^0$, the initial conditions are elaborated from previous thermalizations of the involved samples to steady-states of the \acrshort{ST} with either $\varepsilon=0$ or $\varepsilon=1$. This is to fulfill that one sample must posses much more translational than rotational energy, and the other way around for the other sample. 

\item Finally, \acrshort{DSMC} and \acrshort{EDMD} simulation results confirm the theoretical approaches in the transient and steady states of the system, as well as the protocol designed to observe the \acrshort{ME} in this setup.
\end{itemize}

\subsubsection{Articles 7 and 8}

\begin{itemize}
\item In Article 7 (\refsec{Art7}), the transport properties of a freely evolving monodisperse and dilute granular gas of inelastic and rough \acrshort{HD} and \acrshort{HS} have been computed in a common mathematical framework, by means of the \acrshort{NSF} transport coefficients from the \index{Chapman--Enskog method}\acrshort{CE} method and using a \acrshort{MA} for $f^{(0)}$ (the local version of the \acrshort{HCS} \acrshort{VDF}) and a Sonine-like approximation for $f^{(1)}$. 

\item Explicit expressions of the (reduced) transport coefficients for this model of granular gas have been obtained: the (reduced) shear viscosity $\eta^*$, the (reduced) bulk viscosity $\eta^*_b$, the (reduced) thermal conductivity $\lambda^*$, the (reduced) Dufour-like coefficient $\mu^*$, and the velocity-divergence transport coefficient $\xi$, all of them as functions of the coefficients of restitution, $(\een,\eet)$, the reduced moment of inertia, $\kappa$, and the translational and rotational degrees of freedom, $(\dt,\dr)$.

\item From the expressions of the transport coefficients, we have recovered, in Article 7 (\refsec{Art7}), already known results for purely smooth particles ($\eet=-1$, $\dr\to 0$), as well as novel ones in terms of the degrees of freedom for the quasismooth ($\eet\to-1$) and Pidduck's ($\een=\eet=1$) limits.

\item From the knowledge of the transport coefficients, in Article 8 (\refsec{Art8}), we have analyzed the stability of the \acrshort{HCS} under linear nonhomogeneous perturbations, in which we have recovered known results for \acrshort{HS}, and obtained novel results for \acrshort{HD}. We have computed the critical wave numbers associated with cluster and vortex instabilities, as well as the critical length of the system for instabilities, $L_c$.

\item The theoretical linear stability analysis predicts a region of the parameter space where the \acrshort{HCS} is always unstable due to clustering, referred to as \acrshort{HUR} in this thesis, which would imply a breakdown of the hydrodynamics of the granular gaseous system. However, \acrshort{EDMD} has revealed that there exist stable conditions in contradiction with this strong prediction, although a good agreement is observed outside the \acrshort{HUR}. This could be signaling that, whereas this region seems to be highly unstable, it might not be absolutely unstable. The theoretical strong prediction could fail due to the breakdown of the approximations. This is probably linked to the strong non-Gaussianities of the \acrshort{HCS} \acrshort{VDF} studied in Article 5 (\refsec{Art5}).

\item Finally, the clustering detection in the \acrshort{EDMD} simulations has been addressed via the application of the \acrshort{KLD} of the spatial particle distribution with respect to a homogeneous Poisson distribution as reference. This has been complemented by the control of some non-Gaussianities of the \acrshort{HCS} \acrshort{VDF} computed in Article 5 (\refsec{Art5}). 
\end{itemize}

\section{Outlooks}

In this thesis, different theoretical tools have been developed, accompanied by the implementation of the corresponding simulation methods to check whether the approximations and predictions were reasonable or not. We hope that the knowledge created in the works appearing in this thesis could encourage future works, not only from the most personal point of view, but also, in general, in the granular gas community.

One of the objectives of the thesis has been the study of driven states in granular gases of inelastic and rough particles under the \acrshort{ST}. However, the non-Gaussian properties of the \emph{steady} \acrshort{VDF} have been restricted to the freely evolving case. We are planning to study in deep those properties for the \acrshort{ST}, which absorbs the limiting cases of purely translational ($\varepsilon=0$) or rotational ($\varepsilon=1$) drivings. It would be interesting to determine whether the \acrshort{MA} used in Article 6 (\refsec{Art6}) is a good approach at the steady state for every pair of values of the coefficients of restitution $(\een,\eet)$, as well as for different values of the reduced moment of inertia $\kappa$, as computer simulations in Article 6 (\refsec{Art6}) indicate. The first nontrivial cumulants and the \acrshort{HVT} of the marginals \acrshort{VDF} at the steady state would be again a good measure for that. Moreover, once the homogeneous steady state is described, it would be important the determination of the transport properties by means of the \acrshort{NSF} transport coefficients, as well as checking whether the action of the driving associated with the \acrshort{ST} erases, or not, any sort of instability, as occurs with stochastic thermostats acting on granular gases of inelastic particles~\cite{KG18,KG19,G19}.

Furthermore, a connection between the nonlinear drag analysis and the driving by means of the \acrshort{ST} would be the consideration of nonlinear driving to the translational and rotational degrees of freedom. An extension of Refs.~\cite{CLH00,HCZHL05} with the tools developed in this thesis would be quite an interesting research topic.

Moreover, the results from \refsec{Art8} for the \acrshort{HUR} in a freely cooling granular gas of inelastic and rough \acrshort{HD}, indicates that the approach to the transport properties of this system should be improved. Some methodologies could be, firstly, a modification due to the effect of the Sonine coefficients of the \acrshort{HCS} \acrshort{VDF}, as done in Ref.~\cite{GSM07} for the \acrshort{IHS} model, or the introduction of the predicted \acrshort{HVT} into the computations in a proper way. The investigation of this issue could be complemented by the determination of the \acrshort{NSF} transport coefficients by means of computer simulations. In addition, the application of this knowledge to the rheological properties of a granular gas of rough particles could lead to interesting results, as already studied in the case of inertial suspensions in Ref.~\cite{GG20}.

Whereas the models studied in this thesis seem to describe properly the granular gas dynamics, being the main predicted (translational) properties of the freely evolving and driven systems observed in experiments~\cite{MIMA08,TMHS09,HTWS18,YSS20,GF23}, the hard \dt-sphere and the constant coefficient of normal restitution are, at the end, approaches and simplifications. In fact, in Ref.~\cite{YSS20} is observed an excess of cooling as compared with the \acrshort{IHS}, that could be explained by introducing the role of the rotational degrees of freedom, which has been already commented throughout his memoir. Nevertheless, this latter implementation, might not explain all the discrepancies as commented in Ref.~\cite{PSAS21} for a model for rough particles~\cite{HHZ00} different from the \acrshort{IRHS} one studied in this thesis. Even more, the consideration for $\eet$ to be constant is also an approximation. The effective coefficient of tangential restitution could depend on the impact angle because of friction~\cite{MBF76,LTL97}. Moreover, the theory usually characterizes systems of infinitely many particles without any boundary interaction. But, particle-wall collisions could explain the observed excess of cooling too. All these issues, as well as past and future experimental observation should guide our future theoretical modeling of these systems toward more realistic theories, with the attempt of explaining, in more detail, the real behavior of granular gases.

Another compelling topic, which would extend the present studies, is the application of all these techniques to polydisperse systems. Moreover, in a polydisperse system, its properties would depend on the set of coefficients of restitution, generally different, $\{\een_{ij},\eet_{ij}\}$, the concentrations of the different species $\{x_i\equiv n_i/\sum_k n_k\}$, and the mechanical parameters of the mixture, that is, the masses, $\{m_i\}$, diameters, $\{\sigma_i\}$, and reduced moments of inertia, $\{\kappa_i\}$. Therefore, from this huge number of parameters, we could look for very interesting phenomena. The expressions of the energy production rates have been computed via \acrshort{MA} in previous works within our degrees-of-freedom context~\cite{MS19,MS19b} as a generalization of the results of Ref.~\cite{SKG10}. Moreover, during my visiting period at the \href{https://www.pmmh.espci.fr/?-Home-}{PMMH of the ESPCI} in Paris (France), with Prof. Hans J. Herrmann, we started an analysis of diffusion phenomena in granular gases of inelastic and rough particles, in which we have already computed the diffusion coefficients associated with a binary system described by a \acrshort{IRHS} model, and we are studying their influence into the segregation mechanisms. We expect to publish soon the results, which would generalize previous works in the inelastic but smooth case~\cite{JY02,BRM05,G06a,GV10,G11,BKD11,BKD12,GMV13,VGK14,G19} and would reveal novel phenomena due to the role of roughness into the collision dynamics. Another interesting  effect appearing in polydisperse granular systems is a situation in which intruders, added to a host gas, mimic the thermal behavior of the latter. In fact, during the predoctoral period, some computations about the so-called \emph{mimicry effect}~\cite{S18,LVGS19} in this type of systems, initiated at Ref.~\cite{MS19}, have been carried out and it would be desirable to continue those in the future.

In addition, we have worked in this thesis under the assumptions of instantaneous collisions in granular interactions. However, this could lead to problems, such as the inelastic collapse, as explained, for example, in Ref.~\cite{LM98}. Therefore, the consideration of a finite-time duration of a collision, due to the particle softness, could solve this issue, implying corrections to the transport properties of the granular gases~\cite{ST20,T21,THH22}. In fact, during my stay at \href{https://www.skoltech.ru/en/}{Skolkovo Institute of Science and Technology} in Moscow (Russia), with Prof. Nikolai V. Brilliantov, we started to study how this consideration affects the self-diffusion properties of a granular gas of inelastic \acrshort{HS}. We expect to publish soon the results of this work and extend the analysis to the rest of transport properties.

Finally, it would be interesting to study the homogeneous and transport properties of active matter within the tools described in this thesis. In fact, an interesting topic would be the study of memory effects from a kinetic-theory point of view, such as the \acrshort{ME} or \acrshort{KE}, which have already been of interest in the statistical physics community~\cite{KSI17,SL22}, but from other approaches. Furthermore, an analysis of the transport properties could lead to interesting collective phenomena in active systems.

\appendix 
\addpart{Appendices}
\chapter{Fokker--Planck equation associated with a Langevin-like equation}
\labapp{app_FPLE}

Paul Langevin derived his famous equation in 1908~\cite{L08,LG97}, with the aim to understand the Brownian motion. The \acrshort{LE} was derived after the pioneering Eintein's~\cite{E05} and Smoluchowski's~\cite{S06} works, in 1905 and 1906, respectively, which had already put into manifest the relevance of the action of atomic motion in fluids. On the other hand, the \acrshort{FPE} is named after Adriaan Fokker and Max Planck, who independently derived this equation in 1914~\cite{F14a} and 1917~\cite{P17}, respectively. After its derivation in different contexts, the first application of this equation was, again, to describe statistically the Brownian motion~\cite{QG13}. These two equations, the \acrshort{LE} and the \acrshort{FPE}, became fundamental in the study of diffusion processes.

The aim of this appendix is to summarize the relationship between a Langevin-like equation and a \acrshort{FPE} and finally specify this for the thermostatted states used in this thesis (see \refsubsec{nonlinear_KTGG} and \refsubsec{splitting_KTGG}). Whereas the former is a \acrshort{SDE}, usually written in terms of the velocity, $\vvel$, of a particle, the latter is a partial differential equation on the probability density, i.e., the one-body \acrshort{VDF} in velocity terms. In general, there is not always a relationship between an arbitrary \acrshort{SDE} and a \acrshort{FPE}, but we will work under the proper circumstances where this is guaranteed. 

\section{General relations}

Let us work in a \dt-dimensional velocity space and let us consider a general nonlinear Langevin-like equation, which in differential form reads
\begin{equation}\labeq{general_form_LE}
	\dif \vvel = \mathbf{a}(\vvel)\dif t+\mathsf{b}(\vvel)\cdot\dif\mathbf{W}_t,
\end{equation}
where $\mathbf{W}_t$ is a Wiener process~\cite{R12,R19}, $\mathbf{a}$ is a vector function, and $\mathsf{b}$ is a matrix function. Then, the latter \acrshort{SDE} can be written as~\cite{R12,R19}
\begin{equation}\labeq{app_LE}
	\dot{\vvel} = \mathbf{a}(\vvel)+\mathsf{b}(\vvel)\cdot\bar{\boldsymbol{\eta}}(t),
\end{equation}
with the random vector $\bar{\boldsymbol{\eta}}(t)$ following the properties of a white noise with unit variance,
\begin{equation}\labeq{WN_LE_app}
	\langle \bar{\boldsymbol{\eta}}(t)\rangle =\mathbf{0}, \quad \langle \bar{\boldsymbol{\eta}}(t)\bar{\boldsymbol{\eta}}(t^{\prime})\rangle =\mathbb{1}_{\dt}\delta(t-t^\prime).
\end{equation}

On the other hand, a \acrshort{FPE} is a second order partial differential equation on the \acrshort{VDF}, which reads
\begin{equation}\labeq{app_FPE}
	\frac{\partial f(\vvel,t)}{\partial t} +\frac{\partial}{\partial \vvel}\cdot\left[\mathbf{D}^{(1)}(\vvel,t)f(\vvel,t)\right] - \left(\frac{\partial}{\partial \vvel}\frac{\partial}{\partial \vvel}\right):\left[\mathsf{D}^{(2)}(\vvel,t) f(\vvel,t)\right]=0,
\end{equation}
where $\mathbf{D}^{(1)}$ and $\mathsf{D}^{(2)}$ are usually called the \emph{drift vector} and the \emph{diffusion matrix}, respectively. 

In other words, the \acrshort{FPE} is the truncation of the Kramers--Moyal~\cite{vK07,R12,R19} expansion of the \acrshort{VDF} up to second order terms (higher order truncations are forbidden by Pawula theorem~\cite{P67,R19}). Then, the link between 
both equations become from the identification of the drift vector and the diffusion matrix with the two first coefficients of such expansion. The result depends on the interpretation in which \refeq{app_LE} is solved. The drift and diffusion coefficients associated with \refeq{app_LE} read \cite{vK07,VHBWB10,MM12,R12,R19}
\begin{subequations}
\begin{align}
	D_i^{(1)} \equiv& \left.\lim_{h\to 0}\frac{\langle v_i(t+h)-v_i(t)\rangle}{h} \right|_{\vvel(t)=\vvel}= a_i(\vvel)+\epsilon\sum_{j=1}^{\dt}\sum_{k=1}^{\dt}b_{kj}(\vvel)\frac{\partial}{\partial v_k}b_{ij}(\vvel), \quad i=1,\dots,\dt \\
	D_{ij}^{(2)} \equiv& \left.\lim_{h\to 0}\frac{\langle [v_i(t+h)-v_i(t)][v_j(t+h)-v_j(t)]\rangle}{2h} \right|_{\vvel(t)=\vvel}= \frac{1}{2}\sum_{k=1}^{\dt}b_{ik}(\vvel)b_{jk}(\vvel),\nonumber \\
	& i,j=1,\dots,\dt,
\end{align}
\end{subequations}
for the It\^o ($\epsilon=0$)~\cite{vK07}, Stratonovich ($\epsilon=1/2$)~\cite{vK07}, and Klimontovich ($\epsilon=1$)~\cite{K94} interpretations. Moreover, from It\^o's lemma~\cite{R19}, $D_{i_1,\dots,i_n}^{(n)}=0$, $\forall i_1,\dots,i_n$ and $\forall n\geq 3$. Then, the relationship between the nonlinear \acrshort{LE} defined by \refeq{app_LE} and the \acrshort{FPE} is guaranteed, and higher-order terms for the Kramers--Moyal expansion do not apply.

In particular, the original \acrshort{LE} coincides with the case in which \index{Temperature!of the thermal bath}$\mathbf{a}(\vvel)\equiv -\xi_0 \vvel$, and $\mathsf{b}(\vvel)\equiv \sqrt{2\xi_0 T_b/m} \mathbb{1}_{\dt}$, in which the \acrshort{FDT} has been already assumed [see \refeq{white_noise}]. Then, the latter three interpretations give the same \acrshort{FPE}, reading\index{Temperature!of the thermal bath}
\begin{equation}
	\frac{\partial f(\vvel,t)}{\partial t} -\frac{\partial}{\partial \vvel}\cdot\left(\xi_0\vvel +\frac{\xi_0 T_b}{m}\frac{\partial}{\partial \vvel}\right)f(\vvel,t)=0.
\end{equation}
Let us particularize for the two thermostats considered in this thesis.

\section{Langevin dynamics with a nonlinear drag}\labsec{LD_NLD_FP_app}

Let us work under the conditions of \refsubsec{nonlinear_KTGG}. Here, in the free streaming, the (Brownian) particles of the dilute gas are considered to follow a nonlinear \acrshort{LE} with a velocity-dependent quadratic drag coefficient,\index{Temperature!of the thermal bath}
\begin{equation}\labeq{LE_NLD}
	\dot{\vvel} = -\xi(v)\vvel +\chi(v)\bar{\boldsymbol{\eta}}(t), \quad \xi(v) = \xi_0 \left(1+\gamma\frac{mv^2}{T_b} \right),
\end{equation}
such that $\bar{\boldsymbol{\eta}}(t)$ corresponds to a random variable, describing a white noise with unit variance following \refeq{WN_LE_app}. Comparing \refeq{LE_NLD} with \refeq{app_LE}, one can identify that $\mathbf{a}(\vvel)=-\xi(v)\vvel$, and $\mathsf{b}(\vvel) = \chi(v)\mathbb{1}_{\dt}$. Then, the drift vector and diffusion matrix read
\begin{equation}\labeq{FP_NLD_coeffs}
	\mathbf{D}^{(1)} = -\left[\xi(v)-\frac{\epsilon}{v}\chi(v)\chi^{\prime}(v) \right]\vvel, \quad \mathsf{D}^{(2)} =  \frac{\chi^2(v)}{2}\mathbb{1}_{\dt}.
\end{equation}
From this, we infer that the \acrshort{FPE} depends on the interpretation of the Brownian process, and it reads
\begin{equation}
	\frac{\partial f}{\partial t} - \frac{\partial}{\partial\vvel}\cdot\left[\xi(v)\vvel +\frac{\chi^{2\epsilon}(v)}{2}\frac{\partial}{\partial \vvel}\chi^{2(\epsilon-1)}(v) \right]f =0.
\end{equation}
Thus, the (differential) fluctuation-dissipation relation that can be deduced from the latter \acrshort{FPE} turns out to be\index{Temperature!of the thermal bath}
\begin{equation}\labeq{FDT_NLD}
	\xi(v) +\frac{1-\epsilon}{2v}\frac{\partial \chi^2(v)}{\partial v} = \frac{m\chi^2(v)}{2T_b}.
\end{equation}
Hence, only for $\epsilon=1$, i.e., the Klimontovich interpretation, one recovers the original fluctuation-dissipation relation. Therefore, if we impose that relation to be held, the resulting \acrshort{FPE} is 
\begin{equation}
	\frac{\partial f}{\partial t} - \frac{\partial}{\partial\vvel}\cdot\left[\xi(v)\vvel +\frac{\chi^{2}(v)}{2}\frac{\partial}{\partial \vvel}\right]f =0.
\end{equation}
It can be proved from \refeqs{FP_NLD_coeffs} and~\eqref{eq:FDT_NLD} that the latter \acrshort{FPE} is recovered from the following nonlinear \acrshort{LE} in the It\^o interpretation,\index{Temperature!of the thermal bath}
\begin{equation}\labeq{eff_LE_Ito}
	\dot{\vvel} = -\xi_{\mathrm{eff}}(v)\vvel+\chi^2(v)\bar{\boldsymbol{\eta}}(t), \quad \xi_{\mathrm{eff}}(v)\equiv \xi(v)-2\xi_0\gamma = \xi_0\left(1-2\gamma+\gamma\frac{mv^2}{T_b} \right),
\end{equation}
where we have used the explicit form of the nonlinear drag coefficient written in \refeq{LE_NLD}. The latter nonlinear \acrshort{LE}, \refeq{eff_LE_Ito}, will be helpful in the numerical resolution of this dynamics in the It\^o interpretation (see \refapp{app_AGF}).

\section{Splitting thermostat}\labsec{ST_FP_app}

In the case of the thermostat introduced in \refsubsec{splitting_KTGG} and, according to the stochastic force and torque acting on the system [see \refeqs{ext_stoc_force} and~\eqref{eq:ext_stoc_torque}, respectively], we obtain the following two Langevin-like equations
\begin{equation}\labeq{ST_LE_first}
	\dot{\vvel} = \chi_t \bar{\boldsymbol{\eta}}_{\dt}(t), \quad \dot{\oo} = \chi_r \bar{\boldsymbol{\eta}}_{\dr}(t),
\end{equation}
with $\bar{\boldsymbol{\eta}}_{\dt}$ and $\bar{\boldsymbol{\eta}}_{\dr}$ being \dt- and \dr-dimensional random vectors describing a white noise with unit variance such that, from \refeq{WN_LE_app}, it can be interpreted that
\begin{subequations}\labeq{etas_ST_FP_app}
\begin{align}
	\langle \bar{\boldsymbol{\eta}}_{\dt}(t)\rangle =& \mathbf{0}, \quad \langle \bar{\boldsymbol{\eta}}_{\dt}(t)\bar{\boldsymbol{\eta}}_{\dt}(t^\prime)\rangle = \mathbb{1}_{\dt}\delta(t-t^\prime), \\
	\langle \bar{\boldsymbol{\eta}}_{\dr}(t)\rangle =& \mathbf{0}, \quad \langle \bar{\boldsymbol{\eta}}_{\dr}(t)\bar{\boldsymbol{\eta}}_{\dr}(t^\prime)\rangle = \mathbb{1}_{\dr}\delta(t-t^\prime).
\end{align}
\end{subequations}
In a more compact way, working with the variable $\vom\in\mathbb{R}^{\dt}\times\mathbb{R}^{\dr}$, \refeq{ST_LE_first} can be expressed as
\begin{equation}
	\dot{\vom} = \mathsf{X}\cdot\bar{\boldsymbol{\eta}}(t), \quad \mathsf{X}\equiv \left(
\begin{array}{cc}
\chi_t \mathbb{1}_{\dt} & \mathbf{0}\\
\mathbf{0} & \chi_r \mathbb{1}_{\dr}
\end{array} 
\right), \quad \bar{\boldsymbol{\eta}} \equiv \left(
\begin{array}{c}
\bar{\boldsymbol{\eta}}_{\dt}\\
\bar{\boldsymbol{\eta}}_{\dr}
\end{array} 
\right).
\end{equation}
Therefore, as it occurs with the original \acrshort{LE}, the associated \acrshort{FPE} coincides for It\^o, Stratonovich, and Klimontovich interpretations. The drift vector and diffusion matrix, for this case, are
\begin{equation}
	\mathbf{D}^{(1)} = \mathbf{0}, \quad \mathsf{D}^{(2)} =\frac{1}{2}\mathsf{X}\mathsf{X}^{\top}=   \frac{1}{2}\left(
\begin{array}{cc}
\chi_t^2 \mathbb{1}_{\dt} & \mathbf{0}\\
\mathbf{0} & \chi_r^2 \mathbb{1}_{\dr}
\end{array} 
\right).
\end{equation}
Then, the \acrshort{FPE} is straightforwardly derived, reading
\begin{equation}
	\frac{\partial f}{\partial t}-\frac{\chi_t^2}{2}\left(\frac{\partial}{\partial \vvel}\right)^2f-\frac{\chi_r^2}{2}\left(\frac{\partial}{\partial \oo}\right)^2 f=0.
\end{equation}
Note that this coincides with the Fokker--Planck-related part of the \acrshort{BFPE} in \refeq{FPE_ST_first}, acting on the free-streaming contribution. 
\chapter{Derivation of the nonlinear drag coefficient for a quasi-Rayleigh gas}
\labapp{app_NLD}

Let us work under the conditions introduced in \refsubsec{nonlinear_KTGG}. That is, we consider a \acrshort{QR} gas of Brownian heavy particles with masses $m$, number density $n$, and velocities $\vvel$, in contact with a dilute gas of lighter particles with masses $m_b$, number density $n_b$, and velocities $\vvel_b$ in equilibrium at \index{Temperature!of the thermal bath}temperature $T_b$. The ensemble corresponding to the dilute gas of light particles is assumed to be acting as a thermal bath on the heavy ones. The aim of this section is to obtain formally an expression for the drag coefficient appearing in \refeq{mom_drag} from a power expansion in terms of $m_b/m$. For that, we will follow the derivations developed already in Refs.~\cite{F00b,F07,F14} for \acrshort{HS}, but generalized to hard \dt-spheres.

We know that the average change of momentum of a Brownian particle due to a collision with a bath particle\footnote{Note that primed quantities refer to postcollisional values.} is $m(\vvel^\prime-\vvel)/\tau$, such that $\tau$ is the mean free time, i.e., \phantomsection\label{sym:Qbar}$\tau=\left(n_b \overline{Q}(g) g\right)^{-1}$, where $\mathbf{g}=\vvel_b-\vvel$ is the relative velocity and $\overline{Q}(g)=\int_{\overline{\Omega}_{\dt}} \dif\overline{\Omega}_{\dt} B(g,\overline{\vartheta}_1)$ is the total cross section, $B(g,\overline{\vartheta}_1)$ being the differential cross section and $\overline{\vartheta}_1$ being the angle between $\mathbf{g}^\prime$ and $\mathbf{g}$. We will use that the $\dt$-differential solid angle is~\cite{B60}
\begin{equation}
    \dif\Omega_{\dt} = \prod_{j=1}^{\dt-1}\sin^{\dt-j-1}{\vartheta_j}\dif\vartheta_j,
\end{equation}
where $0\leq\vartheta_{j}\leq \pi$ for $j=1,\ldots,\dt-2$ and $0\leq\vartheta_{\dt-1}\leq 2\pi$. Its integration gives the total solid angle,
\begin{equation}\labeq{solid_angle_dt}
    \Omega_{\dt}=\frac{2\pi^{\frac{\dt}{2}}}{\Gamma\left(\frac{\dt}{2}\right)}.
\end{equation}
Therefore, the average change of momentum of a Brownian particle due to a collision with a background fluid particle per unit time is given by
\begin{align}\labeq{mom_drag_kin}
     m\left(\frac{\dif \vvel}{\dif t}\right)_{\text{Brown}} =& \Omega^2_{\dt-1} \pi^{-\dt/2}n_b \vvel \frac{mm_b}{m+m_b}v_{\mathrm{th},b}^{-\dt} \int_0^\infty \dif v_b v_b^{\dt-1} e^{-\frac{v_b^2}{v_{\mathrm{th},b}^2}}\int_0^\pi \dif \vartheta_1 \sin^{\dt-2}{\vartheta_1} \nonumber \\
     &\times g\left(\frac{v_b}{v}\cos{\vartheta_1}-1 \right)\int_0^\pi \dif\overline{\vartheta}_1 \sin^{\dt-2}{\overline{\vartheta}_1}(1-\cos\overline{\vartheta}_1)B(g,\overline{\vartheta}_1), 
\end{align}
where we have used the collisional rule from the \acrshort{EHS} model for two particles of different masses, $m$ and $m_b$, for the gas and bath particles, respectively,
\begin{equation}
    \vvel^\prime-\vvel = -\frac{m_b}{m+m_b}(\mathbf{g}^\prime-\mathbf{g}),    
\end{equation}
that the background fluid particles are at equilibrium,
\begin{equation}\labeq{equilibrium_VDF_fb}
    f_b(\vvel_b) \to f_{\mathrm{M}}(\vvel_b) = \pi^{-\dt/2}v^{-\dt}_{\mathrm{th},b} e^{-\frac{v_b^2}{v^2_{\mathrm{th},b}}},
\end{equation}
and $g^\prime=g$ due to the elastic collision between gas and background fluid particles, i.e., no relative velocity magnitude is lost in the collision. Moreover, from the definition of the relative velocity, its modulus is given by the expression
\begin{equation}
    g= v_b\sqrt{1-2\frac{v}{v_b}\cos\vartheta_1+\left(\frac{v}{v_b}\right)^2},
\end{equation}
so that $\vartheta_1=\arccos{(|\vvel_b\cdot\vvel|/v_b v)}$. From the \acrshort{QR} gas assumption, one can expand the product  $g B(g,\overline{\vartheta}_1)$ as a MacLaurin series for $v/v_b$ or, equivalently $\sqrt{m_b/m}$, which up to the term of order $(m_b/m)^{3/2}$ is
\begin{align}\labeq{Taylor_gB}
    g B(g,\overline{\vartheta}_1) =& v_b\left\{B\left(c_b,\overline{\vartheta}_1\right)-\left(\frac{m_b}{m}\right)^{1/2}\frac{c}{c_b}\cos{\vartheta_1}\frac{\partial}{\partial c_b}\left[c_b B(c_b,\overline{\vartheta}_1) \right] +\frac{1}{2}\frac{m_b}{m}\left(\frac{c}{c_b}\right)^2\right.\nonumber \\
    &\times\left.\left[\sin^2\vartheta_1B(c_b,\overline{\vartheta}_1)+c_b\left( 1+\cos^2{\vartheta_1}\right)\frac{\partial B(c_b,\overline{\vartheta}_1)}{\partial c_b} +c_b^2\cos^2{\vartheta_1}\frac{\partial^2 B(c_b,\overline{\vartheta}_1)}{\partial c_b^2}\right]\right.\nonumber\\
    &\left.+\frac{1}{2}\left(\frac{m_b}{m}\right)^{3/2}\left(\frac{c}{c_b}\right)^3\left[\sin^2{\vartheta_1}\cos{\vartheta_1}\left(B(c_b,\overline{\vartheta}_1)-c_b\frac{\partial B(c_b,\overline{\vartheta}_1)}{\partial c_b} \right)-c_b^2\cos\vartheta_1\right.\right.\nonumber \\
    &\left.\left.\times\frac{\partial^2 B(c_B,\overline{\vartheta}_1)}{\partial c_b^2} -\frac{1}{3}c_b^3\cos^2\vartheta_1\frac{\partial^3 B(c_B,\overline{\vartheta}_1)}{\partial c_b^3}\right]\right\}+\mathcal{O}\left[ \left(\frac{m_b}{m}\right)^2\right].
\end{align}
Inserting \refeq{Taylor_gB} into \refeq{mom_drag_kin}, and comparing with \refeq{mom_drag}, we obtain for hard $\dt$-hyperspheres, whose differential cross-section is independent of the velocity and the angle, $B(c,\overline{\vartheta}_1)=[(\sigma+\sigma_b)/4]^{\dt-1}$, that\index{Temperature!of the thermal bath}
\begin{equation}\labeq{xi_QR}
    \xi(v) \approx \xi_0\left[1-\frac{m v^2}{T_b} \frac{m_b}{2m}\left(1-\frac{\xi_1}{\xi_0}\right) \right],
\end{equation}
with
\begin{align}\labeq{xi_QR_orders_NLD}
    \xi_0 = \frac{\Omega_{\dt}^2}{2^{\dt-1}\pi^{\dt/2}\dt}\Gamma\left( \frac{\dt+3}{2}\right)n_b\left( \frac{\sigma+\sigma_b}{2}\right)^{\dt-1} v_{\mathrm{th},b}\frac{m_b}{m+m_b}, \quad \xi_1 = \xi_0\frac{\dt+3}{\dt+2}.
\end{align}
Thus, \refeq{xi_QR} can expressed as\index{Temperature!of the thermal bath}
\begin{equation}\labeq{xi_QR_gamma_NLD}
    \xi(v) = \xi_0\left( 1+\gamma\frac{mv^2}{ T_b}\right),
\end{equation}
with
\begin{equation}\labeq{gamma_NLD}
    \gamma=\frac{m_b}{2(\dt+2)m},
\end{equation}
which is expected to be small. The quadratic expression in \refeq{xi_QR_gamma_NLD} for the nonlinear drag coefficient, $\xi(v)$, has been used throughout this thesis, specifically in the development of Article 1 (\refsec{Art1}) and Article 4 (\refsec{Art4}).

\chapter{Tests and numerical schemes for the thermostatted states}
\labapp{app_AGF}

In \refsubsec{LD_EDMD} we introduced the numerical scheme associated with the \acrshort{AGF} algorithm proposed in Refs.~\cite{SVdM07,S12} for the usual \acrshort{LE}. That is, given the \acrshort{LE} for the velocities, positions follow a \acrshort{SDE} coming from a second integration of the \acrshort{LE}. The solutions to these equations are fundamental to reproduce the free-streaming evolution of particles in the absence of collisions. In the case of homogeneous \acrshort{DSMC} simulations, only a first integration of the \acrshort{LE} is enough. However, for the \acrshort{AGF} scheme in \acrshort{EDMD} simulations, the equation for the particle positions becomes fundamental.

The aim of this chapter is two-fold. First, results of the implementation of the \acrshort{AGF} algorithm in the \acrshort{EDMD} computer simulations for some test systems are shown. Then, the numerical schemes used for $N$-particle ensembles of the thermostatted systems introduced in this thesis (see \refsec{therm_models_KTGG}) are exposed.

\section{Test results}

In order to test the implementation of the \acrshort{AGF} algorithm in an \acrshort{EDMD} custom-made program~\cite{M23_github}, two homogeneous granular gaseous systems made of a monodisperse set of inelastic \acrshort{HD} are proposed, whose results have been previously studied in the literature. The first one consists in a model of a granular suspension in a three-dimensional setup, where the \acrshort{IHS} are in contact with an interstitial fluid, acting as a thermal bath at $T_b$. The interaction between the granular gas and the bath particles is described by the \acrshort{LE}. In the second system, the granular gas is driven by a stochastic thermostat.

\subsection{Granular gas of inelastic HD with Langevin dynamics}

Here, the numerical scheme for the solutions of the \acrshort{LE} are the same as the ones derived in \refsubsec{LD_EDMD}. The collisional part of the \acrshort{EDMD} algorithm is described by the \acrshort{IHS} model (see \refsubsec{IHS_KTGG}) in two dimensions. Previous works in Refs.~\cite{CVG12,CVG13} studied this homogeneous granular gaseous system with a linear drag force, $\mathbf{F}_{\mathrm{drag}}=-m\xi_0 \vvel$, plus a stochastic force with variance $2m\xi_0 T_b$ (assuming the fulfillment of the \acrshort{FDT}). Note that this is a particular case of the system analyzed in Article 4 (\refsec{Art4}), concretely, in the linear drag limit ($\gamma\to 0$).

We have computed 14 runs of a system of $10^4$ \acrshort{HD} with $\xi_0/\nu_b\approx 0.89$ and $n\sigma^2=10^{-3}$, for different values of the coefficient of normal restitution\index{Coefficient of normal restitution}, and for a time step $\Delta t\approx 0.02/\nu_b$. The initial velocities follow a Maxwellian distribution, 6 of the runs at \index{Temperature!of the thermal bath}temperature $T(0)=T_b/2$, and the other 6 at $T(0)=2T_b$, for each value of the coefficient of normal restitution\index{Coefficient of normal restitution}. In \reffig{Langevin_IHS_st}, the steady-state values $T^{\mathrm{st}}/T_b$ and $a_2^{\mathrm{st}}$ are compared with the theoretical results approximated by a \acrshort{SA}, consisting in a truncation of the \acrshort{VDF} up to the first nontrivial Sonine coefficient. Then, the theoretical values under this approach are given by the numerical solution of the following system of equations [see \cite{CVG12} and the linear drag limit in Article 4 (\refsec{Art4})]
\begin{subequations}\labeq{Langevin_st_SA}
\begin{align}
	\frac{T_b}{T^{\mathrm{st}}}-1 =& \frac{\mu_2^{(0)}+\mu_2^{(1)}a_2^{\mathrm{st}}}{\dt\xi_0^{*}}\sqrt{\frac{T^{\mathrm{st}}}{T_b}}, \\ a_2^{\mathrm{st}}=&\frac{\mu_4^{(0)}-(\dt+2)\mu_2^{(0)}}{\mu_2^{(0)}-(\dt+2)\left[\mu_2^{(0)}+\mu_2^{(1)}-\dt\xi_0^{*}(T^{\mathrm{st}}/T_b)^{-3/2} \right]},
\end{align}
\end{subequations}
with the coefficients
\begin{subequations}
\begin{align}
	\mu_2^{(0)} =& 1-\een^2, \quad \mu_4^{(0)} = \left(\dt+\frac{3}{2}+\een^2 \right)\mu_2^{(0)},\\
	\mu_2^{(1)}=&\frac{3}{16}\mu_2^{(0)}, \quad \mu_4^{(1)}=\frac{3}{32}\left( 10\dt+39+10\een^2\right)\mu_2^{(0)}+(\dt-1)(1+\een),
\end{align}
\end{subequations}
coming from the linearization of the collisional moments~\cite{GS95,vNE98,MS00,BP04,BP06,BP06b,SM09,G19,MS22c}, introduced also in \refsec{Art4},
\begin{align}
\mu_2 \approx \mu_2^{(0)}+a_2\mu_2^{(1)}, \quad \mu_4 \approx \mu_4^{(0)}+a_2\mu_4^{(1)}.
\end{align}
Furthermore, the thermal evolution is compared, in \reffig{Langevin_IHS_ev}, with the \acrshort{MA} of its evolution equation, which reads~\cite{CVG12,CVG13,MS22c}
\begin{equation}\labeq{T_Langevin_ev}
	\frac{\partial (T/T_b)}{\partial t^*} = 2\left(1-\frac{T}{T_b}\right)\xi_0^{*}-\frac{2\mu_2^{(0)}}{\dt}\left(\frac{T}{T_b}\right)^{3/2},
\end{equation}
with $\xi_0^{*}\equiv \xi_0/\nu_b$, and $t^*=t\nu_b$.

The simulation steady-state values in \reffig{Langevin_IHS_st} come from the average over the 14 simulation runs and the last 50 points, for which the curve is ensured to be stationary. On the other hand, the evolution curves in \reffig{Langevin_IHS_ev} are the result of the average over the 6 different runs, according to the different initial conditions. We can conclude that the simulation results are in very good agreement with the theoretical predictions.
\begin{figure}[h!]
 \centering
 \includegraphics[width=0.47\textwidth]{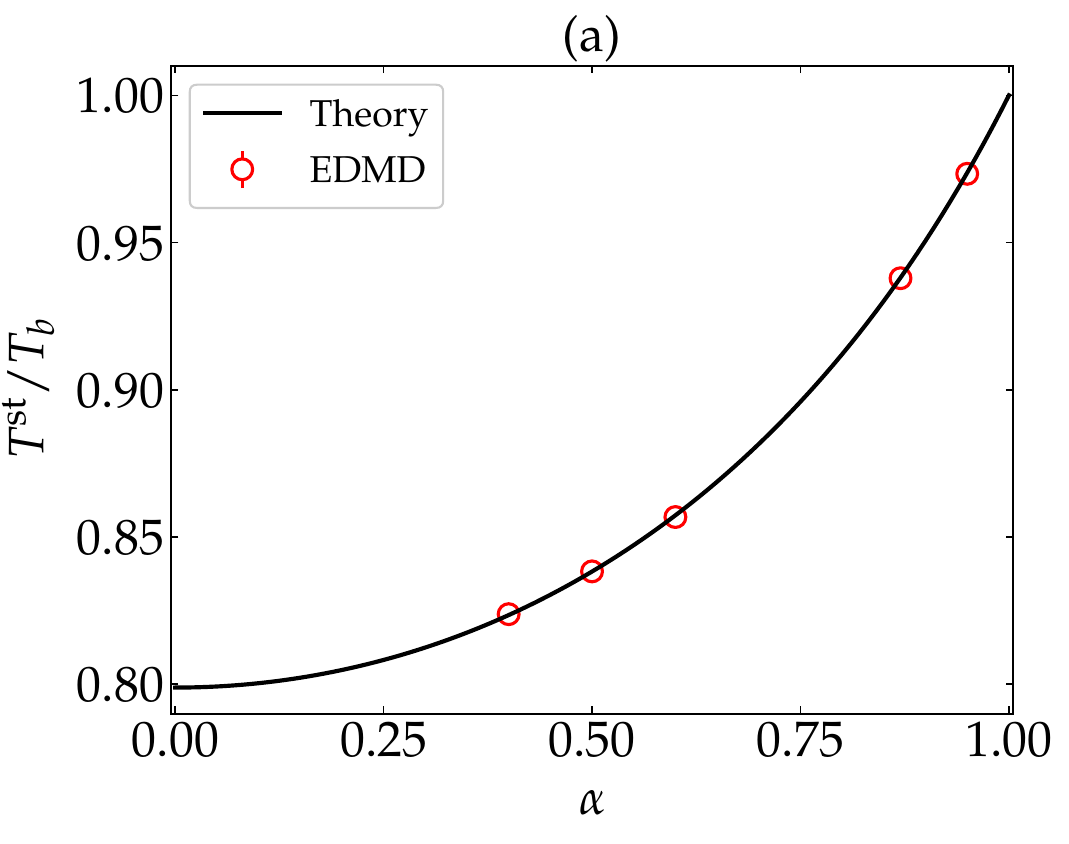}
 \includegraphics[width=0.47\textwidth]{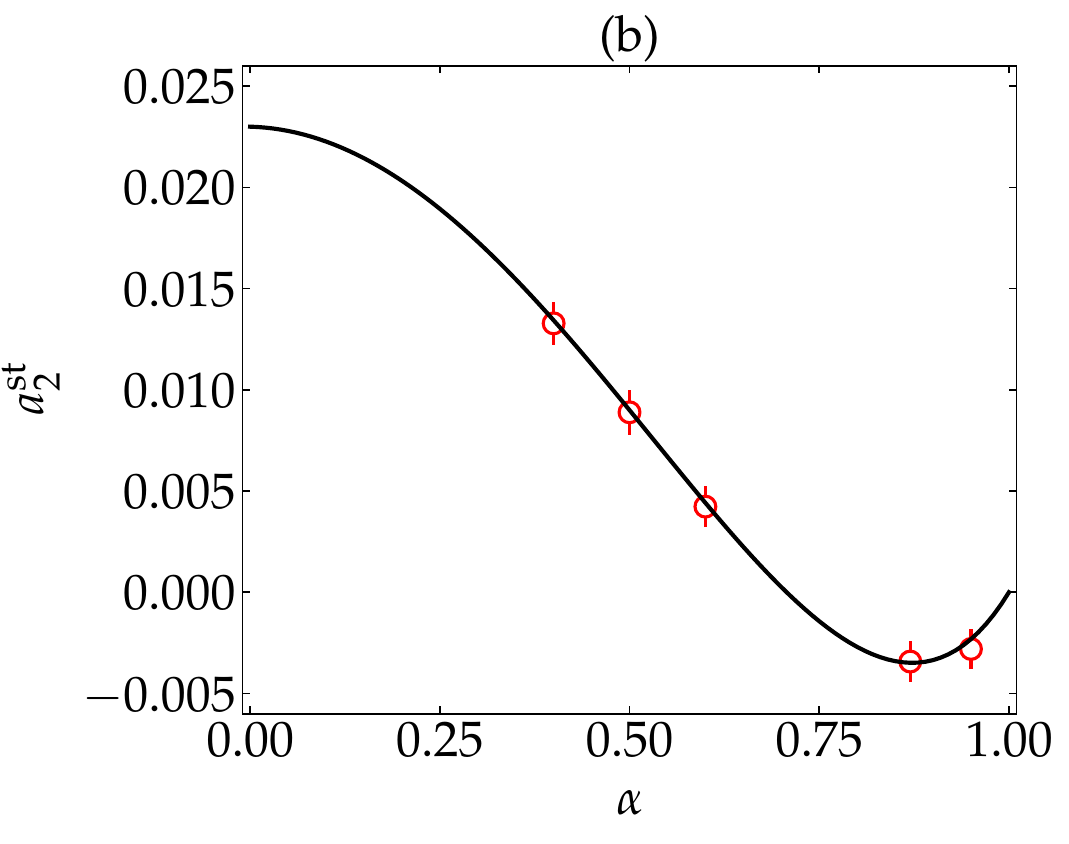}
 \caption{Steady-state values of (a) the granular \index{Temperature!of the thermal bath}\index{Temperature!steady-state}temperature, $T^{\mathrm{st}}/T_b$, and (b) the excess kurtosis $a_2^{\mathrm{st}}$, vs. the coefficient of normal restitution $\een$ for $\xi_0^{*}\approx 0.89$ and \acrshort{HS} ($\dt=2$). The circles ($\circ$) stand for \acrshort{EDMD} simulation results, while the thick black lines (---) refer to the values corresponding to \refeqs{Langevin_st_SA}. The values of the coefficient of normal restitution used in the simulations are $\een=0.95,0.87,0.6,0.5,0.4$.}
 \labfig{Langevin_IHS_st}
\end{figure}
\begin{figure}[h!]
 \centering
 \includegraphics[width=0.47\textwidth]{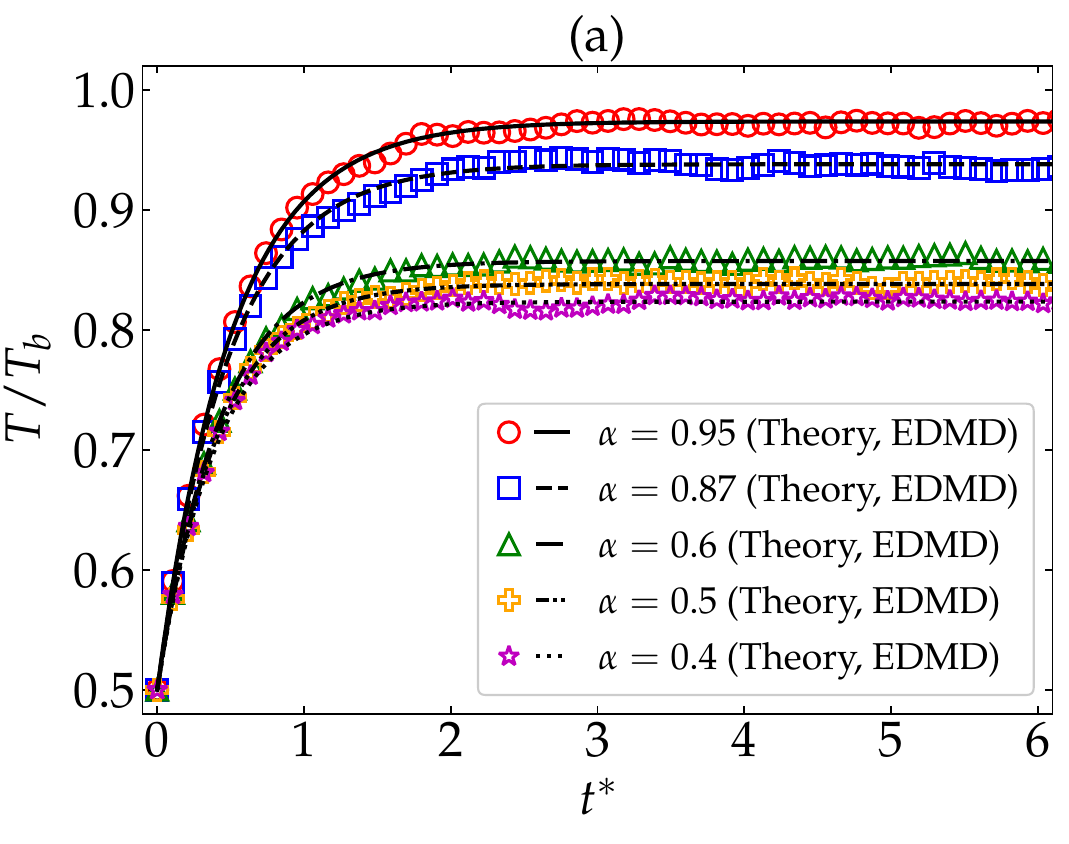}
 \includegraphics[width=0.47\textwidth]{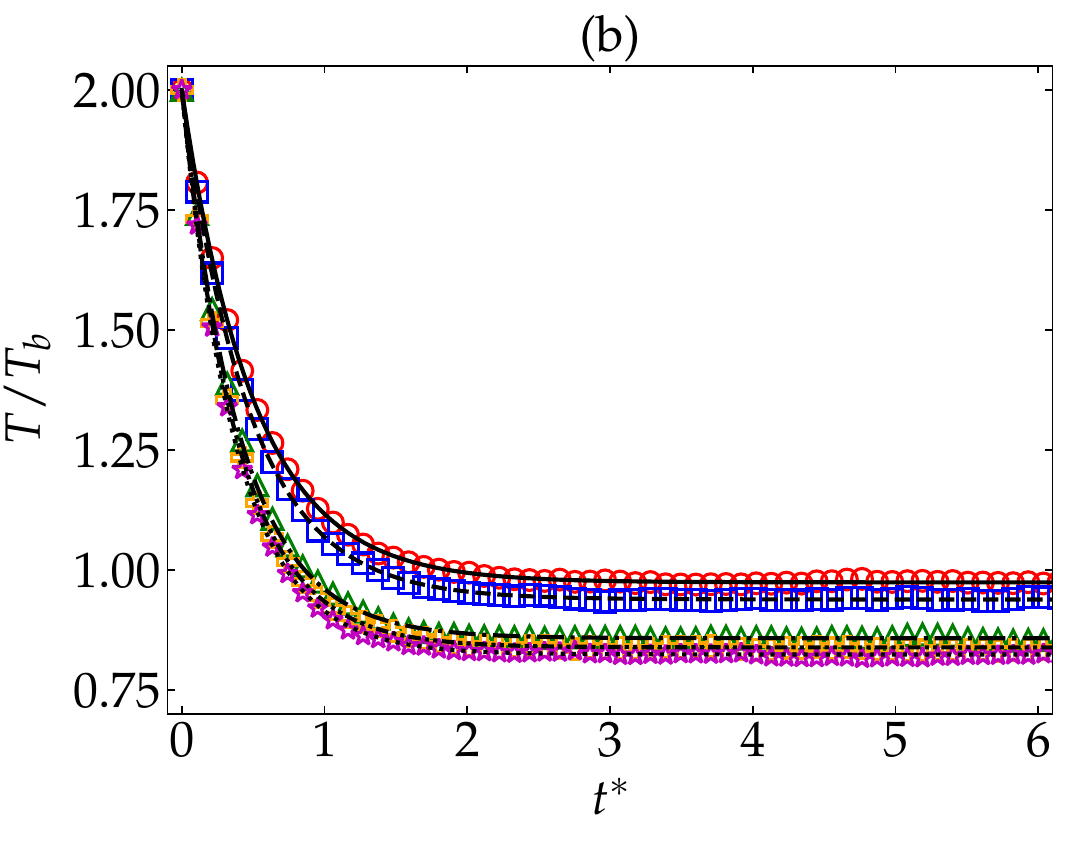}
 \caption{Plot $T/T_b$ vs. $t^*\equiv t\nu_b$ for a system of \acrshort{HD} ($\dt=2$), with $\xi_0^{*}\approx 0.89$, and for (a) $T(0)=T_b/2$ and (b) $T(0)=2T_b$. The symbols ($\circ$, $\square$, $\triangle$, $+$, $*$) stand for \acrshort{EDMD} simulation results, while the lines (---,-- --,$--\, \cdot$,$--\,\cdot\,\cdot$,$\cdots$) refer to the values corresponding to the  differential equation, \refeq{T_Langevin_ev}, for the different values of the coefficient of normal restitution, $\een=0.95,0.87,0.6,0.5,0.4$.}
\labfig{Langevin_IHS_ev}
\end{figure}

\subsection{Granular gas of inelastic HD driven by a stochastic thermostat}

The system under consideration is a monodisperse and homogeneous granular gas, whose binary collisional rules are described by the \acrshort{IHS} model, and the cooling is counterbalanced by a constant energy injection via a stochastic force with variance $m^2\chi^2$. Its associated \acrshort{BFPE} is
\begin{equation}
	\frac{\partial f}{\partial t}+\chi^2\left(\frac{\partial}{\partial \vvel}\right)^2 f = J[\vvel|f,f],
\end{equation}
with the Boltzmann collisional operator given by that of the \acrshort{IHS} model [see \refeq{BCO_IHS}]. Note that this is a particular case of the \acrshort{ST} restricted to translational degrees of freedom ($\varepsilon\to 0$) and acting on an inelastic but smooth system. Then, its associated Langevin-like equations for the free-streaming stage of the particles are
\begin{equation}
 \dot{\rr}(t) = \vvel(t), \quad \dot{\vvel}(t) = \chi \bar{\boldsymbol{\eta}}(t),
\end{equation}
with $\bar{\boldsymbol{\eta}}(t)$ being a random vector, representing a white noise of zero mean and unit variance [see \refeqs{white_noise_eta_bar}]. Then, according to the description in \refsubsec{LD_EDMD}, we obtain the following numerical scheme
\begin{subequations}\labeq{StocT_scheme}
\begin{align}
    \vvel_i(t+\Delta t)=& \vvel_i(t)+\chi\boldsymbol{\mathcal{W}}_i, \\
    \rr_i(t+\Delta t)=& \rr_i(t)+\vvel_i(t)\Delta t+\chi\bar{\boldsymbol{\mathcal{W}}}_i,
\end{align}
\end{subequations}
for $i=1,\dots,N$, with
\begin{align}
    \boldsymbol{\mathcal{W}}_i = \int_0^{\Delta t}\dif t^\prime\, \bar{\boldsymbol{\eta}}_i(t^\prime), \quad
    \bar{\boldsymbol{\mathcal{W}}}_i = \int_0^{\Delta t}\dif t^\prime\,\int_0^{t^\prime} \dif t^{\prime\prime}\, \bar{\boldsymbol{\eta}}_i(t^{\prime\prime}),
\end{align} 
and
\begin{align}\labeq{W_stochastic_therm}
   \langle {\mathcal{W}}^2\rangle=\Delta t, \quad
    \langle \boldsymbol{\mathcal{W}}\cdot \bar{\boldsymbol{\mathcal{W}}}\rangle = \frac{1}{2}\Delta t^2, \quad \langle \bar{\mathcal{W}}^2\rangle=\frac{2}{3}\Delta t^3.
\end{align}
The latter values are recovered from the leading-order term in the approximation $\xi_0\Delta t\ll 1$ in \refeqs{W_values_LE}. Therefore,
\begin{align}
	\boldsymbol{\mathcal{W}} = \sqrt{\Delta t} \boldsymbol{\mathcal{Y}}, \quad \bar{\boldsymbol{\mathcal{W}}} = \frac{\Delta t^{3/2}}{2}\left(\boldsymbol{\mathcal{Y}}+\sqrt{\frac{5}{3}}\bar{\boldsymbol{\mathcal{Y}}} \right).
\end{align}
Here, $\boldsymbol{\mathcal{Y}}$ and $\bar{\boldsymbol{\mathcal{Y}}}$ are \dt-dimensional Gaussian variables [see \refeq{gaussian_Y_and_Ybar}].

We have performed 10 runs for a system of $10^4$ inelastic \acrshort{HD} with a time step $\Delta t= 6\times 10^4/\nuwn$. Those runs were initialized to a Maxwellian distribution for the velocities at a temperature $T(0)\approx 1.46 \Twn$. The simulation outcomes are compared with theoretical results of transient and steady states. For the former, a \acrshort{MA} for the \index{Temperature!nonequilibrium}temperature evolution~\cite{MS00,SM09},
\begin{equation}\labeq{Stochastic_ev_MA}
	\frac{\partial (T/\Twn)}{\partial t^*}=-\zeta^*\left(\frac{T}{\Twn}\right)^{3/2}+\frac{1}{2},
\end{equation}
with $t^*=t\nuwn$. For the steady values, $T^{\mathrm{st}}/\Twn$ and $a_2^{\mathrm{st}}$, we used a \acrshort{SA} based on a truncation of the Sonine expansion of the \acrshort{VDF} up to the fourth cumulant, with those theoretical values computed upon linearization on the $a_2$~\cite{MS00,SM09}, reading 
\begin{align}\labeq{Stochastic_st_SA}
        \frac{T^{\mathrm{st}}}{\Twn} = m\left[\frac{\dt}{2\left(\mu_2^{(0)}+\mu_2^{(1)}a_2^{\mathrm{st}} \right)} \right]^{2/3}, \quad
        a_2^{\mathrm{st}} = \frac{4(1-\alpha)(1-2\alpha^2)}{19+14d-3\alpha(9+2d)+6(1-\alpha)\alpha^2}.
\end{align}

\begin{figure}[h!]
 \centering
 \includegraphics[width=0.47\textwidth]{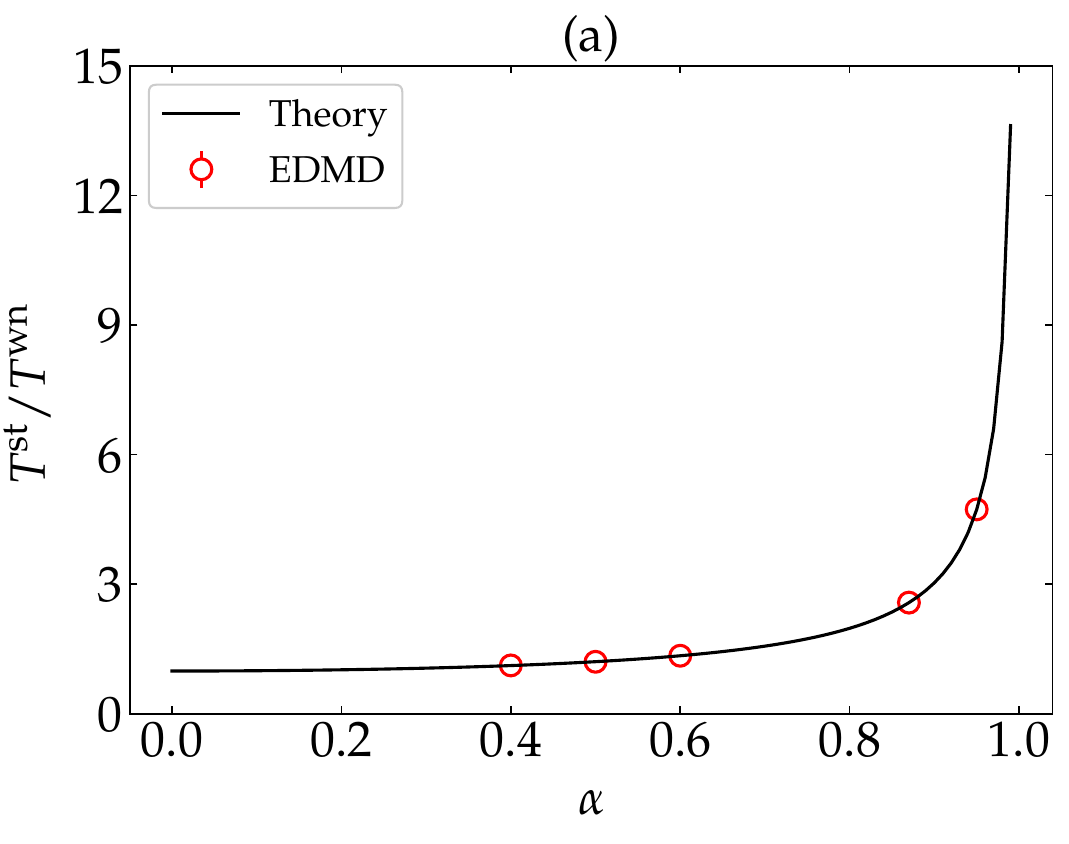}
 \includegraphics[width=0.47\textwidth]{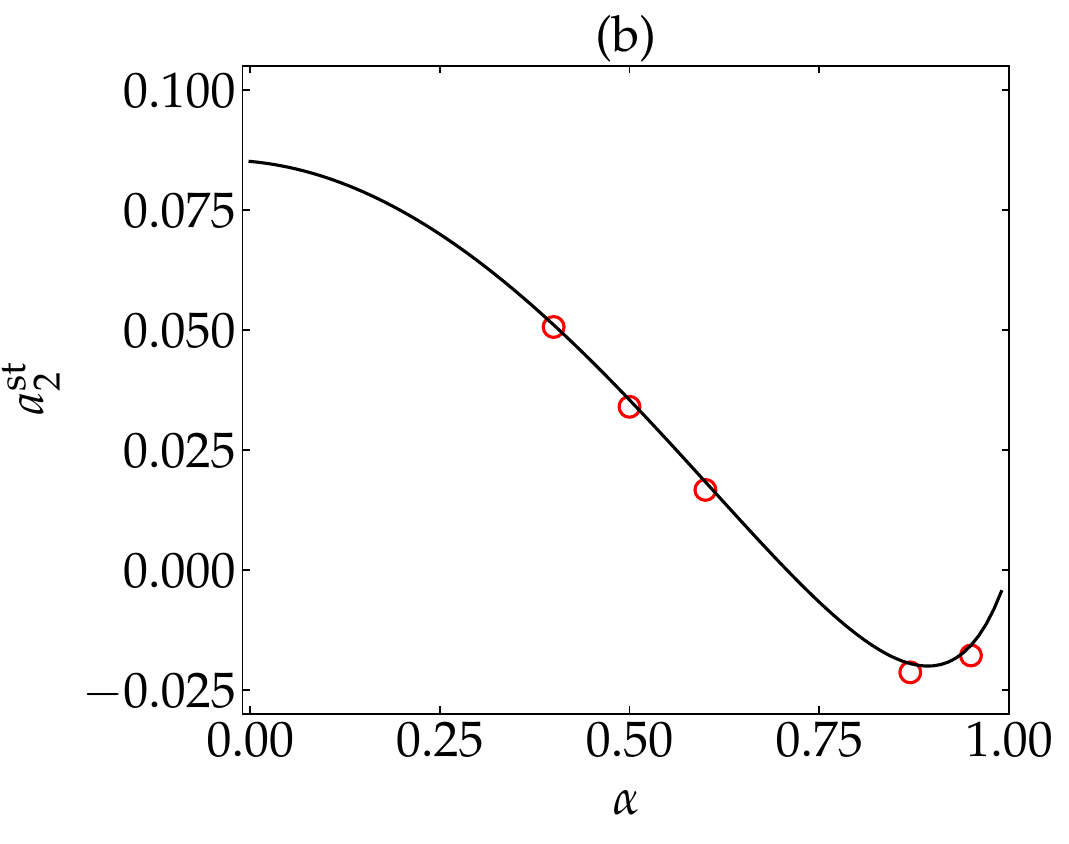}
 \caption{Steady-state values of (a) the granular \index{Temperature!steady-state}temperature, $T^{\mathrm{st}}/\Twn$, and (b) the excess kurtosis, $a_2^{\mathrm{st}}$ vs. the coefficient of normal restitution $\een$, for \acrshort{HS} ($\dt=2$). The circles ($\circ$) stand for \acrshort{EDMD} simulation results, while the thick black lines (---) refer to the values corresponding to \refeq{Stochastic_st_SA}. The values of the coefficient of normal restitution used in the simulations are $\een=0.95,0.87,0.6,0.5,0.4$.}
 \labfig{Stochastic_IHS_st}
\end{figure}

\begin{figure}[h!]
 \centering
 \includegraphics[width=0.47\textwidth]{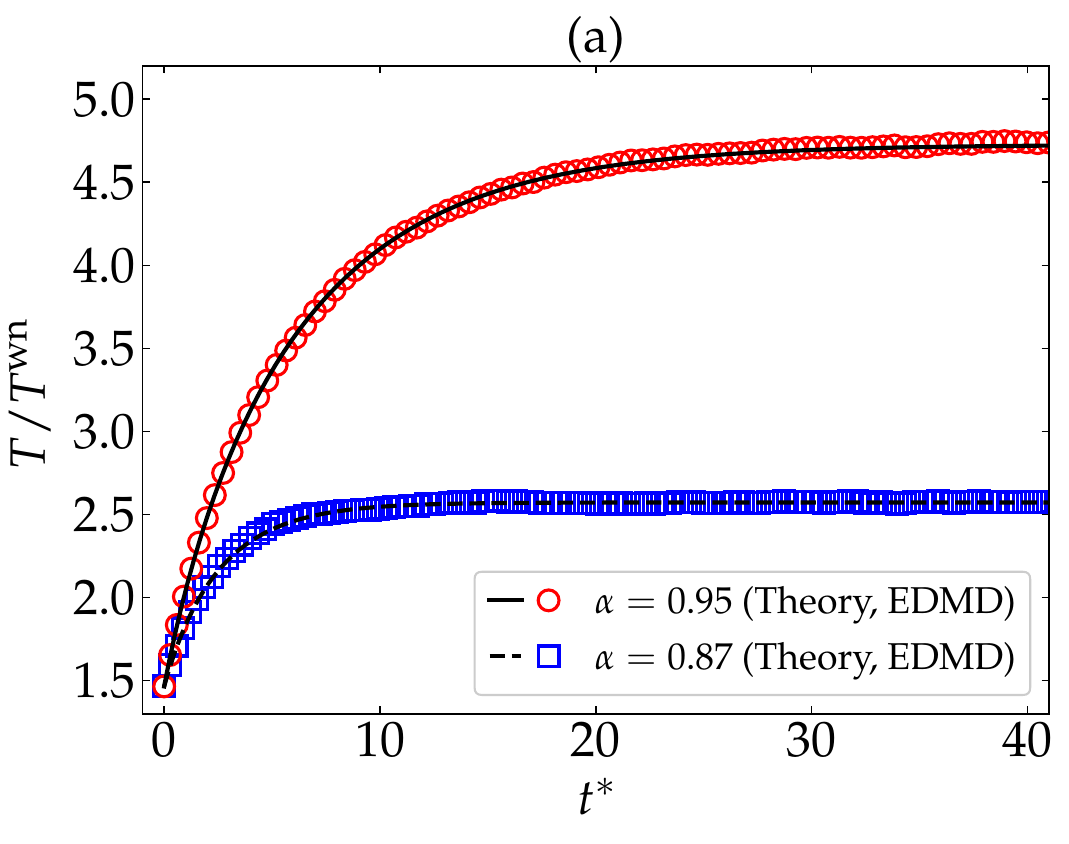}
 \includegraphics[width=0.47\textwidth]{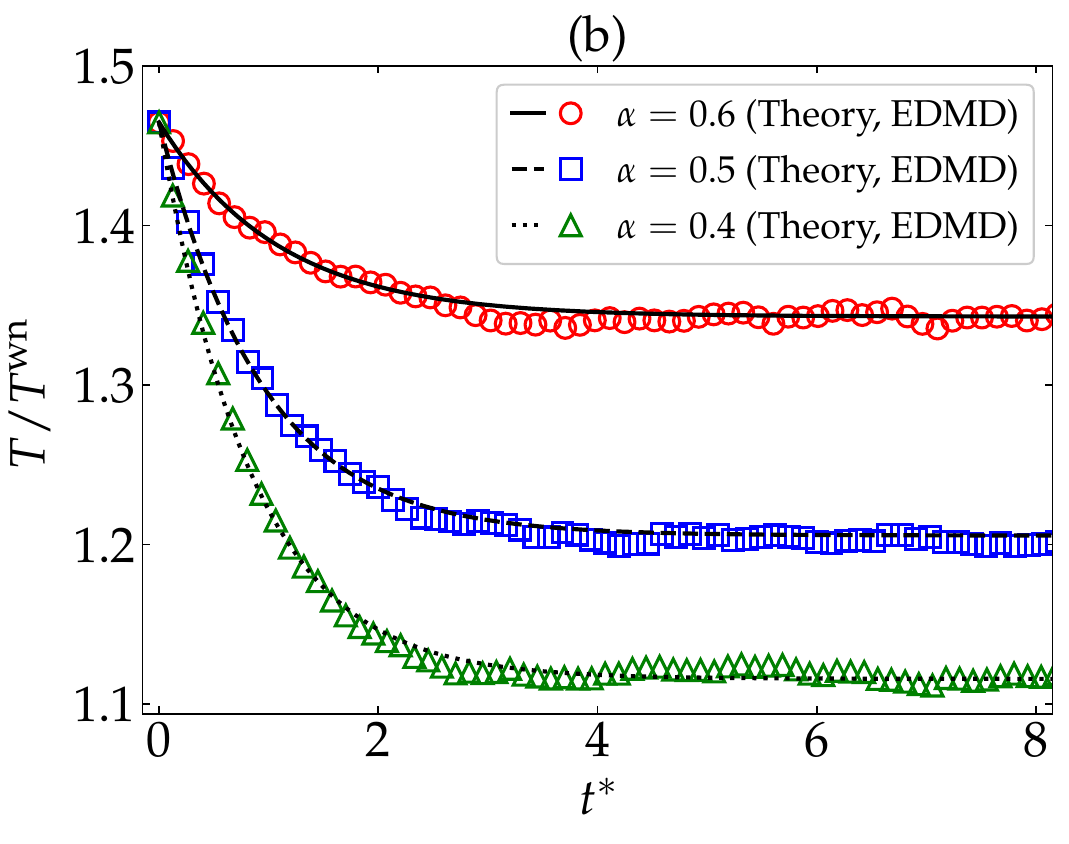}
 \caption{Plot $T/\Twn$ vs. $t^*\equiv t\nu_b$, for a system of \acrshort{HD} ($\dt=2$), for an initial condition $T(0)\approx 1.46\Twn$ and (a) $\een=0.95,0.87$ (whose steady value is greater than the initial condition), and (b) $\een=0.6,0.5,0.4$ (whose stationary temperature is below its initial value). The symbols ($\circ$, $\square$, $\triangle$) stand for \acrshort{EDMD} simulation results, while the lines (---,-- --,$\cdots$) refer to the values corresponding to the differential equation, \refeq{Stochastic_ev_MA}.}
 \labfig{Stochastic_IHS_ev}
\end{figure}

In \reffig{Stochastic_IHS_st}, the theoretical approximation for the steady-state values in \refeq{Stochastic_st_SA} are compared with computer outcomes. The simulation results are computed as the average of the 10 simulation curves and 50 points of the evolution curve, where the stationary state is ensured, and for each value of the coefficient of normal restitution\index{Coefficient of normal restitution}. On the other hand, in \reffig{Stochastic_IHS_ev}, the numerical solutions of \refeq{Stochastic_ev_MA} are plotted, together with the corresponding simulation curves resulting as the average over the 10 runs. We conclude that there is quite a good agreement between theory and simulation results. Therefore, we confirm that the implementation of the algorithm in our computer programs~\cite{M23_github} works properly.

\section{Numerical scheme for the Langevin dynamics with nonlinear drag}

In order to reproduce the simulation results of Article 1 (\refsec{Art1}) and Article 4 (\refsec{Art4}), it is necessary to elaborate the corresponding numerical scheme for the integration of the associated \acrshort{LE} in \refeq{eff_LE_Ito}. That is, from the discussion in \refsec{LD_NLD_FP_app}, it is deduced that the choice of interpretation of the integration of a Wiener process is subtle and plays an important role on the integration for this case. We have chosen the It\^o interpretation to work with~\cite{vK07}. Then, the system will follow the \acrshort{LE} derived in \refsec{LD_NLD_FP_app} [\refeq{eff_LE_Ito}]. The integration of the nonlinear \acrshort{LE}, for the positions and velocities, in a small time interval $\Delta t$, reads~\cite{S12,MSP22,MS22c}
\begin{subequations}
\begin{align}
    \vvel_i(t+\Delta t)=& \vvel_i(t)\left[1-\xi_{\mathrm{eff}}(v(t))\Delta t \right]+\sqrt{\chi^2(v(t))\Delta t}\boldsymbol{\mathcal{Y}}_i+\mathcal{O}(\Delta t^{3/2}), \\
    \rr_i(t+\Delta t)=& \rr_i(t)+\vvel_i(t)\Delta t\left[1-\frac{\xi_{\mathrm{eff}}(v(t))}{2}\Delta t \right]+\frac{1}{2}\sqrt{\chi^2(v(t))\Delta t^3}\left(\boldsymbol{\mathcal{Y}}_i+\sqrt{\frac{5}{3}}\bar{\boldsymbol{\mathcal{Y}}}_i\right)\nonumber \\
    &+\mathcal{O}(\Delta t^{5/2}),
\end{align}
\end{subequations}
for $i=1,\dots,N$, with $\boldsymbol{\mathcal{Y}}$ and $\bar{\boldsymbol{\mathcal{Y}}}$ being Gaussian variables [see \refeq{gaussian_Y_and_Ybar}], and the quantities $\xi_{\mathrm{eff}}(v)$ and $\chi^2(v)$ being defined in \refsec{LD_NLD_FP_app}. Here, we have approximated the random variables $\boldsymbol{\mathcal{W}}$ and $\bar{\boldsymbol{\mathcal{W}}}$ to those of \refeq{W_stochastic_therm}. This approximated scheme coincides with an Euler--Maruyama expansion for the system of \acrshort{SDE} of the Langevin description~\cite{KP92}.

\section{Numerical scheme for the ST}

To derive the proper numerical scheme for the integration of the \acrshort{LE} associated with the thermostatted states of a granular gas driven by the \acrshort{ST}, we use the results derived in \refsec{ST_FP_app}. In the free streaming, the particles admit the following equations of motion,
\begin{align}\labeq{eom_ST_pos_vel_ang}
	\dot{\rr} = \vvel, \quad \dot{\vvel} = \sqrt{\frac{\nuwn\Twn}{m}\frac{1-\varepsilon}{2}}\bar{\boldsymbol{\eta}}_t(t), \quad \dot{\oo} = \sqrt{\frac{\nuwn\Twn}{I}\frac{\varepsilon}{2}\frac{\dt}{\dr}}\bar{\boldsymbol{\eta}}_r(t),
\end{align}
with $\bar{\boldsymbol{\eta}}_{\dt}$ and $\bar{\boldsymbol{\eta}}_{\dr}$ being \dt- and \dr-dimensional random white noise vectors, which follow \refeqs{etas_ST_FP_app}. Note that, to derive \refeq{eom_ST_pos_vel_ang}, we have introduced in \refeq{ST_LE_first} the definitions of $\Twn$, $\nuwn$, and $\varepsilon$, appearing in \refeqs{def_Twn_ST}, \eqref{eq:nuwn}, and \eqref{eq:total_and_varepsilon_ST}, respectively. Thus, the integration of the equations of motion yields
\begin{subequations}\labeq{ST_eqs_scheme}
\begin{align}
    \vvel_i(t+\Delta t)=& \vvel_i(t)+\sqrt{\frac{\nuwn\Twn}{m}\frac{1-\varepsilon}{2}\Delta t}\boldsymbol{\mathcal{Y}}_{t,i}, \\
    \oo_i(t+\Delta t)=& \oo_i(t)+\sqrt{\frac{\nuwn\Twn}{I}\frac{\varepsilon}{2}\frac{\dt}{\dr}\Delta t}\boldsymbol{\mathcal{Y}}_{r,i}, \\
    \rr_i(t+\Delta t)=& \rr_i(t)+\vvel_i(t)\Delta t+\frac{1}{2}\sqrt{\frac{\nuwn\Twn}{m}\frac{1-\varepsilon}{2}\Delta t^3}\left(\boldsymbol{\mathcal{Y}}_{t,i}+\sqrt{\frac{5}{3}}\bar{\boldsymbol{\mathcal{Y}}}_{t,i} \right),
\end{align}
\end{subequations}
for $i=1,\dots,N$, with $\boldsymbol{\mathcal{Y}}_t$ and $\bar{\boldsymbol{\mathcal{Y}}}_t$ being $\dt$-dimensional Gaussian variables, and $\boldsymbol{\mathcal{Y}}_r$ being a \dr-dimensional Gaussian random vector. Note that the translational equations of the scheme in \refeqs{ST_eqs_scheme} coincide with those of \refeqs{StocT_scheme} by changing $\chi\to\chi_t$.
\chapter{Balance equations for the IRHS model}
\labapp{app_balance_equations}

One of the main goals of any fundamental or microscopic theory is to reproduce macroscopic phenomena. From the \acrshort{BE}, one can derive the equations describing the macroscopic behavior of the \index{Hydrodynamic!fields}hydrodynamic fields of the (monodisperse) system: $n$, $\mathbf{u}$, $T$. These equations are the so-called \index{Balance equations}\emph{balance equations}, and they are essential in fluid mechanics. Let us derive them for the particular case of a monodisperse granular gas described by the \acrshort{IRHS} model described by the \acrshort{BE},
\begin{equation}\labeq{BE_free_IHS}
	\frac{\partial f(\vom_1;t)}{\partial t}+\vvel\cdot\nabla f(\vom_1;t) = \sigma^{\dt-1}\int\dif\vom_2\int_{+}\dif\ssab\, 
    (\vvel_{12}\cdot\ssab)\left[\frac{f(\vom^{\prime\prime}_1;t)f(\vom^{\prime\prime}_2;t)}{\een^{2}|\eet|^{2\dr/\dt}}-f(\vom_1;t)f(\vom_2;t)\right].
\end{equation}
Let us start by multiplying an arbitrary one-body hydrodynamic function, $\phi(\vom)$, at both sides of \refeq{BE_free_IHS}, and integrating over all values of $\vom$, yielding
\begin{equation}\labeq{balance}
\frac{\partial}{\partial t}(n\langle \psi \rangle)+\nabla\cdot(n\langle \mathbf{v}\psi\rangle)=\mathcal{J}[\psi|f,f],
\end{equation}
where
\begin{align}\labeq{coll_integral_J_cal}
\mathcal{J}[\psi|f,f]&\equiv\int \dif\vom_1 \medspace\psi J[\vom|f,f]\nonumber\\
&=\sigma^{\dt-1} \int \dif\vom_1 \int \dif\vom_2\int_{+} \dif\ssab\,(\vvel_{12}\cdot\ssab)f_1f_2 \left(\mathfrak{B}_{12,\ssab}-1\right)\psi(\vom_1)
\end{align}
is the collisional production term of $\psi$, corresponding to the nonreduced version of \refeq{coll_integral_I_cal}. Substituting $\psi(\vom)=1,\, m\vvel,\,mV^2/2+I\omega^2/2$ into \refeq{balance}, the \index{Balance equations}balance equations for the mass density, $n$, the flow field, $\mathbf{u}$, and the mean \index{Temperature!mean granular}temperature, $T$, are directly obtained.

\section{Particle (mass) density balance equation}

Let us define $\psi(\vom)=1$ and substitute into \index{Balance equations!of the mass density}\refeq{balance}. Then, one gets
\begin{equation}\labeq{boltzmass}
\frac{\partial}{\partial t}n+\boldsymbol{\nabla}\cdot(n\langle \vvel\rangle)=0,
\end{equation}
where the collision term vanishes because of the conservation of mass. That is, if the system is closed, the total mass of the system must be preserved.

Using the definition of the material derivative, $\mathcal{D}_t$ [see \refeq{mat_time_der}], \refeq{boltzmass} becomes
\begin{equation}\labeq{PMDBE}
\mathcal{D}_t n+n\boldsymbol{\nabla}\cdot \mathbf{u}=0.
\end{equation}

\section{Momentum density balance equation}

If we now consider $\psi(\vom)=m\mathbf{v}$ and we substitute that quantity into \index{Balance equations!of the momentum density}\refeq{balance}, the balance equation associated with the linear momentum reads
\begin{equation}\labeq{boltzmom}
\frac{\partial}{\partial t}(nm\langle \mathbf{v}\rangle) +\boldsymbol{\nabla}\cdot(nm\langle\vvel \vvel\rangle)=0.
\end{equation}
Again, $\vvel$ is a collisional invariant and thus the collisional integral is identically null. Introducing the definition of the \index{Peculiar velocity}peculiar velocity, $\mathbf{V}$,  and the mass density \phantomsection\label{sym:rho_den}$\rho=nm$, \refeq{boltzmom} can be rewritten as
\begin{equation}\labeq{mom2}
\frac{\partial}{\partial t}(\rho\mathbf{u})+\boldsymbol{\nabla}\cdot\left[\rho\langle(\mathbf{u}+\mathbf{V})(\mathbf{u}+\mathbf{V})\rangle\right]=0.
\end{equation}
To elaborate more this expression, we use again \refeq{PMDBE}, the definition of the \index{Material time-derivative}material time-derivative, and the pressure tensor $\mathsf{P}$, defined as
\begin{equation} \labeq{tensor_P}
\mathsf{P}\equiv \rho\langle \mathbf{V}\mathbf{V} \rangle.
\end{equation}
Finally, the momentum density balance equation becomes
\begin{equation}\labeq{MBE}
\mathcal{D}_t\mathbf{u}+\rho^{-1}\boldsymbol{\nabla}\cdot \mathsf{P}=0.
\end{equation}
The latter equation is a conservation equation for $\mathbf{u}$. This is expected due to the linear momentum conservation.

Up to now, in \refeqs{PMDBE} and \eqref{eq:MBE}, there is no explicit signature coming from the \acrshort{IRHS} model. In fact, those equations are the same for both the \acrshort{EHS} and \acrshort{IHS} models. Of course, this is due to the already mentioned laws of conservation of mass and momentum.

\section{Energy density balance equation}

Let us now look for the inhomogeneous equation for the mean \index{Temperature!mean granular}temperature in a monodisperse gas of inelastic and rough \acrshort{HD} or \acrshort{HS}. First of all, we derive the equations for the partial \index{Temperature!translational}\index{Temperature!rotational}temperatures, $\Ttr$ and $\Trot$. We already know that, if $\alpha< 1$ or $|\beta|< 1$, the total kinetic energy is not a collisional invariant any more.

\subsection{Equation for the translational temperature}

Let us first consider that, $\psi=\frac{1}{2}mV^2$, and using \refeq{balance}, we get
\begin{equation}\labeq{boltzTt}
\frac{\partial}{\partial t}\left(\frac{nm}{2}\langle V^2 \rangle\right)+\boldsymbol{\nabla}\cdot\left(\frac{nm}{2}\langle \mathbf{v}V^2\rangle\right)=\frac{m}{2}\mathcal{J}\left[V^2|f,f\right].
\end{equation}
Taking into account the definitions of the \index{Peculiar velocity}peculiar velocity, $\mathbf{V}=\mathbf{v}-\mathbf{u}$, of the translational \index{Temperature!translational}temperature $\Ttr=\frac{\dt}{2}m\langle V^2\rangle$ [see \refeq{tr_rot_temp}], and of $\xi_t$ [see \refeq{P_Rates_t}], after some algebra, one obtains
\begin{equation}
\frac{\dt}{2}\frac{\partial}{\partial t}(n\Ttr)+\frac{\dt}{2}\boldsymbol{\nabla}\cdot(n\mathbf{u}\Ttr+ \mathbf{q}_t)+\mathsf{P}:\boldsymbol{\nabla}\mathbf{u}-n\xi_t\Ttr=0,
\end{equation}
where,
\begin{equation}
\mathbf{q}_t\equiv\frac{\rho}{2}\langle \mathbf{V}V^2\rangle
\label{qtr}
\end{equation}
is the translational contribution of the heat flux.

If we use the mass balance equation, \refeq{PMDBE}, and the definition of material derivative, then, \refeq{boltzTt} becomes
\begin{equation}
\mathcal{D}_t\Ttr+\frac{2}{\dt n}(\boldsymbol{\nabla}\cdot\mathbf{q}_t+\mathsf{P}:\boldsymbol{\nabla}\mathbf{u})+\xi_t \Ttr=0.
\labeq{TtBE}
\end{equation}

\subsection{Equation for the rotational temperature} 

We consider now $\psi=\frac{1}{2}I\omega^2$ in \refeq{balance}. Then, we get\index{Temperature!rotational}
\begin{equation}\labeq{boltzrot}
\frac{\partial}{\partial t}\left(\frac{nI}{2}\langle \omega^2 \rangle\right)+\boldsymbol{\nabla}\cdot\left(\frac{nI}{2}\langle\omega^2\vvel\rangle\right)=\frac{I}{2}\mathcal{J}\left[\omega^2|f,f\right].
\end{equation}
Hence, using $\Trot=\frac{\dr}{2}I\langle \omega^2\rangle$ [see \refeq{tr_rot_temp}], $\xi_r$ [see \refeq{P_Rates_r}], and defining the rotational contribution to the heat flux, $\mathbf{q}_r$, as
\begin{equation}
\mathbf{q}_r\equiv\frac{\rho}{2}\langle \mathbf{V}\omega^2\rangle,
\labeq{qrot}
\end{equation}
it is possible to rewrite \refeq{boltzrot} as
\begin{equation}
\mathcal{D}_t\Trot+\frac{2}{\dr n}(\boldsymbol{\nabla}\cdot\mathbf{q}_r)+\xi_r\Trot=0.
\labeq{TrBE}
\end{equation}

\subsection{Balance equation for the mean temperature} 

In order to obtain the equation for the \index{Balance equations!of the temperature}mean temperature\index{Temperature!mean granular}, we will just use its definition as a weighted sum of the translational\index{Temperature!translational} and rotational\index{Temperature!rotational} partial temperatures, i.e., $T\equiv (\dt\Ttr+\dr\Trot)/(\dt+\dr)$ [see \refeq{def_T_IRHS}]. Then, from the latter definition, and \refeqs{TtBE} and \eqref{eq:TrBE}, one obtains
\begin{equation}
\mathcal{D}_t T+\frac{2}{(d_t+d_r)n}(\boldsymbol{\nabla}\cdot\mathbf{q}+\mathsf{P}:\boldsymbol{\nabla}\mathbf{u})+T\zeta=0,
\labeq{TBE}
\end{equation}
where we have introduced the total heat flux, defined by
\begin{equation}
\mathbf{q}\equiv\mathbf{q}_t+\mathbf{q}_r,
\labeq{heat_flux_total}
\end{equation}
and the cooling rate, $\zeta$, already defined in the context of homogeneous states in \refeq{Haffs_law_IRHS}.

This derivation is equivalent to the one found by directly substituting $\psi(\vom)= \frac{1}{2}mV^2+\frac{1}{2}I\omega^2$ into \refeq{balance}. Finally, it can be stressed that the balance equations, \refeqs{PMDBE}, \eqref{eq:MBE}, and \eqref{eq:TBE}, are formally exact regardless of the use of any approximation to the \acrshort{VDF}.


\backmatter 
\setchapterstyle{plain} 



\printbibliography[heading=bibintoc, title=Bibliography] 

\printindex 




\end{document}